\documentclass{elsarticle}

\usepackage{natbib}
\usepackage{graphicx,epsfig}
\usepackage[normalem]{ulem}
\usepackage[makeroom]{cancel}
\usepackage[T1]{fontenc}
\usepackage{pifont,dsfont}
\usepackage{multirow,array}
\usepackage{empheq}
\usepackage{hyphenat}
\usepackage{lscape}
\usepackage{latexsym,amsmath,amsthm,amssymb,amsfonts,amsbsy}
\usepackage{epsf,wasysym,stmaryrd,slashed,mathrsfs,mathdots}
\usepackage{multicol,xcolor,longtable}
\usepackage{url,color}
\usepackage{verbatim,enumerate}
\usepackage{comment}
\usepackage{titlesec}
\usepackage{hyperref}
\usepackage[width=0.8\textwidth]{caption}
\usepackage[english]{babel}
\usepackage{bbm,amstext,xspace}
\usepackage{bbold}
\usepackage{float}
\usepackage{afterpage}
\usepackage{geometry}
\usepackage{multicol}
\usepackage{varwidth}
\usepackage[titletoc]{appendix}

\titleclass{\subsubsubsection}{straight}[\subsection]

\newcounter{subsubsubsection}[subsubsection]
\renewcommand\thesubsubsubsection{\thesubsubsection.\arabic{subsubsubsection}}

\titleformat{\subsubsubsection}
{\normalfont\normalsize\bfseries}{\thesubsubsubsection}{1em}{}
\titlespacing*{\subsubsubsection}
{0pt}{3.25ex plus 1ex minus .2ex}{1.5ex plus .2ex}

\makeatletter
\renewcommand\paragraph{\@startsection{paragraph}{5}{\z@}%
	{3.25ex \@plus1ex \@minus.2ex}%
	{-1em}%
	{\normalfont\normalsize\bfseries}}
\renewcommand\subparagraph{\@startsection{subparagraph}{6}{\parindent}%
	{3.25ex \@plus1ex \@minus .2ex}%
	{-1em}%
	{\normalfont\normalsize\bfseries}}
\def\toclevel@subsubsubsection{4}
\def\toclevel@paragraph{5}
\def\toclevel@paragraph{6}
\def\l@subsubsubsection{\@dottedtocline{4}{7em}{4em}}
\def\l@paragraph{\@dottedtocline{5}{10em}{5em}}
\def\l@subparagraph{\@dottedtocline{6}{14em}{6em}}
\makeatother

\definecolor{darkred}{rgb}{0.65,0.15,0}
\definecolor{newgreen}{rgb}{0.2,0.62,0.14}
\hypersetup{pdfborder={0 0 0},colorlinks=true,urlcolor=blue,citecolor=blue,linkcolor=darkred,linktocpage=true}

\DeclareFontFamily{U}{matha}{\hyphenchar\font45}
\DeclareFontShape{U}{matha}{m}{n}{
      <5> <6> <7> <8> <9> <10> gen * matha
      <10.95> matha10 <12> <14.4> <17.28> <20.74> <24.88> matha12
      }{}
\DeclareSymbolFont{matha}{U}{matha}{m}{n}

\DeclareMathSymbol{\oleft}{2}{matha}{"68}
\DeclareMathSymbol{\oright}{2}{matha}{"69}

\def\n{\widehat{n}}
\def\ft#1#2{{\textstyle{\frac{\scriptstyle #1}{\scriptstyle #2} } }}
\def\fft#1#2{{\frac{#1}{#2}}}
\def\cO{{\cal O}}
\def\cP{{{\cal P}}}
\def\cG{{{\cal G}}}

\def\p{\partial^{\mu}}
\def\be{\begin{equation}}
\def\ee{\end{equation}}
\def\ba{\begin{array}}
\def\ea{\end{array}}
\def\bea{\begin{eqnarray}}
\def\eea{\end{eqnarray}}
\def\bd{\begin{displaymath}}
\def\ed{\end{displaymath}}

\let\la=\label

\def\nn{\nonumber}

\def\a{\alpha}
\def\al{\alpha}
\def\b{\beta}
\def\g{\gamma}
\def\G{\Gamma}
\def\d{\delta}

\def\e{\epsilon}
\def\ve{\varepsilon}
\def\f{\phi}
\def\vf{\varphi}
\def\vp{\varphi}

\def\p{\psi}
\def\bpsi{{\bar\psi}}
\def\P{\Psi}
\def\k{\kappa}
\def\l{\lambda}
\def\la{\lambda}
\def\ta{\tau}
\def\L{\Lambda}
\def\m{\mu}
\def\n{\nu}
\def\r{\rho}
\def\rh{\rho}
\def\s{\sigma}

\def\Tr{{\rm Tr\,}}
\def\t{\tau}

\def\o{\omega}
\def\om{\omega}
\def\O{\Omega}

\def\tR{{\widetilde R}}

\def\cL{{\mathcal L}}
\def\ad{{\dot{\alpha}}}
\def\bd{{\dot{\beta}}}
\def\gd{\dot{\gamma}}
\def\dd{\dot{\delta}}
\def\bM{{\bar M}}

\def\tG{{\widetilde G}}
\def\tH{{\widetilde H}}
\def\tR{{\widetilde R}}

\def\mR{{\mathbb{R}}}
\def\mZ{{\mathbb{Z}} }
\def\td{{\tilde d}}

\def\del{\partial}
\def\pd{\partial}
\def\rmi{{\rm i}}
\def\R2{\cL_{{\rm Riem}^2}}
\def\bD{{\bar D}}

\def\vE{{\cal E}}

\def\mN{{\mathcal N}}

\newcommand{\grad}[3]{{\scriptscriptstyle (#1 , #2, #3 )}}
\newcommand{\gra}[2]{{\scriptscriptstyle (#1 , #2 )}}
\newcommand{\ord}[1]{{\scriptscriptstyle (#1)}}

\newcommand{\w}[1]{\\[0.#1cm]}
\newcommand{\eq}[1]{(\ref{#1})}
\newcommand{\tr}{\, \textrm{tr}}
\def\tr{\mathop{\rm tr}\nolimits}

\newcommand{\cH}{{\cal H}}

\newcommand{\EW}{{  \widetilde{\cal E}}}
\newcommand{\veccomp}{\mbox{\sf u}\xspace}

\newsavebox{\uuunit}
\sbox{\uuunit}
{\setlength{\unitlength}{0.825em}
\begin{picture}(0.6,0.7)
\thinlines
\put(0,0){\line(1,0){0.5}}
\put(0.15,0){\line(0,1){0.7}}
\put(0.35,0){\line(0,1){0.8}}
\multiput(0.3,0.8)(-0.04,-0.02){12}{\rule{0.5pt}{0.5pt}}
\end {picture}}

\makeatletter
\@addtoreset{equation}{section}
\makeatother

\begin{document}

\begin{frontmatter}



\title{Higher derivative supergravities in diverse dimensions}


\author{Mehmet Ozkan}

\address{Department of Physics, Istanbul Technical University, Maslak 34469 Istanbul, T\"urkiye}

\ead{ozkanmehm@itu.edu.tr}

\author{Yi Pang}

\address{Center for Joint Quantum Studies and Department of Physics, School of Science, Tianjin University, Tianjin 300350, China}
\address{Peng Huanwu Center for Fundamental Theory, Hefei, Anhui 230026, China}

\ead{pangyi1@tju.edu.cn}

\author{Ergin Sezgin}

\address{George P. \& Cynthia Woods Mitchell  Institute for Fundamental Physics and Astronomy, Texas A\&M University, College Station, TX 77843, USA}

\ead{sezgin@tamu.edu}

\begin{abstract}

We survey on-shell and off-shell higher derivative supergravities in dimensions $1\le D\le 11$.
Various approaches to their construction, including the Noether procedure, (harmonic) superspace, superform method, superconformal tensor calculus, $S$-matrix and dimensional reduction, are summarized. Primarily the bosonic parts of the invariants and the supertransformations of the fermionic fields are provided. The process of going on-shell, solutions to the Killing spinor equations, typical supersymmetric solutions, and the role of duality symmetries in the context of $R^4, D^4 R^4$ and $D^6 R^4$ invariants are reviewed.

\end{abstract}

\end{frontmatter}

\newpage

\tableofcontents{}

\newpage

\section{Introduction}
		
It was realized long ago that supergravity theories have improved quantum behavior compared to ordinary gravity. However, it was also found that three-loop counterterms exist for $N=1$ and $N=2$ supergravities, which entail supersymmetrizations of quartic in Riemann tensor terms with suitable contraction of indices\cite{Deser:1977yyz,Deser:1978br,Ferrara:1978wj}. This meant that divergences are not forbidden by local supersymmetry starting at 3-loops in these theories. Nonetheless, it was hoped that supergravities with a higher degree of extended supersymmetry would have improved finiteness behavior. This led to several studies of higher derivative superinvariants. For an early account of higher-order superinvariants in extended supergravity theories, see \cite{Howe:1980th}. These considerations have implications for attempts to build up an effective supergravity theory as well. For a pedagogical introduction to the effective field theory treatment of quantum gravity, see \cite{Donoghue_2012}.

Initially, the focus in the study of higher derivative supergravities was in four dimensions. However, with advances made in string theory and their compactifications, their construction and study in diverse dimensions started to attract attention as well. The interest in higher derivative supergravities began to evolve in a direction not necessarily restricted to their relevance to higher loop divergences but also their properties in relation to higher derivative terms that arise in the low energy limit of string theory. These properties notably include the powerful duality symmetries, and the role of local supersymmetry in the organization of the higher derivative terms in the low-energy effective action.

Given that string theory is the prime candidate for providing a UV complete theory of quantum gravity, one may question the wisdom of investing much effort in the construction and study of higher derivative supergravities in diverse dimensions in their own right. However, it should also be kept in mind that it is not known if string theory is the unique UV complete theory of quantum gravity. Indeed, it is one of the premises of the so-called Swampland program \cite{Vafa:2005ui} to determine the conditions imposed on a UV completion on the basis of physical considerations; see, for example,  \cite{Palti:2019pca, Harlow:2022ich} for a review. The criteria put forward are mostly motivated by string theory but not all of them. In any event, the criteria inspired by string theory are natural to take into account while maintaining freedom in building up an effective theory which is not necessarily identical to string  theory. Among the universal criteria are the requirements of unitarity and causality. These criteria have attracted much attention in recent years, and we shall come back to them briefly in the conclusions. A property which is not necessarily essential but which may certainly be useful is local supersymmetry, and the attendant duality symmetries. Indeed, requiring anomaly freedom of a chiral supergravity naturally requires higher derivative extensions. Higher derivative supergravities, in turn, have led to investigations of their consequences for black hole physics, AdS/CFT correspondence, and cosmology. There is a large body of literature already on the study of the higher derivative corrections to black hole and string solutions, and consequences for the black hole entropy, mostly in four and five dimensions. See \cite{Mohaupt:2000mj,deWit:2007maa, Castro:2008ne} for a review, where an extensive list of references to earlier papers can be found. In addition to their relevance to the black hole entropy,  higher derivative corrections to supergravities, in the context of holography, also provide information on the conformal anomalies of the dual CFT's; see, for example,  \cite{Blau:1999vz, Bobev:2021qxx}. As to the applications in cosmology, the possible shapes for non-gaussianity for gravitational waves in the de Sitter approximation was computed in \cite{Maldacena:2011nz}. In a more general context, the cosmology of modified gravity theories has been reviewed in \cite{Nojiri:2017ncd}. All in all, there is an abundance of motivation for constructing and studying higher derivative supergravities.

Even though several reviews of two-derivative supergravities have appeared in the literature over the years, not that many exist for their higher derivative extensions. Presumably, this is in part due to the fact that their explicit construction is a very complicated task compared to the case of two-derivative supergravities. Our aim here is to remedy this to some extent by providing a relatively extensive survey of presently known higher derivative extensions of supergravities in diverse dimensions. Our focus will be on surveying the known results, rather than describing their detailed derivations, which would deserve separate reviews by themselves. We also summarize briefly their applications.

We shall begin by summarizing briefly different approaches to their constructions, including the Noether procedure, superconformal tensor calculus, (harmonic) superspace, superform method (ectoplasm), dimensional reduction and $S$-matrix. We shall then start with $D=11$ and work our way down to $D=1$, though the focus will be on $D=10,11$ and $3 \le  D \le 6$ as most results available are in these dimensions.
We shall typically give the bosonic parts of the actions and supersymmetry transformations of the fermionic fields. The latter is relevant in finding supersymmetric solutions. Indeed, we shall also survey the known solutions to the Killing spinors equations in diverse dimensions. This will be done for on-shell as well as off-shell supersymmetric theories. We shall give a concise summary of the existing results on the  $R^4, D^4R^4$ and $D^6 R^4$ invariants and their duality symmetries in diverse dimensions. The relevance of these invariants to potential UV divergences and counterterms in supergravities in diverse dimensions will also be reviewed briefly.

Finally a word on conventions. We do not attempt to provide a universal set of conventions for all the supergravity theories that will be reviewed. Instead, we will adhere to the conventions used in the original papers in each case, except in a few cases where we may use slightly different notations.  We refrain from listing the contents of all the sections, as the Table of Contents provided serves that purpose.
\section{Approaches to construction of higher derivative supergravities}
\label{sec2}

Let us recall briefly the approaches that have been employed so far for the construction of higher derivative extensions of supergravities.

\begin{itemize}

\item  \mbox{\it Noether procedure:} In this approach (see, for example, \cite{VanNieuwenhuizen:1981ae}), if an off-shell formalism exists, there is no need to deform the supersymmetry transformation rules since the supersymmetry algebra closes without the need to impose equations of motion that follow from the total action. Thus, the task is to find the partners of the desired higher derivative term that will be supersymmetrized under the fixed off-shell supertransformations. If the supertransformations are known only on-shell, life gets much harder since the closure will depend on the equations of motion, and therefore any deformation of the action will require the deformation of the supertransformations as well, and this will go on order by order in suitable expansion parameters. The difficulty in this case lies in the fact that there are too many structures one can write down, and their variations under supersymmetry produce an even much larger number of terms. This means a large number of variations grouped into a large number of independent structures that need to cancel separately. That is why, even four-derivative extensions of supergravity by this approach are rather rare, let alone the eight derivatives and beyond, as we shall review in the following sections.

The Noether procedure is used in the construction of higher derivative corrections in $11D$, type IIB and heterotic supergravities in $10D$, matter couplings of $N=(1,0), 6D$ supergravity, and $N=1,4,6,8$ supergravities in $3D$.

\item \mbox{\it Superconformal tensor calculus:} Instead of starting with off-shell supergravities, it turns out to be more convenient to work with off-shell conformal supergravities (and their matter couplings) first, and then fix the dilatation, conformal boost and $R$-symmetries to obtain off-shell Poincar\'e supergravity possibly coupled to matter. The convenience is due to the fact that the off-shell conformal supergravity construction is based on the construction of curvatures that follow the structure constants of the underlying superconformal algebra. Nonetheless, there is still quite a bit of work to be done because the system is highly reducible to begin with, and in order to achieve maximum irreducibility of the superconformal gauge field configuration, a maximal set of the so-called conventional constraints need to be imposed on the superconformal curvatures. Final results emerge from the study of the consequences of these constraints. This procedure has been explained in detail in the textbook \cite{Freedman:2012zz}, and summarized in the case of $D=6, N=(1,0)$ in \cite{Bergshoeff:1985mz}, which follows closely the case of $N=2, D=4$ dealt with in detail in \cite{deWit:1984rvr, Lauria:2020rhc}.  For reader's convenience, we shall outline the procedure for obtaining off-shell $N=1, 4D$ supergravity from the superconformal tensor calculus.

The superconformal procedure is based on the $SU(2,2|1)$ superalgebra whose commutation rules can be found, for example, in \cite[(16.2)]{Freedman:2012zz}.
The construction procedure starts with assigning a gauge field to each generator of the superconformal algebra
\begin{eqnarray}
h_\mu = h_\mu^I T_I = \e_\mu{}^a P_a + \omega_\mu{}^{ab} M_{ab} + b_\mu D + f_{\mu}{}^a K_a + A_\mu T + \bar{\psi}_\mu  Q + \bar{\phi}_\mu S \,.
\end{eqnarray}
Here $M_{ab}$ and $P_a$ are the usual Poincar\'e generators while $D$ is the generator for dilatation, $K_a$ generate special conformal boosts and $T$ is the generator of the chiral $U(1)$ symmetry.  Furthermore, $Q$ and $S$ are the generators of supersymmetry and special supersymmetry, respectively.
The linear transformation rules and the curvatures for these gauge fields can be obtained by using the structure constants of the superconformal algebra $f_{IJ}{}^K$, i.e.
\begin{align}
\delta h_\mu{}^I &= \partial_\mu \e^I + \e^K h_\m{}^J f_{JK}{}^I\,, & R_{\mu\nu}{}^I & = 2 \partial_{[\mu} h_{\nu]}{}^I + h_\nu^K h_\mu{}^J   f_{JK}{}^I \ .
\label{StandardRules4dN1}
\end{align}
As certain curvatures depend linearly on specific gauge fields, it becomes possible to impose conventional constraints on these curvatures. Solving these constraints allows for the determination of the corresponding gauge fields, leading to an irreducible multiplet. In the case of four dimensional $N=1$ supersymmetry, the conventional constraints are given by
\begin{align}
R_{\mu\nu}{}^a (P) & = 0 \ , & e^{\mu}{}_a \widehat{R}_{\mu\nu}{}^{ab} (M) & = 0 \ , &  \gamma^\mu R_{\mu\nu}(Q) & = 0 \ ,
\label{ConvConst4dN1}
\end{align}
where in the second constraint a supercovariantization is employed (see \cite[(16.23)]{Freedman:2012zz} for the explicit formula) so that this set of constraints close under supersymmetry. These constraints determine $(\omega_\mu{}^{ab}, f_{\mu}{}^{a}, \phi_\m)$, respectively, in terms of the independent fields $(e_\mu{}^a, b_{\mu}, A_\mu, \psi_\mu)$. In particular, ones finds
\be
f_a^a = -\frac{1}{12} \left( R - \bar\psi_a \gamma^{ab} \phi_b \right) \ .
\label{fe}
\ee
Noting that the field $b_\mu$ has a shift symmetry, these fields form the off-shell Weyl multiplet with 8 bosonic plus 8 fermionic degrees of freedom.

To construct a supergravity action, we now need a compensating multiplet to gauge fix the redundant symmetries $(D, K_a, T, S)$. While $K_a$ is fixed by utilizing the gauge field of dilatation, $b_\mu$,  the remaining ones can be fixed by considering various matter multiplets, giving rise to different off-shell formulations of supergravity. For simplicity, here we demonstrate the procedure by using a chiral multiplet which consists of a complex scalar $Z$, a left-chiral projection of a Majorana spinor $P_L \chi=\frac12(1+\gamma_5) \chi $ and a complex auxiliary scalar field $F$, whose supertransformation rules can be found in \cite[(16.33)]{Freedman:2012zz}. Next, one constructs a superconformal invariant action for the chiral multiplet coupled to the Weyl multiplet for which fixing the redundant symmetries yields off-shell Poincar\'e supergravity. A suitable such action is given by (see, for example, \cite{Freedman:2012zz})
\begin{equation}
e^{-1} \cL = \mathrm{Re} \left[F \bar{F} + Z \Box^c \bar{Z} - \bar{\chi}  P_L \slashed{\mathcal{D}} \chi  + \frac{1}{\sqrt{2}} \bar\p_\m \g^\m \left( P_L\chi F + Z \slashed{\mathcal{D}} P_R \chi  \right) + \frac12 Z \bar{F}  \bar\p_\m \g^{\m\n}P_R \p_\n \right] \,,
\end{equation}
where $\slashed{\mathcal{D}}\chi$ is supercovariantized derivative of $\chi$, and
\begin{eqnarray}
\Box^c Z &=& e^{a\mu} \Big( \partial_\mu \mathcal{D}_a Z - 2 b_\mu \mathcal{D}_a Z + \omega_{\mu a b} \mathcal{D}^b Z + 2 f_{\mu a} Z + \rmi A_\mu \mathcal{D}_a Z \nonumber\\
&& - \frac{1}{\sqrt{2}} \bar\psi_\mu P_L \mathcal{D}_a \chi + \frac{1}{\sqrt{2}} \bar\phi_\mu \gamma_a P_L \chi \Big)\ .
\end{eqnarray}
Note that the term $f_a{}^a Z$ contains the Ricci scalar in view of \eq{fe}. One can now gauge fix this action to obtain an off-shell Poincar\'e supergravity
\begin{align}
\text{D-gauge and T-gauge} &: Z = \sqrt{3}/\kappa\,, &\text{S-gauge} &: \chi = 0\,,
\end{align}
which gives rise to
\begin{equation}
e^{-1} \cL = \frac{1}{2\kappa^2} \left(R - \bar\p_\m \g^{\m\n\r} D_\nu(\omega)\p_\r + 6 A_a A^a\right) + F \bar{F} \ ,
\end{equation}
with the field content $(e_\m{}^a, A_a, F, \psi_\m)$, counting 12 bosonic plus 12 fermionic  degrees of freedom.

The superconformal tensor calculus is also useful for studying matter couplings. For four-dimensional $N=1$ supersymmetry, one can begin with a chiral multiplet action along with the action for the desired matter multiplet. The chiral multiplet can then serve as the compensator for the superconformal symmetries, leading to an off-shell matter-coupled supergravity theory. In the context of extended supersymmetry, possible Weyl multiplets and compensators for $N=2$ supersymmetry can be found in details in \cite{Freedman:2012zz}. For $N=4$, the conformal supergravity framework is discussed in \cite{Bergshoeff:1980is}, although an off-shell formulation with a compensating multiplet, where the only viable option is a vector multiplet, remains unknown. In higher dimensions, the Weyl multiplets of five-dimensional $N=2$ supergravity were introduced in \cite{Bergshoeff:2001hc,Fujita:2001kv,Hutomo:2022hdi}. Possible constructions with a compensating vector or linear multiplet were discussed in \cite{Coomans:2012cf,Ozkan:2013nwa}, while a construction involving hyper and vector multiplets can be found in \cite{Fujita:2001kv,Bergshoeff:2004kh}. In six dimensions, with $N=(1,0)$ supersymmetry, the elements of superconformal tensor calculus, possible Weyl multiplets, and the off-shell construction with a linear multiplet were discussed in \cite{Bergshoeff:1985mz,Coomans:2011ih}.

The superconformal tensor calculus method is used in the description of the $N=(1,0), 6D$, $N=2, 5D$, $N=1,2, 4D$ and $N=1, (1,1), (2,0), 3D$ supergravities.

\item  \mbox{\it Ordinary superspace:} As is well known, supergravity theories can be formulated in superspace in terms of suitably chosen torsion and curvature constraints (see, for example, \cite{Howe:1981gz,Gates:1983nr,Buchbinder:1998qv}). In $4D$, for example, the $N=1$ superspace has in addition to the spacetime coordinates $x^\mu$, anticommuting coordinates $\theta^\alpha$ which have complex conjugates $\theta^\ad$, with $\alpha, \ad=1,2$, that are chiral $SL(2,C)$ spinors. Denoting the entire set of coordinates by $z^M$, and the supervielbein by $E_M{}^A$, the supertorsion 2-form is the exterior derivative of the 1-form $E^A=dz^M E_M{}^A$,
\be
T^A = DE^A = \frac12 E^C\wedge E^B\, T_{BC}{}^A\ ,
\label{st}
\ee
where $DE^A=dE^A +E^B \wedge \Omega_B{}^A$ and $\Omega_A{}^B$ is a connection one-form. Denoting the flat indices by $A=(a,\alpha)$ its nonvanishing components are the Lorentz algebra valued $\Omega^{ab}$, where $\Omega^{\alpha\beta}=\frac12 \Omega^{ab} (\sigma_{ab})^{\alpha\beta}$ and $\Omega^{\ad\bd}=\frac12 \Omega^{ab} (\sigma_{ab})^{\ad\bd}$. Differentiation \eq{st} one finds the Bianchi identity
\be
DT^A= E^B\wedge R_B{}^A\ ,
\label{bi}
\ee
where $R_A{}^B=d\Omega_A{}^B +\Omega_A{}^C \wedge \Omega_C{}^B$. In components, this reads
\be
\sum_{(ABC)} \left( R_{ABC}{}^D-D_A T_{BC}{}^D -T_{AB}{}^E T_{EC}{}^D\right)=0\ ,
\ee
where $\sum_{(ABC)}$ indicates the graded cyclic sum over $ABC$ and $D_A=E_A{}^M D_M$. These identities become nontrivial when constrains are imposed on the components of the torsion and curvature. The description of the supergravity requires the imposition of an appropriate set of constraints. Such a set which gives off-shell $N=1, 4D$ supergravity is given by \cite{Wess:1977fn}
\bea T_{\a\b}{}^C &=&0\ ,\qquad T_{\ad\bd}{}^C=0\ ,
\nn\w2
T_{\a\ad}{}^B&=&-2i\delta_c^B \left(\sigma^c\right)_{\a\ad}\ ,\qquad R_{\a\ad}{}^{cd}=0\ ,
\nn\w2
T_{\a b}{}^c &=&0\ ,\qquad T_{\ad b}{}^c=0\ .
\eea
Recalling that $D_\a$ and $D_a$ have dimensions $1/2$ and $1$ respectively, possible dimensions of the superfield expressions in \eq{bi} run from $1/2$ to $5/2$. A very useful strategy is to analyze them in the order of increasing dimensions. Nonetheless this is still a long and complicated calculation \cite{Grimm:1978ch}; for a sketch, see, for example, Chapter 16 of \cite{West:1990tg}. In summary, Bianchi identities are solved in terms of the superfields $R, G_{\a\bd}, W_{\a\b\gamma}=W_{(\a\b\gamma)}$ in terms of which all torsion and curvature components can be expressed, and they contain spacetime coordinate dependent fields $(e_\mu, \psi_\mu, Z, A_\mu)$ that form an off-shell multiplet in which the complex scalar $Z$ and the gauge field $A_\mu$ are auxiliary fields, arising as the $\theta=0$ components of the superfields $R$ and $G_{\a\ad}$, respectively. Setting $R=0$ and $G_{\a\ad}=0$ gives the equations of motion of $N=1, 4D$ supergravity.
The framework outlined above can be extended to yield conformal supergravity. In a subsequent step the conformal symmetry can be fixed to yield Poincar\'e supergravity. For a detailed account of this procedure, see for example, \cite{Buchbinder:1998qv}, where matter couplings are also covered. For extended supersymmetry in superspace see, for example, \cite{Howe:1981gz}. These can be generalized to conformal superspace to describe conformal supergravities \cite{Howe:1980sy}. As to supergravities in higher dimensions, the most studied ones are in $D=6,10, 11$, for which we shall give appropriate references in the sections where we cover these dimensions.

In general, demanding the Bianchi identities in presence of a given set of torsion and curvature constraints may or may not require the imposition of equations of motion. In the first case, one gets an on-shell supergravity, while the latter case yields an off-shell supergravity, assuming that the solution is nontrivial. Often, the solutions to the torsion Bianchi identities facilitate the construction of certain super p-form field strengths with their own Bianchi identities, and these forms prove to be very useful. This framework is very powerful in determining the two-derivative supergravities, and their matter couplings. In order to go beyond two derivatives, one needs to deform the torsion and possibly the p-form constraints, as discussed further in section \ref{ssa}. However, this complicates the matter because solutions to the Bianchi identities bring in a number of new superfields whose relation to the supergravity multiplet can involve a large number (e.g. thousands) of possible structures, which are hard to analyze; see for example \cite{Cederwall:2004cg}.

The superspace space approach is used in the description of type IIB and heterotic supergravities in $10D$, $N=1,2, 4D$ and $N=4,6,8, 3D$ supergravities.

\item  \mbox{\it Harmonic superspace:}  The $N$-extended superspace has a useful extension known as harmonic superspace, introduced in \cite{Galperin:1984av}, which is the product of ordinary superspace with a coset space $K=H\backslash G$ where $G$ is the $R$-symmetry group and $H$ is a suitably chosen isotropy group such that $K$ is always a compact complex manifold \cite{Galperin:2001seg}. This framework makes it possible to construct action formulae that may not necessarily admit an ordinary superspace formulation. While it is relatively straightforward to construct such actions with linearized supersymmetry, their non-linear generalizations may not always exist, for example, due to the non-integrability of the chirality condition on the integration measure caused by certain nonvanishing supertorsion components; see for example the case of Type IIB supergravity \cite{deHaro:2002vk}. However, in some cases, there may exist non-linear action formulae with appropriate integration measures in which the integral is over the harmonic superspace. For a further discussion of harmonic superspace and its application to the construction of $N=4,5,6,8$ supergravities in $4D$, see section \ref{ordinary invariants}.

\item \mbox{\it The superform method (ectoplasm):}  In a superspace of $D$-dimensional spacetime and $n$ dimensional Grassmannian coordinates, consider a closed superform $J$,
\be
J=\frac{1}{D!} dz^{M_D}\wedge \cdots \wedge dz^{M_1}\,J_{M_1...M_D}\ ,\qquad dJ=0\ ,
\ee
where $z^M$ are the superspace coordinates. Such a superform furnishes a supersymmetric action formula \cite{Voronov:1992,Gates:1997kr,Gates:1998hy}
\be
S= \int_{{\cal M}_D}  i^\star J  \ ,
\label{ep}
\ee
where $i: {\cal M} \to {\cal M}^{D|n}$ is the inclusion map and $i^\star$ is its pullback which effectively embodies the projections $\theta=0$ and $d\theta=0$. A more explicit formula is given in \eq{epa} below. The fact that this action is invariant under supersymmetry can be seen as follows. Under the superdiffeomorphisms generated by the superfield $ \xi= \xi^M\partial_M $, in which the fermionic part is the local supersymmetry transformation, one has
\be
\delta_\xi J= \cL_\xi J = i_\xi dJ + di_\xi J\ ,
\ee
and the first term vanishes due to the closure $dJ=0$, and the second term gives a surface term in the variation of the action, which we neglect. Denoting the supervielbein by $E_M{}^A$, and identifying the spacetime vielbein and the gravitino fields by $e_m{}^a = E_m{}^a\Big|$ and $\psi_m{}^\a = E_m{}^\a\Big|$, where the bar denotes evaluation of a superfield at $\theta=0$, the action formula reads
\bea
S &=& \int d^D x\, \ve^{m_1...m_D} \Big( e_{m_D}{}^{a_D} \cdots e_{m_1}{}^{a_1} \, J_{a_1...a_D} + D\, e_{m_D}{}^{a_D} \cdots e_{m_2}{}^{a_2}\psi_{m_1}{}^{\a_1} \, J_{\a_1a_2...a_D} +...
\nn\w2
&& ...+ \psi_{m_D}{}^{\a_D} \cdots \psi_{m_1}{}^{\a_1}\, J_{\a_1...\a_D} \Big)\ ,
\label{epa}
\eea
where each of the $J$s is evaluated at $\theta=0$. Since $S$ is invariant under the replacement $J_D \to J_D+dK_{D-1}$, the mathematical problem at hand is to compute the d'th de Rham cohomology group. It is understood, however, that $J_D$ is to be constructed out of the physical fields of the supergravity theory at hand. In the case of $D=10$, pure spinors in the sense of Cartan, namely those obeying the condition $\lambda^\a \left(\gamma^a\right)_{\a\b} \lambda^\b =0$, appear in the construction.

In the closed superform method, there are two cases to consider: strict invariants and Chern-Simons invariants. In the first case, the nonvanishing components of the closed form $J_D$ are tensorial. In the latter case, they may involve the gauge potentials and possibly $\theta$s. These lead to Chern-Simons terms in addition to tensorial terms. In that case $J_D$ can be constructed as follows \cite{Voronov:1992,Gates:1997kr,Gates:1998hy, Berkovits:2008qw, Gates:1997ag}. Let $W$ be a closed $(D+1)$-form in the $(D|n)$ dimensional superspace that can be written as $dZ$ where $Z$ is a local $D$-form that involves gauge potentials. If $W$ can also be written as $dK$ where $K$ is a tensorial $D$-form, then
\be
J_D=K-Z
\ee
is a closed $D$-form that can be used to form invariants using the superform method. In this construction, $Z$ gives the Chern-Simons term while $K$ gives the rest of the terms that go with it. We refer the readers to \cite{Berkovits:2008qw} for a review.

The ectoplasm approach is used in the description of higher derivative couplings in maximal supergravities in $D=7,8,9$, $N=(1,0),6D$ and $N=4, 4D$ supergravities.

\item \mbox{\it Dimensional reduction and hidden symmetries:} Given a higher derivative extension of a supergravity theory in $D$-dimensions, it is straightforward to perform a dimensional reduction on $T^d$ to obtain higher derivative extensions of supergravity in $(D-d)$-dimensions. At the two-derivative level, the resulting theories are known to possess hidden symmetries, see Table \ref{dualgr} in section \ref{modular functions}.
In the case of $N=1, 10D$ supergravity coupled to $n_V$ abelian vector multiplets compactified on a torus $T^d$, the hidden symmetry is $O(d,d+n_V)$. A construction of the theory in which this symmetry is manifest by doubling the $10D$ spacetime and introducing additional $n_V$ coordinates is known as double field theory, reviewed in \cite{Aldazabal:2013sca, Hohm:2013bwa, Berman:2013eva}. A four derivative extension of this theory where a $2d+n_V$ dimensional group is gauged has been provided in \cite{Baron:2017dvb}. Dimensional reduction of this result, in particular on tori, gives rise to four-derivative extension of half-maximal gauged supergravities coupled to Yang-Mills.

The dimensional reduction method is used in the description of the higher derivative couplings of $10D$ type IIB supergravity and supergravities in $D=1,2,3,6,7,8,9$.

\item \mbox{\it The $S$-matrix and holography:} The $S$-matrix method relies on extracting the supergravity effective action from the scattering amplitudes. For a textbook treatment of this subject, see \cite{Elvang:2015rqa}. If such amplitudes are available from an $S$-matrix approach, which is a situation encountered in string theory, but less robust in supergravity as a field theory, one can look for an action modulo field redefinitions that produces those amplitudes. Combinations of terms that give vanishing contributions to a given $n$-point amplitude may require the knowledge of $(n+1)$ or higher point amplitudes, which increases the complexity of the problem. See, for example, \cite{Peeters:2005tb} for the situation arising in the context of four graviton amplitudes in $11D$ supergravity. Quantization of the two-derivative supergravity and computation of loop amplitudes instead give information on local higher derivative functionals in terms of divergences that depend on a cut-off. Adding counterterms to remove these divergences leads to undetermined coefficients in front of them. If the theory is a low energy limit of string theory in some background, these coefficients are determined by string theory. Regarding the fate of hidden symmetries in higher derivative extensions of supergravity, at least the order $\alpha'$ corrections in the $O(d,d+n_V)$ covariant setup has been achieved \cite{Baron:2018lve,Lescano:2021guc,Hronek:2022dyr}. Whether this construction can be carried out to higher orders in $\alpha'$ remains to be seen. Indeed, doubt has been cast on whether the $\alpha^{'3} \zeta(3)$ correction in heterotic string theory effective action can be captured in a supersymmetric double field theory framework \cite{Hronek:2020xxi}.

Remarkably, a super conformal field theory in $3D$ known as the ABJM model \cite{Aharony:2008ug}, which is a $3D$ Chern-Simons theory with gauge group U(N)$\times$ U(N) and level $k=1,2$, coupled to bifundamental matter, has been employed to compute successfully the coefficients of the $R^4$ \cite{Chester:2018aca} and $D^4 R^4$ \cite{Binder:2018yvd} terms in $11D$ low energy effective action of M-theory. This is a remarkable result because it provides a framework for going beyond $11D$ supergravity, as can be seen by the fact that it fixes the coefficient of the $R^4$ term which cannot be fixed by $11D$ supersymmetry alone.  The approach used in \cite{Chester:2018aca} is based on studying the flat spacetime limit of the Mellin amplitude associated with the four-point correlation function of scalar operators in the stress tensor multiplet of ABJM theory \cite{Aharony:2008ug}. In this way, the momentum expansion of the M-theory four-graviton S-matrix elements is obtained. In practice, however, so far this approach has been tractable for four-point graviton amplitudes. Similarly, the relation between the large-N expansion of the integrated correlators in $N=4$  supersymmetric Yang-Mills theory and the scattering amplitudes in type IIB superstring theory in AdS$_5 \times S^5$  has been reviewed recently in \cite{Dorigoni:2022iem}.

The S-matrix and holography approach is used in the description of the higher derivative couplings of supergravity in $11D$, type IIA and type IIB supergravities in $10D$, and $N=8, 4D$ supergravity.

\end{itemize}

\section{\texorpdfstring{$D=11$}{D=11}}

The $11D$ supergravity multiplet consists of a metric $g_{\m\n}$, a 3-form potential $C_{\m\n\r}$ and a Majorana gravitino $\psi_\m$. The bosonic part of the Lagrangian is given by \cite{Cremmer:1978km}
\bea
e^{-1}\cL_{CJS} = R(\omega)-\frac{1}{48} F_{\m\n\r\s} F^{\m\n\r\s}
 - \frac{1}{144^2} \varepsilon^{\m_1...\m_{11}} F_{\m_1...\m_4} F_{\m_5...\m_8} C_{\m_9...\m_{11}}\ ,
\label{CSJ}
\eea
where $F_{\m\n\r\s}= 4\partial_{[\m} C_{\n\r\s]}$. The supertransformation of the gravitino, up to leading terms in fermions, is given by
\be
\delta\psi_\m = D_\m \epsilon +\frac{1}{288} \left(\gamma_\m{}^{\n\r\s\tau} -8\delta_\m^\n \gamma^{\r\s\tau} \right) F_{\n\r\s\tau}\ .
\ee
Since $11D$ supergravity emerges in the strong coupling limit of Type IIA string theory, and string theory admits a derivative expansion in its low energy effective action, it is natural to expect a higher derivative deformation of $11D$ supergravity as well. Let us also recall that $11D$ supergravity admits the fundamental membrane and solitonic M5-brane solutions. Studying the local symmetries on the M5-brane worldvolume, one discovers that freedom from anomalies requires the presence of the $ C_3 \wedge R^4$ term in the $11D$ action and the attendant anomaly inflow mechanism that ensures the cancellation of the anomalies, as will be discussed further below. The presence of this term in $11D$ supergravity calls for the restoration of supersymmetry order by order in the Planck length $\ell_p$, and hence the need for uncovering the higher order in derivative terms in the action. The need for the higher derivative terms is also clear from the consideration of loop corrections to $11D$ supergravity. On the account of its nonrenormalizability one needs to put a cut-off and introduce higher derivative counterterms. Despite all these motivations for getting a handle on the higher derivative terms in the effective action, their determination turns out to be an extremely complicated problem. In what follows we shall review those which have been obtained so far, starting with the eight-derivative deformations\footnote{In  \cite{Tsimpis:2004rs}, a five derivative deformation has been discussed but it can simply be obtained from a field redefinition of the 3-form potential. } .

\subsection{Eight-derivative deformations }
In $11D$ supergravity, the first non-topological deformation of the two-derivative
action appears at the eight-derivative level. There exist no quadratic curvature terms in 11D maximal supergravities, as can be seen by the following argument. Assuming that quadratic curvature terms exist in $11D$ supergravity, their reduction on a torus $T^8$ yields quadratic curvature terms in $4D, N=8$ supergravity. This theory admits a maximally supersymmetric Minkowski vacuum about which the modes of perturbations are arranged into irreducible representations, with $s_{\rm max} \le 2$,  of the $N=8$ Poincar\'e  superalgebra. In particular, there will be massive spin-2 modes generated by the combination of the Einstein-Hilbert term and Riemann curvature-squared term. However, from Table \ref{higherNirrep}, one observes that when $N\ge5$, the massive spin-2 state must live in a multiplet with $s_{\rm max} >2$, yielding a contradiction\footnote{The massive modes generated by higher derivative terms are different from the KK modes. Owing to the presence of the central charge, if the mass term originates for the two-derivative action, then $s_{\rm max} \le 2$ even in the case of $N=8$. However, if the mass is generated by the presence of the  Riemann-squared term, then $s_{\rm max}>2$ can arise.}. As for the cubic in curvature terms, it is known they are incompatible with the $10D$ maximal supersymmetry \cite{Metsaev:1986yb}. Therefore, there should not be cubic curvature terms in $11D$ either, since their circle reduction would generate such terms. For these reasons, we will first review attempts to construct the eight-derivative superinvariant in $11D$.
\begin{table}[ht!]
\centering
\begin{tabular}{|r|r|r|r|r|r|}
\hline
spin & $N=1$ & $N=2$ & $N=3$ & $N=4$ & $N=5$ \\ \hline
5/2  &       &       &       &       & 1     \\
2    & 1     & 1     & ~ 1   & 1     & 10    \\
3/2  & 1 2   & 1 4   & 1~~6   & 8     & 44    \\
1    & 1 2 1 & 1 4 6 & 6 15  & 27    & 110   \\
1/2  & 1 2 1~~ & 4 6 4 & 14 20 & 48    & 165   \\
0    & 2 1~~~~~  & 5 4 1 & 14 14 & 42    & 132   \\
\hline
\end{tabular}
\caption{On-shell massive multiplets in $D=4$ super-Poincar\'e algebra without central charge up to $N=5$ \cite{deWit:1978pd}. }
\label{higherNirrep}
\end{table}

\subsubsection*{Noether procedure approach}
		
The most extensive Noether procedure approach to the construction of the eight-derivative extension of $11D$ supergravity was carried out in  \cite{Hyakutake:2006aq,Hyakutake:2007vc,Hyakutake:2007sm}.
The ansatz considered by these authors has $1544$ terms which take the schematic form\footnote{Some of the $R^3\bar{\psi}\psi_2$ terms were obtained by lifting higher derivative corrections of type IIA theory in \cite{Peeters:2000qj}.}
\bea
S^{(6)} &=& \ell_p^6 \int dx^{11}e\Big[ [R^4]_7 + [CR^4]_2 +  [R^3 F^2]_{30} +[R^2 (DF)^2]_{24}
\nn\w2
&& + [R^2\bar{\psi}_2D\psi_2]_{25} + [R^3\bar{\psi}\psi_2]_{92}
+ [R^3 F \bar\psi\psi]_{447}
\nn\w2
&&
+ [R^2 F\bar\psi_{(2)} \psi_2]_{190}
+[R^2 D F \bar\psi\psi_2]_{614} + [RD F \bar\psi_2 D\psi_2]_{113} \Big]\ ,
\label{an2}
\eea
where $[X]_n$ schematically denotes the structure of the terms where the indices need to be contracted in various ways, $n$ denotes the number of independent such terms and $\psi_2$ denotes the gravitino curvature. The explicit form of the terms $[R^3 F^2]_{30}$ can be found in \cite[Eq. (15)]{Hyakutake:2006aq}, and the terms $[R^2 (DF)^2]_{24}$ in \cite[Eq. (A.5)]{Peeters:2005tb}\footnote{The basis for the $R^2(DF)^2$ terms in \cite{Hyakutake:2008} and \cite{Peeters:2005tb} differ from each other. We have determined the exact relation between these two bases but we refrain from giving it here.}.
In this ansatz only a subset of eight-derivative terms are considered because the following assumptions have been made: 1) Terms with covariant derivatives of the Riemann tensor are not considered; 2) the covariant derivative of the gravitino appear only as a field strength; 3) the terms involving $F$ are considered only up to second order; 4) parity invariance under which $x^{10}\to -x^{10}, A\to -A$ and $\psi_\mu \to \gamma^{10} \psi_\mu$ is assumed;  5) terms proportional to the field equations resulting from $S_0$ are left out; 6) higher derivative deformations of the supersymmetry transformation rules are not considered at all, and only the lowest order supersymmetry variations, namely those of ordinary two-derivative $11D$ supergravity, are implemented and 7) only cancellations of terms containing at most one factor of $F$ are sought in  \cite{Hyakutake:2008}.
The 5th and 6th assumptions are connected in the following sense. Supersymmetric variation of terms in the action that contain the lowest order field equations can be canceled by modifying the transformation rules in a way that can be deduced from the variations proportional to the stated field equations \cite{Hyakutake:2006aq}. However, the variation of the higher derivative action with respect to the lowest order supersymmetry transformations $\delta_0$, cannot be canceled in this way, and achieving the cancellations by fixing the parameters is highly nontrivial.

Even under the assumptions listed above, the Lagrangian still has $1544$ parameters, and the subset of variations mentioned above give rise to 4643 structures that have the schematic form  \cite{Hyakutake:2008}\footnote{ We thank Y. Hyakutake for making his unpublished work available to us, in which the counts for the 3rd and 6th structures given in  \cite{Hyakutake:2007vc}, have been corrected.}
\bea
&& [\bar\epsilon \psi R^4]_{116}\ , \quad [\bar\epsilon\psi_2 R^2 DR]_{88}\ , \quad {[\bar\epsilon D\psi_2 R^3]_{51}}\ ,\quad [\bar\epsilon\psi R^2(DR) F]_{1563}\ , \quad [\bar\epsilon\psi R^3F]_{513}\ ,
\nn\w2
&& {[\bar\epsilon \psi_2 R^2 D^2F]_{614}}\ ,\quad [\bar\epsilon \psi R^3 DF]_{995}\ ,\quad [\bar\epsilon\psi_2 RDR DF]_{371}\ ,\quad [\bar\epsilon D\psi_2 R^2 DF]_{332}\ .
\label{v2}
\eea
The $F$-independent terms in \eq{an2} and \eq{v2} are spelled out in  \cite{Hyakutake:2006aq}. Partial results were obtained in \cite{Hyakutake:2006aq,Hyakutake:2007vc}, while a fuller analysis was carried out in  \cite{Hyakutake:2008}, where it was found that the bosonic part of the Lagrangian \eq{an2} is completely fixed in terms of two parameters, $a$ and $b$, as follows
\bea
S^{(6)}_H &=& \ell_p^6 \int d^{11}x\left[a \left( t_8 t_8 e R^4
-\frac{1}{24}\epsilon_{11} \epsilon_{11} e R^4  - \frac{1}{6} \epsilon_{11} t_8 C R^4\right)\right.
\nn\w2
&&\left. +\sum_{i=1}^{24} c_i (a,b) [R^2 (DF)^2]_i
+ \sum_{r=1}^{30} d_i (a,b) [R^3 F^2]_i
\right]\ ,
\label{d11}
\eea
 in which the coefficients $c_i$ and $d_i$ have been determined in \cite{Hyakutake:2008} and the symbols $\e_{11}\e_{11}, \e_{11}t_8$ and $t_8t_8$ are defined in the appendix. It is natural to expect that further variations in the Noether procedure would relate the parameters $a$ and $b$ to each other. Another way to fix them is to reduce the result on a circle and compare it with the results obtained long ago in \cite{Gross:1986iv} in the NS-NS sector.

Finally, we note that the $\epsilon_{11} t_8 C R^4$ term is related to anomaly inflow in the presence of $M5$-brane as follows \cite{Duff:1995wd}. The $M5$-brane which supports chiral fermions and a chiral 2-form has gravitational anomalies which can be obtained by descent from the following anomaly polynomial
\be
X_8 = \frac{1}{(2\pi)^4}\frac1{768}\left[-({\rm tr}R^2)^2+4 {\rm tr }R^4\right]\ .
\ee
The descent equations $X_8=dI_7$, $\d I_7=dA_6$ determine the anomaly ${ A}_6$. It turns out that this anomaly is precisely canceled by the variation of the bulk term $\epsilon_{11} t_8 C R^4 \sim C\wedge X_8$ under the $11D$ local Lorentz transformations restricted to $SO(5,1)$. In other words, in the presence of an M5-brane, we have the so-called anomaly inflow
\be
\d \int_{M_{11}} C\wedge X_8 = \int_{\partial M_{11}} F\wedge \delta I_7  = -\int_{W_6} dF\wedge A_6\ ,
\ee
where we have used the relation $dF=2\pi \delta_5$ with $\delta_5$ representing a 5-form which integrates to one in the directions transverse to the $M5$-brane
and has delta function support on the $M5$-brane. Thus the anomaly inflow from the bulk cancels precisely the $M5$-brane worldvolume gravitational anomaly $A_6$.  For more details, including the cancellation of $R$-symmetry related anomalies, see \cite{deAlwis:1997gq, Freed:1998tg}.  The $C\wedge X_8$ term has also been utilized to compute the Weyl anomaly on the worldvolume of multiple M5-branes using holography  \cite{Henningson:1998gx}. It was shown that the leading order coefficient $4N^3$  \cite{Henningson:1998gx} in the Weyl anomaly is shifted to $4N^3-3N$  \cite{Tseytlin:2000sf} which for $N=1$ indeed matches in the case of a single free tensor multiplet theory.

\subsubsection*{Superparticle approach}

As mentioned above, an alternative approach to computing the higher derivative extension of $11D$ supergravity is to compute the loop corrections. Instead of employing the usual BRST quantization, it turns out that an equivalent but more efficient way to proceed is to use the superparticle vertex operators \cite{Green:1999by}. To this end, we recall that $\kappa$-invariant superparticle propagating in $11D$ target superspace requires $11D$ supergravity field equations. However, since $\kappa$-symmetry is infinitely reducible, its covariant quantization is problematic. This problem is bypassed by quantizing the superparticle in the lightcone gauge. The resulting superparticle vertex operators for the supergraviton were used in \cite{Peeters:2005tb} to determine the $(DF)^2 R^2$ and $(DF)^4$ terms from the four-graviton amplitude. The result schematically takes the form
\be
S^{(6)}_{PPS} = \ell_p^6 \int d^{11}x e \Big[ \alpha t_8 t_8  R^4 +\sum_{i=1}^{24} \beta_i [R^2 (DF)^2]_i
+\sum_{r=1}^{24} \gamma_r [(DF)^4]_r \Big]\ ,
\label{pr}
\ee
where the coefficients $\alpha, \beta_i, \gamma_r$  and all the tensorial structures are given in \cite[Eqs. (3.1-3), (A.2) and (A.5)]{Peeters:2005tb}. Furthermore, as noted in \cite{Peeters:2005tb}, there are 6 linear combinations of $(DF)^2 R^2$ terms and 9 linear combinations of $(DF)^4$ terms in the effective action which lead to vanishing 4-pt amplitudes. These combinations are given in \cite[Eqs. (A.3) and (A.7)]{Peeters:2005tb}. Note also that the $\e_{11}\e_{11} R^4$ doesn't contribute to the four-graviton amplitude. Its determination requires the five-graviton amplitude \cite{Peeters:2005tb}.

Comparing the results of \cite{Peeters:2005tb} with those of \cite{Hyakutake:2008}, the $a$ $t_8t_8R^4$ terms agree. As to the $R^2 (DF)^2$ terms, not only do they agree by relating $(\alpha,\beta_i)$ to $(a,b)$, but the results of \cite{Hyakutake:2008} also provide the coefficients for the 6 linear combinations mentioned earlier that are not fixed by the 4-pt supergraviton amplitude.

It is useful to note that the $R^4$ term in the Lagrangian discussed above also arises as a one-loop divergent part of the four-graviton amplitude \cite{Green:2006gt}. Introducing  a momentum cut-off $\Lambda$ the divergence is given by $\frac{4\pi}{3} \Lambda^3 e t_8t_8R^4$, in conventions of \cite{Green:2008bf}. This can be eliminated by adding the counterterm $\Delta \cL = (a-\frac{4\pi}{3} \Lambda^3)e t_8t_8 R^4$, thereby obtaining the action \eq{d11}.
In \cite{Green:2008bf}, the coefficient $a$ is fixed to be $a=2\pi^3$ by comparing the result with the finite 4-pt graviton amplitude in type II string theory. It has been argued that the $R^4$ term we are discussing does not receive corrections beyond one-loop \cite{Bern:1998ug,Green:2005ba}.

Considering higher loops in $11D$ supergravity, one expects divergences of the form
$D^{2k}R^4$ arising in the four-graviton amplitudes. It turns out that the $D^2 R^4$ term vanishes on-shell, and at two-loops, $D^4 R^4$ and $D^6 R^4$ terms may arise \cite{Green:2006gt}. If nonvanishing, these terms would suggest the existence of their supersymmetric completions. The two-loop amplitude has been studied in detail in \cite{Green:2006gt}. Consistency with string theory indicates that the renormalized value of the $D^4 R^4$ term should vanish and the $D^6R^4$ interaction, together with other terms of the same dimension, are the first nontrivial corrections to the eleven-dimensional M-theory
effective action after $R^4$ \cite{Green:2005ba}. Higher loops in $11D$ supergravity and their consequences for $D^6 R^4, D^8 R^4$ and $D^{10} R^4$ counterterms have also been discussed in \cite{Green:2005ba}.

\subsection{Superspace approach }
\label{ssa}
		
The equations of motion of the $11D$ supergravity \cite{Cremmer:1978km} were reformulated in superspace in  \cite{Cremmer:1980ru,Brink:1980az}. The key to superspace formulations is the nature of the constraints imposed on the superspace torsion. Some of these are called the conventional constraints, which amount to field redefinitions, while other physical (non-conventional) constraints put the theory on-shell in $11D$.  For a self-contained succinct review of these points, see, for example,  \cite{Cederwall:2004cg}. In the original superspace formulation, the minimal set of constraints that put the theory on-shell were not studied. In addition, there is the additional issue of whether to introduce a super 4-form field strength into the formalism from the start, as opposed to working in what is referred to as the geometrical part of the theory, entailing the supervielbein and connection. For a discussion of the relationship between the two approaches, see  \cite{Cederwall:2004cg}, where it has been argued that while they are equivalent in the formulation of the standard two-derivative $11D$ supergravity, it is not known whether that is the case in the higher derivative extension of the theory discussed below in a geometric framework. In the superspace formalism, there is also a choice to be made as to whether to enlarge the structure group to include a Weyl(scale) transformation.  This entails the introduction of a connection which takes its values in the Lie algebra of $Spin(1, 10) \times \mathbb{R}^+$, and superspace with such a connection is called Weyl Superspace \cite{Howe:1997he}.
		
In  \cite{Howe:1997he}, working in Weyl superspace, Howe showed to obtain the $11D$ supergravity equations of motion, it is only necessary to consider the geometric part of the theory and to take the dimension-$0$ component of the supertorsion to be
\be
T_{\alpha\beta}{}^c = -i(\gamma^c)_{\alpha\beta}\ ,
\label{mt}
\ee
where $\alpha,\beta=1,...,32$ are the spinor indices, and $c=0,1,...,10$ is the Lorentz vector index. In a later work  \cite{Howe:2003cy}, it was shown that taking the lowest components of the closed superspace four-form to vanish, one obtains the standard supergravity equations of motion.
		
\subsubsection*{A geometrical framework}

Higher derivative extensions of $11D$ supergravity, referred to as the deformed theory, call for a new set of superspace constraints. This problem has been studied in the geometrical framework, that is, without introducing the superspace four-form and its Bianchi identities, in  \cite{Cederwall:2004cg,Nilsson:1998ca,Cederwall:2000ye}. The deformation problem, which deals with the question of which torsion components are to be subject to physical constraints, very beneficially has been mapped to a problem in {\it spinorial cohomology} introduced in   \cite{Cederwall:2001bt,Cederwall:2001dx}, and developed further in  \cite{Howe:2003cy}.
In this approach, it has been pointed out that the deformations in question can be understood perturbatively in the cohomology of a supersymmetric spinor derivative followed by a projection onto the highest weight representation. For a succinct review of this procedure, see, for example,  \cite{Howe:2003cy}. There also exists a closely related {\it pure spinor cohomology} approach\cite{Berkovits:2000fe,Berkovits:2002zk}. For a discussion of the relation between the "spinor" and "pure spinor" cohomologies, see \cite{Howe:2003cy}.

To summarize the results of  \cite{Cederwall:2004cg}, it was found that by using the conventional constraints, with the structure group taken to be the Lorentz group, the torsion can be brought into a form in which, in particular, the dimension-0 components are given by
\be
T_{\alpha\beta}{}^c = 2\Big( \gamma_{\alpha\beta}{}^c + \frac12 \gamma_{\alpha\beta}{}^{d_1d_2}\,X_{d_1d_2}{}^c + \frac{1}{5!} \gamma_{\alpha\beta}{}^{d_1\cdots d_5}\, Y_{d_1\cdots d_5}{}^c \Big)\ ,
\label{nt}
\ee
where $X$ and $Y$ tensors are in the $(11000)$ and $(10002)$ representations of the $11D$ Lorentz group. The remaining components of the torsion can be found in \cite[Eq. (2.10)]{Cederwall:2004cg}. They span dimensions $1/2$ to $3/2$ and contain $31$ more representations! It has been argued that these representations are general enough to account for any deformation allowed by supersymmetry, and when substituted to the superspace Bianchi identities (SSBI's), they will contain components of the most general stress tensor multiplet. In  \cite{Cederwall:2004cg}, all SSBI's of dimensions $1/2$ and $1$ are solved to linear order in tensor superfields $X$ and $Y$, and the solution is used to obtain the deformed equations of motion at dimensions $3/2$ and $2$. However, to find the explicit form of the equations of motion in terms of physical fields requires the determination of the $X$ and $Y$ tensors in terms of the physical gauge covariant fields such as the Riemann tensor and gravitino curvature. This is a very complicated, and yet to be cracked problem, since there are thousands of possible independent combinations of fields that can be harbored in the representations of $X$ and $Y$, as has been noted in  \cite{Cederwall:2004cg}, and as can be glimpsed from the discussion of the Noether procedure results summarized in the previous section.

\subsection{Lifting from \texorpdfstring{$10D$}{10D}, amplitudes and duality}
		
There have been proposals for the construction of the higher derivative deformation of $11D$ supergravity by lifting the corresponding results in Type IIA theory in $10D$. The lifting ansatz takes the form
\bea
ds_{11}^2&=&e^{-2\phi/3}\left(ds_{10}^2+e^{2\phi}(dx^{11}+A_{\mu}dx^{\mu})\right)\,,
\nn\\
B_{\mu\nu}&=&C_{\mu\nu 11},\quad C_{\m\n\r}=A_{\m\n\r}\ ,
\eea
where $\m,\n=0,\cdots 9$. The 10D/11D physical parameters are related via
\be
2\kappa_{10}^2=(2\pi)^7g_s^2\alpha'^4\,,\quad 2\kappa_{11}^2=(2\pi)^5\ell^9_{11}\,,\quad \kappa_{10}^2=\frac{\kappa^2_{11}}{2\pi R_{11}}\,,\quad R_{11}=g_s\sqrt{\alpha'}\,,
\ee
where $R_{11}$ is the period of $x^{11}$, i.e. $x^{11}\sim x^{11}+R^{11}$. Using this ansatz, the eight-derivative terms in the type IIA one-loop effective action were lifted to $11D$, thereby obtaining the purely bosonic eight-derivative corrections to $11D$ supergravity \cite[Eq. (6.17)]{Liu:2013dna}. The reduction of the $11D$ action to $10D$ apparently gives rise to an expression with complicated dilation dependence. However, one can show that it is on-shell equivalent to the standard one-loop term without the dilaton, as discussed in  \cite{Frolov:2001jh}.
		
The eight-derivative terms in the type IIA tree-level effective action do not survive the M-theory limit ( $g_s\rightarrow \infty$ ) and therefore it gives no contribution to the effective action of $11D$ supergravity \cite{Russo:1997mk}. When M-theory is compactified on a circle of finite radius, one recovers the eight-derivative terms in IIA tree-level effective action, upon the inclusion of the contributions from the KK modes \cite{Russo:1997mk}.
		
The bosonic action of the eight-derivative deformation of $11D$ supergravity, is partially obtained by lifting the deformations of type IIA supergravity \cite{Liu:2013dna}  and partially from the four-point superparticle correlators in $11D$ in the light-cone gauge \cite[Eq. (3.1)-(3.3)]{Peeters:2005tb}, or from the four-point amplitudes in $11D$ supergravity  \cite{Deser:1998jz,Deser:2000xz,Deser:2005kb}. Taking into account all this information, the eight-derivative deformation of $11D$ supergravity, up to an overall constant, is given by\footnote{Prior to  \cite{Liu:2013dna}, partial results on the eight-derivative deformations of $11D$ supergravity were also derived by lifting certain deformations of type IIA supergravity in $10D$  in its NS-NS sector \cite{Peeters:2000qj}. }  \cite{Liu:2013dna}
\bea
\Delta S_2\Big|_{11D} &=& \ell_p^6 \int \Big[ (t_8t_8-\frac1{24}
\epsilon_{11}\epsilon_{11})R^4 +  C\wedge X^8(R) -t_8t_8F^2R^3-\frac1{96}\epsilon_{11}\epsilon_{11}F^2R^3
\nn\\
&&  + \frac13 F\wedge \left( R^{ab}\wedge R^{bc}\wedge F^{cde}\wedge D F^{dae}+2R^{ab}\wedge F^{bce}\wedge R^{cd}\wedge DF^{dae}\right.
\nn\\
&&+R^{ab}\wedge R^{bc}\wedge DF^{cde}\wedge F^{dae}-\frac12 {\rm tr}R^2\wedge F^{abe}\wedge DF^{bae},
\nn\\
&&\left. -3 R^{ab}\wedge F^{bae}\wedge R^{cd}\wedge F^{dce} \right) \Big] + \int {\cal L}\left[(DF)^2 R^2\right] + \int {\cal L}\left[(DF)^4\right ] +\cdots\ ,
\label{LM}
\eea
where the $(DF)^4$ and $(DF)^2 R^2$ terms are the ones in \eq{pr}, and $F^{abc} = dx^\mu F_\mu{}^{abc}/(4\pi)^2$, and the ellipsis denotes yet to be determined couplings, including those of the form $(DF)^3 R$ which are not accessible in the light-cone gauge. For the definitions of $t_8 t_8 R^3$ and $\e_{11}\e_{11}F^2R^3$ see the appendix. A different method based on the four-point tree scattering amplitudes in $11D$ supergravity was used in  \cite{Deser:1998jz,Deser:2000xz,Deser:2005kb}, where $(DF)^4, (DF)^2 R^2$ and $R^4$ terms were computed\footnote{According to  \cite{Peeters:2005tb} their results for the $(DF)^2 R^2$ terms do not agree with those of  \cite{Deser:2000xz}.}. For one-loop computations in $11D$ supergravity, see   \cite{Green:1997as,Russo:1997mk,Green:1997me,Green:1999by}. Note that the $F^2R^3$ terms obtained via lifting from IIA supergravity have not been compared to those obtained by Noether procedure \cite{Hyakutake:2008}, although we expect them to match modulo field redefinitions.

\section{\texorpdfstring{$D=10$}{D=10}}
		
In ten-dimensions, the higher derivative deformation of supergravity theories has been studied in the case of $N=(1,0)$ supergravity coupled to Yang-Mills from supersymmetry alone, and the results have been compared with the $\alpha'$ expansion of heterotic and type I string effective action, as we shall review below. In the case of $N=(1,1)$ and $N=(2,0)$ supergravities, also known as type IIA and type IIB supergravities, respectively, a direct Noether procedure has not been pursued, with the exception of the study of a limited sector in type IIB supergravity in \cite{Green:1998by}, where compatibility with the $SL(2,\mathbb{Z})$ duality symmetry is discussed in detail, as we shall review briefly below. In the case of type II supergravities, the higher derivative extensions have been deduced to some extent from the string amplitudes, mostly from the computation of the four-point graviton amplitude. Less is known about higher point amplitudes in general \cite{Liu:2019ses}, with the exception of the gravitational 5-pt and 6-pt amplitudes worked out in \cite{Stieberger:2009rr}.

\subsection{Type IIB  }

To begin with, let us consider the string theory amplitude approach for finding the higher derivative effective action. Using the pure-spinor formalism  \cite{Berkovits:2000fe,Berkovits:2002zk}, the full {\it four-point} tree amplitudes of both type II strings to all orders in $\alpha'$ were computed in  \cite{Policastro:2006vt}. The result was found to be derivable from the Lagrangian given in  \cite[Eq. (1.2)]{Policastro:2006vt}. Suppressing the terms containing the RR five-form field strength and truncating the result at order $\alpha'^3$, this result was cast into a manifestly  $SL(2,{\mathbb Z})$ invariant form in \cite{Policastro:2008hg}. In this subsection, we shall begin by writing down the two-derivative action, Then, we shall review the work of \cite{Green:1998by} which uses supersymmetry and $SL(2,\mathbb Z)$ duality to get information on the scalar field dependent couplings at order $\alpha'^3$. Next, we shall turn to the amplitude considerations in type IIB and certain relations with $11D$ supergravity on $T^2$.

The bosonic part of the two-derivative classical type IIB effective (pseudo)action in Einstein frame, and including the RR five-form field strength only here, is given by \cite{Polchinski:1998rr}
\be
S^{(0)}_{IIB} = \frac{1}{2\kappa_{10}^2} \int \left( R-2 P_\m {\bar P}^\m- \frac{|G_3|^2}{2\cdot 3!} -\frac{|F_5|^2}{4\cdot 5!}\right) \star \mathbbm{1}+\frac{1}{8i\kappa_{10}^2} \int C_4\wedge G_3 \wedge {\bar G}_3\ ,
\ee
where
\begin{align}
\tau &=C_0+ie^{-\phi}\ ,\quad P_\m= \frac{i}{2} (\tau_2)^{-1} \nabla_\m\tau \ ,\quad G_3= (\tau_2)^{-1/2} \left(F_3-\tau H_3\right)\ ,
\nn\w2
H_3&=dB_2\ , \quad F_3=dC_2\ ,\quad F_5=dC_4-\frac12 H_3\wedge C_2 +\frac12 F_3\wedge B_2\ ,
\end{align}
and $\tau_2=(\tau-\bar\tau)/(2i)$. The equation of motion $F_5=\star F_5$ is to be imposed by hand after varying the action. Strictly from the point of view of supersymmetry, the computation of a higher derivative extension of this action by employing the Noether procedure is notoriously complicated.

\subsubsection*{A no-go at the nonlinear level}

Superspace approach may be simpler but unfortunately, it runs into an obstacle at the nonlinear level as follows. In the type IIB superspace described in \cite{deHaro:2002vk}, it is known that one can construct an analytic superfield whose lowest component is related to the axion-dilaton $\tau$ satisfying the constraint ${\bar D}\Phi=0$. As highlighted in\cite{deHaro:2002vk}, this constraint is integrable even at the non-linear level. A natural ansatz for an  action formula has the form
\be
S= \int d^{10}x\, d^{16}\,\theta {\cal E}\, W(V,U^\star)\ ,
\ee
where $V, U^\star$ are the analytic superfields whose lowest components are $v, u^\star$ that make up the $SU(1,1)$ element $\begin{pmatrix} u & v\\ v^\star & u^\star \end{pmatrix}$, and ${\cal E}$ is the integral measure which is required to be an analytic superfield as well, but with lowest component being equal to the vielbein determinant $e$. It was shown in \cite{deHaro:2002vk} that such a measure does not exist at the nonlinear level, due to the non-integrability of the analyticity constraint on ${\cal E}$ at the nonlinear level.

\subsubsection*{ The use of linearized supersymmetry and $SL(2,\mathbb Z)$ duality}

There are useful building blocks for the higher derivative terms that can be deduced by considering superspace formulation at the level of linearized supersymmetry. To ensure the linearized equations of motion, one works with the analytic superfield $\Phi$ whose lowest component is the axion-dilaton that satisfies the constraints
\be
{\bar D}\Phi=0\ ,\qquad D^4\Phi =0\ ,
\ee
which are integrable at the linearized level. In view of these constraints, $\Phi$ has the  component expansion
\be
\Phi=\tau+ \theta \lambda + \theta^2 G + \dots + \theta^4 \left( W + \partial F_5 \right) +\dots+ \theta^8 \partial^4\bar\tau\ ,
\ee
where $G=dB$ and $W$ is the linearized Weyl tensor. Using this superfield, one can write down the following action
\begin{align}
S^{(3)} &=\alpha'^3 \int d^{10}x\, d^{16}\theta\,F(\Phi) + c.c.
\nn\w2
&= \alpha'^3 \int d^{10}x\, \Big( f^{(-24)} \lambda^{16} + f^{(-22)} G \lambda^{14} +\cdots+f^{(0)} W^4 +\cdots+ f^{(24)}\lambda^{\star 16} \Big)\ ,
\label{GS3}
\end{align}
where $f^{q}(\tau)$ are functions related to $F(\tau)$ through the Taylor expansion needed in the action formula, and $q$ denotes the $U(1)$ charge such that each term in the expansion is $U(1)$ invariant, recalling that $\lambda$ and $G$ have $U(1)$ charges $3/2$ and $+1$, respectively. This action, however, is not invariant under $SL(2,R)$ or $SL(2,Z)$, and $f(\tau)$ is an arbitrary function. Nonetheless, it motivates the following procedure as a step towards the construction of the nonlinear higher derivative extension of type IIB supergravity. Firstly, all functions $f^{(q)}(\tau)$ can be replaced by the non-holomorphic modular forms $f^{(w,-w)}_h(\tau,\bar\tau)$ that carry weight $w=-q/2$, to ensure $SL(2,Z)$ covariance.
These are also discussed further below and defined in \eqref{fh}. These forms are expected to receive contributions from string amplitudes up to {\it $h$-loops.} It turns out that the relevant contribution for the $\alpha'^3$ arises for $h=1$, due to the properties displayed in \eq{ef}. Thus, in \eq{GS3}, we let
\be
f^{(q)}(\tau) \to f^{(-q/2,q/2)}_1(\tau,\bar\tau)\ .
\ee
This step in general does not preserve supersymmetry. To restore it, one natural step is to replace the linearized field strengths with their nonlinear and supercovariant forms. Next, one can examine a subset of variations of the supergravity action $S^{(0)}$ plus $S^{(3)}$ under (possibly $\alpha'$ deformed) supersymmetry transformations. This is what was achieved in \cite{Green:1998by} who considered the following action (in Einstein frame)
\be
S =  \int d^{10} x\,  e \Big[  \frac{1}{256} \left({\bar\lambda} \gamma^{\m\n\r}\lambda^\star \right)\left({\bar\lambda}^\star\gamma_{\m\n\r}\lambda\right) +\dots
+ \alpha'^3 \left( f_1^{(12,-12)} \lambda^{16} -432 f_1^{(11,-11)} \left( \lambda^{15}\gamma^\m \psi_\m^\star\right) +\cdots\right) \Big]\ .
\label{GSA}
\ee
It was shown in \cite{Green:1998by} that introducing the following order $\alpha'^3$ variation of the dilatino
\be
\delta^{(3)} \lambda =  -\frac{i}{16} \alpha'^3  g(\tau,\bar\tau)  \left(\lambda^{14}\right)_{cd} \left(\gamma^{\m\n\r}\gamma^0 \right)_{dc} \left(\gamma_{\m\n\r}\e^\star \right)_a\ ,
\ee
where $g(\tau,\bar\tau)$ is a to be determined function, the zeroth order in $\alpha'$ variation of the $\alpha'^3$ term (reviewed in \cite{Green:1998by}), and the order $\alpha'^3$ variation of the quartic fermion term in \eq{GSA}, give rise to the following $ \left({\bar\e}^\star\gamma^\m\psi_\m^\star\right) \lambda^{16} $ and $ \left({\bar\e} \lambda^\star\right)\lambda^{16}$ terms
\bea
\delta \cL &=& 8i e \alpha'^3 \left( {\bar\e}^\star\gamma^\m\psi_\m^\star\right) \lambda^{16} \Big(f_1^{(12,-12)}+ 108 D_{11} f_1^{(11,-11)} \Big)
\nn\w2
&& -2i \alpha'^3 \left({\bar\e}\lambda^\star\right) \lambda^{16} \Big( {\bar D}_{-12} f_1^{(12,-12)} + 3240 f_1^{(11,-11)} -90 g \Big)\ ,
\label{2v}
\eea
where
\be
D_w = i\left(\tau_2\frac{\partial}{\partial_\tau} -i\frac{w}{2} \right)\ ,
\ee
and $\bar{D}_{w}$ is the complex conjugate of $D_w$. It was also shown that the closure of the supersymmetry algebra on $\lambda^\star$ at order $\alpha'^3$, modulo the $\lambda^\star$ equation of motion, requires that
\be
 D_{11} g= \frac{1}{32} f_1^{(12,-12)}\ .
 \label{c}
\ee
This relation, together with two equations that follow from the requirement of the vanishing of \eq{2v} give
\be
g = -\frac{27}{8} f_1^{(11,-11)}\ ,\quad {\bar D}_{12} D_{11} f_1^{(11,-11)} =-\frac{525}{4} f_1^{(11,-11)}\ ,\quad D_{11} {\bar D}_{12}f_1^{(12,-12)} = -\frac{525}{4} f_1^{(12,-12)}\ ,
\ee
where one uses the fact that $D_{11}g=0$ has no solution. These are Laplace equations whose solutions involve representation functions of $SL(2,\mathbb{R})$ in general. However, motivated by the fact that the low energy effective action of type IIB string has $SL(2,\mathbb{Z})$ instead, we are more interested in properties of the solution under the discrete subgroup $SL(2,\mathbb{Z}) \subset SL(2,R)$. For  general weight $w$, such solutions are the non-holomorphic Eisenstein series defined as
\be
f_h^{(w,-w)} (\tau,\bar\tau)= \sum_{(m,n)\ne (0,0)} \frac{(\tau_2)^{\frac12+h}}{(m+\tau n)^{h+\frac12+w}(m+\bar\tau n)^{h+\frac12-w}}\ .
\label{fh}
\ee
Under $SL(2,Z)$ transformation $\tau \to (a\tau + b) (c\tau +d)^{-1}$ with $ad-bc=1$, a field $\Phi$ with weight $(q,-q)$ is
\be
\Phi\to (c\tau+d)^{q}(c\bar\tau +d)^{-q} \Phi\ .
\ee
It is useful to note that expanding $f_1^{(w,-w)}$ in the large $\tau_2 \gg1$, i.e. small string coupling, regime gives
\be
f_1^{(w,-w)}(\tau,\bar\tau) = 2\zeta(3)\, \tau_2^{3/2} + \frac{2\pi^2 }{3(1-4w^2)}\,\tau_2^{-1/2} + \cO\left(e^{-\tau_2}\right)\ ,
\label{ef}
\ee
where the first term is associated with closed string tree level \cite{Gross:1986iv}, the second term with the 1-loop effects \cite{Green:1981ya,Green:1997as}, and the last terms encode the contributions from non-perturbative D-instanton states \cite{Green:1997tv}.  While the Noether procedure at the nonlinear level has not been carried out to determine the coefficient of $R^4$ term, strong arguments, including the use of linearized supersymmetry, have been given in \cite{Green:1998by} for its being $f_1^{(0,0)}$  (see also \cite{Fleig:2015vky}). It is also useful to note that the function $f_{1}^{(0,0)}$ is often denoted by $\vE^{(10)}_{(0,0)}$, as discussed in section \ref{modular functions}.

The same idea has been applied to the analysis of the modular property of the function in front of the $\l^{16}\widehat{G}^4$ term, where $\widehat{G}$ is the supercovariantized three-form field strength \cite{Sinha:2002zr}, which appears in the IIB low energy effective action at order $\alpha'^5$. To be specific, the tensorial structure of the $\widehat{G}^4$ term considered in \cite{Sinha:2002zr} is $\widehat{G}^4=(\widehat{G}_{\m\n\r}\widehat{G}^{\m\n\r})^2$. There are two other terms mixing with the $\l^{16}\widehat{G}^4$ term under the supersymmetry transformation. Thus to explore the consequences of supersymmetry on the $\l^{16}\widehat{G}^4$ term, the first three terms in the following action at ${\cal O}(\a'^5)$ are considered
\bea
S^{(5)} &=&\a'^5 \int d^{10}x\,e\Big( f_2^{(14,-14)}\l^{16}\widehat{G}^4+f_2^{(13,-13)}\l^{15}\g^\m\psi_{\m}^{*}+\widetilde{f}_2^{(13,-13)}\l^{16}\widehat{G}^2\widehat{G}_{\m\n\r}\widehat{G}^{\m\n\r}\cdots \nn\w2
&&   +\cdots + f_2^{(0,0)} D^4 R^4 +\cdots \Big)\ ,
\label{5orderS}
\eea
where ``\dots" indicates other ${\cal O}(\a'^5)$ terms in the effective action which are beyond the discussion of \cite{Sinha:2002zr}. We have added the $D^4R^4$ term to emphasize that it is expected to arise from a fuller analysis of supersymmetry. Assuming the modified supersymmetry transformation rules for the dilatino and gravitino at order $\alpha'^5$ to be of the form
\be
\d^{(5)}\l^*=\a'^5 g_1 \widehat{G}^4(\l^{14})_{ab}(\g^{\m\n\r})_{ba}\g_{\m\n\r}\epsilon^*\ ,
\quad
\d^{(5)}\psi_\m=\a'^5 g_2 \l^{16}\widehat{G} \widehat{G}_{\n\r\s}(\g^{\n\r\s}\g_\m)\epsilon^*\ ,
\ee
the $\d^{(5)}$ variation of the following two terms from the two-derivative action
\be
S^{(0)} = \int d^{10}x e\Big(\frac1{256}\bar{\l}\g^{\m\n\r}\l^*\bar{\l}^*\g_{\m\n\r}\l-\frac18\bar{\psi}_\m^*\g_{\n\r}\l^*\widehat{G}^{\m\n\r} \Big)
\ee
cancels the $\d^{(0)}$ variation of the terms in \eqref{5orderS} provided that the undetermined functions in the effective action and modified supersymmetry transformations obey certain linear differential relations. Requiring also the closure of supersymmetry algebra at ${\cal O}(\a'^5)$,  one obtains
\bea
4 D_{13}\bar{D}_{14} f_2^{(14,-14)}&=&-\frac{713}4 f_2^{(14,-14)},\quad D_{13} f_2^{(13,-13)}=\frac{11}2 f_2^{(14,-14)}\ ,
\nn\\
2i (g_1+191 g_2)&=& f_2^{(13,-13)}, \quad 192 D_{13}g_1=i f_2^{(14,-14)}, \quad 108 g_1=\widetilde{f}_2^{(13,-13)}\ .
\eea
It follows that $f_2^{(14,-14)}$ is an eigenfunction of the Laplacian defined on the fundamental domain of $SL(2,R)$ transforming as a weight $(14,-14)$ modular form. Regarding the coefficient of the  $D^4 R^4$ term, as for the case of $R^4$ term discussed earlier, at the nonlinear level the requirement of $SL(2, \mathbb{Z})$ symmetry suggests that it should be proportional to $f_2^{(0,0)}$, which is often denoted by $\vE^{(10)}_{(1,0)}$ as discussed in section \ref{modular functions}. In the weak coupling limit, i.e. $\ta_2 \gg1$, one finds that it gets contributions from tree level, two-loop and $D$-instantons \cite{Green:1999pu}.

Type IIB supergravity has an anomaly in composite local $U(1)$ symmetry which also implies an anomaly in the global $SL(2, R)$ transformations; this means that $SL(2, R)$ is not a symmetry of the theory. For a detailed discussion of this anomaly, its cancellation, and restrictions on spacetime background, see \cite{Gaberdiel:1998ui}.

The use of duality symmetry is a powerful approach to studying the higher derivative extensions of supergravity theories, not just in $10D$ but in other dimensions where duality groups $E_{n(n)}$ are present. These are the Cremmer-Julia groups listed in Table \ref{dualgr} in section \ref{modular functions}, where we summarize the basic aspects of this approach. For a textbook exposition of this subject, see \cite{Fleig:2015vky}.

\subsubsection*{$11D$ on $T^2$ and decompactification of type IIB string}

Another avenue for using supersymmetry and duality to get a handle on the higher derivative extension of type IIB supergravity is to reduce the higher derivative extensions of $11D$ supergravity (to the extent they are determined by Noether procedure discussed in the section on $11D$ above) on $T^2$ and compare to an appropriate decompactification limit of type IIB string. We shall come back to this point briefly in section \ref{lastsection}.  $11D$ supersymmetry may also be exploited at the level of amplitude computations at one-loop and beyond,  and this approach has been successfully implemented in a series of papers pioneered in \cite{Gross:1986mw}.

\subsubsection*{String theory amplitudes}
%
Using supersymmetry and duality alone, the story unfortunately does not go much further for type II theory, and there are a large number of terms in the effective action even at the eight derivative level, let alone the higher derivative corrections to the supersymmetry transformation rules, that have not been determined as yet. Given the difficulties in deforming the superspace constraints to accommodate the higher derivative terms, and the obstacles in dealing with the RR sector in the beta function method, the most promising approach that remains is the computation of sting theory amplitudes, and the construction of the effective action that produces them.  The pioneering work along these lines was done long ago in \cite{Gross:1986mw} where 4-pt amplitudes in the NS-NS sector of heterotic string theory, which constitutes a universal sector in all string theories, were computed. In \cite{Stieberger:2009rr}, 5- and 6-pt graviton amplitudes were computed as well. In what follows we shall summarize various results for the higher derivative corrections at order $\alpha'^3$ obtained so far from the computation of the various string theory amplitudes at the tree and one-loop level.

\subsubsection*{$\bullet$  The tree-level 4-point amplitudes to all orders in $\alpha'$ }

Using the pure-spinor formalism, the complete four-point
effective action of both type II superstrings to all orders in $\alpha'$, at tree level in string loops was computed in \cite{Policastro:2006vt}. The result, which includes the fermions as well, is given in \cite[Eq. (5.36)]{Policastro:2006vt}. The explicit tensorial structures are to be deduced by employing a procedure described in \cite{Policastro:2006vt} but the result is not provided. Furthermore, in the case of type IIB, the $SL(2,R)$ invariance (at tree level) is not manifest. In a subsequent paper \cite{Policastro:2008hg}, putting aside the RR 5-form, the terms at order $\alpha'^3$ were put into a form with manifest $SL(2,R)$ invariance (in Einstein frame) as follows  \cite[Eq. (3.3)]{Policastro:2008hg} \footnote{The fourth term has been corrected in \cite{Liu:2019ses}.}
\bea
\cL_{4pt, tree}^{IIB} &=& c \alpha'^3 t_8t_8 \Big[ R^4 +6R^2 \Big( |\partial P|^2 +|\partial G_3|^2 \Big) + 6 |\partial P|^2 |\partial G_3|^2
+ 6R \left( \partial P(\partial {\bar G}_3)^2 +c.c \right) \Big]
\nn\w2
&& \qquad \qquad + \alpha'^3 \zeta(3) \Big[ {\cal O}_1 ( (|\partial P|^2))^2+  {\cal O}_2 ((|\partial G_3|^2)^2) \Big]\ ,
\label{Tsimpis}
\eea
where, $c= \zeta(3)/(3\cdot 2^8)$, and in the linear approximation \cite{Policastro:2008hg}
\bea
\partial G_3 &=& \sqrt2 \left( e^{-D/2} \partial H -ie^{D/2} \partial F_3\right)\ ,\qquad \left(\partial H\right)_{abcd}= \partial_{[a} H_{bcd]}\ ,
\nn\w2
\left(\partial P\right)_{ab}{}^{cd}  &=& \partial\partial D +ie^D \partial\partial\chi\ ,\quad F_3=dC_2\ ,
\eea
with $(D,\chi)$ representing the (dilaton, axion), $(\partial\partial)_{ab}{}^{cd} :=\delta_{[a}^{[c} \partial_{b]} \partial^{d]}$, and $\cO_1$ and $\cO_2$ are operators that are complicated combination of the products of Kronecker deltas, for which we refer the reader to \cite{Policastro:2008hg}. The dilaton exponentials in the above definitions are put in accordance with the field's conformal weights, but the overall dilaton factor in the action has been omitted. The terms involving the Riemann tensor are deduced by general covariance, and therefore they don't require the computation of higher point amplitudes. Note also that the expected $\e_{8}\e_{8} R^4$ term is not visible in the linearized approximation.

The effective action at order $\alpha'^3$  in the NS-NS sector has been computed in \cite{Garousi:2020gio,Garousi:2020lof} solely from the bosonic gauge symmetries, and the requirement that a circle reduction produces a $T$-duality invariant result.  The action at order $\alpha'^3$ found in this way in $10D$ schematically takes the form,
\bea
S_{\rm NS-NS, tree}^{IIB} &=& c \alpha'^3 \int dx^{10} e\Big[ [R^4]_2 + [R^3 H^2]_{22} +[R^2 (\nabla H)^2]_{22} + [R^2 H^4]_7
\nn\w2
&& + [R (\nabla H)^2 H^2 ]_{106} +[R H^6]_1  +    [(\nabla H)^4]_{12}  +[(\nabla H)^2 H^4]_{77}  + [H^8]_2 \Big]\ ,
\label{aga}
\eea
where $[X]_n$ denotes $n$ number of structures with different index contractions of the fields $X$. The structures denoted by $[X]$ can be found in \cite{Garousi:2020lof}.  However, a comparison of the result above with that of \cite{Policastro:2006vt} discussed above remains to be carried out. Given that the action \eq{aga} is for the NS-NS sector, it should be the same for the heterotic string \cite{Moura:2007ks}.
\subsubsection*{ $\bullet$ The one-loop  4-point amplitudes at order $\alpha'^3$}

The dependence of the type IIB effective action on the three-form and five-form RR field strengths at order $\alpha'^3$ was obtained in \cite{Peeters:2003pv} from the 4-pt amplitude at the one-loop level. The result for the $R^2 (DF_3)^2$ terms, upon choosing suitably the 4 parameters that cannot be determined from the 4-pt amplitude at the linearized level, comes out to be compatible with the result \eq{Tsimpis}. In other words, the $\zeta(3)$ term from \eq{Tsimpis} and the one-loop term from \cite{Peeters:2003pv} are compatible with the $SL(2,Z)$ invariant structure $f_1^{(0,0)} R^2 (DG_3)^2$. The first two terms in the weak string coupling limit expansion of  $f_1^{(0,0)}$ given in \eq{ef} describe the tree-level and one-loop contributions. As for the $R^2 (DF_5)^2$ terms, 58 structures with the maximum number of contractions between the indices of $R_{a_1...a_4} R_{a_5...a_8}$ and those of $ \left( \partial F_5\right)^2_{a_1...a_8}$, are given explicitly in \cite[ eqs. (2.13), (A.12-15) and (A.17-18)]{Peeters:2003pv}.

\subsubsection*{ $\bullet$  The $R^3 G_3^2$ terms at order $\alpha'^3$ }

The $G_3^2 R^3$ and $|G_3|^2 R^3$ term at order $\alpha'^3$ have been given in \cite[Eq. (3.36) ]{Liu:2022bfg} as
\bea
\cL_{IIB}\Big|_{ G^2R^3}& =& c\alpha'^3 \Big[ f_1^{(0,0)} (\tau,\bar\tau)\Big(-\frac13 \e_9\e_9 +2t_8t_8 -\frac12 \e_8\e_8-t_{18}\Big) |G_3|^2 R^3
\nn\\
&&+ \Big(\frac32 f_1^{(1,-1)}(\tau,\bar\tau) t_{18} G_3^2 R^3 +c.c.\Big) \Big]\ ,
\label{IIBstuff}
\eea
where
\be
t_{18} G_3^2 R^3 = \frac12 t_8t_8 G^2 R^3 -\frac{1}{24} \e_9\e_9 G^2 R^3 -2\cdot 4! \sum_{i=1}^8 {\tilde d}_i G_{\m_1\m_2\m_3} G_{\m_4\m_5\m_6} {\tilde Q}_i^{\m_1...\m_6}\ ,
\label{t18}
\ee
and ${\tilde Q}_i$ are eight independent $R^3$ structures listed in  \cite[Eq. (A.13) ]{Liu:2022bfg}. Furthermore,
\be
(\td_1,...,\td_8) = 4\left(1,-\frac14, 0, \frac13, 1, 1, \frac14, -2, \frac18 \right)\ .
\ee
The  definitions of various contractions of the symbols in \eq{IIBstuff} are given in the appendix. In view of the definition of the functions $f_1^{(0,0)}$ and $f_1^{(1,-1)}$, the Lagrangian above contains both tree-level, one-loop and instanton contributions.

\subsubsection*{$\bullet$ The tree level 5- and 6-point graviton amplitudes}

The 5- and 6-point graviton amplitudes at tree level were computed in \cite{Stieberger:2009rr} where the absence of $R^5$ terms and certain higher derivatives of  $R^4, R^5$ and $R^6$ terms with zeta valued coefficients was shown.

\subsubsection*{ $\bullet$  One loop 5-pt amplitudes and $B\wedge X_8$ terms }
The one loop 5-pt point amplitude involving NS-NS $B$-field and gravitons were studied in \cite{Richards:2008jg, Richards:2008sa}. The result includes the term, which is of the Green-Schwarz type, though the type IIB theory is anomaly-free, given by
\be
\cL_{\rm GS} =     B\wedge \Big(X_8(R(\o_+))- X_8(R(\o_-))\Big)\ .
\label{GS2B}
\ee
Despite the progress that has been made so far, the eight derivative terms still have not yet been completely determined. For further references in which partial results are obtained, see \cite{Liu:2022bfg}. In particular, for a treatment of the usefulness, as well as limitations,  of the Lorentz connection with bosonic torsion $H_3$, see \cite{Liu:2013dna}.
As for the use of supersymmetry, a study of a particular set of variations, to wit, those involving the 16th and 14th power of the dilatino, have been considered to discover that the modular function $f_0(\tau,\bar\tau)$ is needed in the $R^4$ term \cite{Green:1998by}. However, linearized supersymmetry, while predicting the existence of certain higher derivative terms, cannot produce them all, including the $f_h(\tau,\bar\tau)$ factors. A fruitful approach to obtaining more detailed information is to consider 11D superparticle on $T^2$ of vanishing volume, and integrate out the winding modes  \cite{Green:1999by, Green:1997as,Green:1997me}. Doing so, in  \cite{Liu:2022bfg}, for example, a conjecture is made for specific types of terms, namely
\be
\cL \sim \alpha'^3 \sum_{h=0}^4 c_h f_h(\tau,\bar\tau) G_3^m {\bar G}_3^{m-2h} R^{4+h-m} +h.c.\ ,
\ee
where the coefficients $c_h$ are specified.

\subsection{Type IIA}

The bosonic part of the classical two-derivative action for the type IIA string is given by
\bea
S^{(0)}_{IIA} &=&
-\frac{1}{  \kappa_{10}^2 } \int d^{10}x \sqrt{-g} \Bigg\{ e^{-2\phi} \left[ R-4 (\partial\phi)^2 +\frac12 H^2\right]
\nn\\
&& + \frac12 G^{(2)} \cdot G^{(2)} + \frac12 G^{(4)} \cdot G^{(4)}
-\star \left[ \frac12 dC^{(3)} \wedge dC^{(3)} \wedge B \right]  \Bigg\}\ ,
\label{2A}
\eea
where
\be
G^{(2)} = dC^{(1)}\ ,\qquad G^{(4)} = dC^{(3)} + dB \wedge C^{(1)}\ .
\ee
 From tree and  one-loop level 5-pt point amplitudes, and considerations of known dualities, it is  conjectured in \cite{Liu:2019ses} that the 8-derivative terms in the bosonic part of the type IIA string effective action, not taking into account the $RR$ fields and the $\partial_\mu\phi$ terms (and taking into account some terms that were missing in  \cite{Liu:2013dna}) takes the form,
\bea
S^{(3)}_{IIA} &=&  c\alpha'^3  e^{-2\phi} \Big[   t_8 t_8 R(\omega_+)^4 -\frac14 \epsilon_{8}\epsilon_{8} R(\omega_+)^4 -2t_8 t_8 H^2 R(\omega_+)^3
\nn\\
&& -\frac16 \epsilon_9\epsilon_9 H^2 R(\omega_+)^3
+8\cdot 4! \sum_{i} \td_i H^{\mu\nu\lambda} H^{\rho\sigma\tau} {\widetilde Q}^i_{\mu\nu\lambda\rho\sigma\tau} +\cdots \Big]
\nn\\
&& + \frac{c\alpha'^3\pi^2}{3}\Big[  e \Big( t_8 t_8 R(\omega_+)^4 +\frac14 \epsilon_{8}\epsilon_{8} R(\omega_+)^4 +\frac13 \epsilon_9 \epsilon_9  H^2 R(\omega_+)^3
\nn\\
&& -\frac49\epsilon_9\epsilon_9 H^2 (DH)^2 R(\omega_+) + \cdots \Big) +\cdots \Big]
\nn\w2
&& - \frac{(2\pi)^6}{2} \alpha'^3 B\wedge \Big(X_8(R(\omega_+))+ X_8(R(\omega_-)\Big)\ ,
\label{IIA}
\eea
where $c := \zeta(3)/(3\cdot 2^{11})$,  and
\be
\omega_\pm = \omega(e) \pm \frac12 H\ ,\qquad X_8 (R)= \frac{1}{ (2\pi)^4 3 \cdot 2^6} \left( R^4 -\frac14 (R^2)^2\right)\ ,
\ee
See the appendix for the definitions of various symbol contractions. The terms with the overall $e^{-2\phi}$ factor are tree level, and the remaining terms are one-loop contributions. In the tree level Lagrangian the ellipsis inside the round brackets refers to the terms that have the structure $H^2 (\nabla H)^2 R$ which can be computed from 5-pt amplitudes, and terms such as $H^4 R^2$ that contribute to 6-pt amplitudes which have not been computed so far. In the one-loop level Lagrangian the ellipsis in the round brackets refers to terms of the form $H^4 R^2$, and possibly other terms that would contribute to six- and higher point amplitudes.
The tree level contribution in \eqref{IIA} is obtained by the four graviton amplitude and the sigma-model computation \cite{Schwarz:1982jn,Gross:1986iv}. The first two terms of the one-loop contributions in \eqref{IIA} are found in the sigma model beta function approach in \cite{Grisaru:1986px,Grisaru:1986dk,Grisaru:1986kw, Grisaru:1986vi, Freeman:1986br, Freeman:1986zh}, and by making use of the four graviton amplitude in \cite{Sakai:1986bi}, while the last term is introduced to ensure the string-string duality between type IIA on K3 and heterotic string on $T^4$ \cite{Hull:1994ys,Witten:1995ex,Sen:1995cj,Harvey:1995rn}. Under this duality, the last term is related to the Green-Schwarz anomaly cancellation term in the heterotic string effective action \cite{Vafa:1995fj,Duff:1995wd}.

It should be noticed that when restricted to the NS-NS sector, the type IIA tree level effective action at order $\a'^3$ shares the same form as the one for type IIB \cite{Liu:2013dna}. Thus, as noted before, the $N=(1,0)$ truncation of the NS-NS sector of Type IIA  effective action at tree level is also the same as that of the heterotic effective action \cite{Moura:2007ks}. At one loop level, the terms contracted with two $t_8$ tensors are also the same for IIA and IIB. However, terms involving two epsilon tensors, as well as the $B\wedge X_8$ terms appear with opposite signs.

\subsection{\texorpdfstring{$N=(1,0), 10D$}{N=(1,0), 10D} supergravity coupled to Yang-Mills}
\label{hetsugra}
		
We begin by reviewing a particular deformation of $N=(1,0)$ supergravity coupled to Yang-Mills up to quartic in Riemann curvature terms, strictly from the point of view of local supersymmetry as was obtained by Bergshoeff and de Roo (BdR) \cite{BD1989,Bergshoeff:1989de}. Next, we shall review another deformations that are quartic in Riemann curvature but only in the gravitational sector  \cite{Suelmann:1994vk,deRoo:1992zp,Suelmann:1994qk}. In a subsequent section, we shall compare these results with the low energy limits of $E_8 \times E_8$ and $SO(32)$ heterotic string. Note that higher derivative terms in type I effective action can be obtained from those in $SO(32)$ heterotic string via suitable field redefinitions \cite{Tseytlin:1995fy, Tseytlin:1995bi, Bachas:1996bp}.
		
\subsubsection*{${\bf R}+ {\bf Riem}^2 + ({\bf Riem}^2)^2$ from Noether procedure}
%
The supergravity multiplet and the Yang-Mills multiplet have the fields $(e_\mu{}^a, B_{\mu\nu}, \phi, \psi_\mu, \chi)$ and $(A_\mu, \lambda)$ respectively. The bosonic part of the higher derivative extension of the coupled system up to the fourth power of the Riemann tensor is given by  \cite{Bergshoeff:1989de}:
\be
{\cal L}_{BdR}^{\rm het} = -ee^{-2\phi} \left[ R(\omega(e))+ 4\partial_\mu\phi \partial^\mu \phi -\frac{1}{12} H_{\mu\nu\rho} H^{\mu\nu\rho} -\frac12 T + \frac{\a}2\big( 3 T_{\mu\nu\rho\sigma} T^{\mu\nu\rho\sigma} + T_{\mu\nu} T^{\mu\nu}\big) \right]\ ,
\label{BdR1}
\ee
where
\be
H = dB + \alpha\, X_3(\omega_-)+ \beta\, X_3(A) \ ,\qquad \omega_{\mu\pm}{}^{ab} = \omega_\mu{}^{ab}(e) \pm \frac12 H_\mu{}^{ab}\ ,
\label{HDB}
\ee
and
\bea
&& T_{\mu\nu\rho\sigma}=  \alpha \tr (R_{[\mu\nu}(\omega_-) R_{\rho\sigma]}(\omega_-))+\beta \tr (F_{[\mu\nu} F_{\rho\sigma]})\ ,
\nn\w2
&& T_{\mu\nu}= \alpha\,\tr R_{\mu\lambda}(\omega_-) R^\lambda{}_\nu(\omega_-)+\beta\,\tr F_\mu{}^\lambda F_{\lambda\nu}\ ,\qquad T= g^{\mu\nu} T_{\mu\nu}\ ,
\nn\w2
&& X_3(A) = \tr_{\rm YM} ( A\wedge dA+\frac23 A\wedge A\wedge A)\ ,\qquad X_3(\omega) = \tr_L ( \omega\wedge d\omega+\frac23 \omega\wedge \omega\wedge \omega)\ .
\eea
 The deformation parameters are $\alpha$ and $\beta$, and $\beta = 1/g_{YM}^2$. One can solve for $H$ recursively from \eq{HDB} in powers of $(\alpha, \beta)$, the first few terms taking the form
\bea
H^{(0)}&=& dB\ ,
\nn\\
H^{(1)}&=&\alpha\, X_3(\omega_-)^{(0)}+ \beta\, X_3(A)\ ,
\nn\\
H^{(n)}&=&\alpha\, X_3(\omega_-)^{(n-1)}\,\quad n\ge 2\ .
\label{exp}
\eea
It is worth noting that the BdR action at first order in $(\alpha, \beta)$ is complete including the four-fermion terms, though, it is understood that supersymmetry holds up to the same order.

The action \eqref{BdR1} is invariant up to order ${\cal O}(\alpha^3, \alpha^2\beta, \alpha\beta^2)$ with the supersymmetry transformations up to that order taking the form
\bea
\delta e_\mu^a &=& {\bar\epsilon}\gamma^a \psi_\mu+\cdots\ ,
\nn\\
\delta \psi_\mu &=& D_\mu (\omega_+)\epsilon +\cdots \ ,
\nn\\
\delta B_{\mu\nu} &=& 2{\bar\epsilon}\gamma_{[\mu}\psi_{\nu]} + 2\alpha \tr (\omega_{-[\mu}\delta\omega_{-\nu]} + 2\beta \tr (A_{[\mu}\delta A_{\nu]} +\cdots \ ,
\nn\\
\delta \chi &=& (\gamma^\mu \partial_\mu \phi -\frac{1}{12}\gamma^{\mu\nu\rho} H_{\mu\nu\rho})\epsilon +\cdots\ ,
\nn\\
\delta \phi &=&  \frac12{\bar\epsilon}\chi +\cdots \ ,
\nn\\
\delta A_\mu &=& {\bar\epsilon}\gamma_\mu \lambda +\cdots\ ,
\nn\\
\delta\lambda &=& -\frac14\gamma^{\mu\nu} F_{\mu\nu}\epsilon + \cdots\ ,
\label{st1}
\eea
where the ellipsis for the supergravity multiplet transformations denotes the order $\cO(\alpha^3, \alpha^2\beta)$ terms \cite[Eq. (4.19)]{Bergshoeff:1989de}, and in the case of Yang-Mills multiplet the order $\cO(\alpha^2\beta,\alpha\beta^2)$ terms \cite[Eqs. (4.13) and (4.14)]{Bergshoeff:1989de}.   At order ${\cal O}(\alpha^3, \alpha^2\beta, \alpha\beta^2)$, the quartic fermion terms as well as terms in which $\partial_\mu \phi$ and $H$ may possibly arise have not been determined in \cite{Bergshoeff:1989de}.
In obtaining the result \eq{BdR1}, the convenient trick  \cite{Bergshoeff:1988nn,Bergshoeff:1989de} (employed previously in off-shell $6D$ conformal supergravity context in \cite{Bergshoeff:1987rb})  that uses the fact that $(A_\mu, \lambda)$ transform under supersymmetry at lowest order in $(\alpha,\beta)$ as $(\omega_{\mu -}{}^{ab}, \psi^{ab}(\omega_-))$ do, is employed. In the Lagrangian and supersymmetry transformation rules obtained in this way, the dependence on the parameters $(\alpha,\beta)$ arises either explicitly, or implicitly through the expansion \eq{exp}. It is understood that only terms of order $\cO(\alpha^2,\alpha\beta, \alpha^3, \alpha^2\beta,\alpha\beta^2)$ are to be kept.

It was also shown in  \cite{Bergshoeff:1989de} that no $\alpha^2 R^3$ and $\alpha\beta RF^2$ arise but order $O(\alpha^2,\alpha\beta)$ terms do arise and they involve fermionic bilinears multiplying the $(B_{\mu\nu}, \psi_\mu, \chi)$ field equations. These terms \cite[Eq. (3.17)]{Bergshoeff:1989de} arise as a result of putting the supersymmetry deformations in a form without a differentiated supersymmetry parameter.

The dilaton factor $e^{-2\phi}$ appears as an overall factor, and therefore the BdR action is to be compared with the tree-level contributions to the heterotic string low energy effective action. Since $\tr F^4$ terms, for which these two groups have different properties, do not arise in the BdR action discussed above, it can be compared with the heterotic string effective action with either gauge group. The results from heterotic string theory can be extracted from the amplitude computations, some of the earliest ones being \cite{Bento:1986hx,Nunez:1987ig,Gross:1986mw}, or from the beta function calculations carried out in \cite{Metsaev:1987zx}. Comparisons are not straightforward due to the complicated consequences of the field redefinitions. However, putting aside the Yang-Mills couplings, we can see from the work of \cite{Metsaev:1987zx} that the effective action at order $\alpha'$ obtained from the beta function calculations does agree with the BdR action.

\subsubsection*{ ${\bf Riem}^4$ invariants from Noether procedure}
%
 There are three independent eight-derivative extensions of the two-derivative action for ${\cal N}=1, 10D$ supergravity. Two of these were constructed in  \cite{deRoo:1992zp} where an action with the generic form
\be
S= \int d^{10}x\sqrt{-g} \left( R + \gamma R^4+ \cdots \right)
\label{S1}
\ee
was considered. Here $R^4$ refers to terms that are fourth order in Riemann tensor with indices contracted in all possible ways allowed by supersymmetry. In \cite{deRoo:1992zp}, only terms independent of, or linear in, $H$ and $\partial_\m \varphi$ were considered. Consequently, in the variation of the action, only terms in which $H$ and $\partial_\m \varphi$ are absent were studied. A combination of the two invariants constructed in  \cite{deRoo:1992zp} and a third one found in \cite{Suelmann:1994qk} can be expressed as suitable combinations of the following invariants, in the notation of  \cite{Peeters:2000qj}, taking the schematic form
\bea
I_X&=&(t_8+\frac12\epsilon_{10}B)t_8R^4+\cdots\ ,
\nn\\
I_{Y_1}&=&(t_8+\frac12\epsilon_{10}B)(tr R^2)^2+4HR^2DR+\cdots\ ,
\nn\\
I_{Z}&=&-\epsilon_{10}\epsilon_{10}R^4+4\epsilon_{10}t_8BR^4+\cdots\ ,
\eea
where $\e_{10} B$ denotes the Levi-Civita symbol with two of its indices contracted with those of $B$. These are independent of the 8-derivative invariant expressed in terms of $T^2$ terms in \eq{BdR1}. Up to quadratic fermions, the invariant $I_X$ is given in \cite[Eq. (3.8)]{Peeters:2000qj}. It should be noted that in the expressions above the terms involving bare $B$-field are not invariant under Yang-Mills gauge transformations. Since the commutator of two supersymmetry transformations involves a Yang-Mills gauge transformation as well as local Lorentz transformations, anomalies in these transformations are also expected in a manner in which the Wess-Zumino consistency conditions are satisfied. We are not aware of a detailed study of this phenomenon in $10D$, but it has been displayed in $(1,0), 6D$ supergravity coupled Yang-Mills in considerable detail in \cite{Ferrara:1996wv,Riccioni:1998th}. Note, however, that there is just one combination of the invariants listed above which has no bare $B$-field, and therefore manifestly gauge invariant. As we shall see below, it turns out that it is that combination which appears in the heterotic tree-level effective action, with the celebrated $\zeta(3)$ dependent coefficient.

The invariants (modulo the anomalies just discussed) $I_1\,,I_2$ obtained by de Roo and Suelmann  \cite{deRoo:1992zp} are related to $I_X\,,I_{Y_1}\,,I_Z$ by\footnote{In comparing the results in the literature, we map the two-potential as
$B^{dRS}=\sqrt{2}B^{PVW}=-\frac1{\sqrt{2}}B^T$, where dRS, PVW and T refer to the papers  \cite{deRoo:1992zp,Peeters:2000qj,Tseytlin:1995bi}.}
\be
I_1=\frac1{48}(I_X+\frac38 I_Z)\ ,\quad I_2=-\frac12 I_{Y_1}\ .
\ee
As noted above, the combination $I_X-\frac18 I_Z$ does not contain the term $\e_{10}t_8 BR^4$ and appears in string theory at arbitrary loop order. According to \cite{Peeters:2000qj},
\be
\label{action}
e^{-1}{\cal L}^{\rm het}\vert_{(\alpha')^3}=\Big[ {\cal L}^{\rm het}_{BdR}+ e^{-2\phi}\frac{\zeta(3)\a'^3}{3\cdot 2^{14}} \left( I_X-\frac18 I_Z \right)\Big] + \a'^3J_1 \ ,
\ee
with $\cL^{\rm het}_{BdR}$ from \eq{BdR1} and $J_1$ is a one-loop term explained below. Note that  $ I^{\rm het}_{BdR}$ already contains the overall dilaton factor  $e^{-2\phi}$. Demanding the $O(d)\times O(d)$ symmetry to be present in tree-level string action compactified on $d$-torus, ref. \cite{Wulff:2021fhr} has obtained the $B$-field dependent couplings in the eight-derivative term proportional to $\zeta(3)$ up to fifth order in $H$.  As discussed below \eq{S1}, $I_X$ and $I_Z$ are partially determined by supersymmetry. At order $\alpha'^3$ at the tree level the Yang-Mills dependent terms come solely from $I_{BdR}$ \cite{Gross:1986mw, Cai:1986sa, Kikuchi:1986cz,Kikuchi:1986rk, Ellwanger:1988cc}.
$J_1$ is the one loop term  \cite{Tseytlin:1995bi,Ellis:1987dc,Abe:1988cq,Lerche:1988zy} involving the $B\wedge X_8$ required by the GS mechanism. Its bosonic part is given by
\be
J_1= \left(t_8+\frac12\epsilon_{10} B\right)X_8(R,F) +\cdots\ ,
\ee
where $X_8 (R,F)$ has different forms for $E_8\times E_8$ and $SO(32)$ invariant models\footnote{Note that in the absences of the Yang-Mills field, $J_1=\frac{1}{192} I_X +\frac{1}{16} I_{Y_1} +\cdots$.}. In the first case, $\tr F^4$ factorizes as $(\tr F^2)^2$ but for $SO(32)$ this factorization does not occur. Therefore, an invariant including $\tr F^4$ is needed to make up $X_8$. Finally $X_8$ can be expressed as a combination of $I_X,\, I_{Y_1}$ and 3 other separate invariants involving the Yang-Mills field strength $F$ \cite{ deRoo:1992zp, Suelmann:1994qk, deRoo:1992sm}\footnote{The last invariant is given in \cite{Suelmann:1994qk}.}
\bea
I_3 &=& (t_8+\frac12\epsilon_{10}B)(tr F^2)^2+\cdots\ ,
\nn\w2
I_4 &=&(t_8+\frac12\epsilon_{10}B)tr R^2tr F^2+\cdots\ ,
\nn\w2
I_5 &=& (t_8+\frac12\epsilon_{10}B)tr F^4+\cdots\ .
\eea
The first two invariants in the equation above can be obtained by replacing $R^2$ in $I_{Y_1}$ with $F^2$ and $F^2\times I + I\times R^2$ respectively.
In the $E_8$ model, $\tr F^4$ factorizes into $(\tr F^2)^2$. In that case, $X_8(R,F)$ and the bosonic part of its supersymmetric completion can be obtained as a suitable combination of $I_X, I_{Y_1}, I_3$ and $I_4$. For completeness, we recall here the well-known expressions for $X(R,F)$, in  conventions of \cite{Becker:2006dvp}:
\bea
SO(32):  & \qquad X_8= \frac18 \tr R^4+\frac1{32}(tr R^2)^2-\frac1{240} \tr R^2 \tr F^2+\frac1{24} \tr F^4-\frac1{7200}(trF^2)^2\ ,
\nn\w2
E_8\times E_8: & \qquad X_{8(i)}= \frac18 \tr R^4+\frac1{32}(\tr R^2)^2-\frac1{12}\tr R^2 trF_{(i)}^2-\frac1{3600}(\tr F_{(i)}^2)^2\ .
\eea

At leading order, the low energy effective action of the heterotic string admits a class of Mink$_4\times$CY$_3$ solutions \cite{Candelas:1985en} which play an important role in phenomenological applications of string theory. Whether such solutions continue to exist when $\a'$ corrections are switched on was partially investigated in  \cite{Candelas:1986tz}.  It was found that for tree-level stringy corrections up to order $\a'^3$, the metric of the internal six-dimensional K\"ahler space satisfies
\be
R_{ij}=\frac{\a'^3}{24}\zeta(3)(\nabla_i\nabla_j S-J_i^kJ_j^l\nabla_k\nabla_l S)\ ,
\ee
where $J^i_j$ is the complex structure and
\be
S=R_{ij}{}^{kl}R_{kl}{}^{mn}R_{mn}{}^{ij}-2R_{i~j}^{~k~l}R_{k~l}^{~m~n}R_{m~n}^{~i~j}\ .
\label{a3Ri}
\ee
The corresponding Ricci form $P_{ij}:=J_i^k R_{kj}$ is an exact form implying that the first Chern class is still vanishing. Thus the internal six-dimensional K\"ahler space is still a Calabi-Yau manifold \cite{Candelas:1986tz}. The 3-form flux remains vanishing while field equations determine the dilaton to be
\be
\phi={\rm constant}+\frac{\a'^3}{24}\zeta(3)S\ .
\label{a3di}
\ee
The spin connection is embedded in the Yang-Mills gauge group as in the two-derivative case. The first corrections to the embedding condition begin at $\a'^3$ which comes from the $\a'^4$ part of the action which is currently unknown. For \eqref{a3Ri} and \eqref{a3di} to naturally arise from the integrability condition of the Killing spinor equation, the $10D$ supersymmetry transformation rules of gravitino and dilatino must be modified to include terms proportional to the gradient of $S$  \cite{Candelas:1986tz}. Based on this analysis, it was concluded that with the tree level $\a'^3$ terms taken into account, the $E_8\times E_8$ heterotic string still admits compactification on a six-dimensional Calabi-Yau which gives rise to four-dimensional models that have an $E_6$ gauge group with four standard generations of fermions.

\subsection{Dualization of the Riemann-squared action}

Starting from the BdR Lagrangian in $10D$, the dualization of the two-form to a six-form was achieved in  \cite{BdR90}, and it was investigated in greater detail in  \cite{Chang:2022urm}. Earlier results were also obtained in superspace \cite{Saulina:1996vn}, on which we shall comment further below.

Here we shall describe the bosonic sector of the Lagrangians involved, and the supertransformations, up to quadratic fermions terms. The bosonic part of  the BdR Lagrangian, already given above, but in a slightly different notation, is given by \cite{Chang:2022urm}
\begin{align}
\cL_{BdR}^{\rm het} =& e e^{2\vp} \Big[
\frac{1}{4} R(\omega) + g^{\m\n} \partial_\m \vp \partial_\n \vp
- \frac{1}{12} \cG_{\m\n\rh} \cG^{\m\n\rh}
 - \frac14 \alpha' R_{\m\n ab}(\omega_-) R^{\m\n ab}(\omega_-) \Big]\ ,
\label{BdR6}
\end{align}
where
\begin{align}
\omega_{\pm \m ab} =& \omega_{\m ab} \pm G_{\m ab}\ , \qquad G_{\m\n\rh} = 3\partial_{[\m} C_{\n\rh]}\ ,
\nn\w2
\cG_{\m\n\rh} =& G_{\m\n\rh}  -6\alpha'\, X_{\m\n\rh}(\omega_-(G))\ ,
\nn\w2
X_{\m\n\rh}(\omega_-) =& \tr \left(
\omega_{-[\m} \partial_\n \omega_{-\rh]} + \frac23 \omega_{-[\m} \omega_{-\n} \omega_{-\rh]} \right)\ .
\label{dn}
\end{align}
It is understood that the term proportional to $\al'^2$ coming from $\cG^2$ is to be dropped, since we are considering the Lagrangian to first order in $\alpha'$. The action of the Lagrangian \eq{BdR6} is invariant under the following supersymmetry transformation rules up to ${\cal O}(\alpha')$ and higher order fermion terms,
\begin{align}
\delta e_\m{}^r
=& {\bar\e} \gamma^r \psi_\m\ ,
\qquad
\delta\psi_\m = D_\m(\omega_+(\cG)) \epsilon \ ,
\nn\w2
\delta C_{\m\n}
=& - {\bar\e} \gamma_{[\m} \psi_{\nu]} + 2 \alpha' \,\big( \omega_{-[\mu}{}^{rs}\delta \omega_{-\nu] rs}\big) \ ,
\nn\w2
\delta \chi =& \frac12 \gamma^\m \epsilon \partial_\m \vp -\frac{1}{12} \cG_{\m\n\rh} \gamma^{\m\n\rh} \epsilon\ ,
\nn\w2
\delta \vp =& {\bar\e}\chi\ .
\label{6dsuper}
\end{align}
To dualize the two-form potential, one adds the following Lagrange multiplier term to the Lagrangian
\begin{align}
\Delta \cL^{10D} (B,C) =& \frac{1}{6\times 7!}  \e^{\m\n\rh\s_1...\s_7} H_{\s_1...\s_7} G_{\m\n\rh}
= \frac16 e\, \tH^{\m\n\rh} \big( \cG_{\m\n\rh} + 6\alpha' X_{\m\n\rh} (\omega_-) \big)\ ,
\end{align}
where
\be
H_{\m_1...\m_7} = 7\partial_{[\m_1} B_{\m_2...\m_7]}\ ,\qquad \tH^{\m\n\rh} = \frac{1}{7!} \ve^{\m\n\rh\s_1...\s_7} H_{\s_1...\s_7}\ ,
\label{defs10d}
\ee
and integrates over $\cG_{\m\n\rh}$. To this end, it is convenient to write the total Lagrangian as
\begin{align}
& \cL_{BdR} +\Delta\cL^{10D}(B,C) = \cL_{01}+ \alpha' \cL_1\ ,
\label{L1}\w4
& \cL_{01} =  ee^{2\vp} \Big[
\frac{1}{4} R(\omega) + g^{\m\n} \partial_\m \vp \partial_\n \vp
- \frac{1}{12} \cG_{\m\n\rh}\big( \cG^{\m\n\rh} -2e^{-2\vp} \tH^{\m\n\rh} \big) \Big]\ ,
\label{L2}\w2
& \cL_1 =  - \frac14 ee^{2\vp} R_{\m\n ab}(\omega_-) R^{\m\n ab}(\omega_-)
+ e \tH^{\m\n\rh}  X_{\m\n\rh}({\om}_-) \ .
\label{L3}
\end{align}
The $\cO(\al')$ terms are collected in $\cL_1$ where the dependence on $\cG$ arises through the torsionful connection $\omega_-$. We are treating $\cG$ as an independent variable, while $H=dB$. Thus, the field equation for $B$ gives the relation $d\cG = -\alpha' \tr (R \wedge R)$, which can be solved to yield $\cG$ given in \eq{dn}. The field equation for $\cG$ at $\cO(\al')$ following from $\int d^{10}x (\cL_{01}+\alpha' \cL_1)$ is given by
\begin{align}
& \cG_{\m\n\rh} = e^{-2\vp} \tH_{\m\n\rh}  + 6 \alpha' e^{-2\vp} \frac{\delta \cL_1}{\delta G^{\m\n\rh}} \ .
\label{de}
\end{align}
This equation is readily solved for $\cG$ in terms of $H$, again at $\cO(\alpha')$, as
\begin{align}
& \cG_{\m\n\rh} = e^{-2\vp} \tH_{\m\n\rh}  + 6 \alpha' e^{-2\vp} \frac{\delta \cL_1}{\delta G^{\m\n\rh}}\Big|_{G=e^{-2\vp} \tH} \ .
\label{de3}
\end{align}
As shown in  \cite{BdR90}, the term $\delta \cL_1 / \delta \omega_{-\m ab}$ is proportional to field equations. Substituting for $\cG$ in the action \eq{L1} by using this equation, and going over to the brane frame  \cite{Duff:1990wv} by rescaling the metric as
\be
g_{\m\n} \to  g'_{\m\n} = e^{-2\vp/3} g_{\m\n}\ ,
\label{rs}
\ee
the following dual Lagrangian is obtained
\begin{align}
\cL^{\rm het, \rm dual}_{BdR} =&\, e e^{- 2\vp/3} \Big[ \frac14 R - \frac{1}{2 \times 7!} H_{\m_1 \cdots \m_7} H^{\m_1 \cdots \m_7} \Big]
\nn\w2
&\, + \alpha'\, e\, \Big[ - \frac14 R^{\m\n\rh\s} R_{\m\n\rh\s} + 6 H^2_{\m\n} R^{\m\n} - 3 H^2 R - \frac32 R^{\m\n\rh\s} H_{\m\n, \rh\s}
\nn\w2
&\, - \frac23 H^2_{\m\n} H^{2 \m\n} - \frac16 H_4 + \frac{14}{3} ( H^2 )^2 + \frac{2}{7!} ( D_\m H_{\n_1 \cdots \n_7} ) D^\m H^{\n_1 \cdots \n_7}
\nn\w2
&\, + 6 \vp^\m D_\m H^2  + \frac49 H^2_{\m\n} \vp^\m \vp^\n + \frac{16}{3} H^2_{\m\n} \vp^{\m\n} + \frac{22}{9} H^2 \vp^2
\nn\w2
&\, - \frac29 R_{\m\n} \vp^\m \vp^\n - \frac23 R_{\m\n} \vp^{\m\n} + \frac19 R \vp^2
- \frac{16}{27} \vp_{\m\n} \vp^\m \vp^\n
\nn\w2
&\, - \frac89 \vp_{\m\n} \vp^{\m\n} - \frac49 ( \vp^2 )^2 + \frac{16}{27} \vp^2 \vp^\m{}_\m - \frac19 ( \vp^\m{}_\m )^2
\nn\w2
&\,  + \frac{2}{6!} \tH^{\m\n\rh} H_\m{}^{\s_1 \cdots \s_6} D_\n H_{\rh \s_1 \cdots \s_6} - \frac43 \tH^{\m\n\rh} \vp^\s H_{\m\n, \rh\s} + \tH^{\m\n\rh} X_{\m\n\rh}(\omega) \Big]\ ,
\label{BdRDual10D}
\end{align}
where
\be
H_{\m\n, \rh\s} := \frac{1}{5!} H_{\m\n\la_1 \cdots \la_5} H_{\rh\s}{}^{\la_1 \cdots \la_5} , \quad H^2_{\m\n} := \frac{1}{6!}H_{\m\la_1 \cdots \la_6} H_\n{}^{\la_1 \cdots \la_6} , \quad H^2 := \frac17 H^2_{\m\n} g^{\m\n} \ .
\ee
Using the duality relation \eq{de3} and performing the rescaling \eq{rs}, the supersymmetry transformations take the form
\begin{align}
\delta e_\m{}^a
=& {\bar\e} \gamma^a \psi_\m\ ,
\nn\w2
\delta\psi_\m
=& D_\m(\omega)\e + \frac{1}{72} \tH_{abc} \big( 3\gamma^{abc}\gamma_\m +\gamma_\m \gamma^{abc}\big)\e + {\rm EOMs}\ ,
\nn\w2
\delta B_{\m_1...\m_6}
=& 3 {\bar\e} \gamma_{[\m_1...\m_5} \psi_{\m_6]} +{\rm EOMs}\ ,
\nn\w2
\delta \chi =& \frac12 \gamma^\m \epsilon \partial_\m \vp -\frac{1}{12} \tH_{\m\n\rh} \gamma^{\m\n\rh} \e +{\rm EOMs}\ ,
\nn\w2
\delta \vp =& {\bar\e}\chi\ .
\label{10dsuper2}
\end{align}

\subsection{Superspace approach}

In a superspace approach to the construction of the higher derivative extension of heterotic supergravity in $10D$, the key equations are the superspace Bianchi identities
\be
DT^A= R^A{}_B \wedge E^B\ ,\qquad D\cG = \alpha' \tr (R\wedge R)\ .
\label{10SG}
\ee
With a particular set of constraints these were solved in \cite{Bonora:1986ix, Bonora:1987xn, DAuria:1987tdr, Raciti:1989je, Bonora:1990mt, Bonora:1992tx,Fre:1991ef, Pesando:1992pa}, where the consistency of the BI's was proven to all orders in $\alpha'$. In this approach, the dimension zero torsion component is taken to be $T_{\al\beta}^a= \gamma^a_{\al\beta}$ but certain other components are deformed by $\al'$ dependent terms. In particular, the following relation (in our notation) arises
\be
\cG_{abc} = e^{-2\vp} T_{abc} + \al' W_{abc} (T)\ ,
\label{GBI}
\ee
where $W_{abc}$ is a nonlinear function of the torsion superfield $T_{abc}$ which can be found in papers referred to above. To obtain the deformed equations of motion, one solves for $T_{abc}$ in terms of $\cG_{abc}$ order by order in $\alpha'$, and uses the result in the supertorsion BI's. The resulting equations of motion were obtained at $\cO(\al')$ in\cite{Fre:1991ef, Pesando:1992pa}. These equations apparently have not been compared with those which arise from the BdR action. While they are expected to agree at $\cO(\alpha')$, it is an open question whether equivalence holds to all orders in $\alpha'$. This approach has been updated in \cite{Lechner:2008uz} where the relationship to another approach by \cite{Bellucci:1988ff, Bellucci:1990fa} which focuses on order by order in $\al'$ analysis (without addressing fully the question of the consistency of the entire procedure) was clarified. Interestingly, the formulation of \cite{Lechner:2008uz} is such that the Gauss-Bonnet action appears as part of the bosonic action. The full four-derivative action in this framework has not been worked out but it is expected to be related to the result that follows from \cite{Bonora:1986ix, Bonora:1987xn, DAuria:1987tdr, Raciti:1989je, Bonora:1990mt, Bonora:1992tx} by field redefinitions.

The solution of the Bianchi identities \eq{10SG} will yield the deformed equations of motion to any order in $\al'$. However, this framework does not capture the most general supersymmetric deformation of heterotic supergravity. For example, at order $\cO(\al'^3)$, deformations involving $(\tr R^2)^2$, but not $\tr (R^4)$, will arise. To get the latter, one can either deform the constraint on $T^a_{\al\beta}$ to include a tensor in $1050$ dimensional representation of $SO(9,1)$ \cite{Bellucci:2006cx,OReilly:2006eeg,Nilsson:1986cz,Nilsson:1986rh,Candiello:1994ew,Howe:2008vb,Lechner:2010ti} or take $\cG_{\alpha\beta\gamma}$ to be nonvanishing\cite{Lechner:1987ip}.

Heterotic supergravity in the six-form formulation in superspace including $\alpha'$ corrections was studied in \cite{Gates:1985wh, Gates:1986is, Gates:1986tj, Nishino:1986mj, Nishino:1990ky} where partial results were obtained. A more complete treatment which builds especially on the results of \cite{Nishino:1990ky} appeared in \cite{Terentev:1993wm, Terentev:1994br, Zyablyuk:1994xk, Saulina:1995eq, Saulina:1996vn}, where the dualization phenomenon in superspace, suggested in \cite{DAuria:1987qjh}, was spelled out. Here we shall focus on the key results of \cite{Saulina:1996vn} where the equations of motion deduced from superspace were also integrated into an action for the bosonic fields, and we shall compare the result with ours.

The super BIs for supertorsion  $T_{MN}^A$ and the super seven-form $H_7=DB_6$ are given by
\begin{align}
DT^A &= R^A{}_B \wedge E^B\ , \qquad  DH_7 =0\ .
\label{DBI}
\end{align}
Note that the BI for $H_7$ does not acquire $\alpha'$ deformation, unlike the BI for $\cG$ in \eq{GBI}. The BI's \eq{DBI} are solved by (see \cite{Saulina:1996vn} and references therein)
\begin{align}
& T_{\alpha\beta}{}^c = \gamma^c_{\alpha\beta}\ ,\qquad  T_{a\beta}{}^\gamma = \frac{1}{(72)^2} T_{bcd} \left(\gamma^{bcd}\gamma^a \right)_\beta{}^\gamma\ , \qquad  T_{\alpha b}{}^c =0\ ,\qquad  T_{\alpha\beta}{}^\gamma=0\ ,
\nn\w2
& H_{a_1...a_5 \alpha\beta} = -\left(\gamma_{a_1...a_5}\right)_{\alpha\beta}\ ,\qquad H_{a_1...a_7}= \frac{1}{6!} \e_{a_1...a_7}{}^{abc} T_{abc} \ ,
\nn\w2
& \mbox{other components of}\ \  H_7=0\ ,
\label{ssc}
\end{align}
together with a scalar superfield $\phi$ satisfying
\be
D_\alpha \phi = \chi_\alpha\ ,\qquad D_\alpha \chi_\beta  = \frac12 \gamma^a_{\alpha\beta} D_a \phi +\left(-\frac{1}{36} \phi T_{abc} +\alpha' A_{abc} \right)\left(\gamma^{abc}\right)_{\alpha\beta}\ ,
\label{dchi}
\ee
where $D_a$ is the covariant derivative with bosonic torsion, and $A_{abc}$ is a crucial superfield which governs the $\alpha'$ deformation given by~\cite{Saulina:1996vn}\footnote{Certain terms for $A_{abc}$ and their implications for the $\alpha'$ corrections were considered in \cite{Gates:1985wh, Gates:1986is, Gates:1986tj, Nishino:1986mj, Nishino:1990ky}.}
\begin{align}
A_{abc}& = \Big[-\frac{1}{18} \Box T_{abc} +\frac{1}{36} D^d T_{da,bc} -\frac{1}{36} T^{de}{}_a D_b T_{c de}
-\frac{5}{1944} T^2 T_{abc}
\nn\w2
& -\frac{5}{108} T^2_{da} T_{bc}{}^d +\frac{5}{54} T^3_{abc} -\frac{1}{3888} \e_{abc}{}^{a_1...a_7} T_{a_1...a_3} D_{a_4} T_{a_5...a_7}
\nn\w2
& -\frac{1}{48} T_{a_1 a_2}{}^\alpha \left( 2\gamma_{abc} \eta^{a_1b_1} \eta^{a_2 b_2} + \gamma^{a_1}\gamma_{abc} \gamma^{b_1} \eta^{a_2 b_2} +24 \gamma^{a_1} \gamma_c \gamma^{b_1} \delta_a^{a_2} \delta_b^{b_2} \right)_{\a\b} T_{b_1 b_2}{}^\beta \Big]_{[abc]}\ ,
\end{align}
where $T_{abc}= T_{[abc]}$ and $T_{ab}{}^\alpha$ is the gravitino curvature, and
\be
T_{ab,cd} := T_{ab}{}^e T_{cde}\ ,\ \ T^2_{ab}:= T_a{}^{cd} T_{bcd}\ ,\ \  T^3_{abc} := T_{a d_1d_2} T_b{}^{d_2d_3} T_{cd_3}{}^{d_1}\ ,\ \  T^2 := T_{abc} T^{abc}\ .
\ee
It is noteworthy that the solution is an exact one, even though there is an $\alpha'$ dependent deformation. The EOMs that result from the analysis of the superspace BI's are also given in \cite{Saulina:1996vn} in terms of superfields whose lowest order components in $\theta$ expansion are the supergravity multiplet of fields.  For a more detailed explanation of how the EOMs are obtained in superspace, see \cite{Terentev:1994br}. These equations imply an action with $\alpha' {\rm Riem}^2$ term, and yet their supersymmetry is realized exactly. No higher than first order in $\alpha'$ terms arise in supersymmetric variations of these EOMs since, as can be seen in \cite{Saulina:1996vn}, the $\alpha'$ dependent terms do not involve the dilatino $\chi$ which is the only field that develops $\alpha'$ deformation; see \eq{dchi}.

A bosonic Lagrangian which yields these EOMs can only be determined up to terms proportional to the lowest order (i.e. two-derivative) EOMs. Such terms can always be removed by field definitions (see, for example, \cite{Chang:2023pss} for a detailed explanation).  With this understood, the resulting bosonic Lagrangian is found to be \cite{Saulina:1996vn}
\footnote{In converting the conventions of \cite{Saulina:1995eq} ours, we first let $\omega_\m{}^{ab} \to - \omega_\m{}^{ab}$, and then let $\eta_{ab} \to -\eta_{ab}, \e_{a_1...a_{10}} \to -\e_{a_1...a_{10}},  M_{\m\n\rh} \to 2 \tH_{\m\n\rh}$, $\tilde{\phi} \to e^{- \frac23 \vp}$, $k_g \to \alpha'$, and $\cL \to 4 \cL$. Note also that the term $\frac{1}{162} ( M^2 )^2$ term in \cite[Eq. (4.10)]{Saulina:1995eq} should be absent, as noted later in\cite{Saulina:1996vn} as well.}
\begin{align}
\cL_{STZ} =&\, e e^{- \frac23 \vp} \Big[ \frac14 R - \frac{1}{2 \times 7!} H_{\m_1 \cdots \m_7} H^{\m_1 \cdots \m_7} \Big]
\nn\w2
&\, + \frac14 e \alpha' \Big[ - R^{\m\n\rh\s} R_{\m\n\rh\s} + 2 R^{\m\n} R_{\m\n} + 4 R^{\m\n} H^2_{\m\n} - 4 R H^2
\nn\w2
&\, - \frac{4}{7!} ( D_\m H_{\n_1 \cdots \n_7} ) D^\m H^{\n_1 \cdots \n_7} + \frac{16}{3} H^2_{\m\n} H^{2 \m\n} - \frac23 H_4 - \frac{40}{3} ( H^2 )^2
\nn\w2
&\, + 4 \tH^{\m\n\rh} D^\s H_{\m\n, \rh\s} + 4 \tH^{\m\n\rh} \omega^L_{\m\n\rh}(\omega) \Big] \ .
\end{align}
The supertransformation  resulting from the constraints \eq{ssc} are \cite{Zyablyuk:1994xk} (up to cubic fermions here)
\begin{align}
\delta e_\m{}^a =& {\bar\e} \gamma^a \psi_\m\ ,
\nn\w2
\delta \psi_\m =& D_\m \e -\frac{1}{72} T_{abc} \big( 3 \gamma^{abc}\gamma_\mu +\gamma_\mu \gamma^{abc} \big) \e\ ,
\nn\w2
\delta B_{\m_1...\m_6} =& 3 {\bar\e} \gamma_{[\m_1...\m_5} \psi_{\m_6]}\ ,
\nn\w2
\delta\chi =& \frac12 \gamma^\m \e \partial_\m \phi + \left( -\frac{1}{36} \phi T_{abc} + \alpha' A_{abc}\right) \gamma^{abc} \e\ ,
\nn\w2
\delta \phi =& {\bar\e} \chi\ ,
\label{STZsusy}
\end{align}
where it is understood that $\phi \to e^{-2\vp/3}$ and $\chi \to e^{-2\vp/3} \chi$. These are also understood to be valid up to the lowest order EOMs. It has been shown in \cite{Saulina:1996vn} that the algebra closes on-shell, and that the closure functions are $\alpha'$ independent. Thus, the closure of the algebra is not a statement up to order $\alpha'$ but an exactly valid statement. The fact that $A_{abc}$ obeys the relation  $DA_{abc}= \gamma_{abc}{}^{de} X_{de}$ where $X_{de}{}^\alpha$ is an arbitrary function \cite{Saulina:1996vn} is behind this property.

Comparison of the Lagrangian $\cL_{STZ}$ with the bosonic sector of the dual of the BdR Lagrangian in $10D$ \eq{BdRDual10D}, which was obtained by solving the duality equation to order $\alpha'$, was carried out in \cite{Chang:2022urm}. It was shown that the difference amounts to field redefinition of the dilaton. However, it was argued that a full comparison of the action as well as supersymmetry transformations require the dualization of the BdR action in the two-form formulation to all orders in $\alpha'$, and it was conjectured that a solution may be obtained as follows.

In superspace, leaving the solution of the BIs reviewed above intact, one can also construct a super three-form $\cG$ which obeys the super BI \eq{10SG} as \cite{Saulina:1996vn}
\begin{align}
\cG_{\al\beta\gamma} =& 0\ ,
\nn\w2
\cG_{\al\beta a}=& \phi \left(\gamma_a\right)_{\al\beta} +\alpha'\,U_{\alpha\beta a}\ ,
\nn\w2
\cG_{\al bc} =& -\left(\gamma_{bc} \chi\right)_\alpha +\al'\,U_{\al bc}\ ,
\nn\w2
\cG_{abc}=& -\phi T_{abc} +\al'\, U_{abc}\ ,
\label{Geq}
\end{align}
where \cite{Saulina:1996vn}
\begin{align}
U_{abc}=& \Big[-2 \Box T_{abc} -6 D^d T_{da,bc} -6 T^{de}{}_a D_b T_{cde}
-6 {\cal R}^{de}{}_{ab} T_{cde} -6 {\cal R}_{da} T_{bc}{}^d +4T^3_{abc}
\nn\w2
& - T_{a_1 a_2}{}^\alpha \Big( \gamma_{abc} \eta^{a_1b_1} \eta^{a_2 b_2} + \gamma^{a_1}\gamma_{abc} \gamma^{b_1} \eta^{a_2 b_2} +12 \gamma^{a_1} \gamma_c \gamma^{b_1} \delta_a^{a_2} \delta_b^{b_2}
\nn\w2
& + 12 \delta_a^{a_1} \delta_b^{b_1} \eta^{a_2 b_2} \gamma_c + 6 \delta_a^{a_1} \delta_b^{b_1} \delta_c^{a_2} \gamma^{b_2} \Big)_{\a\b} T_{b_1 b_2}{}^\beta \Big]_{[abc]}\ ,
\end{align}
the expressions for $U_{\al\beta a}$ and $U_{\al bc}$, which are functions of $T_{abc}$ and $T_{ab}{}^\al$, can be found in \cite{Saulina:1996vn}, and ${\cal R}_{abcd}$ is the supercovariant curvature (calculated with the
torsion full spin-connection). The last equation in \eq{Geq} is expected to be equivalent to \eq{GBI} upon field redefinitions, and it also represents the duality relation between the two-form and six-form formulations as can be seen by  substituting $T_{abc} = \tH_{abc}$ from \eq{ssc} into this relation, which now takes the form
\be
\cG_{abc}= -\phi \tH_{abc} +\al'\, U_{abc}\Big|_{T_{def}=\tH_{def}}\ .
\label{STZduality}
\ee
Solving for ${\widetilde H}_{abc}$ order by order in $\alpha'$ and substituting the result into the EOMs obtained by STZ in \cite{Saulina:1996vn} is expected, though not proven, to generate the EOMs of BdR to all orders in $\alpha'$, just as solving for $T_{abc}$ in \eq{GBI} is expected, but not proven, to lead to the same result up to field redefinitions, as explained above. For a further discussion of these issues, see \cite{Chang:2022urm}.

\subsection{\texorpdfstring{ Killing spinors in $N=(1,0), 10D$}{N=(1,0), 10D} supergravity with higher derivatives}

The Killing spinor equation in the heterotic string with leading $\a'$ corrections was analyzed in  \cite{Fontanella:2019avn}. These corrections come from the Chern-Simons modification of the three-form field strength, as in \eq{exp}. Up to the first order in $\a'$, the  supersymmetric solutions of this theory imply the existence of a Killing spinor $\epsilon$ satisfying
\bea
0&=&\left(\nabla_\m-\frac14\o_{+\m ab}\gamma^{ab}\right)\epsilon\ ,
\nn\\
0&=&\left(\g^{\m}\partial_\m\phi-\frac1{12}H_{\m\n\r}\gamma^{\m\n\r}\right)\epsilon\ ,
\nn\\
0&=& F_{\m\n}\gamma^{\m\n}\epsilon\ ,
\eea
where $H=dB+\a' X(\omega_+)$, with $\omega_+ = \omega + dB$, and $F=dA$, thus switching on only one vector field. Using the Killing spinor, one can build spinor bilinears
\be
\ell_{\m}=\bar{\epsilon}\gamma_{\m}\epsilon,\quad W_{\m_1\cdots\m_5}=\bar{\epsilon}\g_{\m_1\cdots\m_5 }\epsilon\ ,
\ee
which satisfy
\be
\ell_\m\ell^\m :=0,\quad W_{\m_1\cdots\m_5 } :=5\ell_{[\m_1}\O_{\m_2\cdots\m_5]}\ .
\ee
Since $\ell_\m$ is null, one can pair it with another null vector $n_\m$ obeying $n_\m\ell^\m=-1$. Using the Killing spinor equations above, one can show that $\ell_\m$ is a Killing vector of the solution. Thus one can choose a coordinate system in which $\ell^\m\partial_\m=\partial_v$.  The fact that the metric admits a null Killing vector alone implies that the metric can be parametrized as follows
\be
ds^2=-2f(du+\b)(dv+Kdu+\overline{\o})+h_{mn}dx^mdx^n\ ,
\label{metric}
\ee
where $\b=\b_mdx^m$ and $\overline{\o}=\overline{\o}_m dx^m$, and all the functions are independent of $v$. Algebraic manipulations of the Killing spinor equations lead to
\bea
\partial_v\phi&=&0,\quad F_{\m\n}\ell^\n=0\ ,
\nn\\
F_{pq}\Pi^{-pq}{}_{mn}&=&0,\quad H_{-pq}\Pi^{-pq}{}_{mn}=0\ ,
\nn\\
H_{qrs}\Pi^{-qrs}{}_{mnp}&=&\frac17(2\partial_q\phi-H_{+-q})\O^q{}_{mnp}\ ,
\label{KSlemmas}
\eea
where certain projection operators $\Pi^-$ in the space orthogonal to $\ell_\m$ and $n_\m$ have been defined, and they can be found in the appendix of  \cite{Fontanella:2019avn}. It was shown that \cite{Fontanella:2019avn} the necessary condition for a Killing spinor to exist is in fact sufficient for its existence. As is well known, the integrability conditions of the Killing spinor equations imply a subset of the equations of motion. To state this subset, let us define the frames
\be
e^+=\ell_\m dx^\m,\quad n_\m dx^\m=e^-\ .
\ee
One finds that all equations motion follow from the integrability conditions of the Killing spinor equations, except the following components \footnote{The integrability conditions can be studied by either commuting the differential operators that define the Killing spinors equations, or perhaps more conveniently by using the method of Killing spinor identities \cite{Fontanella:2019avn}, developed in \cite{Kallosh:1993wx}.}
\be
\vE_{g}^{--},\quad \vE_{B}^{+-},\quad \vE_{B}^{-m},\quad
\widehat{\vE}_A^-
\label{subset}
\ee
where $\vE$ refer to the EOMs with self-explanatory notation, except that $\widehat{\vE}_A^\m$ is a particular combination of the vector field equations which can be simplified to read
\be
\widehat{\vE}_A^\m=\nabla_{\n}(\omega_+)(e^{-2\phi} F^{\m\n})\ .
\ee
Of course, the Bianchi identity for $H_{(3)}$ must also be satisfied. Thus, to find a supersymmetric solution, one may make an ansatz consistent with \eq{metric}, and use the equations \eq{KSlemmas} as well, to solve the equations listed in \eq{subset}, and the Bianchi identity for $H_{(3)}$. Based on the results above, a $\frac14$-BPS black hole solution in heterotic supergravity with leading stringy corrections has been obtained in \cite{Cano:2018qev}, and it takes the form (setting $\k^2=1$)
\bea
ds^2&=&-\frac2{Z_-}du(dv-\frac12Z_+ du)+Z_0(d\rho^2+\rho^2d\Omega^2_{(3)})+dy^mdy^m,\quad m=1,\cdots,4\ ,
\nn\\
H&=&dZ_+\wedge du\wedge dv- \rho^3\partial_\r Z_0 \Omega_{(3)}\ ,
\nn\\
A&=&\frac2{\r^2(1+\r^2)}M^-_{ij}x^idx^j,\quad e^{-2\phi}=e^{-2\phi_\infty}\frac{Z_1}{Z_0}\ ,
\eea
where $\r^2=x^ix^i$, $\Omega_{(3)}$ is the volume form of the 3-sphere with unit radius, $M^-_{ij}$ is the anti-self dual part of the SO(4) generators (i.e. the constant 't Hooft symbol) and $Z_0, Z_{\pm}$ are functions depending on $\r$ only and they are determined in \cite{Cano:2018qev}.  A particular solution is of the form \cite{Cano:2018qev}
\bea
Z_0&=& 1+8\a'\frac{\r^2+2}{(\r^2+1)^2}+{\cal O}(\a'^2)\ ,
\nn\\
Z_-&=& 1+\frac{Q_-}{\r^2}+{\cal O}(\a'^2)\ ,
\nn\\
Z_+&=&1+\frac{Q_+}{\r^2}-16\a'\frac{Q_-Q_+}{\r^4(\r^2+Q_-)}+{\cal O}(\a'^2)\ ,
\eea
where $Q_\pm$ are integration constants. See \cite{Cano:2018qev} for an interpretation of this particular solution.  Generalizations to non-supersymmetric black holes were investigated in \cite{Cano:2019oma,Cano:2019ycn}. It is found that for fixed mass and charges, higher derivative corrections to black hole entropy are always positive \cite{Cheung:2018cwt}.

\section{\texorpdfstring{$D=9,8,7$}{D=9,8,7}}
\label{sec987}

Results on higher derivative extensions of supergravities in $D=9,8,7$ are scarce. One such rare result is provided in \cite{Eloy:2020dko}, where the bosonic part of half-maximal supergravities in $D$ dimensions is obtained by a toroidal reduction of the heterotic supergravity with its four-derivative extension, but without the Yang-Mills sector. (See also \cite{Chang:2021tsj} where such a reduction is carried out for the 4-torus reduction, including the fermions.) In a double field theory approach, in which the Yang-Mills sector is also considered, the bosonic sector in $D$ dimensions is also obtained in \cite{Baron:2017dvb}. Putting aside the Yang-Mills sector, the reduction on a $d$-torus $T^d$ gives half-maximal  supergravity in $D=10-d$ dimensions coupled to $d$ vector multiplets. The full bosonic field content is
\be
(g_{\mu\nu}\ , B_{\mu\nu}\ , \phi\ , M_{mn}\ , A_\mu^m )\ ,
\ee
where $m=1,...,2d$. Furthermore, the scalar fields other than the dilaton $\phi$ parametrize the coset $SO(d,d)/(SO(d)\times SO(d))$, and the scalar matrix  $M_{mn}$ is built out of the coset representative $L_m{}^A= \left( L_m{}^a, L_m{}^{\bar a}\right)$ with $a,\bar a =1,...,d,$ as $M_{mn}= L_m{}^a L_n{}^a + L_m{}^{\bar a} L_n{}^{\bar a}$. The Maurer-Cartan form associated with this coset is defined as
\bea
(L^{-1}\partial_{\mu}L)_A{}^{B} \equiv \begin{pmatrix}  Q_{\mu a}{}^{b} & P_{\mu a}{}^{\bar{b}} \\[0.7ex]
\bar{P}_{\mu \bar{a}}{}^{b} & \bar{Q}_{\mu \bar{a}}{}^{\bar{b}}
\end{pmatrix}\ ,
\label{mc1}
\eea
where $Q_{\mu a}{}^b$ and $\bar Q_{\m \bar a}{}^{\bar b}$ are the composite connections.
Up to order $\alpha'$, the bosonic part of the action in $D$ dimensional spacetime takes the form \cite{Eloy:2020dko}
\begin{align}
I=&\int d^{D}x\,\sqrt{-g}\,e^{-\phi}\Bigg\{R+\partial_{\mu}\phi\,\partial^{\mu}\phi-\frac{1}{12}\,H_{\mu\nu\rho} H^{\mu\nu\rho}+\frac{1}{8}\,\Tr{\partial_{\mu}S\,\partial^{\mu}S}-\frac{1}{4}\,F_{\mu\nu}^m\,{S_m}^n\,F^{\mu\nu}_n
 \nn\\
&\quad+\frac{1}{8}\,\alpha' \,\Big[-2H^{\mu\nu\rho}\Big(\, \Omega_{\mu\nu\rho}(\omega) +\Omega_{\mu\nu\rho}(Q) - \Omega_{\mu\nu\rho}({\bar Q}) \Big)
\nn\\
&\quad +R_{\mu\nu\rho\sigma}R^{\mu\nu\rho\sigma}
-\frac{1}{2}\,R_{\mu\nu\rho\sigma} \left( {H}^{\mu\nu,\rho\sigma}+F^{\mu\nu\,m}F^{\rho\sigma}{}_m
+F^{\mu\nu\,m}{S_m}^n{F^{\rho\sigma}}_n\right)
\nn\\
&\quad + \frac{1}{8}\Tr{S\nabla_{\mu}S\nabla^{\mu}S\nabla_{\nu}S\nabla^{\nu}S}
+\frac{1}{16}\,\Tr{\nabla_{\mu}S\nabla_{\nu}S\nabla^{\mu}S\nabla^{\nu}S} -\frac{1}{32}\, \Tr{\nabla_{\mu}S\nabla_{\nu}S} \Tr{\nabla^{\mu}S\nabla^{\nu}S}
\nn\\
& \quad   + \frac{1}{24}\, H_{\mu\nu\rho}{H}^{\mu\ \, \lambda}_{\ \,\sigma}{H}^{\nu\ \,\tau}_{\ \,\lambda}{H}^{\rho\ \,\sigma}_{\ \,\tau} -\frac{1}{8}\,{H}^{2}_{\mu\nu}{H}^{2\,\mu\nu} +\frac{1}{8}\,H^{2}_{\mu\nu}\Tr{\nabla^{\mu}S\nabla^{\nu}S}
\nn\\
&\quad-\frac{1}{2}\,H^{2}_{\mu\nu}{{F^{\mu}}_{\rho}}^m{S_m}^n{F^{\nu\rho}}_n +\frac{1}{4}\, {H}^{\mu\nu,\rho\sigma}\left( {F_{\mu\rho}}^m{S_m}^n F_{\nu\sigma\,n} +2F_{\mu\nu}^mF_{\rho\sigma\,m}\right)
\nn\\
&\quad-\frac{1}{2}\,H^{\mu\nu\rho}{F_{\mu\sigma}}^m{\left(S\nabla_{\nu}S\right)_m}^n{{F_{\rho}}^{\sigma}}_n +\frac{1}{4}\,H^{\mu\nu\rho}F_{\mu\sigma}{}^m\nabla^{\sigma}S_m{}^n F_{\nu\rho\,n}
\nn\\
&\quad+\frac{1}{8}\, {F_{\mu\nu}}^m{S_m}^n F_{\rho\sigma\,n} F^{\mu\rho\,p} {S_p}^q{F^{\nu\sigma}}_q -\frac{1}{2}\,{F_{\mu\nu}}^m{S_m}^n{F^{\mu\rho}}_n F^{\nu\sigma\,p}{S_p}^q F_{\rho\sigma\,q}
\nn\\
&\quad+\frac{1}{8}\,{F_{\mu\nu}}^m{F_{\rho\sigma}}_m F^{\mu\rho\,n}{F^{\nu\sigma}}_n  + \frac{1}{8}\,F_{\mu\nu}{}^mF_{\rho\sigma\,m}\,F^{\mu\nu\,p}S_p{}^q F^{\rho\sigma}{}_q
\nn\\
&\quad-\frac{1}{2}\,{F_{\mu\nu}}^m{\left(S\nabla_{\rho}S\nabla^{\nu}S\right)_m}^n {F^{\mu\rho}}_n + \frac{1}{4}\,F^{\mu\rho\,m}{S_m}^n{F^{\nu}}_{\rho\,n}\Tr{\nabla_{\mu}S\nabla_{\nu}S}
\nn\\
&\quad    -\frac{1}{4}\,F_{\mu\nu}{}^m\left(\nabla_{\rho}S\nabla^{\rho}S\right)_m{}^nF^{\mu\nu}{}_n -\frac{1}{8}\,F_{\mu\nu}{}^m\left(\nabla^{\mu}S\nabla_{\rho}S\right)_m{}^nF^{\nu\rho}{}_n \Big)\Big]\Bigg\}\ ,
\label{PreD2}
\end{align}
where
\bea
S_m{}^n &=& M_{mp} \eta^{pn}\ ,\quad F_{\mu\nu}^m = 2\partial_{[\mu} A_{\nu]}^m\ , \quad H_{\mu\nu\rho} = 3\partial_{[\mu} B_{\nu\rho]}\ ,
\nn\\
\Omega_{\mu\nu\rho}(\omega) &=& \Tr \left(\omega_{[\mu} \partial_\nu \omega_{\rho]}
+\frac23  \omega_{[\mu} \omega_\nu \omega_{\rho]} \right)\ ,\quad \Omega_{\mu\nu\rho}(Q) = \Tr \left(Q_{[\mu} \partial_\nu Q_{\rho]}
+\frac23 Q_{[\mu} Q_\nu Q_{\rho]} \right)\ ,
\label{ehsd}
\eea
and similarly for $\Omega_{\mu\nu\rho}(\bar Q)$.

Gauged version of the half-maximal supergravity coupled to $n$ vector multiplets in $D$ dimensions with higher derivative extension terms, has been obtained from double field theory (DFT) in \cite{Baron:2017dvb}, where the potential and the implications of the $\alpha'$ corrections for the vacua of the theory are also discussed. In this case, the scalar fields other than the dilaton $\phi$ parametrize the coset $SO(d,d+N)/(SO(d)\times SO(d+N))$, and the scalar matrix  $M_{mn}$ is built out of the coset representative $L_m{}^A= \left( L_m{}^a, L_m{}^{\bar a}\right)$ with $a =1,...,d,\ {\bar a}= 1,...,d+N$,  as $M_{mn}= L_m{}^a L_n{}^a + L_m{}^{\bar a} L_n{}^{\bar a}$, where $d=10-D$. It is important to note that here, and in the rest of this section \mbox{\it we are now using the index $m$ which takes the values $m=1,...,2d+N$} for simplicity in notation, while  the same index is used in eqs. (5.1), (5.3) and (5.4) to take the values $m=1,...,2d$. The $SO(d,d+N)$ invariant matrix $\eta_{mn}$ and its inverse $\eta^{mn}$ are used to lower and raise indices. The result found in \cite{Baron:2017dvb} for the bosonic part of the action in $D=10-d$ dimensions is given by
\begin{align}
S = \int d^D X \sqrt{-g} e^{-2 \phi} &\Bigg\{R + 4 \nabla_\mu \nabla^\mu \phi - 4 \nabla_\mu \phi \nabla^\mu \phi - \frac 1 {12} H_{\mu \nu \rho} H^{\mu \nu \rho}
\nn\\
& - \frac 1 4 F_{\mu \nu}{}^m F^{\mu \nu n} M_{m n} + \frac 1 8 \nabla_\mu M_{mn} \nabla^\mu M^{mn} - V_0
\nn\\
& +\alpha' \Big[ H^{\mu\nu\rho} \Big( -\frac12 \Omega_{\mu\nu\rho}(A)-\frac14 \Omega_{\mu\nu\rho}(\omega_-) -\frac14 \Omega_{\mu\nu\rho}(Q) -\frac14 \Omega_{\mu\nu\rho}({\bar Q}) \Big)
\nn\\
&
+\frac{1}{8}\, {\tilde{R}{}^{(-)}}^{\mu \nu \rho \sigma} {\tilde{R}{}^{(-)}}_{\mu \nu \rho \sigma}
+ {\mathcal L}_{ungauged} + {\mathcal L}_{gauged} - {\cal V} \Big] \Bigg\}\ ,
\label{na}
\end{align}
where $H=dB$ and the potential associated with the gauged two-derivative theory is
\be
V_0 = \frac 1 {12} f_{m p}{}^r f_{n q}{}^s M^{mn} M^{pq} M_{r s} + \frac 1 4 f_{m p}{}^q f_{n q}{}^p M^{m n} + \frac 1 6 f_{mnp} f^{mnp} \ ,
\label{V0}
\ee
and $f_{mnp}$ are the structure constants of the gauge group with $2d+N$ generators. The order $\alpha'$ contribution to the potential is found to be \cite{Baron:2017dvb}
\bea
{\cal V}&=&
\left( P_{m m'} P_{n n'} {\bar P}_{p p'}
- {\bar P}_{m m'} {\bar P}_{n n'} P_{p p'} \right)
P_{q q'} P_{r r'} {\bar P}_{s s'}
f^{m p q} f^{n p' q'} f^{m' r s} f^{n'r's'}
\label{calV}
\nn\\
&+& \left(  P_{m m'} P_{n n'} {\bar P}_{p p'}
+ \frac43 {\bar P}_{m m'} {\bar P}_{n n'} P_{p p'} \right)
P_{q q'} P_{r r'} {\bar P}_{s s'}
f^{m n s} f^{m' p r} f^{n' p' q} f^{q' r' s'}\ . \ \ \ \ \
\label{efpot}
\eea
Further definitions used above are
\bea
\tilde{R}_{\mu \nu \rho \sigma}(\omega_-) &=& R_{\mu \nu \rho \sigma}(\omega_-) - \frac12 F_{\mu \nu m} F_{\rho \sigma n} (P^{mn}-2\, {\bar P}^{mn})\ ,\quad \omega_\pm = \omega \pm \frac12 H\ ,
\nn\\
P_{mn} &^=& \frac12 \left(\eta_{mn} -M_{mn}\right)\ ,\quad {\bar P}_{mn} = \frac12 \left(\eta_{mn} +M_{mn}\right)\ ,
\nn\\
\Omega_{\mu\nu\rho} (A) &=& A_{[\mu} \partial_\nu A_{\rho]} -\frac13 f_{mnp} A_{[\mu}{}^m A_\nu{}^n A_{\rho]}{}^p\ .
\eea
 The Chern-Simons forms $\Omega(\omega), \Omega(Q)$ and $\Omega(\bar Q)$ are as defined in \eq{ehsd} except that in the latter one the composite connection $\bar Q$ is now valued in $SO(d+N)$. The Lagrangians ${\mathcal L}_{ungauged}$ and $ {\mathcal L}_{gauged}$ have very complicated forms, and they are given as ${\mathcal L}_{ungauged}^{(-)}$ and $ {\mathcal L}_{gauged}^{(-)}$ in eqs. (3.66) and (3.67) of \cite{Baron:2017dvb}. In fact, the {\it total} ungauged Lagrangian can be obtained from \eq{na} by setting $f_{mn}{}^p=0$, and the result is expected to agree with \eq{PreD2}, which was obtained from the toroidal reduction of the higher derivative extended heterotic supergravity, upon letting $m=1,...,2d$ as well. As for the potential $V_0$ given in \eq{V0}, we have checked that it agrees with that found in \cite{Bergshoeff:1985mr} in $7D$, where the couplings were obtained directly by Noether procedure.

In general, some of the higher derivative extensions of supergravity are relevant for potential ultraviolet divergences. As we shall see in section \ref{divergences}, the $D^8 R^4, D^6R^4,D^4 R^4 $ and $R^4$ terms are relevant for counterterms in dimensions $D=9,8,7$, respectively. Certain action integrals, including those in $D=8,7$, are formulated in the ectoplasm approach, as will be summarized in section \ref{ectoplasm}.

\section{\texorpdfstring{$D=6$}{D=6}}

Higher derivative deformations of $6D$ supergravities with $N=(2,2), N=(1,1)$ and $N=(2,0)$ apparently have not been studied directly but some results have been obtained in the latter two cases from the compactifications of the heterotic string on $T^4$ and of Type II on $K3$, respectively \cite{ Liu:2013dna, Liu:2019ses}. In the case of $N=(2,2)$ supergravity, resulting from a toroidal compactification of type II supergravities, the consequences of the duality symmetry for the higher derivative extension, albeit at the level of the leading terms in $R^4,\,D^4R^4,\,D^6 R^4$ have been explored by various authors, as we shall review in section \ref{lastsection}.

In the case of $N=(1,0)$ supergravity, the existence of off-shell formulations makes the construction of higher derivative supersymmetric invariants more manageable, and we shall review the results obtained in this way. We shall also review aspects of results obtained from the compactification of heterotic string on $K3$.

\subsection{\texorpdfstring{$N=(1,1), 6D$}{N=(1,1), 6D} supergravity from Type IIA on \texorpdfstring{$K3$}{K3}}
		
The $N=(1,1)$ supergravity multiplet consists of the fields
\bea
(e_\mu{}^a, \psi_{\mu}^i, B_{\mu\nu}, A_\mu, A_{\mu i}{}^j, \chi^i, \phi )\ ,
\eea
where $i=1,2$ is the $SU(2)_R$ doublet index. The two-derivative Lagrangian was constructed by Romans  \cite{Romans:1985tw}. The bosonic Lagrangian of the ungauged $N=(1,1)$ supergravity without mass deformation is given by \cite{Romans:1985tw}
\bea
e^{-1}{\cal L}_{6D, (1,1)}&=&\frac14 R-\frac12\partial_\m\phi\partial^\m\phi-\frac14e^{-\sqrt{2}\phi}(f_{\m\n}f^{\m\n}+F_{\m\n i}{}^jF^{\m\n}{}_j{}^i)-\frac1{12}e^{-2\sqrt2\phi}H_{\m\n\r}H^{\m\n\r}
\nn\\
&&-\frac18\epsilon^{\m\n\r\l\t\s}B_{\m\n}(f_{\r\l}f_{\t\s}+F_{\r\l i}{}^jF_{\t\s j}{}^i)\ ,
\eea
where $f_{\m\n}$ and $F_{\m\n i}{}^j$ are the abelian field strengths of the $A_\m$ and $A_{\m i}{}^j$, respectively. The supersymmetry transformations of the fermions are given by
\bea
\delta \psi_{\m\,i}&=& \nabla_\m\epsilon_i-\frac{1}{24}e^{\sqrt{2}\phi}\g_7\slashed{H}\g_\m\epsilon_i-\frac{1}{4\sqrt2}(\g_\m{}^{\n\r}-6\delta_\m{}^\n\g^\r)e^{-\sqrt{2}\phi}(\frac{1}{2}f_{\n\rho}\delta_i^j+\g_7 F_{\nu\rho\,i}{}^j) \epsilon_j\, ,
\nn\\
\delta\chi_i&=&\frac1{\sqrt2}\g^\m\partial_\m\phi\epsilon_i-\frac{1}{12}e^{\sqrt{2}\phi}\g_7\slashed{H}\epsilon_i+\frac{1}{2\sqrt2}\g^{\m\n} e^{-\sqrt{2}\phi}(\frac{1}{2}f_{\m\n}\delta_i^j+\g_7 F_{\m\n\,i}{}^j)\epsilon_j\ .
\eea
Coupling to an arbitrary number of vector multiplets was obtained in \cite{Giani:1984dw}.
Type IIA string on $K3$ yields $N=(1,1)$ supergravity coupled to $20$ vector multiplets. Focusing only on the NS-NS sector, the effective action up to four-derivatives has been investigated in  \cite{Liu:2019ses}. The $K3$ compactification of the eight-derivative action for type IIA string \eq{IIA} was reconstructed from the 5-pt string amplitudes, which upon the use of the lowest order equations of motion in six-dimensions, reads
\cite{Liu:2019ses}
\bea
e^{-1} {\mathcal L}_{6D, (1,1)} &=& e^{-2\phi}\alpha'\Big[\frac5{12}H^4+\frac16(H^2_{\mu\nu})^2+\frac{19}{36}(H^2)^2-8H^2_{\mu\nu}\partial^\mu\phi\partial^\nu\phi+\frac83H^2\partial\phi^2-16(\partial\phi^2)^2\Big]
\nn\\
&& + \fft{\alpha'}{16}\Big[(t_4t_4+\frac14\epsilon_4\epsilon_4)R(\Omega_+)^2+\frac16\epsilon_5\epsilon_5H^2R(\Omega_+)+\frac1{36}\epsilon_4\epsilon_4H^4
\nn\\
&& \qquad\qquad +2B_2\wedge\left(\tr R(\Omega_+)^2\Big|_{{\rm even\ in}\ B_2\, {\rm terms}}\,\right)\Big]\ ,
\label{6D11}
\eea
where in the formula above $R(\O_+)$ is a shorthand notation for the Riemann tensor defined with respect to the torsionful spin connection
\be
\Omega_{\pm\m}{}^{ ab} =\Omega_\m{}^{ab} \pm \frac12H_\m{}^{ab}\ ,
\ee
and various contractions are given by
\bea
t_4t_4R(\O_+)^2&=&R_{\m\n}{}^{\a\b}(\O_+)R^{\m\n}{}_{\a\b}(\O_+)\ ,
\nn\\
\frac14\epsilon_4\epsilon_4 R(\O_+)^2&=&R_{\m\n}{}^{\a\b}(\O_+)R_{\a\b}{}^{\m\n}(\O_+)-4R_\m{}^\a(\O_+)R_{\a}{}^\m(\O_+)+R(\O_+)^2\,
\nn\\
-\frac16\epsilon_5\epsilon_5H^2R(\O_+)-\frac1{36} \e_4\e_4 H^4&=&
-4R^{\m\n\r\s}R_{\m\r\a}H_{\n\s}{}^{\a}+4R^{\m\n}H^2_{\m\n}-\frac23RH^2
\nn\w2
&& +\frac1{18}(H^2)^2-\frac13H^4\ ,
\eea
in which we have defined
\bea
R_{\m}{}^\a(\O_+)&=& R_{\m\r}{}^{\a\r}(\O_+)\,,\quad R(\O_+)=R_{\m}{}^\m(\O_+)\ ,
\nn\\
H^2&=&H_{\m\n\r}H^{\m\n\r}\ ,\quad H_{\m\n}^2=H_{\m\r\s}H_{\n}{}^{\r\s}\,,\quad H^4=H_{\m\n\s}H_{\r\l}{}^{\s}H^{\m\r\d}H^{\n\l}{}_{\d}\ .
\label{RHH}
\eea
In arriving at the four-derivative couplings above, one focuses on the
factorized (four derivatives) $\times R^2$ terms in ten dimensions and makes use of the topological data of $K3$
\bea
&& \frac{1}{24\pi^2}\int_{K3}R\wedge R = 16\ ,
\nn\\
&& \frac{1}{36\pi^2}\int_{K3}d^4x\sqrt{g} (R_{\mu\nu\rho\sigma}R^{\mu\nu\rho\sigma}-
4R_{\mu\nu}R^{\mu\nu}+R^2) = 24\ .
\eea
Note that the Riemann squared terms, which would naively come from the reduction of the  quartic in Riemann curvature terms in $10D$, have canceled out in the tree level part of the Lagrangian. Furthermore, the tree-level terms given in \eq{6D11} are expected to be canceled by reduction of the eight-derivative action for $10D$ IIA, reconstructed from 6- and higher-point string amplitudes \cite{Liu:2019ses}. This is consistent with the expected duality between type IIA string on $K3$ and heterotic string on $T^4$ \cite{Liu:2013dna}, which maps the one-loop terms in the former to the tree level of the latter.
However, to demonstrate the full cancellation of four-derivative terms in the tree level action requires knowledge of 6-pt and higher-point amplitudes in $10D$ which are currently unavailable.

\subsection{\texorpdfstring{$N=(2,0), 6D$}{N=(2,0), 6D} supergravity from Type IIB on \texorpdfstring{$K3$}{K3}}
	
The $N=(2,0)$ supergravity and tensor multiplets are
\be
(e_\mu{}^a, \psi_{\mu +}^I, B_{\mu\nu}^{IJ} )\ ,\qquad (B_{\m\n}, \chi_{-}^I, \phi^{IJ} )\ ,
\ee
where $I=1,...,4$ is the $USp(4)_R$ vector index, the 2-form potentials $B_{\mu\nu}^{IJ}$ and the scalars $\phi^{IJ}$ are anti-symmetric, and $\pm$ denote chirality. The equations of motion of the two-derivative supergravity were given in \cite{Romans:1986er}. Its coupling to an arbitrary number of tensor multiplets was obtained in  \cite{Bergshoeff:1999db,Riccioni:1997np}. In particular,
higher derivative extensions of $(2,0)$ supergravity with or without matter couplings do not seem to be available. One may consider obtaining such extensions from the $K3$ compactification of type IIB string. At the level of Kaluza-Klein spectrum, the $K3$ compactification gives an anomaly-free model with 21 tensor multiplets. Some aspects of the reduction involving higher derivatives in which only a $(1,0)$ subsector, which furthermore keeps only the NS-NS fields $(e_\m^a, B_{\m\n}, \phi)$ are considered in \cite{Liu:2019ses}. The results for this sector resemble those in  \eqref{6D11}.

\subsection{\texorpdfstring{$N=(1,0), 6D$}{N=(1,0), 6D} off-shell \texorpdfstring{$U(1)_R$}{U(1)R} gauged supergravity and curvature-squared invariants}

The $6D$ off-shell ${\cal N}=(1,0)$ supergravity was constructed in  \cite{Bergshoeff:1985mz} and the Poincar\'e supermultiplet consists of the fields
\be
\Big(\,e^a_\m(15),\, V_{\m}^{'ij}(12),\,V_\m(5),\, B_{\m\n}(10),\, L(1),\, E_{\m\n\r\s}(5),\,\psi^i_\m(40),\,\chi^i(8)\,\Big)\ ,
\label{6dcontent}
\ee
where $i=1,2$ and $V_\m$ is a gauge field of the $R$-symmetry group $U(1)_R$, while $V^\prime_{\m ij}=V^\prime_{\m ji}$ is traceless and it has no gauge symmetry.  This multiplet is obtained by coupling the dilaton Weyl multiplet to a linear multiplet and making the gauge choices given in \cite[Eq. (3.1)]{Bergshoeff:1985mz}. The off-shell supergravity obtained in this way was also coupled to an off-shell vector multiplet
\be
\Big( A_\m(5), \lambda^i(8), Y^{ij}(3) \Big)\ ,
\ee
where $Y^{ij}$ are the triplet of auxiliary fields. The bosonic part of the  resulting Maxwell-Einstein supergravity Lagrangian is given by  \cite{Bergshoeff:1985mz}
\bea
e^{-1}{\cal L}_{ME} &=&\frac12LR+\frac12L^{-1}\partial_{\m}L\partial^{\m}L-\frac1{24}LH_{\m\n\r}H^{\m\n\r}+LV_{\m}^{'ij}V^{\prime \m}_{ij}-\frac14L^{-1}E^{\m}E_{\m} +\frac1{\sqrt{2}}E^\m V_{\m}
\nn\w2
&& + \frac12 g E^\m A_\m + Y^{ij} Y_{ij} +\frac1{\sqrt2}gL\delta^{ij}Y_{ij}-\frac14 F_{\m\n} F^{\m\n} -\frac{1}{16} \e^{\m\n\r\s\la\ta} B_{\m\n} F_{\r\s} F_{\la\ta}\ ,
\label{SG}
\eea
where
\be
H_{\m\n\r}=3\partial_{[\m}B_{\n\r]}\ ,\quad F_{\m\n} = 2\partial_{[\m} A_{\n]}\ , \quad E^{\m}=\frac1{24}e^{-1}\varepsilon^{\m\n_1\cdots\n_5}\partial_{[\n_1}E_{\n_2\cdots\n_5]}\ .
\ee
The $B$-field is inert under the Maxwell gauge transformations, and therefore, the $B\wedge F\wedge F$ term respects the gauge symmetry. The complete off-shell supersymmetry transformation rule for the supergravity multiplet can be found in  \cite{Bergshoeff:2012ax}. Here we only give the supertransformations of the fermions which take the form
\begin{align}
\delta \psi_\m^i &= \bigg( \partial_{\mu} +\frac14\omega_{\mu ab}\gamma^{ab} \bigg) \epsilon^i-\frac12{V}_\m\d^{ij}\epsilon_j +{V^\prime}_\mu{}^i{}_j\epsilon^j +\frac18 H_{\mu\nu\rho}\gamma^{\nu\rho}\epsilon^i  \ ,
\nn\\
\delta \chi^i &= \frac{1}{2\sqrt 2} \gamma^\mu
\delta^{ij}\partial_\mu L \epsilon_j -\frac14 \gamma^\mu E_\mu\epsilon^i
+\frac{1}{\sqrt2}\gamma^\mu {V^\prime}{}_\mu^{(i}{}_k \delta^{j)k} L
\epsilon_j - \frac{1}{12\sqrt2}L\delta^{ij}\gamma_{\m\n\r} H^{\m\n\r} \epsilon_j  \ ,
\nn\w2
\delta \lambda^i &=\frac18 \gamma^{\m\n} F_{\m\n}\e^i -\frac12 Y^{ij} \e_j\ .
\label{6dsusyt}
\end{align}

Using Noether procedure in which certain properties of the spin connection with $H$-torsion were utilized, the supersymmetric Riemann-squared action was obtained in \cite{Bergshoeff:1986wc}\footnote{The stated property was exploited in \cite{Bergshoeff:1987rb} where it was observed that the torsionful spin connection and the gravitino curvature transform like Yang-Mills supermultiplet.} To present the results, we first define the torsionful connection
\be
\omega_{\pm\m}{}^{ ab} =\omega_\m{}^{ab} \pm \frac12H_\m{}^{ab}\ .
\ee
The bosonic part of the supersymmetric Riemann squared invariant takes the form \cite{Bergshoeff:2012ax}.
\bea
e^{-1}\R2 &=& R_{\m\n}{}^{ab}(\omega_-)R^{\m\n}{}_{ab}(\omega_-)-2F_{\m\n}(V)F^{\m\n}(V)-
4F_{\m\n}^{ij}(V')F^{\m\n}_{ij}(V')
\nn\w2
&& +\frac14 \e^{\m\n\r\s\l\t} B_{\m\n} R_{\r\s}{}^{ab} (\omega_-) R_{\l\t ab}(\omega_-)\ ,
\label{R2}
\eea
where
\be
F(V)_{\m\n} = 2\partial_{[\m}V_{\n]} +2V_{\m}^{\prime i}{}_k V_\n{}^{\prime jk}\delta_{ij}\ ,\qquad
F^{ij}(V')_{\m\n} = 2\partial_{[\m} V_{\n]}^{\prime ij} -2\delta^{k(i} V_{[\m} V_{\n]}{}^{\prime j)}_{~k}\ .
\label{defF}
\ee
Combining the Maxwell-Einstein supergravity with $\R2$ as
\be
\cL=\cL_{ME} +\frac18 \alpha \R2\ ,
\ee
the elimination of the auxiliary field to go on-shell requires order by order in derivative solution for the vector auxiliary field, as explained in \cite{Bergshoeff:2012ax}.  At the lowest order in $\alpha$, the bosonic part of the resulting Lagrangian is \cite{Bergshoeff:2012ax}
\bea
e^{-1}\cL_{BCSV} &=& \frac12 LR-\frac14 g^2L^2
+\frac12 L^{-1}\partial_{\mu}L\partial^{\mu}L-\frac{1}{24} L H_{\mu\nu\rho}H^{\mu\nu\rho}
\nn\w2
&& -\frac14 \left(1+\frac12 \alpha g^2 \right)F_{\mu\nu}F^{\mu\nu}
-\frac{1}{16} \varepsilon^{\mu\nu\rho\sigma\lambda\tau}B_{\mu\nu}
F_{\rho\sigma} F_{\lambda\tau}
\nn\w2
&&+\frac{\alpha}{8}\Bigl[ R_{\mu\nu}{}^{ab}(\omega_-)R^{\mu\nu}{}_{ab}(\omega_-)
+\frac14 \epsilon^{\mu\nu\rho\sigma\lambda\tau}B_{\mu\nu}
R_{\rho\sigma}{}^{ab}(\omega_-)R_{\lambda\tau}{}_{ab}(\omega_-)\Bigr]\,.
\label{totalbos2}
\eea
Note that the critical coupling $\alpha=-2/g^2$ observed in  \cite{Bergshoeff:2012ax} falls into the non-unitary regime.

The off-shell Gauss-Bonnet invariant has been constructed by utilizing a gauged 3-form multiplet  \cite{Butter:2016qkx,Butter:2017jqu} in the intermediate steps. Using a gauged 3-form multiplet composed from the dilaton-Weyl multiplet, one can also obtain a new curvature-squared invariant. One can combine this invariant with $\cL_{{\rm Riem}^2}$ to form the off-shell supersymmetric Gauss-Bonnet combination  with the bosonic part given by \cite{Novak:2017wqc, Butter:2018wss}
\bea
e^{-1}\mathcal{L}_{\mathrm{GB}}&=&R_{\mu\nu\rho\sigma}R^{\mu\nu\rho\sigma}-
4R_{\mu\nu}R^{\mu\nu}+R^2+\frac{1}{6}RH^2
-R^{\mu\nu}H^2_{\mu\nu}+\frac{1}{2}R_{\mu\nu\rho\sigma}H^{\mu\nu\lambda}H^{\rho\sigma}_{\ \ \lambda}
\nn\w2
&&+\frac{5}{24}H^4+\frac{1}{144}\left(H^2\right)^2-\frac{1}{8}\left(H^2_{\mu\nu}\right)^2
-\frac{1}{4}\epsilon^{\mu\nu\rho\sigma\lambda\tau}B_{\mu\nu}R_{\rho\sigma ab}(\o_+)R_{\lambda\tau}{}^ {ab}(\o_+)\,
\nn\w2
&&+\frac12\epsilon^{\mu\nu\rho\sigma\lambda\tau}B_{\mu\nu}F_{\r\s}(V)F_{\l\t}(V)+ \epsilon^{\mu\nu\rho\sigma\lambda\tau}B_{\mu\nu}F_{\r\s}^{ij}(V') F_{\l\t ij}(V')\ ,
\label{GB}
\eea
where $H^2, H_{\m\n}^2$ and $H^4$ are as defined in \eqref{RHH}.

The bosonic part of the last four-derivative invariant we review here is given by
\bea
e^{-1} \cL_{R^2} &=&\frac{1}{16}
 \Big( R
+ \frac1{12} H_{\m\n\r} H^{\m\n\r}
+2 L^{-1}\Box L
- L^{-2} \partial_\m L \partial^\m L
+ 4 Z^\m \bar{Z}_\m
- \frac{1}{4} L^{-2} { E}^\m {E}_\m   \Big)^2
 \nn\\
&&
- \frac{1}{4} \Big(2 L^{-1} D^\m (L Z_\m)
- i L^{-1} {E}^\m Z_\m \Big)
\Big(2 L^{-1} D^\n (L \bar Z_\n)
+ i L^{-1} {E}^\n \bar{Z}_\n \Big )
\nn\\
&&
+\frac{1}{8} \ve_{\m\n\l\r\s\d} B^{\m\n} \partial^{\l} ( V^\r + \frac1{2} L^{-1} { E}^\r )
\partial^{\s} ( V^\d+ \frac1{2} L^{-1} {E}^\d )
 \nn\\
&&
- \frac{1}{2} \partial_{[\m} ( V_{\n]} + \frac1{2} L^{-1} {E}_{\n]} )
\partial^{\m} ( V^{\n} + \frac1{2} L^{-1} { E}^{\n} )
\ ,
\label{scalar-squared-Lagrangian}
\eea
where $Z_\m=V^{'12}_\m+iV^{11}_\m$. The field equations of auxiliary fields allow us to set them to zero. Once this is done, the action becomes proportional to the leading order equation of motion of $L$, and therefore it can be removed by a redefinition of the field $L$. If the auxiliary fields are not set to zero, it can be shown that when combined with two-derivative supergravity Lagrangian, the model still admits a maximally supersymmetric Minkowski vacuum, around which the spectrum consists of a massless supergravity multiplet and a massive vector multiplet which is unitary when the coefficient in front of the Ricci scalar squared is positive.

\subsection{\texorpdfstring{$N=(1,0), 6D$}{N=(1,0), 6D} on-shell curvature-squared invariants and their dualizations}

Given that $\cL_{R^2}$ on-shell can be removed by field redefinitions, and putting aside the vector multiplet coupling, let us consider the two parameter Lagrangian
\be
\cL_{\alpha,\gamma} = \cL_{SG} -\frac18 \alpha \R2 -\frac18 \gamma \cL_{GB}\ ,
\label{ML}
\ee
with $\cL_{SG} = \cL_{ME}\big|_{A=0}$ from \eq{SG}, $\R2$ from \eq{R2}, and $\cL_{GB}$ from \eq{GB}.  Going on-shell and redefining the appropriate fields as detailed in  \cite{Chang:2022urm}, the bosonic part of the resulting on-shell supersymmetric (up to order $\alpha$ and $\gamma$) Lagrangian is given by\footnote{For convention changes compared to those of  \cite{Bergshoeff:2012ax} and  \cite{Butter:2017jqu}, see  \cite{Chang:2022urm}. }
\begin{align}
e^{-1}\cL_{\alpha,\gamma} & =  e^{-2\vp} \Big[\, \frac14 R + \pd_\m \vp \pd^\m \vp - \frac1{12} H^{\m\n\rh}H_{\m\n\rh} \Big]
\nn\w2
& -\frac18  \alpha \Big[\, R^{\m\n ab}(\omega_-) R_{\m\n ab}(\omega_-) +\frac12 \ve^{\m\n\rh\s\la\ta} B_{\m\n} R_{\rh\s}{}^{ab}(\omega_-) R_{\la\ta ab}(\omega_-) \Big] \nn\w2
& -\frac18  \gamma \Big[\, R^{\m\n\rh\s} R_{\m\n\rh\s} - 4 R^{\m\n} R_{\m\n} + R^2
+ 2 R_{\m\n\rh\s} H^{\m\n,\rh\s} - 4 R^{\m\n} H^2_{\m\n} +\frac23 R H^2
\nn\w2
&\qquad + \frac{10}{3} H_4 + \frac19 ( H^2 )^2-2( H^2_{\m\n} )^2 + \frac12 \e^{\m\n\rh\s\la\ta} B_{\m\n} R_{\rh\s}{}^{ab}(\omega_+) R_{\la\ta ab}(\omega_+) \Big] \ .
\end{align}

It is useful to consider the dualization of $B_{\m\n}$ to a dual potential $C_{\m\n}$. To this end,  one adds to the Lagrangian $\cL_{\al,\gamma}$ a total derivative Lagrange multiplier term
\be
\Delta \cL (B,C) =  \frac{1}{2\times 3!} \varepsilon^{\m\n\rh\s\ta\la} H_{\m\n\rh}\partial_\s C_{\ta\la}\ .
\ee
Dualization proceeds by treating $H$ as an independent field and integrating over it. After a considerable amount of calculation, and at the end letting $g_{\m\n} \to g^\prime_{\m\n}= e^{2\vp} g_{\m\n}$ in order to pass to the string frame, to first order in $\alpha$ and $\gamma$ one finds that \cite{Chang:2022urm}
\begin{align}
e^{-1}\cL_{\al,\gamma}^{\rm dual} =&\,  e^{2\vp} \Bigg[ \frac14 R + \pd_\m \vp \pd^\m \vp - \frac1{12} G^{\m\n\rh} G_{\m\n\rh} -\frac12 ( \alpha - \gamma ) G^{\m\n\rh} \pd_\m ( \omega_\n{}^{ab} \tG_{\rh ab} )
\nn\w2
&\, +( \alpha - \gamma ) G^{\m\n\rh} \pd_\m ( \vp^\s \tG_{\n\rh\s} ) + ( \alpha + \gamma ) G^{\m\n\rh} \partial_\m ( \omega_\n{}^{ab} e_{\rh a} \vp_b )
\nn\w2
& +\frac12 ( \alpha + \gamma ) G^{\m\n\rh} \omega^L_{\m\n\rh}(\omega) - \frac18 ( \alpha + \gamma ) R_{\m\n\rh\s} R^{\m\n\rh\s} -\frac14 \gamma( - 2 R_{\m\n} R^{\m\n} + \frac12 R^2 )
\nn\w2
& -\frac14( \alpha - \gamma ) R_{\m\n\rh\s} G^{\m\n,\rh\s} +\frac12 \alpha R^{\m\n} G^2_{\m\n} - \frac{1}{12} \alpha R G^2 - ( \alpha - 3 \gamma ) R_{\m\n} \vp^\m \vp^\n
\nn\w2
&  + ( \alpha - 3 \gamma ) R_{\m\n} \vp^{\m\n} +\frac12 ( \alpha + 3 \gamma ) R \vp^2 +\frac32 \gamma R \vp^\m{}_\m - ( \alpha - \gamma ) G^2_{\m\n} \vp^\m \vp^\n
\nn\w2
& - ( \alpha + \gamma ) G^2_{\m\n} \vp^{\m\n} + (\frac56 \alpha -\frac12 \gamma ) G^2 \vp^2 +\frac13 \alpha G^2 \vp^\m{}_\m -2 ( \alpha - 3 \gamma ) \vp^{\m\n} \vp_{\m\n}\nn\w2
&  -\frac12 ( \alpha + 12 \gamma ) ( \vp^\m{}_\m )^2 +4 ( \alpha - 3 \gamma ) \vp_{\m\n} \vp^\m \vp^\n  -2 ( 2 \alpha + 9 \gamma ) \vp^2 \vp^\m{}_\m -5 ( \alpha + 3 \gamma ) ( \vp^2 )^2
\nn\w2
&+\frac16 \alpha (D_\m G_{\n\rh\s} )(D^\m G^{\n\rh\s}) -\frac12\alpha (D^\la G_{\la\m\n})(D_\ta G^{\ta\m\n}) \nn
\end{align}
\begin{align}
& -\frac23 \alpha G^{\m\n\rh} \vp^\s D_\s G_{\m\n\rh} - \alpha \vp^\rh G_{\rh\m\n} D_\la G^{\la\m\n}
\nn\w2
& -\frac{1}{12}(-3\alpha+5\gamma)  G^{\m\n,\rh\s} G_{\m\rh,\n\s} -\frac14 (\alpha-\gamma) G^2_{\m\n}G^{2\m\n} -\frac{1}{72} \gamma\left(G^2\right)^2
\nn\w2
&  -\frac16 (\alpha-\gamma) \tG^{\m\n\rh} \Big(G^2_{\m\alpha} G_{\n\rh}{}^\alpha -3 G_{\m\n}{}^\alpha \big( R_{\rh\alpha}-4\vp_{\rh\alpha} +4\vp_\rh\vp_\alpha\big)\Big)  \Bigg]\ ,
\label{genDL}
\end{align}
where
\bea
G_{\m\n\rh} &=& 3\partial_{[\m} C_{\n\rh]}\ ,\quad
\tG^{\m\n\rh} = \frac{1}{3!} \e^{\m\n\rh\s\la\ta} G_{\s\la\ta}\ ,\quad G^2_{\m\n} = G_{\m\r\s} G_\n{}^{\r\s}\ ,
\nn\w2
G^2 &=& G_{\m\n\r} G^{\m\n\r}\ ,\quad \vp_\m = \partial_\m \vp\ , \quad \vp_{\m\n}= \nabla_\m \vp_\n \ ,\quad \vp^2= \vp_\m \vp^\m\ .
\eea
While several terms can be removed by field redefinitions, such a step will modify the simple supersymmetry transformations by introducing the corresponding $\alpha$ or $\gamma$ dependent higher derivative terms. The supertransformations of $e_\m{}^a$ and $\vp$ remain the same in the dualized theory.  In the supertransformations of $\psi_\m$ and $\chi$, the duality equation \cite[Eq. (3.6)]{Chang:2022urm} needs to be used to replace $H_{\m\n\r}$ in terms of its dual. Derivation of the supertransformation of $C_{\m\n}$, which is more involved, is given in  \cite{Chang:2022urm}.

It is interesting to note that if we set $\alpha=\gamma$ in the Lagrangian above, it gives the $6D$ BdR action up to field redefinitions  \cite{Chang:2022urm}. Conversely, the dualization of the $6D$ BdR action as obtained from the dimensional reduction of the heterotic string on $T^4$ with the focus on the NS-NS sector was performed in  \cite{Liu:2013dna} where the occurrence of the Riemann-squared and Gauss-Bonnet invariants with equal coefficients was noted.

Turning to the two parameter dual Lagrangian \eq{genDL}, the question of whether the two invariants separately admit a lift to $10D$ was addressed in \cite{Chang:2022urm}, with the motivation that this would give the $10D$ analog of the $6D$ Gauss-Bonnet invariant. It was found that there are obstacles in performing such a lift. This strengthens the expectation that the BdR action in $10D$ is the unique four-derivative extension of $(1,0)$ supergravity, up to field redefinitions. Finally, we note that a superspace approach to the higher derivative extension of heterotic supergravity in $6D$, and its dualization have been analyzed in superspace in \cite{DallAgata:1997yqq}.

\subsection{\texorpdfstring{$N=(1,0), 6D$}{N=(1,0), 6D} on-shell supergravity coupling to higher derivative hypers from dimensional reduction}

 The lack of an off-shell hypermultiplet with a finite number of auxiliary fields makes the construction of its higher derivative couplings directly by a Noether procedure a formidable task simply because there are many structures one can write down, and furthermore, the variation of multitudes of structures gives rise a much larger set of independent variations\footnote{In the superconformal tensor calculus approach explained in great detail in \cite{Bergshoeff:1985mz}, the coupling of the on-shell hypermultiplets to conformal supergravity, which is off-shell by construction, is described, yielding a two-derivative action. However, the generalization of these results to higher derivative couplings has not been explored.}. Nonetheless, recently this task has been accomplished for the case of hyperscalars parametrizing the coset $Sp(n,1)/Sp(n)\times Sp(1)$ \cite{Chang:2023pss}. We shall review these results in the next subsection.

 Another approach, which we shall review below, is the construction of higher derivative matter couplings by dimensional reduction of higher derivative invariants in $10D$. In particular, the case of hyperscalars parametrizing the coset $SO(n,4)/(SO(n)\times SO(4))$  may be obtained in this way since it has been proven in \cite{Sen:1991zi} that the dimensional reduction of heterotic supergravity with gauge fields truncated to the Cartan subalgebra must exhibit at string tree level, and to all orders in $\alpha'$, a continuous $O(d, d + 16;R)$ global symmetry, related to the $O(d, d + 16;Z)$ T-duality of heterotic strings on a $d$-torus.
At the two-derivative level, and in the bosonic sector, some time ago it was shown \cite{Maharana:1992my} that reduction on $T^d$ does give an $O(d, d + 16;R)$ invariant result. More recently, it was shown that the effective action for the bosonic string, as well as the bosonic sector of the heterotic string at the four-derivative level, in the absence of Yang-Mills fields, do yield $O(d, d;R)$ invariant action upon reduction on $T^d$  \cite{Eloy:2020dko}. Soon after, the Yang-Mills were taken into account to obtain $O(d, d+16;R)$ invariant result  \cite{Ortin:2020xdm}, where, however, the fermionic sector was not considered.
The dimensional reduction of the BdR Riemann-squared action for heterotic supergravity, in the absence of Yang-Mills coupling, on $T^4$ and truncation to $(1,0)$ supersymmetry was carried out in \cite{Chang:2021tsj}, including the fermion terms and local supertransformations\footnote{In another approach, the $\alpha'$ extended double field theories   \cite{Hohm:2013jaa,Hohm:2014xsa,Marques:2015vua,Baron:2017dvb,Lescano:2021guc} were used to obtain the bosonic sector of $O(d,d)$ invariant higher derivative couplings in  \cite{Baron:2017dvb}.}.  Here shall summarize the main results of \cite{Chang:2021tsj}.

Let us consider the ordinary dimensional reduction on $T^4$. Putting hats on all the fields and indices of $10D$ fields, and decomposing the indices as $\widehat\m = (\m, \alpha)$ and $\widehat r= (r, a)$ where $\m, r=0,1,...,5$ and $\alpha, a =1,...,4$. As we truncate supersymmetry from $(1,1)$ to $(1,0)$, we take the $10D$ vielbein and the two-form potential to be
\be
\widehat{e}_{\widehat\mu}{}^{\widehat r}
= \left(
\begin{array}{cc}
e_\mu{}^r & 0 \\
0 & E_\alpha{}^a
\end{array}
\right)\ ,  \qquad \widehat{B}_{\widehat{\mu}\widehat{\nu}}
= ( B_{\mu\nu}, \ B_{\mu\alpha}=0, \ B_{\alpha\beta})\ .
\label{ve}
\ee
For the truncation of the fermions, see \cite{Chang:2021tsj}. The ansatz \eq{ve}  only gives manifest $GL(4)$ symmetry but not the expected $SO(4,4)$ duality symmetry. To uncover this symmetry, it is convenient to introduce the $SO(4,4)$ valued field
\be
W = \rho^T \left(
\begin{array}{cc}
E_a{}^\alpha & -2E_a{}^\beta B_{\beta\alpha} \\
0 & E_\alpha{}^a
\end{array}
\right) \rho\ ,\qquad    \rho= \frac{1}{\sqrt 2} { \left(\begin{array}{cc} 1& -1 \\ 1 & 1 \end{array} \right)}\ ,
\ee
which satisfies $W^T \eta W = \eta$, where  $\eta = \left(  \begin{array}{cc}
0 & 1 \\ 1 & 0  \end{array} \right)$. The Maurer-Cartan form is
\begin{equation}
W \partial_\mu W^{-1} = \left(
\begin{array}{cc}
Q_{+ \mu ab} & - P_{- \mu ab} \\
- P_{+ \mu ab} & Q_{- \mu ab}
\end{array}
\right)\ , \qquad  P_{-ab} \equiv P_{\m ab}\ ,\qquad P_{+\mu ab} := P^T_{-\mu ab} = P_{\m ba}\ ,
\label{wpq}
\end{equation}
where $Q_{\pm \mu ab} = - Q_{\pm \mu ba}$ are the composite connections associated with $SO(4)_\pm$. It is also important to note $P_{\m ab}$ transforms under $SO(4)_\pm$ as
\be
\delta P_{\m ab} = \Lambda_{+a}{}^c P_{\m\,cb} + \Lambda_{-b}{}^c P_{\m\,ac}\ .
\label{gt}
\ee
Using the ingredients summarized above, after a considerable amount of computation, one finds that the dimensional reduction of the BdR action in $10D$ on $T^4$ yields (for the bosonic part) the result \cite{Chang:2021tsj}
\bea
e^{-1}\cL &=& e^{2\varphi} \Big[ \, \frac{1}{4} R
+ g^{\mu\nu} \partial_\mu \varphi \partial_\nu \varphi
- \frac{1}{12} H_{\mu\nu\rho} H^{\mu\nu\rho} - \frac{1}{4} P_{\mu ab} P^{\mu ab}
\nn\w2
&& + H^{\m\n\rh} \big( \omega^L_{\m\n\rh}(\omega) + \omega^Q_{\m\n\rh} (Q_-)\big)
- \frac14 R_{\mu\nu mn}(\Omega_-) R^{\mu\nu mn}(\Omega_-) - \frac14  Q_{+\mu\nu ab} Q_+{}^{\mu\nu ab}
\nn\w2
&& \quad - \frac14  Q_{-\mu\nu ab} Q_-{}^{\mu\nu ab}
-D_\mu(\Gamma_+) P_{\n ab}  D^\mu (\Gamma_+) P^{\n ab}
-\frac12 Y_{\m\n} Y^{\m\n} +\frac12 Z_{\m\n ab} Z^{\m\n ba}
\nn\w2
&&  \quad -P^{\mu ab } D_\m Y_{ab}  - X^{ab} Y_{ab} \Big] \ ,
\label{h2r2}
\eea
where $\omega^Q_{\m\n\r}(Q_-)$ is the Chern-Simons form for the composite connection $Q_{-\m ab}$, $Q_{\pm\m\n ab}$ denote the standard curvature of the composite connections $Q_{\pm\m ab}$, and
\begin{align}
X_{\m\n ab} :=& P_{\m a}{}^c P_{\n cb}\ , & Y_{\m\n ab}:=& P_{\m a}{}^c P_{\n bc}\ , & Z_{\m\n ab} :=& P_{\m c a} P_\n{}^c{}_b\ ,
\nn\w2
X_{\m\n} := & \delta^{ab} X_{\m\n ab}\ , & Y_{\m\n} :=& \delta^{ab} Y_{\m\n ab} \ ,
\nn\w2
X_{ab} := & g^{\m\n} X_{\m\n ab}\ ,& Y_{ab} :=&  g^{\m\n}  Y_{\m\n ab}\ .
\label{def3}
\end{align}
A key point is that only the last two terms in the Lagrangian are not invariant under $SO(4)_+\times SO(4)_-$, but rather they break that symmetry down to the diagonal $SO(4)$ subgroup. These are removed by a field redefinition under which all but the last two terms in the Lagrangian \eq{h2r2} are invariant. The last two terms turn into the $SO(4)_+\times SO(4)_-$ invariant result $-ee^{2\vp}  Z^{\m\n ab} Z_{\m\n ab}$. Thus the now manifestly duality symmetry invariant Lagrangian is given by \cite[Eq. 6.2]{Chang:2021tsj}
\bea
e^{-1}\cL &=&  e^{2\varphi} \Big[ \, \frac{1}{4} R
+ g^{\mu\nu} \partial_\mu \varphi \partial_\nu \varphi
- \frac{1}{12} H_{\mu\nu\rho} H^{\mu\nu\rho} - \frac{1}{4} P_{\mu ab} P^{\mu ab}
+ H^{\m\n\rh} \big( \omega^L_{\m\n\rh} + \omega^Q_{\m\n\rh} \big)
\nn\w2
&&
-\frac14 R_{\mu\nu mn}(\Omega_-) R^{\mu\nu mn}(\Omega_-) - \frac14  Q_{+\mu\nu ab} Q_{+}^{\mu\nu ab} - \frac14  Q_{-\mu\nu ab} Q_{-}^{\mu\nu ab}
\nn\w2
&& \quad
-D_\mu(\Gamma_+) P_{\n ab}  D^\mu (\Gamma_+) P^{\n ab}
-\frac12 Y^{\m\n} Y_{\m\n} +\frac12 Z^{\m\n ab} \left(Z_{\m\n ba} -2Z_{\m\n ab}\right)\Big]\ ,
\eea
where $\Omega_\pm = \omega \pm H$ and $\Gamma_\pm= \Gamma \pm H$ with $\Gamma$ representing the Christoffel symbol, and $H$ is the three-form field strength $H=dB$. As shown in \cite{Chang:2021tsj}, this result can be written as
\begin{align}
e^{-1}\cL &=  e^{2\varphi} \Bigg\{ \, \frac{1}{4} R
+ g^{\mu\nu} \partial_\mu \varphi \partial_\nu \varphi
- \frac{1}{12} H_{\mu\nu\rho} H^{\mu\nu\rho} - \frac{1}{4} P_{\mu ab} P^{\mu ab}
\nn\w2
& +\Big[ H^{\m\n\rh} \big( \omega^L_{\m\n\rh}(\omega) + \omega^Q_{\m\n\rh} (Q_-)\big)- \frac14 R_{\mu\nu mn}(\omega) R^{\mu\nu mn}(\omega)
+\frac12 R_{\m\n\rh\s} H^{\m\n,\rh\s}
\nn\w2
&  +\frac12 H^2_{\m\n} H^{2 \m\n} - \frac16 H_{\m\n,\rh\s} H^{\m\rh,\n\s}  +  H^{2\m\n} \tr (P_\m P_\n^T)   +  \frac12 \tr (P_\m P^T_\n) \tr (P^\m P^{T\n})
\nn\w2
& -\frac12\tr \big(P_\m P^T_\n P^\m P^{T\n}\big) +\frac12 \tr \big( P_\m^T P^\m P^T_\n P^\n \big)  -\frac12 \tr \big(P^\m P_\m^T P^\n P_\n^T \big) \Big] \Bigg\}\ .
\label{LB3}
\end{align}
This result is given in different forms in \cite{Baron:2017dvb, Eloy:2020dko}, but after some algebra and taking into account the convention differences, the results agree. The supertransformations of the fermions after appropriate field redefinitions are given by \cite{Chang:2021tsj}
\begin{align}
\delta \psi_\mu &= D_\mu(\Omega_+) \epsilon  -\frac32 \alpha' \big[ \omega^L_{\m\n\rh}(\Omega_-) + \omega^Q_{\m\n\rh}(Q_-) \big] \gamma^{\n\rh} \e   - \alpha' P_{\n a}{}^c D_\m P^\n{}_{bc} \Gamma^{ab} \e\  ,
\nn\w2
\delta\chi
&= \frac{1}{2} \gamma^\mu \epsilon \partial_\mu  \varphi
- \frac{1}{12} H_{\mu\nu\rho} \gamma^{\mu\nu\rho} \epsilon  +\frac12 \alpha' \big[ \omega^L_{\m\n\rh}(\Omega_-) + \omega^Q_{\m\n\rh}(Q_-) \big] \gamma^{\m\n\rh} \e \ ,
\nn\w2
\delta \psi_a
&=   - \frac12 \g^\m \Gamma^b \e  P_{\m ba} - \al'\g^\m \Gamma^b \e  P_\m{}^c{}_a  Y_{bc}\ .
\label{susy3}
\end{align}
%

\subsection{Higher derivative couplings of \texorpdfstring{$N=(1,0), 6D$}{N=(1,0), 6D} on-shell supergravity to Yang-Mills and hypermultiplets by Noether procedure }

The construction of the higher derivative couplings of matter multiplets to supergravity is notoriously complicated due to the fact that many structures in the action and transformation rules are possible, and the number of independent structures that arise upon supersymmetry variations grows very rapidly. One may perform the dimensional reduction of the well-established Bergshoeff-de Roo extension of heterotic supergravity action on $T^4$ \cite{Eloy:2020dko,Chang:2021tsj}, and consistently truncate the result to $N=(1,0)$ supersymmetry which gives couplings of hypermultiplets parametrizing the quaternionic K\"ahler (QK) coset $SO(n,4)/(SO(n)\times SO(4))$.  In $K3$ compactification of heterotic string, the low energy effective theory is also a $6D$ (1,0) supergravity coupled to vector and hypermultiplets, in which the NS-NS sector gives rise to scalars which parametrize the QK coset $SO(20,4)/(SO(20)\times SO(4))$ \cite{Seiberg:1988pf,Aspinwall:1994rg,Forste:1996yd}\footnote{For a review where the YM sector is also discussed, see \cite{Forste:1996yd}. Embedding the instanton in one of the $E_8$ groups in a way that $E_8$ is completely broken, for example, gives a $N=(1,0)$ model with a single tensor multiplet, $492$ hypermultiplets, and the remaining $E_8$ gauge group. However, the nature of the QK manifold they parametrize apparently has not been determined.}.  The same coset also arises from $T^4$ compactification of the NS-NS sector of heterotic supergravity and its truncation to (1,0). Note, however, that this leaves open the question of how to construct the higher derivative couplings of more general QK manifolds allowed by $(1,0)$ supersymmetry.

In a rare such calculation that employs the Noether procedure, the higher derivative couplings of hypermultiplets that parametrize a quaternionic projective space $Hp(n)= Sp(n,1)/(Sp(n)\times Sp(1)_R)$ to $N=(1,0)$ supergravity were constructed in \cite{Chang:2023pss}. There are several details for which we refer the reader to \cite[Eqs. (5.1) and (5.2)]{Chang:2023pss} where the final results are summarized. Here we shall give the bosonic part of the action, and the supertransformations of the fermions. To explain the notations, let us begin by noting that using the $(2n+2) \times (2n+2)$ matrix $L$ of $Sp(n,1)$ as a representative of the coset, the Maurer-Cartan form can be written as
\begin{equation}
L^{-1} dL = \left(
\begin{array}{cc}
Q_a{}^b & P_a{}^B \\
P_A{}^b & Q_A{}^B
\end{array}
\right),
\label{oneform}
\end{equation}
where $ Q_{ab} = Q_{ba}, \ Q_{AB} = Q_{BA},\ P_{Ab} = -P_{bA}$  with $a,b=1,...,2n, A,B=1,2$ and
\be
P_\m^{aA} = \partial_\m \phi^\alpha\,V_\alpha^{aA}\ ,\qquad Q_\m^{AB}
= \partial_\m \phi^\alpha\, Q_\alpha^{AB}\ ,\qquad
Q_\m^{ab} = \partial_\m \phi^\alpha\, Q_\alpha^{ab}\ .
\ee
Here $\phi^\alpha$ denote the hyperscalars, $V_\alpha^{aA}$ is the vielbein, $Q_\alpha^{ab}$ and $Q_\alpha^{AB}$ are the $Sp(n)$ and $Sp(1)$ connections on $Hp(n)$. Furthermore, we have the curvatures
\begin{align}
Q_{\mu\nu a}{}^b
&:= 2 \partial_{[\mu} Q_{\nu]a}{}^b + 2 Q_{[\mu|a}{}^c Q_{|\nu]c}{}^b
= 2 P_{[\mu|a}{}^C P_{|\nu]}{}^b{}_C\ ,
\nn\w2
Q_{\mu\nu A}{}^B
&:= 2 \partial_{[\mu} Q_{\nu]A}{}^B + 2 Q_{[\mu|A}{}^C Q_{|\nu]C}{}^B
= 2 P_{[\mu|}{}^c{}_A P_{|\nu]c}{}^B\ .
\label{mc}
\end{align}
The bosonic part of the Lagrangian is given by \cite{Chang:2023pss}
\begin{align}
e^{-1}\cL = e^{2 \vp} \Big[&\, \frac{1}{4} R(\omega) + \vp_\m \vp^\m - \frac1{12} \cH_{\m\n\rh} \cH^{\m\n\rh} -\frac12 P^{\m a A} P_{\m a A}
- \frac{1}{4} \beta F_{\mu\nu}^I F^{I\mu\nu}
\nn\w2
& +\alpha \left( - \frac14 R_{\m\n}{}^{rs}(\Omega_-) R^{\m\n}{}_{rs}(\Omega_-) + Q^{\m\n AB} Q_{\m\n AB}\right)
\nn\w2
& + \gamma \left( \frac14 Q^{\m\n AB} Q_{\m\n AB} - (P^2)^{\m\n} (P^2)_{\m\n} + \frac14 (P^2)^2\right)  \Big] \ ,
\end{align}
and the supertransformations by \cite{Chang:2023pss}
\begin{align}
\del \psi_{\m A} =&\, D_\m \e_A + \frac14 \cH_{\m\n\rh} \g^{\n\rh} \e_A - 8 \al \e^B (PDP)_{\m AB} + 8 \al \e^B P_\m{}^a{}_{(A|} \EW_{a |B)} \ ,
\nn\w2
\del \chi_A =&\, \frac12 \g^\m \e_A \pd_\m \vp - \frac1{12} \cH_{\m\n\rh} \g^{\m\n\rh} \e_A + 4 \al \g^\m \e^B (PDP)_{\m AB} - 4 \al \g^\m \e^B P_\m{}^a{}_{(A|} \EW_{a |B)} \ ,
\nn\w2
\del \psi_a =&\, \g^\m \e^A P_{\m a A} + \g ( \g^\m \e^A ) P^\n{}_{a A} (P^2)_{\m\n}
\nn\w2
&\, - \frac14 \g ( \g^\m \e^A ) P_{\m a A} P^2 - \frac14 \g ( \g^{\m\n} \g^\rh \e^A ) P_{\rh a}{}^B Q_{\m\n AB} \ ,
\nn\w2
\delta A_\mu^I =&\, - \bar{\epsilon} \gamma_\mu \lambda^I \ ,
\nn\w2
\delta \lambda^I
=&\, \frac{1}{4} F_{\mu\nu}^I \gamma^{\mu\nu} \epsilon \ ,
\end{align}
where $\alpha, \beta, \gamma$ are arbitrary constants, $P^2_{\m\n} := P_\m^{aA} P_{\n aA}$, $P^2:= g^{\mu\nu} P^2_{\mu\nu}$, $(PDP)_\mu^{AB} := P^{\nu a(A} D_\mu P_{\nu a}{}^{B)}$, the hyperscalar equation of motion $\EW^{a A} :=  e^{-1} (\del \cL_0/\del \phi^\al) V^{\al a A}$ with $\cL_0$ representing the two-derivative part of the Lagrangian,  and we have the following further definitions:
\begin{align}
\cH_{\m\n\rh} &= 3 \partial_{[\mu} B_{\nu\rh]} - 6 \beta \omega^{YM}_{\m\n\rh} - 6 \alpha\,\omega^L_{\m\n\rh}
- 6 \gamma \,\omega^Q_{\m\n\rh}\ ,
\nn\w2
\omega^{YM}_{\m\n\rh} &=  \tr \Big( A_{[\m}\partial_\n A_{\rh]} + \frac23 A_{[\m} A_\n A_{\rh]}\Big)\ ,
\nn\w2
\omega^L_{\m\n\rh} &=  \tr \Big( \Omega_{-[\m}\partial_\n \Omega_{-\rh]} + \frac23 \Omega_{-[\m} \Omega_{-\n} \Omega_{-\rh]}\Big)\ ,
\nn\w2
\omega^Q_{\m\n\rh} &= \Big( Q_{\m A}{}^B \partial_\n Q_{\rho B}{}^A +\frac23 Q_{\m A}{}^B Q_{\nu B}{}^C Q_{\rho C}{}^A \Big)_{[\m\n\rho]}\ ,
\nn\w2
\Omega_{\pm \mu rs} &= \omega_{\mu rs} \pm  H_{\mu rs}\ ,
 \label{CSQ}
\end{align}
where $A_\m := A_\m^I T^I$ with $\tr (T^I T^J) = -\delta^{IJ}$, $r,s=0,1...,5$ is the Lorentz vector index, and $H=dB$.

It is instructive to truncate the $Hp(n)$ model summarized above to $Hp(1)$, in which case the QK coset is $Sp(1,1)/(Sp(1)\times Sp(1)_R)$, and compare it with the result obtained from the BdR higher derivative heterotic supergravity on $T^4$ followed by a consistent truncation to $N=(1,0)$. The result for the bosonic part of the four-derivative terms obtained by the Noether procedure described above is given by \cite{Chang:2023pss}
\begin{align}
e^{-1}\cL_{\rm Bos.}\Big|_{\al,\gamma}  =&\, e^{2\varphi} \biggl[  H^{\m\n\rh} \big( \al\, \omega^L_{\m\n\rh}(\Omega_-)  + \gamma\, \omega^Q_{\m\n\rh} \big) - \frac14 \al\,R_{\m\n rs}(\Omega_-) R^{\m\n rs}(\Omega_-)
\nn\w2
& -(2\alpha +\frac32 \gamma) (P^2)_{\mu\nu} (P^2)^{\mu\nu} + (2\alpha+\frac34\gamma) (P^2)^2 \biggr]\ .
\label{HB}
\end{align}
On the other hand, the $10D$ BdR heterotic supergravity on $T^4$  gives $N=(1,1)$ supergravity with hyperscalars parametrizing the coset $SO(4,4)/(SO(4)_+\times SO(4)_-)$. Truncation to $N=(1,0)$ was carried out in \cite{Chang:2023pss} where it was observed that while $Sp(1)_R \in SO(4)_+$, there are two distinct ways of embedding the other $Sp(1)$ factor in $SO(4)_+\times SO(4)_-$.  Each of these two embeddings gives a result consistent with the general Noether procedure result \eq{HB} for particular values of the constant parameter $\gamma$. In one case we embed $Sp(1)\subset SO(4)_+$, and  truncate by setting the $SO(4)_-$ vector index to one value, say $a=1$ \footnote{ See \eqref{wpq} for the notations of different components of Maurer-Cartan form associated with the coset $SO(4,4)/SO(4)\times SO(4)$. In the rest of this subsection,  $a, b$ label the $SO(4)$ vector indices.}. Then $P_{\mu ab} \to P_{\mu a1}\equiv P_{\mu a}$ and $Q_{-\mu ab} \to 0$, and as shown in \cite{Chang:2023pss} the bosonic part of the order $\alpha$ Lagrangian obtained from the truncation of the reduction on $T^4$ becomes
\begin{equation}
e^{-1}\cL_{\rm Bos.}\Big|_\al = \al e^{2 \vp} \Big[ H^{\m\n\rh} \omega^L_{\m\n\rh}(\Omega_-) - \frac14 R_{\m\n rs}(\Omega_-) R^{\m\n rs}(\Omega_-) + Q^2 \Big] \ ,
\end{equation}
where $Q^2\equiv Q_{\mu A}{}^B Q^\mu{}_B{}^A$ with $Q_{\m A}{}^B \equiv\frac14 Q_{+ \m a b} ( \s^{a b} )_A{}^B$. This result agrees with the general Noether result \eq{HB} for $\gamma=0$, upon using the second identity in \cite[Eq.(6.1)]{Chang:2023pss}.

In the second way of truncation, again $Sp(1)_R \subset SO(4)_+$ but the remaining $Sp(1)$ factor is now embedded into $SO(4)_-$, instead of $SO(4)_+$ considered above.
Using van der Wardeen symbols, the vector indices of $SO(4)_+$ and $SO(4)_-$ are converted into spinor indices $(A, A')$ and $(\bar{A}, \bar{A}')$ respectively, for instance,
$P_{\m ab}=\ft1{\sqrt2}(\s_a)_{AA'}(\s_b)_{\bar{A}\bar{A}'} P_\m^{A'A\bar{A}'\bar{A}}$.
It turns out that what survives the truncation is $ P_\m{}^{1' A \bar{1}' \bar{A}} =\, P_\m{}^{2' A \bar{2}' \bar{A}} \equiv - P_\m{}^{\bar{A} A}/{\sqrt 2}$. In this case, as shown in \cite{Chang:2023pss}, the bosonic part of the order $\alpha$ Lagrangian obtained from the truncation of the reduction on $T^4$ now becomes
\begin{align}
e^{-1}\cL_{\rm Bos.}\Big|_\al = \al e^{2 \vp} \Big[&\, H^{\m\n\rh} \big( \omega^L_{\m\n\rh}(\Omega_-) - 2 \omega_{\m\n\rh}^Q (Q^{AB})\big)
\nn\w1
&\, - \frac14 R_{\m\n rs}(\Omega_-) R^{\m\n rs}(\Omega_-) + (P^2)_{\m\n} (P^2)^{\m\n} + \frac12 (P^2)^2 \Big] \ .
\label{ae}
\end{align}
For notational details see \cite{Chang:2023pss}. This result agrees with the bosonic part of the Noether result for the $Hp(1)$ model displayed in \eq{HB}, for $\gamma = -2\alpha$. Thus, perhaps not surprisingly, we see that the direct Noether construction in $6D$ gives a more general result for higher derivative couplings compared to that obtained from dimensional reduction.

\subsection{\texorpdfstring{$N=(1,0), 6D$}{N=(1,0), 6D} off-shell superconformal curvature-cubed invariants from ectoplasm approach}

The curvature-cubed terms are known to be absent from
superstring effective action  \cite{Metsaev:1986yb}. There is no proof but is likely that algebraically they cannot accommodate 16 or more supercharges. However, they do exist in $6D$, ${ N}=(1,0)$ supergravity with 8 supercharges  \cite{Butter:2017jqu}. The construction was motivated by supersymmetrizing type B conformal anomalies in six dimensions. Denoting the leading term in each of the $6D$ type B anomalies as
\be
{\cal L}_1=C_{abcd}C^{aefd}C_{e}{}^{bc}_f\,,\quad {\cal L}_2=C_{ab}{}^{cd}C_{cd}{}^{ef}C_{ef}{}^{ab}\ ,\quad
{\cal L}_3=C_{abcd}(\delta_e^a\Box-4R_e^a+\frac65\delta_e^a R)C^{abcd}\ ,
\ee
where $C_{abcd}$ is the Weyl tensor, it is claimed that there exist only two superconformal invariant curvature-cubed terms  \cite{Butter:2017jqu}. Schematically, one of them is the superconformal completion of\footnote{This is not in contradiction with the result of \cite{Deser:1977yyz} where it is shown that $N=1$ supersymmetry in $4D$ does not allow a Riemann-cubed invariant. Reduction of \eqref{c3inv}  from 6 to 4 dimensions leads to curvature-cubed terms, composed of Ricci tensor and Ricci scalar but not the Riemann tensor, which can be removed by field redefinitions.}
\be
-\frac18\varepsilon^{abcdef}\varepsilon_{a'b'c'd'e'f'}C_{ab}{}^{a'b'}C_{cd}{}^{c'd'}C_{ef}{}^{e'f'}=8{\cal L}_1+4{\cal L}_2\ ,
\label{c3inv}
\ee
while the other one contains ${\cal L}_3$. Apparently $\cL_1$ and $\cL_2$ do not admit separately supersymmetric completions \cite{Butter:2016qkx}. The results are obtained by using the ectoplasm approach in off-shell superconformal superspace and the invariants are based solely on standard Weyl multiplet with the field content \cite{Butter:2016qkx}
\be
(e^a_\m,\,b_\m,\, V_{\m}^{'ij},\,V_\m,\, T^-_{\m\n\r},\, D,\, \psi^i_\m,\,\chi^i)\,.
\ee
We now briefly explain how the two curvature-cubed superinvariants were obtained by using this technique. The supersymmetric completion of ${\cal L}_1+\frac12{\cal L}_2$ utilized a primary superfield $A_\a^{ijk}$ of dimension 9/2 obeying the constraint
\be
\nabla^{(i}_{(\a} A_{\b)}^{jkl)}=0\ .
\ee
The components of the superfield $A_\a^{ijk}$ consist of the bosonic fields
\be
(S^+_{abc}{}^{ij},\, E_a^{ij},\, F,\,C_{[ab]},\,A^{ijk},\, A_a{}^{ijk})\ .
\ee
Using $A_\a^{ijk}$, a closed 6-form superfield $J$ was constructed  in \cite{Butter:2016qkx}  whose nonvanishing components are\footnote{The pair of indices $(\alpha i)$ label a symplectic Majorana spinor in $6D$, where $\alpha=1,..4$ is chirally projected spinor index, which cannot be raised and lowered, and $i=1,2$ labels the doublet of the $R$-symmetry group $Sp(1)_R$, which can be raised and lowered by the antisymmetric $\epsilon_{ij}$ and $\epsilon^{ij}$.}
\be
J_{abc\a\b\g}^{~~~~ijk},\,\quad J_{abcd\a\b}^{~~~~~~ij},\,\quad J_{abcde\a}^{~~~~~~i},\,\quad
J_{abcdef}\ .
\ee
The explicit form of these components can be found in  \cite{ Butter:2016qkx}. In particular, one has \cite[Eq. (4.15)]{Butter:2016qkx}
\be
J_{abcdef} = -\ve_{abcdef} F \ .
\ee
Using this form in \eq{ep} yields the superconformal invariant action, whose bosonic part is thus given by
\be
I_A=  \int d^6 x\, e\, F\ .
\ee
Next, substituting into this action an expression for $F$ built out of the standard Weyl multiplet, one obtains the invariant
\be
I_{C^3}=\int d^6 x\sqrt{-g}\left(\cL_1+\frac12 \cL_2 +\mbox{susy completion}\right)\ .
\ee
The supersymmetric completion of ${\cal L}_3$ utilized a different primary superfield, $B_a^{~ij}=B_a^{~(ij)}$, of dimension 3, satisfying the reality condition $(B_a^{~ij})^\star=B_{aij}$ and the constraint
\be
\nabla_\a^{(i|}B^{\b\g |jk)}=\frac{2}3\d^{[\b}_{\a}\nabla_{\d}^{(i|}B^{\g]jk) }\ .
\ee
The bosonic components of $B_a^{~ij}$ are
\be
(B_{a}^{~ij},\,C^{ijkl},\, C_{ab}^{~~ij},\, C_{ab},\, E_{a}^{~ij},\, F)\ .
\ee
Again, a closed super 6-form can be constructed using $B_{a}^{~ij}$ and takes the form  \cite{Butter:2016qkx}
\be
J=J_0+\o(S)^i_\a\wedge J^\a_{Si}+\o(K)_a\wedge J^a_K
\ee
where $\o(S)^i_\a$ and $\o(K)_a$ are the connection 1-forms associated with special supersymmetry and the special conformal symmetry, respectively. The explicit form of the 6-form $J_0$ and 5-forms
$J^\a_{Si},\, J^a_K $ were given in  \cite{Butter:2016qkx}.
The superconformal invariant action is given by the
spacetime component of the 6-form $J$, whose bosonic part takes the form
\be
I_{B}=\int d^6x\sqrt{-g}\left(F+4B_{aij}\nabla_b R(V)^{abij}+\frac23 C_{abij}R(V)^{abij}-16f^{ab}C_{ab}\right)\ ,
\ee
where in the notation of  \cite{Butter:2017jqu}, the bosonic part of $f^{ab}$ is given by
\be
f^b_a=-\frac18R^b_a+\frac1{80}\d^b_a R+\frac14\nabla^cT_{ca}^{-~b}-\frac18T^{-}_{acd}T^{cdb}-\frac1{60}\d^b_aD\,.
\ee
One can then form a composite $B_a^{ij}$ using standard Weyl multiplet and substitute to the action above resulting in the supersymmetric completion of ${\cal L}_3$, namely
\be
I_{C\Box C}=\int d^6x\sqrt{-g}\left(C_{abcd}(\delta_e^a\Box-4R_e^a+\frac65\delta_e^a R)C^{ebcd}+\mbox{susy completion}\right)\ .
\ee
If the composite $B_a^{ij}$ is constructed by using vector multiplet instead, one obtains the supersymmetric completion of $F\Box F$ denoted by $I_{F\Box F}$ .

Based on the supersymmetric completion of conformal anomaly preserving (1,0) supersymmetry, it was proposed that the conformal anomaly of (2,0) theory denoted by ${\cal A}$ can be decomposed into a combination of the (1,0) invariants   \cite{Butter:2017jqu}
\be
{\cal A} = I_{C\Box C} + \frac12 I_{C^3}+I_{F\Box F}\ ,
\ee
where $F$ is the field strength of an extra SU(2) gauge vector, and the $N=(1,0)$ gravitino multiplets
have been truncated.
It should also be interesting to obtain curvature-cubed invariant under local Poincar\'e supersymmetry. This requires coupling the superconformal curvature-cubed invariants with certain compensating matter multiplets and fixing redundant gauge symmetries, which has not been worked out so far.

\subsection{\texorpdfstring{Killing spinors in $N=(1,0), 6D$}{N=(1,0), 6D} supergravity  with higher derivatives}

The Killing spinors in the off-shell (1,0) supergravity were analyzed in  \cite{Chow:2019win, Pang:2019qwq}, and on-shell in \cite{Gutowski:2003rg}. Assuming the fields respect a U(2)$\times\mathbb{R}^2$ isometry, the most general ansatz with vanishing auxiliary fields is given by
\bea
\label{6dansatz}
ds_6^2&=&-a_1^2(r)(dt+\overline{\omega}\s_3)^2+a_2^2(r)(dz+A_{(1)})^2+b(r)^2dr^2+\frac14c^2(r)(\s_3^2+d\theta^2+\sin^2\theta d\phi^2)\ ,\nn\\
B_{(2)}&=&2P\o_2+d(r)dt\wedge dz+f_1(r)dt\wedge\s_3+f_2(r)dz\wedge\s_3\,,\quad A_{(1)}=A_0(r)dt+A_3(r)\s_3\ ,\nn\\
L&=& L(r)\,,\quad V_\m=V_\m^{'ij}=0\ ,
\eea
where in our notation $\s_3=d\psi-\cos\theta d\phi$ and $d\o_2={\rm Vol}(S^3)$. We have turned off the auxiliary fields since the solutions we are interested in have vanishing auxiliary fields. The structure of $N=(1,0)$ off-shell invariants also allows us to truncate the auxiliary fields that appear quadratically in the action. Thus the solution to the off-shell Killing spinor equations is valid beyond the two-derivative supergravity even if the auxiliary fields are set to zero. Supersymmetry of the solution requires the undetermined functions in \eqref{6dansatz} to obey certain relations that follow from the Killing spinor equations
\bea
\label{susyt}
0&=& (\partial_{\mu} +\frac14\omega_{\mu
	\a\b}\gamma^{\a\b})\epsilon^i+\frac18
H_{\mu\nu\rho}\gamma^{\nu\rho}\epsilon^i\ ,
\nn\\
0&=& \frac{1}{2\sqrt 2} \gamma^\mu
\delta^{ij}\partial_\mu L \epsilon_j  - \frac{1}{12\sqrt2}L\delta^{ij}\gamma_{\m\n\r} H^{\m\n\r} \epsilon_j \ .
\eea
For convenience, we introduce the
complex Weyl spinor
\be
\epsilon=\epsilon_1+{\rm i}\epsilon_2\,,
\ee
and assume the Killing spinor to have the form
\be
\label{ks1}
\epsilon=\Pi(r)\epsilon_0\,,
\ee
where $\epsilon_0$ is the standard Killing spinor on a round 2-sphere embedded in the $6D$ spinor obeying the projection conditions
\be
\g^{012345}\epsilon_0=-\epsilon_0\,,\quad \g^{01}\epsilon_0=-\epsilon_0\,.
\ee
Plugging the ansatz for the bosonic fields and Killing spinor into \eqref{susyt},
the necessary and sufficient conditions for the existence of a Killing spinor are given in a set of 9 equations that can be found in  \cite{Pang:2019qwq}, where they are fully solved provided the following relations are satisfied
\bea
\label{kstot}
A_0+\frac{a_1}{a_2}&=&0\ ,\quad d+a_1a_2=0\ ,\quad c^2=r^2+P\ ,\quad c=br\ ,
\nn\\
|d|Lc^2&=&r^2\ ,\quad f_1=0\ ,\quad f_2=-d\overline{\omega}\ ,\quad A_3=A_0\overline{\omega}\ ,\quad \Pi(r)=\sqrt{a_1}\ .
\eea
Thus the solution is determined up to three undetermined functions which we choose to be $d$, $\overline{\o}$ and $A_0$. Furthermore, the equation $|d|Lc^2 = r^2$ indicates that the horizon value of $L$ is determined by the near-horizon behavior of $d$.

\subsection{Exact solutions of \texorpdfstring{$N=(1,0), 6D$}{N=(1,0), 6D} higher derivative supergravities}

Next, we consider solutions of an off-shell theory described by the Lagrangian
\be
\cL = \cL_{ME}\big|_{A=0} +\frac{1}{32}\a \cL_{{\rm Riem}^2}+ \frac{1}{32} \beta \cL_{GB}\ ,
\label{LS}
\ee
with $\R2$ and $\cL_{GB}$ from equations \eq{R2} and \eq{GB}, respectively, and admitting Killing spinors discussed above. The field equations coming from $\R2$ are given in  \cite{Bergshoeff:2012ax}, and those of $\cL_{GB}$ in  \cite{Butter:2018wss}. In summary, we describe the following solutions:
\begin{itemize}
\item
There exists an AdS$_3\times S^3$ solution preserving full supersymmetry, which can be put in the form \cite{Bergshoeff:2012ax}
\bea
 &&R_{\m\n}{}^{\r\s}=-2c^2\d_{[\m}^{\r}\d_{\n]}^{\s},\quad
 R_{\m\n}{}^{\r\s}=-2c^2\d_{[\m}^{\r}\d_{\n]}^{\s},\quad L=L_0\ ,
 \nn\\
&& H_{\m\n\r}=2c\epsilon_{\m\n\r},\quad H_{mnr}=2c\epsilon_{mnr}\ .
\eea
This solution is well studied in the two-derivative case and retains the same form
when the supersymmetric Rimann-squared and Gauss-Bonnet invariants are added. Note that the constants $c$ and $L_0$ are arbitrary.

\item Next, we consider the supersymmetric rotating string solution preserving half of supersymmetry. This solution possesses a U(2)$\times\mathbb{R}^2$ isometry and is captured by the ansatz \eqref{6dansatz}. We saw in the previous section that the solution to Killing spinor equations boiled down to three undetermined functions, namely, $A_0, d$ and $\overline{\o}$. Solving the field equations coming from the Lagrangian \eq{LS} determines these functions. Up to first order in $\alpha$ and $\beta$, they are given by
\bea
&&d=(1+\frac{Q_1}{r^2})^{-1}-\frac{Q_1^2 r^2}{(Q_1+r^2)^4} \a+\frac{P^2r^2(r^2-Q_1)-2 PQ_1 r^4}{2 (P+r^2)^2 (Q_1+r^2)^3}(\b-\a)\ ,
\nn\\
&&\overline{\omega} =\frac{J}{2 r^2}+\frac{J Q_1}{2 r^2 \left(Q_1+r^2\right){}^2}(\b+\a)\ ,
\nn\\
&&A_0=(1+\frac{Q_2}{r^2})^{-1}+\frac{P Q_2 (Q_1+r^2)-2 J^2}{(P+r^2) (Q_1+r^2)^2 (Q_2+r^2)^2}r^2 (\b+\a)\ .
\eea
The near horizon geometry of the solution is extremal BTZ$\times S^3$, in other words, locally AdS$_3\times S^3$, with full supersymmetry.   Plugging the above solution back to  \eqref{kstot}, one finds that near the horizon, namely as $r \to 0$, the value of $L$ is given by
\be
L\quad  \to  \quad    L_0=\frac{Q_1+\frac12(\a+\b)}{P}\ .
\ee

\item Non-supersymmetric solutions of the form $M_1\times M_2$ were studied in  \cite{Bergshoeff:2012ax}, with $\b=0$. These solutions either have no fluxes turned on, or they have either the 2-form or 3-form fluxes turned on. The split of $6D$ spacetime is of the form $4+2$, $3+3$ or $2+4$ in these solutions. Non-supersymmetric dyonic black string solutions with curvature-squared corrections originating from $K3$ compactification of IIA string theory have also been constructed
in \cite{Ma:2021opb,Ma:2022nwq}. These solutions possess interesting applications
in Weak Gravity Conjecture \cite{Harlow:2022ich}. For instance, in \cite{Ma:2022gtm}, it was found that the leading higher derivative corrections to the rotating dyonic black string entropy at fixed conserved charges can be negative, which contradicts the standard expectation.

\item Let us next consider a half-BPS solution of the Riemann-squared extension of the $U(1)_R$ gauged $N=(1,0)$ supergravity, which has the Lagrangian \eqref{totalbos2}. The solution has the form ${\rm Mink}_4 \times S^2$ with a constant dilaton, and a nonvanishing 2-form flux as follows \cite{Bergshoeff:2012ax}
\bea
&&R_{\m\n}=0,\quad R_{mn}=\frac12 g^2L_0 g_{mn},\quad L=L_0\ ,
\nn\\
&&F_{mn}(W)=\pm\frac{gL_0}{\sqrt{2}}\epsilon_{mn},\quad F_{mn}(V)=\mp\frac12g^2L_0\epsilon_{mn}\ .
\eea
The spectrum of perturbations around this vacuum was studied in  \cite{Pang:2012xs}.

\section{\texorpdfstring{$D=5$}{D=5}}

In this section, we shall review off-shell Poincar\'e supergravity and its coupling to vector multiplets, the curvature-squared invariants, as well as the procedure for going on-shell. We shall also discuss the off-shell Killing spinors and exact solutions in the presence of off-shell curvature-squared invariants. A convenient method for constructing these invariants in component expressions turns out to be the superconformal tensor calculus. Thus in this section, we will mainly focus on results obtained from this approach. Readers interested in superspace expressions of curvature-squared invariants are referred to \cite{Butter:2014xxa, Gold:2023dfe, Gold:2023ykx}.  In the case of constructing a Riemann-squared invariant,  a  convenient trick is to dimensionally reduce the Yang-Mills multiplet on a circle from $6D$, and then using the analogy between supergravity and Yang-Mills \cite{Bergshoeff:2011xn}. Invariants with higher than four derivatives are discussed in section \ref{lastsection}.

\subsection{Superconformal approach}
		
In five dimensions with ${ N}=2$ supersymmetry, it is possible to realize curvature-squared invariants off-shell. Superconformal tensor calculus, which is based on the exceptional superalgebra $F^2(4)$ \cite{VanProeyen:1999ni}\footnote{In the notation of \cite{VanProeyen:1999ni}, $F^p(4)$ is a real form of $F(4)$ with bosonic subalgebra $SO(7-p, p) \oplus SU(2)$.} has been one of the main techniques for the construction of the five dimensional higher derivative models \cite{Bergshoeff:2001hc}. It is important to note that on dimensional grounds, in five dimensions the curvature-squared terms do not have the right dimension to be scale invariant on their own, and therefore their constructions require a compensating scalar field with scaling dimension 1. There exist two types of off-shell Weyl multiplets in ${ N}=2,\, 5D$ supergravity  with the following field contents:
\begin{eqnarray}
\text{Standard Weyl Multiplet} &:&  \{e_\mu{}^a\ , \psi_\mu^i\ , V_{\mu}{}^{ij}\ , T_{ab} \ , \chi^i \ , D\ , b_\mu \} \ ,
\nn\\
\text{Dilaton Weyl Multiplet} &:&  \{e_\mu{}^a\ , \psi_\mu^i\ , V_{\mu}{}^{ij}\ ,C_\mu, B_{\mu\nu} \ , \psi^i \ , \sigma\ , b_\mu \}\ .
\label{5DN2WeylMultiplets}
\end{eqnarray}
Here, $e_\mu{}^a$ is the f\"unfbein, $b_\mu$ is the gauge field for dilatation, $V_\mu{}^{ij}$ is the
$SU(2)$ $R$-symmetry gauge field and $\psi_\mu^i$ is the gravitino, which gauges the $Q-$supersymmetry. These fields are common to both Weyl multiplets. The matter content of the standard Weyl multiplet includes a real auxiliary scalar $D$, an anti-symmetric tensor $T_{ab}$ and a symplectic Majorana spinor $\chi^i$.
For the dilaton Weyl multiplet, the matter content is given by a vector field $C_\mu$, a two-form gauge field $B_{\mu\nu}$, a dilaton field $\sigma$ and a dilatino field $\psi^i$. While the dilaton Weyl multiplet has a scalar field $\s$ which can be used as a compensator, the Standard Weyl multiplet lacks such a field (the conformal weight of $D$ is 2). Thus, the curvature-squared models may not need an extra compensating multiplet when the dilaton Weyl multiplet is utilized. However, a vector multiplet, which contains a weight-1 scalar field $\rho$ is essential if the Standard Weyl multiplet is used. Finally, as shown in \cite{Coomans:2012cf} that while gauged supergravity models based on the standard Weyl multiplet can be constructed by combining certain off-shell models \cite{Ozkan:2013nwa}, the same procedure cannot be accomplished in the dilaton Weyl basis without extending it with an additional matter content given by the linear multiplet
\be
\mbox{Linear multiplet:} \quad  \{L_{ij}\ , E_{a}\ , N\ , \vf^i\}\ .
\ee
In the rest of this section, we will investigate the five-dimensional $N=2$ higher derivative invariants separately, depending on the choice of the Weyl multiplet.

\subsection{\texorpdfstring{$N=2, 5D$}{N=2, 5D} off-shell invariants from the standard Weyl multiplet}

\subsubsection{Off-shell (and on-shell) Poincar\'e supergravity and couplings to vector multiplets}

 The off-shell Poincar\'e multiplet can be constructed by coupling the standard Weyl multiplet to a vector multiplet and a linear multiplet, and fixing the redundant gauge symmetry. It has the field content \cite{Bergshoeff:2001hc,Coomans:2012cf}
\begin{table}[H]
\centering
\begin{tabular}{|c|ccccccccccccc|}
\hline
Field & $e^a_\m$  & $V_\m^{'ij} $ & $V_\m$ & $T_{ab}$ & $E_\mu$ & $D$ & $N$ & $A_\m$ & $Y_{ij}$ & $\r$&  $\psi^i_\m$ & $\chi^i$& $\l^i$ \\
\hline
D.o.fs  &   10   &  10   &  4   &   10   & 4  &  1  & 1 & 4 & 3 & 1 & 32 &8&8 \\
\hline
\end{tabular}
    \caption{The field content of $N=2, 5D$ off-shell  Poincar\'e supergravity multiplet based
    on the standard Weyl multiplet.  The degree of freedom count is off-shell.}

\end{table}

In the case of two-derivative supergravity, the on-shell multiplet fields are simply $(e_\mu{}^a, \psi_\mu^i, A_\mu)$. Note that it is not known how to construct a healthy two-derivative supergravity model based on the standard Weyl multiplet in six dimensions \cite{Bergshoeff:1985mz,Coomans:2011ih}. Thus, the off-shell matter content provided here is not related to the six-dimensional field content \eqref{6dcontent}. The off-shell multiplet can be coupled to $n$ off-shell vector multiplets, each containing the fields $(A_\mu, \lambda^i,\rho, Y^{ij})$.  Defining $(\rho^I,\, A_\mu^I,\, \lambda^{Ii},\,Y_{ij}^I)$ where  $I=0,1,...,n$ with ``$0$" representing the fields $\{\rho^0 \equiv \rho, A_\mu^0 \equiv A_\mu\,, \lambda^{0 i} \equiv \lambda^i\,, Y_{ij}^0 \equiv Y_{ij}\}$ coming from the off-shell Poincar\'e multiplet, the bosonic part of the off-shell gauged Poincar\'e supergravity  coupled to $n$ vector multiplets is given by  \cite{Ozkan:2013nwa}
\bea
e^{-1} \cL_{0,S} &=& \frac18 ({\mathcal C}+3)R + \frac13 (104 {\mathcal C} - 8) T^2 + 4 ({\mathcal C}-1)D - N^2- E_\m E^\m + V_\m^{'ij} V^{'\m}_{ij}
\nn\\
&&- \sqrt2 V_\m E^\m+ \frac34 C_{IJK} \r^I F_{\m\n}^J F^{\m\n\, K}+ \frac32 C_{IJK} \r^I \partial_\m \r^J \partial^\m \r^K
\nn\\
&& - 3 C_{IJK} \r^I Y_{ij}^J Y^{ij\,K}- 12 C_{IJK} \r^I \r^J F_{\m\n}^K T^{\m\n} + \frac18 \e^{\m\n\r\s\l} C_{IJK} A^I_\m F_{\n\r}^J F_{\s\l}^K\nn\\
&&    - \frac3{\sqrt2} g_I Y^I_{ij} \d^{ij} - 3 g_I E^\m A_\m^I - 3 g_I \r^I N \ ,
\label{SWSUGRA}
\eea
where $I = 0,1,\ldots n$ and $T^2 := T_{ab}T^{ab}$. The constant coefficient $C_{IJK}$ is symmetric in $I,J,K$  and determines the coupling of $n$ vector multiplets, $\mathcal{C} \equiv C_{IJK} \rho^I \rho^J \rho^K$, $E_\mu$ is a constrained vector $\nabla^\mu E_\mu = 0$, $N$ is an auxiliary scalar  and $V_{\mu}^{\prime ij}$ is the traceless part of the $SU(2)$ $R$-symmetry gauge field $V_{\mu}{}^{ij}$. Note that the $SU(2)$ $R$-symmetry is broken to $U(1)$ due to the gauge fixing from superconformal to super-Poincar\'e, and it is gauged by $V_\mu$ defined by
\bea
V_\mu^{ij} = V_\mu^{\prime ij} + \frac12 \delta^{ij} V_\mu \,, \qquad \text{with} \qquad  V_\mu^{\prime ij} \delta_{ij} = 0 \,.
\label{splitV}
\eea
The supersymmetry transformation rules of the fermionic fields are given by
\begin{eqnarray}
 \d \psi_\m^i   &=& (\partial_\m +\frac{1}{4}\omega_\m{}^{ab}\g_{ab})\e^i - V_\m^{ij}\e_j  + \rmi \g\cdot T \g_\m
\e^i - \rmi \g_\m
\eta^i  \nn\, ,\\
\d \chi^i &=&  \frac14 \e^i D - \frac1{64} \g \cdot {R}^{ij}(V) \e_j + \frac18 \rmi \g^{ab}\slashed{D}T_{ab}\e^i - \frac18 \rmi \g^a D^b T_{ab} \e^i \nn\\
\d\l^{iI} &=&- \frac14 \g \cdot {F}^I \e^i -\frac12\rmi
\slashed{D}\r^I \e^i + \r^I \g \cdot T \e^i - Y^{ij I} \e_j +
\r^I \eta^i \ ,
\label{KSSW}
\end{eqnarray}
where the parameter $\eta^i$ is defined in terms of the $Q$-supersymmetry parameter $\epsilon^i$ as
\begin{eqnarray}
\eta_k &=& \frac13 \Big( \g \cdot T \e_k - \frac1{\sqrt2} N \d_{ik} \e^i + \frac1{\sqrt2} \rmi \slashed{E} \d_{ik} \e^i + \rmi \g^a V_a^{'(i}{}_l \d^{j)l} \d_{ik} \e_j \Big)  \label{ETA} \,.
\end{eqnarray}
The supercovariant curvatures and composite fields are defined as
\begin{eqnarray}
R_{\m\n}{}^{ij}(V)&=&2\partial_{[\m} V_{\n]}{}^{ij} -2V_{[\m}{}^{k( i} V_{\n ]\,k}{}^{j)} - 3\rmi {\bar\phi}^{(i}_{[\m}\psi^{j)}_{\n]} - 8 \bar{\p}^{(i}_{[\m}
\g_{\n]} \chi^{j)} - \rmi \bar{\p}^{(i}_{[\m} \g\cdot T \psi_{\n]}^{j)} \,,\nn\\
D_\m T_{ab}&=&\partial_\m T_{ab} -2\omega_\mu{}^c{}_{[a}T_{b]c}-\frac{1}{2}i\bar{\p}_\m\g_{ab}\chi+\frac{3}{32}i\bar{\p}_\m {R}_{ab}(Q)\,,\nn\\
F_{\m\n}^I &=& 2 \partial_{[\mu} A_{\nu]}^I - \bar{\p}_{[\m} \g_{\n]} \l^I + \frac 12 \rmi
\r^I \bar{\p}_{[\m} \p_{\n]}\,,\nn\\
D_\mu \r^I &=& \partial_\mu \r^I - \frac12\, \rmi\bar{\psi}_\mu \l^I \,.\nn\\
R_{\mu\nu}{}^i(Q) &=& 2\partial_{[\m}\p_{\n]}^i+\frac{1}{2}\o_{[\m}{}^{ab}\g_{ab}\p_{\n]}^i-2V_{[\m}^{ij}\p_{\n] j}-2i\g_{[\m}\phi_{\n]}^i+2i\g\cdot T\g_{[\m}\p_{\n]}^i\,, \nn\\
\phi_\mu &=& \frac13\rmi \g^a {R}^\prime _{\m a}{}^i(Q) - \frac1{24}\rmi
\g_\m \g^{ab} {R}^\prime _{ab}{}^i(Q)\,,
\end{eqnarray}
where we used the notation ${R}_{ab}'(Q)$ to indicate that the expression is obtained from ${R}_{ab}(Q)$ by omitting the $\phi_\m^i$ term. The supersymmetry transformation rules for the bosonic fields can be found by imposing a gauge fixing condition in superconformal transformation rules \cite[Eqs. (2.1), (2.11) and (2.14)]{Ozkan:2013nwa}. See also in  \cite[Eqs. (6.2) - (6.4) ]{Ozkan:2013nwa} for the gauge fixing condition and the decomposition of the superconformal transformation parameters.

This construction of Poincar\'e supergravity utilizes the vector and the linear multiplets. Alternatively, one may replace the linear multiplet with a hypermultiplet, which gives rise to an off-shell supergravity with a different field content \cite{Hutomo:2022hdi}. However, there is no known off-shell higher curvature model with a hypermultiplet compensator, therefore, this option will not be discussed here. We refer to \cite{Fujita:2001kv,Kugo:2000hn,Bergshoeff:2002qk} for the technical details of the construction of five-dimensional ${N} = 2$ off-shell supergravity with vector and hypermultiplets.

If we consider the Lagrangian $\cL_{0,S}$  by itself, going on-shell amounts to eliminating the auxiliary fields by means of their algebraic equations of motion. These are given by (see \cite{Lauria:2020rhc} for a review).
\bea
0&=& V^{'ij}_\m
\nn\\
0&=& V_\m + \frac{3}{\sqrt2} g_I A_\m^I
\nn\\
0 &=& {\cal C} - 1,
\nn\\
0 &=& \frac23 (104 {\cal C} - 8) T_{ab} - 4 c_I F_{ab}^I,
\nn\\
0 &=& 2N + 3 g_I \r^I,
\nn\\
0 &=& {\cal C}_{IJ} Y_{ij}^J + \frac3{\sqrt2} g_I \d_{ij}.
\label{aux}
\eea
where
\bea
\mathcal{C} = C_{IJK} \r^I \r^J \r^K, \quad  \mathcal{C}_{IJ} = 6 C_{IJK} \r^K\ .
\eea
The equation ${\cal C}= C_{IJK} \rho^I \rho^J\rho^K=1$ defines the so-called Very Special Real manifold; for a textbook exposition, see \cite{Lauria:2020rhc}.
Substituting the results above into $\cL_{0,S}$ gives the following result, which agrees with that obtained long ago by the Noether procedure \cite{Gunaydin:1983bi,Gunaydin:1984ak}, and reads
\begin{eqnarray}
\cL_{0,S}^{\rm on-shell} &=& \frac12 R + \frac18 (\mathcal{C}_{IJ} - \mathcal{C}_I \mathcal{C}_J) F_{\m\n}^I F^{\m\n\,J} + \frac14 \mathcal{C}_{IJ} \partial^\m \r^I \partial^\m \r^J
\nn\\
&& + \frac18 \e^{\m\n\r\s\l} C_{IJK} A^I_\m F_{\n\r}^J F_{\s\l}^K -V\ ,
\end{eqnarray}
where
\be
V= -\frac94 (g_I \r^I)^2 - \frac92 \mathcal{C}^{IJ} g_I g_J\ ,\qquad \mathcal{C}_I = 3 C_{IJK} \r^J \r^K,
\label{potential1}
\ee
and  $\mathcal{C}^{IJ}$ is the inverse of $\mathcal{C}_{IJ}$. Note that $\r^I$s are no longer independent fields due to the constraint ${\cal C}=1$. In summary, this Lagrangian describes on-shell $N=2,5D$ supergravity coupled to $n$ vector multiplets.

 The on-shell model can be truncated to a minimal model by setting $\rho^I = \bar\rho^I$=const,  $A_\m^I = \bar\rho^I A_\mu $ and $g_I=\ft16g\,{\cal C}_{IJ}\bar\r^J$ with $C_{IJK}\bar\r^I\bar\r^J\bar\r^K=1$. In this case, we obtain the standard minimal on-shell supergravity in five dimensions
\begin{eqnarray}
 e^{-1} \cL_{\rm min} = \frac12 R - \frac38 F_{\m\n} F^{\m\n} + \frac18 \e^{\m\n\r\s\l} A_\m F_{\n\r} F_{\s\l} + 3 g^2 \ .
 \label{Minimal5DSugra}
\end{eqnarray}

\subsubsection{Off-shell curvature-squared invariants and going on-shell}
\label{SWR2}

When the standard Weyl multiplet is utilized, the off-shell models that have been constructed so far are the Weyl-squared  \cite{Hanaki:2006pj} and the Ricci scalar-squared  \cite{Ozkan:2013nwa} invariants. The bosonic part of the Weyl-squared action is given by  \cite{Hanaki:2006pj}
\bea
e^{-1} \cL_{W^2} &=&  c_{2I} \Big[ \frac18\r^I {C}^{\m\n\r\s} {C}_{\m\n\r\s}+ \frac{64}3 \r^I {D}^2 + \frac{1024}9 \r^I {T}^2 {D} - \frac{32}3 {D} \, {T_{\m\n}} F^{\m\n\,I}   \nn\\
&&  - \frac{16}3 \r^I {C}_{\m\n\r\s} {T}^{\m\n} \, {T}^{\r\s} + 2{C}_{\m\n\r\s} {T}^{\m\n} F^{\r\s\,I} + \frac1{16} \e^{\m\n\r\s\l}A_\m^I {C}_{\n\r\t\d} {C}_{\s\l}{}^{\t\d}    \nn\\
&& -\frac1{12} \e^{\m\n\r\s\l} A_\m^I {V}_{\n\r}{}^{ij} {V}_{\s\l\, ij} +  \frac{16}3 Y^I_{ij} {V}_{\m\n}{}^ {ij} {T}^{\m\n} - \frac{1}3 \r^I {V}_{\m\n}{}^{ij}{V}^{\m\n}{}_{ij} \nn\\
&& +\frac{64}3 \r^I  \nabla_\n T_{\m\r} \nabla^\m T^{\n\r} - \frac{128}3 \r^I {T_{\m\n}} \nabla^\n \nabla_\r {T}^{\m\r} - \frac{256}9 \r^I R^{\n\r} T_{\m\n} T^\m{}_\r \nn\\
&& + \frac{32}9 \r^I R T^2 - \frac{64}3  \r^I \nabla_\m T_{\n\r}\nabla^\m T^{\n\r} + 1024 \r^I \, {T}^4- \frac{2816}{27} \r^I  ({T}^2)^2   \nn\\
&&- \frac{64}9 {T_{\m\n}} F^{\m\n\,I} {T}^2 - \frac{256}3 {T_{\m\r}} {T}^{\r\l} {T_{\n\l}} F^{\m\n\,I}  - \frac{32}3   \e_{\m\n\r\s\l}  {T}^{\r\t} \nabla_\t {T}^{\s\l} F^{\m\n\,I}  \nn\\
&& - 16   \e_{\m\n\r\s\l} {T}^\rho{}_\t \nabla^\s {T}^{\l\t} F^{\m\n\,I}  - \frac{128}3 \r^I \e_{\m\n\r\s\l} {T}^{\m\n} {T}^{\r\s} \nabla_\t {T}^{\l\t}\Big] \ ,
\label{pregb}
\eea
where $c_{2I}$ are arbitrary constants and the five-dimensional Weyl tensor reads
\bea
C_{\m\n\r\s} &=& R_{\m\n\r\s} - \frac13 (g_{\m\r} R_{\n\s} - g_{\n\r} R_{\m\s} - g_{\m\s} R_{\n\r} + g_{\n\s} R_{\m\r})
\nn\w2
&& + \frac{1}{12} (g_{\m\r} g_{\n\s} - g_{\m\s} g_{\n\r}) R\ .
\eea
We have also introduced the following notations,
\bea
T^4 \equiv T_{ab} T^{bc} T_{cd} T^{da}, \qquad (T^2)^2 \equiv (T_{ab} T^{ab})^2.
\eea
The bosonic part of the off-shell Ricci-scalar squared action is given by  \cite{Ozkan:2013nwa}
\bea
e^{-1} \cL_{R^2} &=& c_I \Big( \r^I Y_{ij} Y^{ij} + 2 \r Y^{ij} Y_{ij}^I-\frac18\r^I \r^2 R - \frac14 \r^I  F_{\m\n} F^{\m\n} - \frac12 \r\, F^{\m\n} F_{\m\n}^I  \nn\\
&& \quad+ \frac12 \r^I \partial_\m \r \partial^\m \r + \r^I \r \Box \r - 4 \r^I \r^2 (D + \frac{26}3 T^2) + 4 \r^2 F_{\m\n}^I T^{\m\n} \nn\\
&& \quad + 8 \r^I \r\, F_{\m\n} T^{\m\n} - \frac18 \e_{\m\n\r\s\l} A^{\m I} F^{\n\r} F^{\s\l} \Big)\ ,
\label{R2SW}
\eea
where $(\r, Y^{ij}, F_{\m\n})$ represent the following composite expressions
\bea
\r &=& 2N,
\nn\\
Y^{ij} &=&  \frac1{\sqrt2} \d^{ij}\Big(- \frac3{8} R -  N^2 - E^a E_a + \frac83 T^2 + 4D - V_{a}^{'kl} V_{kl}^{'a} \Big)
\nn\\
&& + 2 E^a V'_{a}{}^{ij}  - \sqrt2 \nabla^a V'_{a}{}^{m(i} \d^{j)}{}_{m} ,
\nn\\
F_{\m\n} &=& 2\sqrt{2} \partial_{[\m} \Big( V_{\n]} + \sqrt2 E_{\n]} \Big) \,.
\label{fixedmap}
\eea	
For the time being,  an off-shell Gauss-Bonnet invariant coupled to $n$ vector multiplets in the standard Weyl basis has not been constructed.

It is worthwhile to note that the model
\be
\cL_{SG+\frac16 W^2} = \frac1{16\pi G_5}{\cal L}_{SG}|_{g=0}+\frac{1}{96\pi^2}\cL_{W^2}
\label{Memb}
\ee
is expected to arise from M-theory compactified on Calabi-Yau threefold where $C_{IJK}$ denotes the triple intersection number of 4-cycles and $c_{2I}$s are the second Chern class numbers of the Calabi-Yau threefold.

The Weyl-squared action \eqref{pregb} has been utilized in AdS/CFT correspondence and black hole physics. In particular, it was found that \cite{Cremonini:2009sy} the four-derivative interactions modify the value of $\eta/s$, namely the ratio of shear viscosity to entropy density, such that the classical bound $\eta/s\ge\frac{1}{4\pi}$ is violated. Applications of the Weyl-squared action \eqref{pregb} in black hole physics will be discussed in section \ref{5dapp}.

We conclude this subsection with remarks on how to obtain the on-shell versions of the higher derivative invariants added to $\cL_{0,S}$.  Schematically, the off-shell action  takes the form
\bea
S_{\text{off-shell}} [\phi] &=& S_0 [\phi] + \a S_1 [\phi] \ .
\label{expS}
\eea
where $S_0$ and $S_1$ denote the two- and four-derivative actions, respectively, and $\alpha$ is a constant with mass dimension -2.  It follows that the auxiliary field equations must receive corrections proportional to $\a$. The solution to those equations can be expressed in terms of a series expansion in $\a$
\bea
\phi &=& \phi_0 + \a \phi_1 + \a^2 \phi_2 + \cdots \ ,
\eea
where $\phi_0$  is the solution to the zeroth order equation given in the previous section. As a consequence, the on-shell action has the form
\bea
S_{\text{on-shell}} [\phi] &=& S_0 [\phi_0] + \a \Big( S_1 [\phi_0]+\phi_1 S'_0[\phi_0]\Big)+ \cdots\,.
\eea
In the above equation, $S'_0[\phi_0]=0$ when $\phi_0$ is an auxiliary field or a Lagrangian multiplier. Thus, we eliminate the auxiliary fields by plugging their zeroth order solutions into the action. For the Lagrangian $\cL_{0,S}$, these equations are listed in \eqref{aux}.

\subsection{\texorpdfstring{$N=2, 5D$}{N=2, 5D} off-shell invariants from the dilaton Weyl multiplet}

\subsubsection{Off-shell gauged Poincar\'e supergravity coupled to vector multiplets and going on-shell}

The off-shell Poincar\'e multiplet based on the dilaton Weyl multiplet has the field content   \cite{Bergshoeff:2001hc}

\begin{table}[hbtp]
\centering
\begin{tabular}{|c|cccccccccc|}
\hline
Field & $e^a_\m$  & $V_\m^{'ij} $ & $V_\m$ &  $L$ &   $C_\m$ & $B_{\m\n}$ & $E_\mu$ & $N$ &  $\psi^i_\m$ & $\varphi^i$ \\ \hline
D.o.fs  &   10   &  10   &  4   &   1   & 4  &  6  & 4 & 1 & 32 &8 \\
\hline
\end{tabular}
\caption{The field content of $N=2, 5D$ off-shell  Poincar\'e supergravity multiplet obtained from the dilaton Weyl multiplet.  The degree of freedom count is off-shell.}

\end{table}
As we shall see below, on-shell this multiplet is reducible, consisting of the supergravity multiplet $(e_\m{}^a, \p_\m^i, C_\m )$, and a vector multiplet $(\tilde{C}_\m, \vf^i, L)$ where $d\tilde C$ is dual to $dB$.
Unlike the standard Weyl multiplet, the Poincar\'e supergravity multiplet obtained from the dilaton Weyl multiplet originates from its six-dimensional counterpart. The bosonic part of the off-shell, gauged Poincar\'e supergravity coupled to $n$ off-shell vector multiplets, $(A_\mu^I, \rho^I, \lambda^{iI}, Y^{I ij})$, where $I=1,...,n$ (unlike the index $I$ used in the previous subsection), is given by \cite{Ozkan:2013nwa}\footnote{ Generalization to Yang-Mills couplings has been obtained \cite{Bergshoeff:2011xn} in the absence of $U(1)_R$ gauging, i.e. for $g=0$.}
\bea
e^{-1} \cL_{0,DW} &=&  L\left(R-\frac12G_{ab}G^{ab}-\frac13H_{abc}H^{abc}+2V_{a}^{'ij}V^{'a}_{ij}\right) +L^{-1}\partial_{a} L\partial^{a} L-2L^{-1}E_{a}E^{a}
	\nonumber\\
	&&-2\sqrt{2}E_a V^a -2N^2L^{-1}  -4g C_{a}E^{a}-2gNL-4gN-\frac12 g^2L^3+2g^2L^2 \nn\\
 && + a_{IJ} \Bigg[ Y_{ij}^I Y^{J ij } - \frac12 D_\m \r^I D^\m \r^J -\frac14 (F_{ab}^I - \r^I G_{ab}) (F^{ab J} - \r^J G^{ab}) \nn\\
&&  - \frac18 \e^{abcde} (F^I_{ab} - \r^I G_{ab})(F_{cd}^J - \r^J G_{cd}) C_e  - \frac12 \e^{abcde} (F_{ab}^I - \r^I G_{ab}) B_{cd} \nabla_e \r^J \nn\\
&& - \frac{1}{\sqrt{2}} g \r^I Y_{ij}^J \d^{ij} - g \r^I \r^J N - \frac14 g^2 \r^I \r^J L^2 \Bigg]\ ,
 \label{DWMultiVector}
\eea
where the bosonic part of supercovariant curvatures are defined as
\be
F_{\mu\nu}^I = 2\partial_{[\mu} A^I_{\nu]}\ ,\qquad G_{\mu\nu} = 2 \partial_{[\mu} C_{\nu]},\qquad H_{\mu\nu\rho} = 3 \partial_{[\mu} B_{\nu\rho]} + \frac{3}{2} C_{[\mu} G_{\nu\rho]} +\frac12 g E_{\m\n\r}.
\ee
The  constrained vector $E_a$ and the three-form gauge field $E_{abc}$ are related to each other via $E_a = -\frac{1}{12} \e_{abcde} \nabla^b E^{cde}$. The supersymmetry transformation rules for the fermionic fields, up to cubic fermions, are given by
\begin{eqnarray}
  \delta \psi_\mu^i &=& \left( \partial_\mu  + \frac{1}{4} \omega_{\mu -}{}^{ab} \gamma_{ab} \right) \epsilon^i  - V_\m{}^{ij} \e_j  - \frac12 \rmi {G}_{\mu\nu} \gamma^\nu \epsilon^i + \frac{1}{2\sqrt{2}} g L \gamma_{\m} \delta^{ij} \e_j \,,
 \nn\\
 \delta \varphi^i &=& - \frac1{2\sqrt2} \rmi \slashed{\partial} L \d^{ij} \e_j - \frac1{\sqrt2} \rmi V'_{\m}{}^{(i}{}_k \d^{j)k} L \e_j  - \frac12 \rmi \slashed{E} \e^i  + \frac12 N \e^i + \frac1{4\sqrt2} L \g \cdot {G} \d^{ij} \e_j  \, \nn\\
&&  - \frac1{6\sqrt2} \rmi L \g \cdot {H} \d^{ij} \e_j + \frac34 g L^2 \e^i \,,
\nn\\
\d\l^{iI} &=&- \frac14 \g \cdot {F}^I \e^i -\frac12\rmi
\slashed{D}\r^I \e^i + \r^I \g \cdot T \e^i - Y^{ij I} \e_j +
\r^I \left( - \gamma \cdot T + \frac14 \g \cdot G \right) \e^i
\nn\\
&& - \frac{1}{2\sqrt2} g L \r^I \d^{ij} \e_j  \,,
\label{Fermi5dN2DW}
\end{eqnarray}
where $T_{ab} = \frac18 \,G_{ab} + \frac1{48}\, \ve_{abcde} H^{cde}$ and $\omega_{\pm\mu ab}=\omega_{\mu ab}\pm H_{\m ab}$. The off-shell supersymmetry transformation rules for the bosonic fields can be found in  \cite[Eq.(2.44)]{Ozkan:2013nwa}. The transformation rules for the vector multiplet can be found by imposing the gauge fixing condition  \cite[Eq.(4.3)]{Ozkan:2013nwa} in the transformation rules given in \cite[Eq.(4.11)]{Ozkan:2013nwa}.

To go on-shell, we eliminate the auxiliary fields by using their equations of motion given by
\bea
N&=&-\frac12 g L(2+L)-\frac14 g L a_{IJ}\r^I\r^J\ ,\qquad E_a=0,\qquad V^{'ij}_\m=0\ ,
\nn\\
Y^{I ij} &=& \frac{1}{2\sqrt 2} g \rho^I \delta^{ij}\ , \qquad
V_{\mu\nu} =-{\sqrt2}g G_{\m\n}-\frac{\sqrt2}{6}g\epsilon_{\m\n\r\s\l}H^{\r\s\l}\ ,
\label{aux2}
\eea
where $V_{\mu\nu} = 2\partial_{[\mu} V_{\nu]}$. Substituting \eqref{aux2} back into the Lagrangian $\cL_{0,DW}$ we obtain
\bea
e^{-1} \cL_{0,DW}^{\rm on-shell} &=&  L\left(R-\frac12G_{ab}G^{ab}-\frac13H_{abc}H^{abc}\right) +L^{-1}\partial_{a} L\partial^{a} L
\nn\\
&& + a_{IJ} \Bigg[ - \frac12 D_\m \r^I D^\m \r^J -\frac14 (F_{ab}^I - \r^I G_{ab}) (F^{ab J} - \r^J G^{ab}) \nn\\
&&  - \frac18 \e^{abcde} (F^I_{ab} - \r^I G_{ab})(F_{cd}^J - \r^J G_{cd}) C_e  \r^J\Bigg]
\nn\\
&& - \frac12 a_{IJ} \e^{abcde} (F_{ab}^I - \r^I G_{ab}) B_{cd} \nabla_e\r^J  -V\ ,
\eea
where
\be
V =2g^2L(1+2L) -\frac14 g^2 (1 -4L-L^2) a_{IJ}\r^I\r^J\ .
\ee
In summary, this Lagrangian describes $N = 2, 5D$ supergravity coupled to a single tensor multiplet and $n$ abelian vector multiplets.  Note that the scalars $\rho^I$ are not constrained, unlike the vector scalars described in the previous subsection. Presumably, there is an underlying $n+2$ dimensional very special real geometry in which the $(L, \rho^I)$ are the intrinsic coordinates.

Truncating the $n$ vector multiplets, and dualizing $B_{\m\n}$ to $\widetilde{C}_\m$ with field strength $\widetilde{G}=d\widetilde{C}$, one obtains the on-shell Einstein-Maxwell theory
\begin{eqnarray}
e^{-1}{\cL} &=&  L R + L^{-1} \partial_\m L \partial^\m L- \frac12 L G_{\m\n} G^{\m\n} - \frac14 L^{-1} \widetilde{G}_{\m\n}  \widetilde{G}^{\m\n} \nn\\
&& + \frac14  \e^{\m\n\r\s\l} C_\m G_{\n\r}  \widetilde{G}_{\s\l} + 2 g^2 L (1+2L) \ ,
\end{eqnarray}
where, as mentioned earlier,  $(e_\m{}^a, \p_\m^i, C_\m )$ are the fields of the supergravity multiplet while $(\tilde{C}_\m, \vf^i, L)$ comprises the Maxwell multiplet. To compare with \eqref{R2SW}, one first passes to the Einstein frame and then makes the identification
\be
C_{112}=C_{121}=C_{211}=\frac13,\quad \r^{(1)}=L^{-\ft13},\quad \r^{(2)}=L^{\ft23},\quad A^1_\m=C_\m,\quad A^2_\m =\widetilde{C}_\m\ .
\ee
The model can be further truncated to the minimal theory by setting $L=1, \tilde{C}_\m = C_\mu$ which is identical to \eqref{Minimal5DSugra} with $A_\m$ and $F_{\m\n}$ now replaced by $C_\m$ and $G_{\m\n}$. To recover the standard convention of minimal
supergravity, we need to rescale the graviphoton $C_\m\rightarrow C_\m/\sqrt3$.

\subsubsection{Off-shell Riemann-squared, Ricci-squared, and Ricci Scalar-squared invariants and going on-shell}
\label{WeylVector}

When the dilaton Weyl multiplet is utilized, all off-shell curvature squared invariants are known in the literature. In fact, these invariants exist in two forms: the minimal actions with no external matter couplings and the models that are coupled to $n$-vector multiplets. Here, we provide the $n$-vector multiplet coupled models.
The off-shell Weyl-squared action coupled to $n$-vector multiplets was constructed in  \cite{Ozkan:2013nwa}, and it has the same form as \eqref{pregb} but with the following definitions for $T_{ab}$ and $D$ \cite{Coomans:2012cf}
\begin{align}
D &= -\frac1{32}R-\frac1{16}G^{ab}G_{ab}-\frac{26}3 {T}^{ab} {T}_{ab}+2{T}^{ab}G_{ab} + \frac{1}{4} g N   + \frac{1}{16} g^2 L^2 +{\rm f.t.} \,,\nn\\
T_{ab} &= \frac18 G_{ab} + \frac1{48} \e_{abcde} H^{cde}+{\rm f.t.}\,.
\label{StoDMap}
\end{align}
Note that as $D$ contains the Ricci scalar, the Weyl-squared invariant in dilaton Weyl multiplet is modified by an $R^2$ contribution, and the leading curvature-squared term is given by $C_{\m\n\r\s} C^{\m\n\r\s} + 1/6 R^2$. Similarly, the Ricci-scalar squared action constructed in \cite{Ozkan:2013nwa} has the same form as \eqref{R2SW} where the composite fields are as defined in \eqref{StoDMap}.

While the Weyl-squared and the Ricci scalar-squared action can be combined with \eqref{DWMultiVector}, the third invariant, namely the Riemann-squared action, requires that we set $g=0$. The construction of this invariant is based on a map between the Yang-Mills multiplet and the off-shell, ungauged Poincar\'e multiplet \cite{Bergshoeff:2011xn}. When coupled to $n$-vector multiplets, the bosonic sector of the Riemann-squared action is given by  \cite{Ozkan:2013nwa}
\bea
&&e^{-1} \cL_{{\rm Riem}^2}= \a_I \Big[ -\frac14 \r^I ( R_{\m\n ab}(\o_+) - G_{\m\n} G_{ab}) (R^{\m\n ab}(\o_+) - 3 G^{\m\n} G^{ab})
\nn\\
&& \qquad\qquad\quad - \frac12 \Big( R^{\m\n ab}(\o_+) - G^{\m\n} G^{ab}\Big) F^I_{\m\n} G_{ab} + \r^I V_{ij}{}^{\m\n} V^{ij}{}_{\m\n} - 2 G_{\m\n} V_{ij}{}^{\m\n} Y^{ij\,I}
\nn\\
&& \qquad\qquad\quad  + \r^I G_{ab} \nabla_\m(\o_+) \nabla^\m(\o_+) G^{ab}  + \frac12 \r^I \nabla_\m (\o_+) G^{ab} \nabla^\m (\o_+) G_{ab}
\nn\\
&& \qquad\qquad\quad  + \frac1{12} \e^{\m\n\r\s\l}\Big(F_{\m\n}^I - 2\r^I G_{\m\n}\Big) H_{\r\s\l} G_{ab} G^{ab} + \frac16 \r^I \e^{\m\n\r\s\l} R_{\m\n ab}(\o_+) G^{ab}H_{\r\s\l}
\nn\\
&& \qquad\qquad\quad  - \frac18 \e^{\m\n\r\s\l} R_{\m\n ab}(\o_+) R_{\r\s}{}^{ab}(\o_+) A_\l^I \Big].
\label{vriem2}
\eea
The Riemann-squared invariant can be combined with the two-derivative action \eqref{DWMultiVector} and the other curvature-squared models as long as $g=0$.

The minimal Weyl-squared and the Ricci scalar-squared models for the gauged dilaton Weyl multiplet are obtained by considering a single vector multiplet, $(A_\mu,\rho,\lambda, Y_{ij})$, and mapping these fields to the fields of the dilaton Weyl multiplet according to \cite{Gold:2023ymc},
\be
  \rho \to \sigma,\quad A_\mu \to C_\mu \,,\quad  \lambda  \to  \psi \,, \quad Y_{ij}  \to \frac14 {\rm{i}} \sigma^{-1} \bar\psi_i \psi_j  - \frac{g}{2} \s^{-1} L^{ij} \ .
  \label{MinimalMap}
\ee
%
Upon fixing the redundant superconformal symmetries by choosing $\sigma = 1$ and $\psi^i = 0$, the bosonic sector of the $\text{Weyl}^2 + 1/6 R^2$ invariant is given by  \cite{Ozkan:2013uk}

\begin{eqnarray}
&&e^{-1} \cL_{W^2+\frac16R^2} =
    - \frac{1}{4}R_{{a} {b} {c} {d}} R^{{a} {b} {c} {d}}
    +\frac{1}{3}R_{{a} {b}} R^{{a} {b}}
    - \frac{1}{12}{R}^{2} - \frac{1}{8} \epsilon^{{a} {b} {c} {d} {e}} C_{{a}} R_{{b} {c} f g} R_{{d} {e}}{}^{f g}
\label{pregb2}\\
&& \qquad\quad
    + \frac{1}{6} \epsilon^{{a} {b} {c} {d} {e}} C_{{a}} V_{{b} {c}}\,^{i j} V_{{d} {e} i j}  +\frac{2}{3} V^{{a} {b} i j} V_{{a} {b} i j}  + \frac{1}{3}R_{{a} {b} {c} {d}} ({{G}}^{{a} {b}} G^{{c} {d}}
    - 2 H^{{a} {b}} H^{{c} {d}}
      \nn\\
&& \qquad\quad -3 {{H}}^{{a} {b}} {{G}}^{{c} {d}})  - \frac{4}{3} R_{{a} {b}} {{H}}^{{a} {c}} G^{{b}}{}_{{c}} +\frac{16}{3}R^{{a} {b}} {{H}}^{2}_{{a} {b}}
    + \frac{1}{3}R {{H}}_{{a} {b}} G^{{a} {b}}
     -\frac{4}{3}R H^2 - 4(H^2)^2
     \nn \\
&&\qquad \quad
    -8H^4  -\frac{16}{3} H^2 H_{{c} {d}} G^{{c} {d}} -\frac{40}{3}H^{2}_{{a} {d}} H^{a}{}_{ {c}}  G^{{c} {d}}
    +\frac{8}{3}H^2 G^2
    +\frac{2}{3} H_{{a} {b}} H_{{c} {d}} G^{{a} {b}} G^{{c} {d}}
    \nn \\
&& \qquad\quad
   + \frac{1}{12}(G^2)^2  - \frac{16}{3}H^{2}_{{a} {b}}  G^{2 {a} {b}}   - \frac{4}{3} H_{{a} {b}} H_{{c} {d}} G^{{a} {c}} G^{{b} {d}}
    - \frac{1}{3}H_{{a} {b}} G^{{a} {b}} G^{2}
    +2 G^{2 {a} {b}} G^{{c}}{}_{{b}} H_{{c} {a}}
     \nn\\
 &&\qquad\quad
     -\frac{1}{3} (\nabla^{{a}}{G_{{b} {c}}}) \nabla_{{a}}{G^{{b} {c}}} +\frac{8}{3} (\nabla^{{a}} H_{{b} {c}} )\nabla_{{a}} H^{{b} {c}}
    - \frac{1}{2}G^{4}
    +\frac{4}{3}\epsilon^{{a} {b} {c} {d} {e}} H_{{a} {b}} H_{{c} {d}} \nabla^{{f}}{G_{{e} {f}}}\nn
   \\
&&\qquad\quad
    - 2 \epsilon^{{a} {b} {c} {d} {e}}  H_{{b} {f}} (\nabla_{{a}}H_{{c}}{}^{{f}}) G_{{d} {e}} - \frac{2}{3}\epsilon^{{a} {b} {c} {d} {e}} H_{{a} {b}} (\nabla^{{f}}{G_{{c} {f}}} ) G_{{d} {e}}
    -\frac{1}{24}\epsilon^{{a} {b} {c} {d} {e}} (\nabla^{{f}}{G_{{a} {f}}}) G_{{b} {c}} G_{{d} {e}}
    \nn \\
&&\qquad\quad  -g \big( -\frac{4}{3} N H_{a b} G^{a b}  + \frac{4}{3} N G^{2} -8 N H^{2}
     +\frac{4}{3} V_{a b}{}^{ij} L_{ij} H^{a b} - \frac{2}{3}V_{a b}{}^{ij} L_{ij} G^{a b} - \frac{2}{3} R N \big)
    \nn\\
 &&\qquad\quad  + g^2 \big( \frac{1}{3}L^2 H^{a b} G_{a b} -\frac{1}{3}L^2 G^{2} +2 L^2 H^{2} -\frac{8}{3}N^2 + \frac{1}{6} R L^2\big) -\frac{4}{3} g^3 N L^2 -\frac{1}{6} g^4 L^4, \nn
\end{eqnarray}
where $V_{ab}{}^{ij} = 2 \partial_{[a} V_{b]}{}^{ij} - 2 V_{[a}{}^{k(i} V_{b]k}{}^{j)}$ and $H_{ab}$ is defined as $H^{ab}  =- \frac{1}{12}\epsilon^{abcde}H_{cde}$, and we have used the notations,
\begin{align}
H^2 & = H^{ab}H_{ab}\,, &G^2 &= G^{ab}G_{ab}\,, & H^4 &= H^{2 ab}H^{2}_{a b}\nn\\
    H^2_{ab}&=H_{a}{}^cH_{b c}\,, & G^2_{ab} & =G_{a}{}^c G_{b c}\,, & G^4 &= G^{2 ab}G^{2}_{a b}\,.
\end{align}
Similarly, the bosonic part of the Ricci scalar-squared action is given by  \cite{Gold:2023ymc}
\bea
e^{-1}\cL_{R^2} &=&  {\cal Y}^{ij} {\cal Y}_{i j} -  2 \nabla^{{a}}(N L^{-1})\nabla_{{a}}(NL^{-1}) -\frac{1}{8}  \e_{{a} {b} {c} {d} {e}}C^{{a}} {\cal G}^{bc} {\cal G}^{de}
\nn \\&&
+ \frac{1}{4} {\cal G}_{ab} \Big( {\cal G}^{ab} -4\e^{abcde} B_{de}  \partial_{c}(NL^{-1})
\Big) +  g^2  (    \frac{1}{4} L^{ij} \nabla^{a} \nabla_{a}{L_{ij}} -  \frac{1}{4}R L^2
- H^{2} L^2
\nn \\&&
+  \frac{1}{8} G^{2} L^2 - \frac{5}{2}  N^2
 -  \frac{1}{2} E^{a}E_{a}   - \frac{1}{2} \nabla^{a}{L} \nabla_{a}{L} )
 - 4 g   N^3 L^{-2}  + \frac{1}{16} g^4   L^4 \ ,
\eea
where
\bea
{\cal G}_{{a} {b}}
&=&   4 \nabla_{[a} ( L^{-1} E_{b]}) + 8 L^{-1} L_{ij}(V_{ab}{}^{ij})
-2 L^{-3}  L_{ij}  \big(\nabla_{[a} L^{ik} \big) \nabla_{b]} L_k{}^{j} - 2NL^{-1} G_{ab}\ ,
\nn\\
    {\cal Y}^{ij}
&=&   \frac{1}{4} L^{-1} \left( 4 \nabla^{a} \nabla_{a}{L^{ij}} - 2 R {L^{ij}}
- 8
H^{2} {L^{ij}} +   G^{2} {L^{ij}}
\right)
\nn\\
&& +  L^{-3} \left( - N^2 {L^{ij}} -  E^{a}E_{a} {L^{ij}} - 2 E^{a}L^{k(i} \nabla_{a} L_k{}^{j)}- {L_{kl}} \nabla^{a}  {L^{k(i}} \nabla_{a} {L^{j)l}}
\right) \ .
\eea
The minimal, ungauged Riemann-squared action can be obtained by using the map \eqref{MinimalMap} with $g=0$ in \eqref{vriem2} with just one vector multiplet and fixing the redundant superconformal symmetries. For the gauged model, the Riemann-squared action can be obtained as
\begin{eqnarray}
     \cL_{{\rm Riem}^2} = \cL_{W^2 + \frac16 R^2} + 2 \cL_{{\rm Ric}^2} \,,
\end{eqnarray}
where the ${\rm Ric}^2$ invariant is only obtained recently and takes the form \cite{Gold:2023ymc}\footnote{This is also referred to as the Log invariant in the literature for reasons explained in detail in \cite{Gold:2023ymc}.}
\bea
e^{-1}\cL_{{\rm Ric}^2}
&=&
- \frac{1}{6} R_{{a} {b}} R^{{a} {b}} +\frac{1}{24}R^2
+\frac{1}{6}R^{{a} {b}}  G^{2}_{{a} {b}} + \frac{1}{3}R H_{{a} {b}} G^{{a} {b}}
-\frac{4}{3}R_{{a} {b}}  H^{{a} {c}} G^{{b}}{}_{{c}} -\frac{1}{3}R H^2
\nn\\
&&  - \frac{1}{12} \epsilon^{{a} {b} {c} {d} {e}} C_{{a}} V_{{b} {c}}\,^{i j} V_{{d} {e} i j}  +\frac{1}{6} V^{{a} {b} i j} V_{{a} {b} i j}
- 2 ({{H}}^2)^2  +\frac{16}{3}H^{2}_{{a} {b}}  H^{{a}{c}} G^{b}{}_{c} - \frac{4}{3}H^2 H_{{a} {b}} G^{{a} {b}}, \nn\\
&& +\frac{2}{3} H_{{a}{b}} H_{{c} {d}} \big( G^{{a} {b}} G^{{c} {d}}  -2  G^{{a} {c}} G^{{b} {d}} \big)   +\frac{2}{3}H^2 G^2
-\frac{4}{3} H^{{2}{a} {b}}G^{2}_{{a} {b}}  - \frac{1}{3}H_{{a} {b}} G^{{a}{b}} G^2
\nn\\
&& +G^{2}_{{a} {b}} H^{a c} G^{b}{}_{c} - \frac{1}{48}(G^2)^2 -\frac{1}{24} G^4
-\frac{1}{6} \nabla_{{c}}G^{{a} {c}}  \nabla^{{b}}G_{{a} {b}} +2 \nabla_{{a}}H_{{b} {c}}   \nabla^{[{a}}H^{{b} {c}]}\nn\\
&& + \frac{1}{48}\epsilon^{{a} {b} {c} {d} {e}} \nabla^{{f}}G_{{e} f}  (4 H_{{a} {b}} - G_{ {a} {b}}) (4 H_{{c} {d}} - G_{ {c} {d}})  +\frac{2}{3}N {L}^{2} {g}^{3}+\frac{5}{24}{L}^{4} {g}^{4}
\nn\\
&& + \frac{g}{6}  \Big(   R N  -4 N H_{{a} {b}} G^{{a} {b}} - 2 N G^2 +  V_{{a} {b}}\,^{i j} L_{ij} ( G^{{a} {b}} + 4 H^{{a} {b}} )
+12 N H^2 - 6 \nabla^{{a}} \nabla_{{a}}N
\Big)\nn\\
&&-  \frac{g^2}{24} \Big( 2 R L^2 -  L^2  (G^{2} -4 G^{{a} {b}}H_{{a} {b}} - 24  H^2 )  +4 N^2 +6 \nabla^{{a}}L^{ij}\nabla_{{a}}L_{ij}  \Big) \ .
\label{log-gauged}
\eea
In Summary, for the ungauged theories, there are three independent higher derivatives invariant actions described above, which can be chosen from the set including $\cL_{{\rm Riem}^2}, \cL_{W^2+\frac16 R^2}, \cL_{{\rm Ric}^2}$ and $\cL_{R^2}$. In the case of $U(1)_R$  gauged theory, the invariants given above are the last three in this list.

We conclude this subsection by describing the procedure for going on-shell in the presence of the higher derivative invariants. As explained at the end of section \ref{SWR2}, we need to apply the auxiliary field equations of motion
\eqref{aux2} in the four-derivative part of the total action. In what follows, we shall give the on-shell result for the higher derivative extension of $U(1)_R$ gauged minimal $N=2, 5D$ supergravity. To this end, one dualizes
 the two-form $B_{\mu\nu}$ to a vector field $\widetilde{C}_\mu$. Next, one truncates the extra vector multiplet by setting  $L = 1$ and  ${\tilde C}_\mu = C_\mu$. As explained before, to get the canonical kinetic term for $C_\mu$ one also rescales  $C_\mu \to C_\mu/{\sqrt 3}$.

When higher-derivative terms are considered, using \eqref{pregb2}, and after some field redefinitions applied to the metric and $C_\mu$, the on-shell Weyl-squared action can be recast into the following form \cite{Gold:2023ymc}
\begin{align}
e^{-1}\cL_{W^2+\frac16R^2} &= R_{\mu\nu\rho\sigma} R^{\mu\nu\rho\sigma} -4R_{\mu\nu}R^{\mu\nu}+R^2
\nn\w2
&  - \frac12 W_{abcd}G^{ab}G^{cd} +\frac{1}{8} G^4 +\frac1{2\sqrt3}\epsilon^{abcde}C_aR_{bc}{}^{fh}R_{de fh}
\nn\\
&-\frac{8 g^2 G^2}{9}+\frac{14 g^2 R}{3}+\frac{50 g^4}{3}-\frac{g^2}{2\sqrt3}\epsilon^{abcde}C_aG_{bc}G_{de} \ .
\label{eq:invariant1}
\end{align}
The on-shell Weyl-squared action \eqref{eq:invariant1} is identical for both the standard and the dilaton Weyl multiplets when the vector multiplet couplings are truncated.

In the ungauged case,  i.e. $g=0$, the on-shell Riemann-squared action based on the dilaton Weyl multiplet is equivalent to the on-shell Weyl-squared action \cite{Liu:2022sew}. For the gauged supergravity, the difference between these two actions is proportional to the on-shell Ricci-squared action of the form
\be
e^{-1}{\cal L}_{{\rm Ric}^2}= g^2 \Big( \frac{23}{6} G^2 -10 R-28 g^2 -\frac{\sqrt{3}}{2}\epsilon^{abcde}C_aG_{bc}G_{de} \Big) \ .
\label{eq:actionRelation}
\ee
The Ricci-scalar squared actions based on different Weyl multiplets are also equivalent on-shell and they can be eliminated by using the lowest-order field equations in the ungauged case when the vector multiplets are truncated. If gauged supergravity is considered, the on-shell Ricci-scalar squared action is given by
\begin{eqnarray}
  e^{-1}\cL_{R^2} =    g^2 \Big(52 g^2+\frac{19 G^2}{6}-2 R-\frac{\sqrt{3}}{2} \e^{abcde} C_{a} G_{bc} G_{de}\Big) \ ,
  \label{eq:R2os}
\end{eqnarray}
which shifts the coefficients of various terms in the two-derivative action.
While the on-shell results are compatible with those of \cite{Cassani:2022lrk}, they do not match with the results presented in \cite{Liu:2022sew} for $g \neq 0$. This is due to the fact that \cite{Liu:2022sew} neglected the $g$-dependent parts in the map between the standard and the dilaton Weyl multiplets \eqref{StoDMap}, leading to missing terms and erroneous coefficients.

\subsection{Off-shell Killing spinors in \texorpdfstring{$N=2, 5D$}{N=2, 5D} supergravity in standard Weyl formulation}

The off-shell Killing spinors in  ${ N}=2,\,5D$ supergravity were classified by  \cite{Castro:2008ne, Bonetti:2018lfb} where the multiplets involved are the standard Weyl multiplet $(e^a_\m,\,V^{AB}_\m,\,b_\m,\,T_{ab},\, D,\,\psi_\m^A,\,\chi^A)$, and abelian vector multiplets $(\rho^I,\,A^I_\m,\, Y_{AB}^I,\,\Omega_A^I)$. For the purpose of this section, to avoid confusion with the index notation that will be used below, we have denote the $R$ symmetry doublet index by $A=1,2$. We also use the conventions of   \cite{Castro:2008ne, Bonetti:2018lfb} for convenience. The full map between the conventions of  \cite{Ozkan:2013nwa} and  \cite{Castro:2008ne, Bonetti:2018lfb} can be found in the Appendix B of  \cite{Ozkan:2013nwa}. In the ungauged case, the structure of all known curvature-squared invariants allows one to set SU(2)-triplet fields $V^{AB}_\m$ and $Y_{AB}^I$ equal to zero. Once this is done, the supersymmetry transformations agree with those in the on-shell theory \cite{Gauntlett:2002nw}. On the other hand, in the gauged case, not all the auxiliary fields can be consistently set to zero. We will thus separate these cases.

In the ungauged case, one can again separate the discussions into two cases based on
the property of the Killing vector built from Killing spinor bilinear.
\begin{itemize}
\item[$\bullet$] When the Killing vector is time-like, the metric can be parameterized as
\be
ds^2=e^{4U}(dt+\o)^2-e^{-2U}\tilde{g}_{mn}dx^mdx^n\ ,
\ee
where $U$ is a time-independent function. Accordingly, one can introduce the f\"unfbeins
\be
e^0=e^{2U}(dt+\o),\quad e^i=e^{-U}\tilde{e}^i\ , \quad i=1,...,4\ ,
\ee
where $\tilde{e}^i$ is the vierbein of $\tilde{g}_{mn}$. For timelike Killing vectors,  it was shown in \cite{Bonetti:2018lfb} that the Killing spinor equations provide a solution for $T_{ab}, D, V_\mu^{AB}$ and the self-dual field strength $ F_{\mu\nu +}^I$ in terms of $U$ and the scalars $\r^I$, and that all equations of motion follow from the integrability of the Killing spinor equation except the following ones:
\be
{\cal E}_D =0,\quad {\cal E}_{V_{-AB}}=0,\quad {\cal E}_{M^I}=0\ ,
\ee
where the field equation of $V_{\mu AB}$ is projected to the lightcone direction.

\item[$\bullet$] When the Killing vector is null, the metric can be parameterized as
 \be
 ds^2=e^{2U}({\cal F}(dy^-)^2+2dy^+ dy^-)-e^{-4U}\d_{ij}(dx^i+a^idy^-)(dx^j+a^jdy^-)\ ,
 \ee
 where all the functions are independent of $y^+$. Choosing the f\"unfbeins
 \be
 e^+=e^U(dy^+ + \frac12 {\cal F}dy^-),\quad e^i=e^{-2U}(dx^i+a^i dy^-)\ ,\quad i=1,2,3\ ,
 \ee
the Killing spinor is of the form
 \be
 \g_+\epsilon=0,\quad \epsilon=e^{U/2}\epsilon_0\ .
 \ee
In this case, the Killing spinor equations determine $T_{+i}, T_{+-},T_{ij}, F^I_{ij}, D$ in terms of $U$ and the scalars $\r^I$, and the only equations of motion which need to be solved are
\be
{\cal E}_D=0,\quad {\cal E}_{-I}=0,\quad  {\cal E}_{-i}=0,\quad {\cal E}_{++}=0\ ,
\ee
\end{itemize}
which are the equations of motion for $D, A^{-I}, T^{-i}$ and $g^{++}$, respectively. In the ungauged case, it is shown by  \cite{Bonetti:2018lfb} that the Ricci scalar-squared invariant does not modify supersymmetric solutions.

In the gauged case, Killing spinor equations together with the integrability condition coming from the vanishing  gravitino supertransformation turn out to imply that
\be
N T_{ab}=0 \ .
\ee
Taking into account all Killing spinor equations and their integrability conditions, one finds that it suffices to solve the following field equations,
\be
{\cal E}(D)=0,\quad {\cal E}(P_a)=0,\quad  {\cal E}(Y)^{AB}_I=0\ .
\ee

In the case $T_{ab}=0$, from the same integrability condition it also follows that \cite{Bonetti:2018lfb}
\be
R_{abcd}=-\frac{N^2}9(g_{a[c}g_{d]b})\ .
\ee
On the other hand, when $T_{ab}$ is non-zero, $N$ must be zero and the integrability condition
of the Killing spinor equation coming from the gravitino supertransformation reduces to that of the ungauged minimal supergravity  \cite{Gauntlett:2002nw} with $V_\m=0$.

In the gauged theory, unlike the ungauged case, the Ricci scalar-squared invariant does not vanish on
the supersymmetric configurations, modifying both the very special geometry satisfied by the real scalars from the vector multiplets and the AdS$_5$ radius  \cite{Ozkan:2013nwa}. Consequently, $5D$ Ricci scalar-squared invariant plays a role in black hole physics and AdS$_5$ holography  \cite{Bobev:2021qxx, Liu:2022sew, Cassani:2022lrk, Baggio:2014hua}.

\subsection{Exact solutions of \texorpdfstring{$N=2, 5D$}{N=2, 5D} higher derivative supergravities}\label{5dapp}

For the ungauged model described by supergravity plus a Weyl-squared invariant in the standard Weyl formulation, solutions preserving maximal supersymmetry are as follows:
\begin{itemize}
\item[$\bullet$] Five-dimensional Minkowski space. All the gauge fields and auxiliary fields vanish. The very special real (VSR) geometry condition remains to be\footnote{In this section, we set $\rho^I=-M^I$ and $T_{ab} =\frac14 v_{ab}$, in order to follow the notation and conventions of \cite{Bonetti:2018lfb}.}
\be
\frac16C_{IJK}M^IM^JM^K=1\ .
\ee
\item[$\bullet$] The G\"odel-type solution  \cite{Gauntlett:2002nw}. All the auxiliary fields vanish, and the metric is of the form
\be
ds^2=k^{-2}(dt+\frac{kr^2}{4}c^{(i)}\s^{(i)}_L)^2-k(dr^2+r^2d\O_3^2)\ ,
\ee
where $k$ and $c^{(i)}, i=1,2,3$ are constants and $\sigma_L^{(i)}$ are the left invariant one-forms on $S^3$. The scalar fields $M^I={\rm const}$, the U(1) field strength $F^I_{\mu\nu}$ and $v_{\mu\nu}$ are anti-self dual and proportional to a linear combination of the hyper-complex structure of the base manifold $\mathbb{R}^4$. The VSR condition is modified to be  \cite{Bonetti:2018lfb}
\be
\frac16C_{IJK}M^IM^JM^K=1-c_{2I}M^I\frac{c^{(i)}c_{(i)}}{12k^2}\ ,
\ee
where $c_{2I}$ is the coefficient in front of the Weyl-squared invariant.

\item[$\bullet$] AdS$_2\times$S$^3$, in which the metric and electric fluxes are
\bea
ds^2&=&\frac{r^4}{4k^2}dt^2-\frac{2k}{r^2}[dr^2+r^2d\O_3^2]\ ,
\nn\\
F^I&=&\frac{1}{2k}M^I dt\wedge dr\ ,
\eea
where scalar field $M^I$ is constant and all the other fields vanish.
The VSR geometry condition is modified to be  \cite{Bonetti:2018lfb}
\be
\frac16 C_{IJK}M^IM^JM^K=1-\frac{c_{2I}M^I}{144k}\ .
\ee
AdS$_2\times$S$^3$ arises as the near horizon limit of electrically charged supersymmetric black holes. The black hole entropy can be computed by extremizing the entropy function and the result turns out to be \cite{Castro:2008ne}
\be
S_{bh}=2\pi\sqrt{Q^3}(1+\frac{c_{2I}q^I}{16Q^{3/2}}+\cdots)\ ,
\label{bhent}
\ee
where $q^I$ is the electric charge and $Q^3 :=\frac16C_{IJK}q^I q^J q^K$.

\item[$\bullet$] AdS$_3\times$S$^2$, in which the metric and fluxes are  \cite{Castro:2007sd}
\bea
ds^2&=&\ell^2_A ds^2_{AdS_3}+\ell_S^2d\O_2^2\ ,\quad \ell_S=\frac12\ell_A\ ,
\nn\\
F^I&=&\frac{p^I}2\epsilon_2,\quad v=-\frac38\ell_A\epsilon_2\ ,
\nn\\
M^I&=&\frac{P^I}{\ell_A}\ ,\quad D=\frac{12}{\ell_A^2}\ ,
\eea
where the modified VSR geometry condition can be expressed as
\be
\ell_A^3=C_{IJK} p^I p^J p^K+\frac1{12}c_{2I}p^I\ .
\ee
AdS$_3\times$S$^2$ arises as the near horizon limit of supersymmetric black strings and captures their entropy. When the excitation energy of the black string is large, the entropy is given by the Cardy formula,
\be
S_{bs}=2\pi\left[\sqrt{\frac{c_L}6 h_L}+\sqrt{\frac{c_R}6 h_R}\, \right]\ ,
\ee
where $h_L,\,h_R$ are eigenvalues of the AdS$_3$ energy generators $L_0,\,\bar{L}_0$ and $c_{L,R}$ are the central charges associated with the CFT residing on the boundary of AdS$_3$. Thus computing black string entropy boils down to deriving the two central charges. The sum of the central charges $c=\frac12(c_L+c_R)$ can be obtained by extremizing the on-shell action over all the parameters in the solution while keeping the magnetic charges $p^I$ fixed \cite{Castro:2008ne}. The result is given by
\be
c=6P^3+\frac34c_{2I}p^I,\quad P^3 :=\frac16 C_{IJK} p^I p^J p^K\ .
\label{cp}
\ee
The difference of the two central charges is obtained from the coefficient of the induced $3D$ Lorentz Chern-Simons term from $S^2$ compactification of the supersymmetric Weyl-squared action. Using $p^I=-\frac1{2\pi}\int _{S^2}F^I$, the term $A\wedge {{\rm Tr}( R\wedge R)}$ yields the following Lorentz Chern-Simons term in the $3D$ effective action
\be
-\frac{c_{2I}p^I}{192\pi}\int_{M_3}{\rm Tr}(\G d\G+\frac23 \G^3)\ .
\ee
From this formula one can read off $c_L-c_R$ using the formula derived in \cite{Kraus:2005zm}
\be
c_L-c_R=-\frac12c_{2I}p^I\ .
\label{cm}
\ee
Combining \eqref{cp} with \eqref{cm}, one finds
\be
c_L=6 P^3+\frac12c_{2I}p^I,\quad c_R=6 P^3+c_{2I}p^I\ .
\ee
As pointed out in \cite{Castro:2008ne}, the central charges obtained from the gravity side match with those expected from the dual CFT \cite{Maldacena:1997de}.

\item[$\bullet$] The near horizon geometry of the rotating BMPV black hole \cite{Breckenridge:1996is} itself turns out to be a maximally supersymmetric solution, in which the metric and fluxes take the form
\bea
ds^2&=&\frac{r^4}{4k^2}(dt+\frac{2k}{r^2}c^{(i)}\s^{(i)}_R)^2-\frac{2k}{r^2}[dr^2+r^2d\O^2_3]\ ,
\nn\\
F^I&=&\frac1{2k}M^I dt\wedge dr+M^I\frac{c^{(i)}}{r^2}\s^{(i)}_R\wedge dr\ ,
 \eea
where $M^I={\rm const}$ and all the auxiliary fields vanish.  For $c^{(i)}=0$, it becomes AdS$_2\times S^3$. The very special geometry condition is modified to be  \cite{Bonetti:2018lfb}
\be
\frac16C_{IJK}M^IM^JM^K=1-\frac{c_{2I}M^I}{36}\left(\frac1k+\frac{3}{k^2} c^{(i)}c_{(i)}\right)\ .
\ee
The entropy of the BMPV black hole is governed by its near horizon geometry above and was studied in \cite{deWit:2009de}. The result is of the form
\be
S_{bh}=2\pi\sqrt{Q^3-J^3}(1+\frac{c_{2I}q^I}{16Q^{3/2}} +\cdots)\ ,
\ee
where $J$ is the angular momentum and the ellipsis denotes corrections from higher derivative terms beyond the current setup.
\end{itemize}

The half-BPS black holes with Weyl-squared correction have been studied in  \cite{Castro:2007hc, Castro:2007ci} where the full solution is given numerically. One feature
is that for singular supersymmetric black string or black hole solutions with vanishing entropy in the two-derivative theory, the four-derivative corrections yield a
non-zero entropy and shield the singularity behind a smooth event horizon \cite{Kraus:2005vz}.
Finally, it is noted that the relation between $5D/4D$ supersymmetric solutions with higher derivative corrections has been studied in \cite{Castro:2007ci, deWit:2009de}.

In the gauged case, there is the maximally supersymmetric AdS$_5$ solution, on which
the VSR receives corrections from the Ricci scalar-squared invariant  \cite{Ozkan:2013nwa}. The half-BPS solutions of the form AdS$_3\times S_g$ with $S_g$ being a genus-$g$ Riemann sphere were studied in  \cite{Bobev:2021qxx}.

\section{\texorpdfstring{$D=4$}{D=4}}

There are several results that have been obtained in the construction and study of higher derivative supergravities in $4D$ over the years, the ones on $N=1$ supergravity going back to the mid-eighties \cite{Cecotti:1985nf,Cecotti:1985mf,Cecotti:1987mr}. We shall survey the existing results for $1\le N\le 8$, most of which are for the $R$-symmetry ungauged supergravities. While explicit higher derivative invariants have been constructed for $N=4$ conformal supergravity, and $N=1,2$ supergravities, thanks to the existence of their off-shell formulations, such results are very difficult to come by for $N>4$. In this section we shall primarily survey the cases of $N\le 4$,  and their four derivative extensions. As a six derivative extension involving Riemann-cubed term, it was shown long ago that it does not exist for $N=1, 4D$ supergravity \cite{Deser:1977yyz}. Regarding $4D$ supergravities with $N>4$, we shall make one exception by considering \cite{Freedman:2011uc} on the $D^{2k} R^4$ invariants for $N=8$ ungauged supergravity, in view of the fact that it is one of the rare papers in which the superpartners of $D^{2k} R^4$ are constructed. In this section, we shall also survey the construction of Killing spinors, and discuss the cosmological applications of $R+R^2$ type supergravities.

\subsection{\texorpdfstring{$D^{2k} R^4$}{D2k R4} invariants for \texorpdfstring{$N=8, 4D$}{N=8, 4D} supergravity from superamplitudes}

In the spin-helicity formulation, the 4-point maximum helicity violating (MHV) superamplitude in ungauged $N=8, D=4$ supergravity, corresponding to supersymmetric completion of $R^4$, is of the form \cite{Freedman:2011uc}
\be
{\cal M}_4^{\rm MHV}= \frac{1}{256} \prod_{i=1}^8 \Big( \sum_{a,b=1}^4 \langle ab\rangle  \eta_{ai} \eta_{bi} \Big) \frac{[34]^4}{\langle12\rangle^4}\ ,
\ee
where $i=1,...,8$ labels the $SU(8)$ fundamental representation, $a=1,...,4$ labels the scattering particles, $\eta_{ai}$ are Grassmann bookkeeping variables and
\bea
P^{(a)}_{\a\ad} &=& \lambda^{(a)}_\a {\widetilde \lambda}^{(a)}_{\ad},\qquad \langle ab\rangle = \e^{\a\b} \lambda^{(a)}_{\a}\lambda^{(b)}_{\b}\ ,\quad [ab] = \e^{\ad\bd} {\widetilde\lambda}^{(a)}_{\ad}{\widetilde \lambda}^{(b)}_{\bd}\ .
\eea
From the superamplitude, one can extract the matrix elements of all independent 4-point amplitudes which carry certain irreps of $SU(8)$. The matrix element for any desired set of four external particles is obtained by applying a specific Grassmann derivative of order $16$, see \cite{Bianchi:2008pu}. The bosonic part of the action constructed in this way is found to be \cite{Freedman:2011uc}
\bea
{\cal L}_{R^4}&=&\frac14R_{\ad\bd\gd\dd}R^{\ad\bd\gd\dd}R_{\a\b\g\d}R^{\a\b\g\d}
+\frac1{24}R_{\ad\bd\gd\dd} R_{\a\b\g\d}\, \partial^{\ad\a}\partial^{\bd\b}\phi^{ijkl}\partial^{\gd\g}\partial^{\dd\d}\phi_{ijkl}
\nn\w2
&&-\frac12 R_{\ad\bd\gd\dd} R_{\a\b\g\d}  F^{\ad\bd ij}\partial^{\gd\g}\partial^{\dd\d}F_{ij}^{\a\b}
+\frac18 R_{\ad\bd\gd\dd} F_{\a\b ij}F_{\g\d kl}\, \partial^{\ad\a}\partial^{\bd\b}\partial^{\gd\g}\partial^{\dd\d}\phi^{ijkl}
\nn\w2
&&+\frac18  R_{\a\b\g\d} F_{\ad\bd}^{ij} F_{\gd\dd}^{kl}\partial^{\ad\a}\partial^{\bd\b}\partial^{\gd\g}\partial^{\dd\d}\phi_{ijkl}
+\frac12 F_{\ad\bd}^{ij}\partial_\m\partial_\n F^{\ad\bd kl}\partial^\m\partial^\n F^{\a\b}_{ij} F_{\a\b kl}
\nn\w2
&&+2F_{\ad\bd}^{ij}\partial_\m\partial_\n F^{\ad\bd kl}\partial^\m F^{\a\b}_{ik}\partial^\n F_{\a\b jl}
-\frac1{24}F^{ij}_{\ad\bd} \partial^\m\partial^\n F_{\a\b ij}\, \phi_{\m\n}^{klmn}\partial^{\ad\a}\partial^{\bd\b}\phi_{klmn}
\nn\w2
&&-\frac13 F_{\ad\bd}^{ij} \partial^\n F_{\a\b jn}\, \phi_{\m\n}^{klmn}\partial^\m\partial^{\ad\a}\partial^{\bd\b}\phi_{iklm}
-\frac14 F_{\ad\bd}^{ij} F_{\a\b mn}\, \phi_{\m\n}^{klmn}\partial^{\ad\a}\partial^{\bd\b}\phi^{\m\n}_{ijkl}
\nn\w2
&&+\frac18\phi_{\m\s}^{ijkl}\phi_{\n\r}^{mnpq}\phi^{\m\r}_{ijmn}\phi^{\n\s}_{klpq}
+\frac19\phi_\m^{ijkl}\phi_{\n\r\s}^{mnpq}\phi^{\n\r}_{ijkm}\phi_{lnpq}^{\m\s}
+\frac1{288}\phi_{\m\n}^{ijkl} \phi_{\r\s} ^{mnpq}\phi^{\m\n}_{mnpq}\phi_{ijkl}^{\r\s}\ ,
\nn\\
\eea
where $\phi_{\mu_1...\m_n}^{ijkl} := \partial_{\mu_1}\cdots \partial_{\mu_n} \phi^{ijkl}$, $\a,\,\b$ and $\ad,\,\bd$ are the indices of the 2-spinors, and $R_{\alpha\beta\gamma\delta}$ is the linearized Weyl tensor, and Yang-Mills curvature is decomposed as,
\be
F_{\m\n} = \frac14 \left({\bar\sigma}_\m\right)^{\ad\a}\left({\bar\sigma}_\n\right)^{\bd\b}\big( \e_{\ad\bd} F_{\a\b} +\e_{\a\b} F_{\ad\bd} \Big)\ .
\ee
There are $30$ additional terms that are quadratic in fermions in the Lagrangian above, and $8$ more terms that are quartic in fermions. The $R^4$ invariant in $N=8$ supergravity was also obtained using the linearized on-shell superspace formalism in \cite{Kallosh:1980fi}.

For the $D^{2k}R^4$ superinvariant the corresponding superamplitude can be conveniently written as \cite{Freedman:2011uc}
\be
{\cal M}_k=P_k(s,\,t,\,u){\cal M}^{\rm MHV}_4\ ,
\label{d2kr4}
\ee
where $P_k$ is a totally symmetric $k$-th order polynomial in Mandelstam variables. For instance \cite{Elvang:2010jv}
\be
P_2=(s^2+t^2+u^2),\quad P_3=(s^3+t^3+u^3) ,\quad P_4=(s^2+t^2+u^2)^2\ ,
\ee
where $s,t,u$ are the standard (dimensionful) Mandelstam variables. Using \eqref{d2kr4} one can build the linearized $D^{2k}R^4$ invariant by distributing the partial derivatives on the four fields involved in the $R^4$ invariant. To be more specific, denoting any term from the $R^4$ invariant as $A(x)B(x)C(x)D(x)$, one applies the replacement rule \cite{Freedman:2011uc}
\bea
s{\cal M}^{\rm MHV}_4& \rightarrow & 2\partial_\m A\partial^\m B CD,\quad
t{\cal M}^{\rm MHV}_4 \rightarrow  2\partial_\m A B \partial^\m CD
,\quad
u{\cal M}^{\rm MHV}_4 \rightarrow  2\partial_\m A B C\partial^\m D,\quad
\nn\\
s^2{\cal M}^{\rm MHV}_4& \rightarrow & 4\partial_\m\partial_\n A\partial^\m\partial^\n B CD,\quad st{\cal M}^{\rm MHV}_4 \rightarrow  4\partial_\m \partial_\n A \partial^\m B \partial^\n CD \ ,
\eea
and so on.

Note that although allowed by maximal supersymmetry, the  $R^4$ term does not appear in $N=8$ supergravity because otherwise its nonlinear supersymmetry completion leads to nonvanishing 6-pt matrix element in the single soft limit \cite{Beisert:2010jx} which is incompatible with the continuous $E_{7(7)}$ symmetry at the perturbative level \cite{Bianchi:2008pu,Bossard:2010dq,Arkani-Hamed:2008owk,Kallosh:2008rr}. On the other hand, the $R^4$ term does appear in the low-energy effective action of string theory, with the continuous $E_{7(7)}$ symmetry broken down to its discrete version by non-perturbative effects.  Consequently, constraints on the matrix element implied by the low energy theorem associated with the continuous $E_{7(7)}$ no longer hold. Similar statements apply to the $D^4R^4$ and $D^6R^4$ in $N=8$ supergravity versus string theory.

The low energy effective action of string theory implies that
the $D^{2k}R^4$ invariants should come with scalar dependent functions which are invariant under $E_{7(7)}(\mathbb{Z})$. In particular,  the moduli dependent functions in front of $R^4$ and $D^4R^4$ satisfy Laplace equations \eq{E1} and \eq{E2} on $E_{7(7)}/SU(8)$,
\be
(\Delta^{(4)}+42)\vE^{(4)}_{(0,0)}=0,\qquad (\Delta^{(4)}+\frac{70}3)\vE^{(4)}_{(1,0)}=0\ .
\ee
Solutions to these equations are \cite{Green:2010wi,Pioline:2015yea}
\bea
R^4:  &&\qquad \vE^4_{(0,0)}= E^{E_{7}}_{[1000000];\frac32}\ ,
\nn\w2
D^4 R^4: &&\qquad \vE^{(4)}_{(1,0)}  = \frac12 E^{E_{7}}_{[1000000];\frac52}\ .
\eea
These functions are defined in section \ref{modular functions}. The expression for $\vE^{(4)}_{(0,0)}$ has a perturbative part in $4D$ string coupling constant expansion, and a nonperturbative part. The perturbative part consists of a tree level and an one-loop level term.
The $D^4R^4$ coupling also has a similar expansion but this time the perturbative parts consist of tree-level, one-loop, and two-loop contributions. On the supergravity side, explicit computation \cite{Bern:2009kd} shows that the $4D$ maximal supergravity is finite at four-loop level. Analysis based on continuous $E_{7(7)}$ symmetry rules out UV divergences at five and six loops \cite{Beisert:2010jx}. It is expected that the first divergence appears at seven loop level and is of the form $D^8R^4$ \cite{Howe:1980th}.

\subsection{\texorpdfstring{$N=4, 4D$}{N=4, 4D} off-shell conformal supergravity in ectoplasm approach}

Off-shell $N=4, 4D$ \mbox{\it Poincar\'e supergravity} is not expected to exist. The two-derivative on-shell version exists \cite{Cremmer:1977tc, Das:1977uy, Cremmer:1977tt} but its higher derivative extension has not been studied, to the our best knowledge. On the other hand, $N=4, 4D$ off-shell conformal supergravity has been constructed \cite{Bergshoeff:1980is}, and it turns out to be of considerable interest (see \cite{Fradkin:1985am} for a nice review ). To name a few, off-shell $N=4, 4D$ conformal supergravity coupled to SU(2)$\times$U(1) or U(1)$^4$ gauge theory was conjectured to be finite to all loop orders and free of conformal anomaly \cite{Fradkin:1985am}. The model also arises from the twistor string theory according to Berkovits and Witten \cite{Berkovits:2004jj}.

The complete $N=4, 4D$ conformal supergravity was obtained recently in  \cite{Butter:2016mtk,Butter:2019edc} using the ectoplasm approach. The fields in the superconformal Weyl multiplet are summarised in  \ref{N=4cs}.

\begin{table}[ht!]
\centering
\begin{tabular}{|c|c|c|c|c|c|c|c|c|c|c|}
\hline
Field & $e^a_\m$ & $\psi^i_\m$ & $b_\m$ & $V_\m^{i}{}_j$ & $\phi_\a$ & $\L_i$ & $E_{ij}$ & $T^-_{[ab]}{}^{ij}$ & $\chi^{ij}{}_k$ & $D^{ij}{}_{kl}$ \\ \hline
SU(4)  &   1    &    4   &   1    &   15    & 1  & $\overline{4}$ & $\overline{10}$ & 6 & $20_c$ & $20_r$    \\
Weyl weight & -1     & -$\frac12 $    & 0   & 0     & 0 & $\frac12$ & 1 & 1& $\frac32$ & 2    \\
\hline
\end{tabular}
\caption{The field content of  $N=4$ superconformal Weyl multiplet. $20_c$ denotes the 20-dimensional complex representation of SU(4) while $20_r$ refers to the real one.  }

\label{N=4cs}
\end{table}
The complex scalar field parametrizes the $SU(1,1)/U(1)$ coset
\be
\phi^\a\phi_\a=1,\quad \phi^\a=\eta^{\a\b}(\phi_\b)^*,\quad \eta^{\a\b}={\rm diag}(1,\,-1)\ .
\ee
The construction of the conformal supergravity action is carried out in ectoplasm approach based on a closed super 4-form $J$. The lowest Weyl weight term $J_{\a\b\d\g}$
is restricted to contain only Lorentz scalars. The other fields that appear in
the closed super 4-form are listed below.
\begin{table}[ht!]
\centering
\begin{tabular}{|c|c|c|c|c|c|c|c|c|c|c|}
\hline
Field & $A^{ij}{}_{kl}$ & $C^{ij}{}_{kl}$ & ${\cal E}^{ij}$ & ${\cal E}_{(ab)}{}^{ij}$ & ${E}_a^i{}_j$ & $F$ & $\r_{ij}^k$ & $\k_{ij}{}^{k}$ & $\O^{i}$ & $\O_a^i$ \\ \hline
SU(4)  &   $20_r$    &    $20_r$   &   10    &   6    & 15  & 1 & $\overline{20}_c$ & $\overline{20}_c$ & 4 & 4   \\
Weyl weight  & 2     & 2     & 3   & 3     & 3 &4 & $\frac52$ & $\frac52$& $\frac72$ & $\frac72$   \\
\hline
\end{tabular}
\caption{Fields that appear in the spacetime component of the closed 4-form. }
\end{table}
 Readers are referred to  \cite{Butter:2019edc} for the detailed properties satisfied by these fields. The superconformal invariant action
is given by the spacetime component $J_{abcd}$ whose bosonic part is simply
\be
S=\int d^4x\, e F\ .
\ee
One can then construct a composite $F$ using Weyl multiplet and substitute the result into the action above to obtain the action for the $N=4$ conformal supergravity \cite{Butter:2016mtk}
\bea
e^{-1}{\cal L}_{CSG}&=& H ( \phi_\a)\Big(\frac12 C_{\m\n\r\s}C^{\m\n\r\s}+F_{\m\n}(V)^i_jF^{\m\n}(V)^j{}_i+\frac14 E_{ij}D^2E^{ij}-4T_{ab}{}^{ij}D^aD_c T^{cb}{}_{ij}
\nn\\
&&-\bar{P}_\m D^{\m}D_\n P^{\n}+\cdots \Big)+c.c.\ .
\eea
The complete bosonic action was given in \cite{Butter:2016mtk} and the fermionic terms can be found in \cite{Butter:2019edc}.
Here we only give kinetic terms of fields, so that one can see which of them are dynamical.  $P_\m$ and $\bar{P}_\m$ are given by
\be
P_\m=\phi^\a\varepsilon_{\a\b}D_\m\phi^\b,\quad \bar{P}_\m=-\phi_\a\varepsilon^{\a\b}D_\m\phi_\b\ .
\ee
In terms of three left-invariant vector fields associated with the group $SU(1,1)$, defining
\be
D^0=\phi^\a\frac{\partial}{\partial\phi^\a}-\phi_\a\frac{\partial}{\partial\phi_\a},\quad D^+=\phi_\a\varepsilon^{\a\b}\frac{\partial}{\partial\phi^\b},\quad D^-=\phi^\a\varepsilon_{\a\b}\frac{\partial}{\partial\phi_\b}\ ,
\ee
$H(\phi_\a)$ is homogeneous of zeroth degree in the holomorphic variables so that it satisfies
\be
D^+H=0,\quad D^0H=0\ .
\ee
In the previous sections, we have discussed how to obtain models preserving Poincar\'e supersymmetry from those which are invariant under superconformal transformations. One has to couple the conformal supergravity to a certain compensating matter multiplet. In $N=4$ supersymmetry, the only matter multiplet is the vector multiplet. At this stage, an off-shell formulation of $N=4$ vector multiplet is still unknown. However, we can still discuss how many vector multiplets are needed at the linearized level \cite{Fradkin:1985am}. It turns out that six $N=4$ abelian vector multiplets with a rigid SO(4) group are needed to fix the dilatation, local SU(4) and special supersymmetry \cite{Howe:1981qj}.

The $N=3$ conformal supergravity can be obtained from the $N=4$ case by decomposing the $N=4$ supermultiplets under $N=3$ superconformal group and truncating out a $N=3$ gravitino multiplet. We refer to  \cite{Hegde:2021rte} and  \cite{Hegde:2022wnb} for the details of this truncation.

\subsection{\texorpdfstring{$N=2, 4D$}{N=2, 4D} off-shell supergravity invariants from the standard Weyl multiplet}

\subsubsection*{$\bullet$\ Off-shell $N=2, 4D$ Poincar\'e theory}

The first way to construct $N=2, 4D$ Poincar\'e supergravity is via coupling standard Weyl multiplet to suitable compensating matter multiplets, followed by fixing redundant gauge symmetries including local conformal boosts, dilatation, SU(2), U(1)$_A$ and special supersymmetry. In this case, it is well known that in order to write down a meaningful two-derivative action, at least two compensating multiplets are needed which are chosen to be a Maxwell multiplet $(X,\,\O^i,\,W_\m,\,Y_{ij})$ and a tensor multiplet $(L_{ij},\,\varphi^i,\,G,\, E_{\m\n})$. After choosing the gauge fixing conditions,
\be
{\rm SU}(2):\, L_{ij}=\d_{ij}\frac{L}{\sqrt2},\quad D:\, L=1, \quad K:\, b_\m=0,\quad
{\rm U}(1)_A:\, X=\bar{X},\quad S_\a:\, \varphi^i=0\ ,
\label{4dswgauge}
\ee
one obtains the off-shell $N=2, 4D$ Poincar\'e supergravity multiplet displayed in Table \ref{4dn2sw}.
\begin{table}[ht!]
\centering

\begin{tabular}{|c|cccccccccccccc|}
\hline
Field & $e^a_\m$  & $V_\m^{'ij} $ & $V_\m$ & $A_\m$ & $T_{ab}$ & $D$ & $X$ & $W_\m$ & $Y^{ij}$ & $G$& $E_{\m\n}$ & $\psi^i_\m$ & $\chi^i$ & $\O^i$ \\
\hline
D.o.f.s  &   6   &  8   &  3   &   4   & 6  &  1  & 1 & 3 & 3 & 2 & 3 &24&8&8 \\
\hline
\end{tabular}

\caption{The field content of off-shell $N=2, 4D$ Poincar\'e supergravity obtained from the standard Weyl multiplet.}

\label{4dn2sw}
\end{table}
The two-derivative Poincar\'e supergravity Lagrangian is the following sum
\be
e^{-1}{\cal L}_{SG}=e^{-1}{\cal L}_{V}-e^{-1}{\cal L}_{T}\ ,
\ee
where $\cL_V$ and $\cL_T$ are the Lagrangians which describe off-shell $N=2, 4D$ Poincar\'e supergravity coupled to a vector \cite{deWit:1980lyi} and a tensor multiplet \cite{deWit:2006gn}, respectively, and they are given by \footnote{Here we use the notations of  \cite{Bobev:2021oku}.}

\bea
e^{-1}{\cal L}_{V}&=&-4X^2(\frac16 R-D)+4D_\m X D^\m X+\frac12 F_{\m\n}^{-}(W)F^{\mu\n-}(W)+\frac12 F_{\m\n}^{+}(W)F^{\m\n+}(W)
\nn\\
&&-\frac14X(F_{\m\n}^{+}(W)T^{\m\n+}+F_{\m\n}^{-}(W)T^{\m\n-})-\frac12 Y^{ij}Y_{ij}
-\frac1{32}X^2(T^{+}_{\m\n}T^{\m\n+}+T^{-}_{\m\n}T^{\m\n-})\ ,
\nn\\
e^{-1}{\cal L}_{T}&=&\frac13R+D+E_{\m}E^{\m}+|G|^2+\frac{1}{2\sqrt 2}E_\m V^\m-\frac14V_{\m}^{'ij}V^{\m'}_{ij}+g X(G+\bar{G})-\frac{g}2E_\m W^\m
\nn\\
&&-\frac{g}{2\sqrt 2}Y^{ij}\d_{ij}\ ,
\eea
where
\be
D_\m X=(\partial_\m-iA_\m)X,\quad E^\m=\frac12\epsilon^{\m\n\r\l}\partial_\n E_{\r\l},\quad V_{\m}^{ij}=V_{\m}^{'ij}+\frac12\d^{ij}V_\m\ .
\ee

The equations of motion  for $D$, $X$, $G$, $Y_{ij}$, $E_\m$, $V_\m$, $V_\m^{'ij}$ and $T_{ab}$ lead to
\bea
&&X=\frac12\ ,\quad  A_\m=0\ , \quad  G=\bar{G}=-\frac12 g\ , \quad Y_{ij}=\frac{g}{2\sqrt 2}\d_{ij}\ ,
\nn\\
&&E_\m=0\ , \quad V_\m=\sqrt2 g W_\m\ , \quad V_{\m}^{'ij}=0\ , \quad T_{\m\n}=-8F_{\m\n}\ .
\eea
Substituting the solutions back to ${\cal L}_{SG}$, one obtains the on-shell two-derivative minimal supergravity
\be
e^{-1}{\cal L}_{SG}=-\frac12 R-\frac12 F_{\m\n}F^{\m\n}+\frac38 g^2\ .
\ee

\subsubsection*{$\bullet$\ Weyl-squared invariant}

The first curvature-squared invariant was constructed in \cite{Bergshoeff:1980is} using superconformal tensor calculus.
We denote it by ${\cal L}_{W^2}$ as it contains the Weyl-squared term. After gauging fixing, the bosonic part of the complete invariant is given by
\bea
e^{-1}{\cal L}_{W^2}&=& C_{\m\n\r\s}C^{\m\n\r\s}+2 F_{\m\n}(A)F^{\m\n}(A)+6D^2+\frac12F_{\m\n}^{ij}(V')F^{\m\n}_{ij}(V')+\frac14F_{\m\n}(V)F^{\m\n}(V)
\nn\\
&&-\frac14T^{\m\n-}\nabla_\m\nabla^\r T^{+}_{\r\n}-\frac14T^{\m\n+}\nabla_\m\nabla^\r T^{-}_{\r\n}-\frac1{512}T_{\m\n}^-T^{\m\n-}T_{\r\s}^+T^{\r\s+}\ ,
\eea
where $F_{\m\n}(V)$ and $F_{\m\n}^{ij}(V')$ are defined as in \eq{defF}.

\subsubsection*{$\bullet$\ Gauss-Bonnet invariant}

The second curvature-squared invariant was obtained in  \cite{Butter:2013lta} using superconformal superspace technique. The construction utilizes a pair of  chiral and anti-chiral superfield $\Phi_{\pm}$ of weight $w\neq 0$. Then one builds the nonlinear version of the kinetic multiplet denoted by ${\mathbb T}[\ln \Phi_{\pm}]$. It is of weight 2 and it is obtained by acting on $\ln\Phi_{\pm}$ with four superspace derivatives. The density formula for the kinetic multiplet gives rise to
\be
\int d^4x\, {\cal L}_{{\rm Ric}^2} =  -\frac1{2w}\int d^4x\, d^4\theta\, {\cal E}_+{\mathbb T}[\ln \Phi_+]-\frac1{2w}\int d^4x d^4\theta {\cal E}_-{\mathbb T}[\ln \Phi_-]\ ,
\ee
where ${\cal E}_\pm$ are (anti)chiral measures, and
\bea
e^{-1} {\cal L}_{{\rm Ric}^2} &=& \frac23 R_{\m\n}R^{\m\n}-2R^2-2F_{\m\n}(A)F^{\m\n}(A)-6D^2-\frac12F_{\m\n}^{ij}(V')F^{\m\n}_{ij}(V')
\nn\\
&& -\frac14F_{\m\n}(V)F^{\m\n}(V) +\frac14T^{\m\n-}\nabla_\m \nabla^\r T^+_{\r\n}+\frac14T^{\m\n+}\nabla_\m \nabla^\r T^-_{\r\n}
\nn\w2
&& + \frac1{512}T_{\m\n}^-T^{\m\n-}T_{\r\s}^+T^{\r\s+}-\frac1{2w}\nabla_\m S^\m\Big]\ .
\eea

The dependence on $\Phi_\pm$ only appears in the total derivative term. Using the expression for Weyl tensor in four dimensions, one can see that the Gauss-Bonnet invariant is given by
\be
{\cal L}_{GB}={\cal L}_{W^2}+{\cal L}_{{\rm Ric}^2}\ .
\ee

\subsubsection*{$\bullet$\ Ricci scalar-squared invariant}

The supersymmetric Ricci scalar-squared action has been constructed by employing superconformal tensor calculus, and the bosonic part of the result is given by \cite{deWit:2006gn,Bobev:2021oku}
\begin{align}
e^{-1}\mathcal{L}_{R^2}= &-\frac12\Bigl(\frac13 R + D\Bigr)^2 + E^2\Bigl(\frac13 R + D\Bigr) + |G|^2\Bigl(\frac16 R + 2D\Bigr)
\nn\w2
&- \mathcal{D}_\m E_\n\Bigl[F(V)^{\m\n} - \frac12\bigl(T^{-\m\n} G + T^{+\m\n}\bar{G}\bigr)\Bigr] + \frac18\Bigl[F(V)_{\m\n} - \frac12\bigl(T^-_{\m\n} G + T^+_{\m\n}\bar{G}\bigr)\Bigr]^2
\nn\w2
&- \frac1{64}\Bigl(T^-_{\m\n} G + T^+_{\m\n}\bar{G}\Bigr)^2 + |\mathcal{D}_\mu G|^2 + 2\,(\partial_{[\m}E_{\n]})^2	- 4\,\bigl(|G|^2 + E_\m E^\m\bigr)^2 \ .
\label{eq:R2}
\end{align}
and the supertransformations of the fermions by
\bea
\d\psi^i_\m &=&2(\partial_\m+\frac14\g^{ab}\o_{\m ab}-\frac{i}2 A_\m)\epsilon^i
-V_{\m}^{ij}\epsilon_j-\frac1{16}\g\cdot T^-\g_\m\epsilon^i-\g_\m\eta^i\ ,
\nn\\
\d\chi^i&=&-\frac1{24}\g\cdot \slashed{D}T^-\epsilon^i -\frac16 \g\cdot F(V)^{ij}\epsilon_j+\frac{i}3\g\cdot F(A)\epsilon^i+D\epsilon^i+\frac1{24}\g\cdot T^-\eta^i\ ,
\nn\\
\d\O_i&=&2\slashed{D}X\epsilon_i+\frac12\g\cdot {\cal F}^-\epsilon_i+Y_{ij}\epsilon^j+2X\eta_i\ ,
\eea
where
\be
{\cal F}_{\m\n}^-=F_{\m\n}-\frac12 XT^-_{\m\n}\ ,\qquad
\eta_i=\frac1{\sqrt 2}G\epsilon^{j}\d_{ij}-\frac1{\sqrt 2}\slashed{E}\epsilon^{j}\d_{ij}+\slashed{V}_m{}^{(j}\d^{k)m}\epsilon_k\d_{ji}\ .
\ee
In the expressions above, the $i\,,j$ indices are raised and lowered using $\varepsilon^{ij}$ and $\varepsilon_{ij}$.

The four-derivative supergravity invariants have been applied to study
holography and black hole physics in \cite{ Bobev:2021oku,Charles:2016wjs, Charles:2017dbr, Bobev:2020egg}. To be specific, \cite{ Bobev:2021oku,Bobev:2020egg} considered the model
\be
{\cal L}_{HD}=(16\pi G_N)^{-1}{\cal L}_{SG}+(c_1-c_2){\cal L}_{W^2}+c_2{\cal L}_{GB}\ .
\label{bob4d}
\ee
It was found that every solution of the two-derivative theory also solves the field equations derived from the four-derivative Lagrangians. Nonetheless, the four-derivative Lagrangians do contribute to the on-shell action and to the black hole entropy. For asymptotically  AdS$_4$ solutions listed in Table 2 of \cite{Bobev:2020egg}, the on-shell Euclidean action takes the form
\be
I_{HD}=\left[1+\frac{64\pi G_N}{\ell^2}(c_2-c_1)\right]\frac{\pi\ell^2}{2G_N}{\cal F}+32\pi^2c_1\chi\ ,
\ee
where $\ell$ denotes the AdS$_4$ radius,  ${\cal F}$ is a constant resulting from the evaluation of the two-derivative action on ${\cal M}_4$ and $\chi$ is the Euler characteristic of ${\cal M}_4$, whose values are given in Table 2 of \cite{Bobev:2020egg}. If the theory \eqref{bob4d} arises from $S^7$ compactification of 11D or $CP^3$ compactification of 10 IIA supergravity with higher derivative corrections, it is conceivable that \eqref{bob4d} provides
a gravitational description of the dynamics involving only the stress tensor sector of ABJM theory \cite{Bobev:2020egg}.
So far, coefficients $c_1$ and $c_2$ are not determined from the gravity side since the
$S^7$ compactification of 11D or $CP^3$ compactification of 10 IIA supergravity with
 higher derivative corrections
has not been worked out. However, they may be computed via a combination of holography, conformal bootstrap and localization techniques \cite{Alday:2021ymb, Alday:2022rly}.
In \cite{Bobev:2020egg} some attempts for fixing $c_1$ and $c_2$ were carried out by comparing the gravitational on-shell action to the free energy of ABJM theory on squashed 3-sphere. Furthermore, one also requires that the stress tensor 2-pt function computed from the gravity side matches with those in the CFT side. In this way, it is possible to fix just the coefficient $c_1$ due to a possible unknown shift in the relation between $\ell^2/G_N$ and $N$.

In \cite{Bobev:2020egg}, the entropy of black hole is also computed. Although the solutions are not modified by the  four-derivative interactions, the entropy does receive corrections. Using the Wald entropy formula, one obtains \cite{Bobev:2020egg},
\be
S_{bh}=\left[1+\frac{64\pi G_N}{\ell^2}(c_2-c_1)\right]\frac{A}{4G_N}-32\pi^2c_1\chi(H)\ ,
\ee
where $A$ and $\chi(H)$ denote the area and Euler number of the horizon.

\subsubsection*{$\bullet$\ Quartic in Weyl tensor invariant}

An off-shell Weyl$^4$ invariant was introduced in superspace in \cite{Moura:2002ip} with the Lagrangian
\be
\mathcal{L} = \int   d^4 \bar{\theta} E \Big\{ \nabla^{Aa}\nabla^{b}_{A} \left(\nabla^{B}_a \nabla_{Bb} + 16X_{ab} \right)
 - \nabla^{Aa}\nabla^{B}_a \left( \nabla^b_A \nabla_{Bb} - 16 \rmi Y_{AB} \right) \Big\}  W^2 \overline{W}^2  + \text{h.c.}\ ,
\ee
where $a=1,2$ is the $U(2)$ $R$-symmetry index, $A=1,2$ is the Weyl spinor index, and the superfields $X_{ab}$ and $Y_{AB}$ are certain components of the torsion and curvature in superspace. The  off-shell $N=2$ Poincar\'e supermultiplet involved is the $40+40$ component multiplet displayed in Table \ref{4dn2sw}.  We will not present the details of the superspace construction here, as they can be found in \cite{Moura:2002ip}. It is pertinent to mention, however, that \cite{Moura:2002ip} focuses exclusively on the superspace formulation, and does not present the component formulation of the Weyl$^4$ invariant. Nevertheless, it has been shown that the above action does provide the off-shell supersymmetric completion of the $C_+^2C_-^2$, where $C_\pm$ is the (anti)self-dual part of the Weyl tensor.

\subsection{\texorpdfstring{$N=2, 4D$}{N=2, 4D} off-shell supergravity invariants from the dilaton
Weyl multiplet}

\subsubsection*{$\bullet$\ Off-shell supergravity}

Another formulation of off-shell  $N=2, 4D$ Poincar\'e supergravity can be obtained by coupling the dilaton Weyl multiplet to a tensor multiplet  \cite{Mishra:2020jlc}.
After fixing local SU(2), U(1)$_A$, dilatation, conformal boosts and special supersymmetry, the off-shell supergravity multiplet is given in Table \ref{4dn2dw} in which the auxiliary fields are $ V^{'ij}_\m,\,V_\m,\, C,\, E_{\m\n}$.
\begin{table}[ht!]
\centering

\begin{tabular}{|c|ccccccccccc|}
\hline
Field & $e^a_\m$  & $V_\m^{'ij} $ & $V_\m$ & $W_\m$ & $\widetilde{W}_\m$ & $B_{\m\n}$ & $L$ & $E_{\m\n}$ & $C$ & $\psi^i_\m$ & $\phi^i$  \\
\hline
D.o.f.s  &   6   &  8   &  3   &   3  & 3  &  3  & 1 & 3 & 2 &24&8 \\
\hline
\end{tabular}

\caption{The field content of off-shell $N=2, 4D$ Poincar\'e supergravity obtained from the dilaton Weyl multiplet.}

\label{4dn2dw}
\end{table}
 This multiplet can also be obtained by reducing the $5D$ off-shell supergravity multiplet and truncating out a vector multiplet.
The bosonic two-derivative supergravity action based on the multiplet above is given by  \cite{Mishra:2020jlc}
\bea
&&e^{-1}{\cal L}_{LR}=\frac12LR-\frac12L^{-1}\partial_\m L\partial^\m L+\frac18LF_{\m\n}(W)F^{\m\n}(W)+\frac18LF_{\m\n}(\widetilde{W})F^{\m\n}(\widetilde{W})
\nn\\
&&\quad +\frac{L}{24}H_{\m\n\r}H^{\m\n\r}+|C|^2L^{-1}-E_{\m}E^{\m}L^{-1}+\frac1{2\sqrt2}\epsilon^{\m\n\r\l}E_{\m\n}F_{\r\l}(V)+LV^{'ij}_\m V^{'\m}_{ij}\ ,
\eea
where
\be
E^\m=\frac12\epsilon^{\m\n\r\l}\partial_\n E_{\r\l},\quad H_{\m\n\r}=3\partial_{[\m}B_{\n\r]}\ .
\ee
On-shell, the field equations imply that the auxiliary fields can be set to zero, and the off-shell multiplet decomposes into the on-shell supergravity multiplet plus the on-shell vector multiplet. One of the scalars in the vector multiplet is obtained by dualizing the massless two form $B_{\mu\nu}$ to a pseudoscalar.
We shall consider the coupling of $n$ off-shell Maxwell multiplets to the off-shell $N=2, 4D$ supergravity. The $n$ vector multiplets will be denoted by
\be
\big( A_\m^I,\ \lambda^{iI},\ X^I, Y^{ij I} \big)\ ,\qquad I=1,...,n\ ,
\ee
where $X^I$ are the complex scalars and $Y^{ij I}$ are the auxiliary scalars. Further useful ingredients are defined as follows
\bea
T_{\m\n}^- = K_{\m\n}^-\ ,\qquad K_{\m\n} & :=& F(W)_{\m\n}
+i F(\widetilde{W})_{\m\n}\ ,\qquad  {\cal P} :=  T^{-}_{\m\n} T^{-\m\n}\ ,
\nn\w2
{\cal A} &:=& -\frac{1}{16} K^{\m\n} K_{\m\n}\ ,\qquad {\cal B} :=C^*L^{-1}-\frac1{\sqrt2}\bar{\phi}^i\phi^j\d_{ij}L^{-2}\ .
\eea

\subsubsection*{$\bullet$\ Riemann-squared invariant and coupling to $n$ vector multiplets}

Off-shell $N=2, 4D$ supergravity coupled to $n$ Maxwell multiplets is described by the Lagrangian \cite{Mishra:2020jlc}
\bea
e^{-1}{\cal L}&=&4D_{\m}G_ID^\m \bar{X}^I+\frac13\epsilon_{\m\n\r\l}H^{\n\r\l}G_I\overleftrightarrow{D}^\m X^I+4D_{\m}(G_{\cal A}K^{\m\n})D^\l\bar{K}_{\l\n}
\nn\\
&&+\frac13\epsilon_{\m\n\r\l}H^{\n\r\l}(G_{\cal A}K^{\s\d})\overleftrightarrow{D}^\m K_{\s\d}+32D_{\m}(G_{\cal P}T^{-\m\l})D^{\n}T^{+}_{\n\l}
\nn\\
&&-8H^{\m\n\l}(G_{\cal P}T^-_{\n\l})\overleftrightarrow{D}^\s T^{+}_{\s\m}+4D^\m G_{\cal B}D_\m\bar{\cal B}+\frac13\epsilon_{\m\n\r\l}H^{\n\r\l}G_{\cal B}\overleftrightarrow{D}^\m\bar{\cal B}
\nn\\
&&+G{\cal L}_0+G_I{\cal L}_1^I+G_{\cal A}{\cal L}_2+G_{\cal P}{\cal L}_3+G_{{\cal A}I}{\cal L}_4^I+G_{{\cal P}I}{\cal L}_5+G_{IJ}{\cal L}_6^{IJ}+G_{\cal AA}{\cal L}_7
\nn\\
&&+G_{\cal AP}{\cal L}_8+G_{\cal PP}{\cal L}_9+G_{\cal B}{\cal L}_{10}+G_{\cal BB}{\cal L}_{11}+G_{{\cal B}I}{\cal L}_{12}+G_{{\cal BA}}{\cal L}_{13}+G_{\cal BP}{\cal L}_{14}+{\rm h.c.}\ ,\nn\\
\label{4dR2DW}
\eea
where $G(X^I, {\cal B}, {\cal A}, {\cal P})$ is a prepotential, $G_{I}=\partial G/\partial X^I$,  $G_{{\cal A}I}=\partial^2 G/\partial X^I\partial {\cal A}$ and similar notation is used for the derivatives of $G$ with respect to other fields.
 Expressions for ${\cal L}_i$ can be found in  \cite{Mishra:2020jlc}. In particular, choosing $G={\cal A}$, one obtains the Riemann-squared invariant
\bea
 && e^{-1}{\cal L}_{{\rm Riem}^2}=16R_{abcd}(\o_-)R^{abcd}(\o_-)+ 4D_\m(\o_-)K^{ab}D^\m(\o_-)K_{ab}
 \nn\w2
 && +\frac13\epsilon_{abcd}H^{bcd}K_{ef}\overleftrightarrow{D}^a(\o_-)\bar{K}^{ef}
-F(V')^{ij}_{\m\n}F(V')_{ij}^{\m\n}-\frac12F_{\m\n}(V)F^{\m\n}(V)
\nn\w2
&& +\frac12K^{ab}\bar{K}_{ab}(F_{\m\n}(W)F^{\m\n}(W)+F_{\m\n}(\widetilde{W})F^{\m\n}(\widetilde{W}))-4K^{ab}R_{cdab}(\o_-)\bar{K}^{cd+}+{\rm h.c.}\ ,
\eea
where $\o_{-\m}^{ab}=\o^{ab}_\m-\frac12 H_{\m}{}^{ab}$. A more general curvature-squared invariant can be obtained by noting that $\cL_2, \cL_3, \cL_{11}$ contain independent curvature-squared structures, and therefore, as a special case, choosing the prepotential as
\be
G(X^I, {\cal A}, \cP, {\cal B}) = \frac{1}{16} \alpha\,{\cal A} +\beta\cP + \gamma {\cal B}^2
\ee
gives
\be
e^{-1}\cL : = \alpha R^{-abcd} R^{-}_{abcd} +8\beta C^{-abcd} C^{-}_{abcd} +\left(\frac23 \beta -\frac{1}{128}\gamma \right) R^2 +\cdots + (h.c.) \ ,
\ee
where $C_{abcd}$ is the Weyl tensor and the ellipsis denotes the remaining bosonic terms which can be read off from \eqref{4dR2DW}. Up to cubic fermions, the supersymmetry transformation rules of fermions are given by
 \cite{Mishra:2020jlc}
\bea
\d\psi^i_\m&=& 2D_\m(\o_+)\epsilon^i-\frac12\g^\n(F_{\m\n}(W)+iF_{\m\n}(\widetilde{W}))\varepsilon^{ij}\epsilon_j\ ,
\nn\\
\d\phi^i&=&\frac1{\sqrt 2}\slashed{D}L\delta^{ij}\epsilon_j +\slashed{E}\varepsilon^{ij}\epsilon_j-G\epsilon^i+\sqrt{2} L \d^{ij}\eta_j\ ,
\nn\\
\d\l^I_i&=& 2\slashed{D}X^I\epsilon_i+\frac12\g^{\m\n}\widehat{F}^{I-}_{\m\n}\epsilon_{ij}\epsilon^j+Y^I_{ij}\epsilon^j+2X^I\eta_i\ ,
\eea
with the following definitions
\bea
\widehat{F}_{\m\n}&=& F^I_{\m\n}-\frac14X^I T^+_{\m\n}-\frac14\bar{X}^I T^-_{\m\n}\ ,
\nn\\
\eta_i&=&-\frac1{12}\g^\m\epsilon_{\m\n\r\s}H^{\n\r\s}\epsilon_i-\frac18\g^{\m\n}(F_{\m\n}(W)+i F_{\m\n}(\widetilde{W}))\epsilon_i\ .
\eea

 The $N=2, 4D$ supergravity coupled to $n$ vector multiplets, extended by Weyl-squared invariant, has been employed to derive the higher derivative corrections to the near horizon geometry of asymptotically flat supersymmetric black holes. These results, when combined with the Wald formula \cite{Wald:1993nt}, yield the macroscopic entropy of these black holes \cite{LopesCardoso:1999fsj,Sahoo:2006rp,Behrndt:1998eq,LopesCardoso:1998tkj,LopesCardoso:1999cv,LopesCardoso:1999xn,LopesCardoso:2000qm,LopesCardoso:2000fp}. For models arising from type IIA string theory compactified on a CY3, the prepotential, up to linear order in ${\cal P}$, takes the form \cite{LopesCardoso:1999fsj}
 \be
 G(X^I,{\cal P})=-\frac16 C_{IJK}\frac{X^IX^JX^K}{X^0}-\frac1{24}\frac1{64}c_{2I}\frac{X^I}{X^0}{\cal P}\ ,
 \ee
 where $C_{IJK}$ is the intersection numbers of the four-cycles of the CY3 and $c_{2I}$s denote its second Chern-class numbers, with $I=1,...,b_2$ where $b_2$ is the second Betti number. The resulting macroscopic entropy of the supersymmetric black holes carrying electric/magnetic charges $(q_0,q_I,\,p^0=0, p^I)$ is given by
 \bea
 S&=&2\pi\sqrt{\frac16\widehat{q}\,(C_{IJK}p^Ip^Jp^K+c_{2I}p^I)}\ ,
 \nn\\
 \widehat{q}_0&=&q_0+\frac1{12}D_{IJ}q^Iq^J,\quad D_{IJ}=-\frac16C_{IJK}p^K\ .
 \label{4dentropy}
 \eea
This result is in agreement with the microscopic entropy formula computed in \cite{Maldacena:1997de,Vafa:1997gr}. A comprehensive review of $D=4$ supersymmetric black holes with stringy higher derivative corrections can be found in
\cite{Mohaupt:2000mj}.

A special class of supersymmetric black holes in $N=2, 4D$ supergravities arising from Calabi-Yau compactifications satisfies
\be
C_{IJK}p^Ip^Jp^K=0\ ,
\ee
corresponding to the so-called small black holes with vanishing classical horizon area, indicating the existence of a null singularity.  However, when higher derivative corrections are taken into account, analysis based on near horizon geometry suggests the singularity is smoothed out by a horizon with a non-vanishing area arising at order $\alpha'$ \cite{Hubeny:2004ji, Dabholkar:2004dq},
\be
A=8\pi\sqrt{\frac{|\widehat{q}_0|c_{2I}p^I}{24}}\ ,
\label{sbh}
\ee
which is proportional to $\a'$.  Therefore this class of black holes are given the name ``small black holes". From \eqref{4dentropy}, its entropy is deduced to be
\be
S=4\pi\sqrt{\frac{|\widehat{q}_0|c_{2I}p^I}{24}}=\frac{A}2\ ,
\ee
differing from the Bekenstein-Hawking entropy formula obeyed by the large black hole by a factor of 2. This difference is natural given that this result for the entropy is not merely a leading order result but encodes significant contributions from $\alpha'$ correction.

The resolution of the singularity, however, has been questioned in \cite{Chimento:2018kop, Cano:2018brq, Cano:2018hut} for certain two-charge black hole solution arising in the 6-torus compactification of heterotic string which is $S$-dual to IIA on $K3\times T^2$. First of all, general four-charge black holes with Riemann-squared corrections were constructed in \cite{Chimento:2018kop, Cano:2018brq} which illustrated that the two-charge black holes still contain a curvature singularity. Based on these solutions, \cite{Cano:2018hut} further argued that the small black holes are in fact already regular in the zeroth-order supergravity approximation and $C_{IJK}p^Ip^Jp^K=0$ does not necessarily imply a singular horizon of vanishing area. The reason is that the charges $p^I$ in \eqref{4dentropy} were obtained from the near horizon geometry and can differ from the genuine conserved charges in the presence of higher curvature couplings. Thus in the context of supergravity with stringy corrections, from $p^I=0$ one cannot deduce that the corresponding conserved charges vanish. However, it is the vanishing of the genuine conserved charges that leads to a singular horizon. Further discussions of small black holes in heterotic string can be found in \cite{Cano:2021dyy, Massai:2023cis}.

\subsection{\texorpdfstring{$N=1, 4D$}{N=1, 4D} off-shell supergravity in \texorpdfstring{$U(1)$}{U(1)} extended superspace and four-derivative invariants}

The construction of higher derivative off-shell $N=1, 4D$ supergravity was carried out long ago within frameworks employing different sets of auxiliary fields. A convenient way to list  possible sets of auxiliary fields is to start from a $20+20$ off-shell multiplet consisting of \cite{Muller:1985vga}
\be
\{ e_\m^a(6), M(2), b_\mu(4), A_{\m\n}(3), C(1), a_\m(3), D(1);\, \psi_\m(12), \chi(4), \lambda(4)\}\ ,
\label{master}
\ee
where $M$ is a complex scalar, $b_\m$ is a non-gauge vector field, $C$ and $D$ are real scalars, $A_{\m\n}$ and $a_\m$ are 2-form and 1-form gauge fields, and $\psi_\mu, \chi, \lambda$ are Majorana spinors. Setting to zero the fields $(A_{\m\n}, C,\chi)$  gives the off-shell $16+16$ multiplet, which we refer to as Type I \cite{deWit:1978ww},
\be
16 +16\ , \mbox{ Type I:} \qquad
\{ e_\m^a(6), M(2), b_\mu(4), a_\m(3), D(1);\, \psi_\m(12), \lambda(4)\}\ .
\ee
Setting to zero $(a_\m, D, \lambda)$ in \eq{master} instead gives another off-shell $16+16$ multiplet, which we refer to as Type II,
\be
16 +16\ , \mbox{Type II:} \qquad
\{ e_\m^a(6), M(2), b_\mu(4), A_{\m\n}(3), C(1);\, \psi_\m(12), \chi(4)\}\ .
\ee
Setting to zero $(A_{\m\n}, C,\chi; a_\m, D, \lambda)$ in \eq{master} instead gives an off-shell $12+12$ multiplet, known as ``old minimal" \cite{Ferrara:1978em},
\be
\mbox{Old minimal:} \qquad
\{ e_\m^a(6), M(2), b_\mu(4);\, \psi_\m(12)\}\ .
\label{om}
\ee
Finally, setting to zero  $(M, b_\mu, C, D; \chi, \lambda )$ gives another off-shell $12+12$ multiplet known as ``new minimal" \cite{Stelle:1978ye, Sohnius:1981tp},
\be
\mbox{New minimal:} \qquad
\{ e_\m^a(6), A_{\m\n}(3), a_\m(3);\, \psi_\m(12)\}\ .
\ee
Here, we shall begin by reviewing the known curvature-squared invariants in the framework of the Type I off-shell formalism \cite{LeDu:1997us}. Later, we shall consider its truncation to the old minimal off-shell framework, and separately we shall also consider certain higher derivatives invariants in the new minimal off-shell framework \cite{deRoo:1990zm}.

In Type I off-shell formalism, the bosonic part of the off-shell invariants discussed in \cite{LeDu:1997us}, where references to earlier literature can be found, are
\bea
e^{-1} \cL_1 &=& -\frac12 R -\frac13 \left( M\bM -b^a b_a\right) +2gD\ ,
\nn\w2
e^{-1} \cL_2 &=& \frac12 D^2 + f^{\m\n} f_{\m\n}\ ,
\nn\w2
e^{-1} \cL_{W^2} &=& \frac18 C^{abcd} C_{abcd} + \frac13 F^{\m\n} F_{\m\n}\ ,
\nn\w2
e^{-1}\cL_{{\rm Ric}^2} &=& -\frac18 \tR^{ab} \tR_{ab}+\frac{1}{96} R^2 -\frac16 D^2 -\frac16 \big( F^{\m\n}F_{\m\n} +2 f^{\m\n} f_{\m\n} \big)\ ,
\nn\w2
e^{-1} \cL_{R^2} &=& -\frac34 (R-2D)^2 +\left(b^\m b_\m +\frac12 |M|^2 \right) R -2\left(b^\m b_\m +2 |M|^2 \right)D
\nn\w2
&& +3 D^\m M D_\m \bM  -3\left( e^\m_a D_\m b^a \right)^2 +ib^\m \left( \bM D_\m M -M D_\m \bM\right)
\nn\w2
&& -\frac13 \Big( \big( |M|^4 + |M|^2 b^\m b_\m + (b^\m b_\m)^2 \Big)\ ,
\label{StarobinskyLR2}
\eea
where ${\tR}_{ab} =R_{ab}-\frac14 \eta_{ab} R$ and
\bea
&& f_{\m\n} = 2\partial_{[\m} a_{\n]}\ , \qquad  F_{\m\n}=f_{\m\n} +i \partial_{[\m} b_{\n]}\ ,
\nn\w2
&& D_\m M =\del_\m M +2g a_\m M\ ,\qquad  D_\m b_a = \partial_\m b_a -  \omega_{\m a}{}^c b_c\ .
\eea
The off-shell supertransformations of the fermions, up to cubic fermionic terms, are
\bea
\delta \psi_\m &=& 2(\partial_\m-\frac14\o_\m^{ab}\g_{ab}+ga_\m) \epsilon-i b_\m\g_5\epsilon-\frac{i}3\g_\m\slashed{b}\g_5\epsilon-\frac13\g_\m({\rm Re}M+i{\rm Im} M\g_5)\epsilon\ ,
\nn\w2
\delta \lambda &=& if_{\m\n}\g^{\m\n}\epsilon+i\g_5 D\epsilon\ .
\eea
It is worth noting that the combination $2\cL_{{\rm Ric}^2} + \cL_{W^2}$ gives the Gauss-Bonnet term, which is a total derivative in $D=4$. Considering $\cL_1+\cL_2$ alone, it has unusual properties \cite{Freedman:1976uk}. Firstly, the gauge field $a_\m$ couples not only to the gravitino but also to the gaugino $\lambda$. Furthermore,  eliminating the auxiliary field $D$ using its field equations gives a positive cosmological term with a fixed value proportional to the square of the $U(1)$ coupling constant. The elimination of $D$ also gives a homogeneous supersymmetry transformation of the gaugino, thereby triggering a super-Higgs effect \cite{Freedman:1976uk}.

One can truncate the Type I multiplet to the old minimal formulation by setting
\be
a_\m=0\ , \quad D=0\ , \quad \lambda=0 \ .
\label{truncation1}
\ee
This is a consistent truncation at the level of supersymmetry transformations, and it is to be implemented in the off-shell action. Performing this truncation,
one obtains off-shell invariants in the old-minimal formulation from \eqref{StarobinskyLR2}. In this case, given that we need not worry about  the local $U(1)$ symmetry gauged by $a_\mu$, a new off-shell invariant  becomes possible, and it is given by
\be
{e}^{-1}{\cal L}_3=M+\bar{M}-\bar{\psi}_\m\g^{\m\n}\psi_\n\ .
\ee
Considering the combination ${\cal L}_1+{\cal L}_3$, the elimination of $M$ generates a negative cosmological constant. The spectrum of the model ${\cal L}_1+a{\cal L}_3+b{\cal L}_{W^2}|_{a_\m=0}$ was studied in \cite{Lu:2011mw}. For generic choices of the coefficients $a$ and $b$, it was found that
excitations of the theory around an AdS$_4$ background
consist of the massless supergravity multiplet, and a single massive spin 2 multiplet. The latter one consists of the following AdS$_4$ irreps
\be
D\big(E_0 +\frac12, 2\big) \oplus D\big(E_0, \frac32 \big) \oplus D\big(E_0+1, \frac32\big) \oplus D\big(E_0+\frac12, 1\big)\ , \qquad E_0 > \frac52\ ,
\ee
with
\be
E_0= 1+\frac12\sqrt{1-\frac1{a^2b}} > \frac52  \ \ \implies \ \ \ -\frac{1}{8a^2} <b <0 \ .
\label{UC}
\ee
Here $D(E,s)$ denotes a UIR of AdS$_4$ with the lowest weight state with energy $E$, and spin $s$. The condition $E > s+1$ is needed for the unitarity of the representation. However, this does not guarantee ghost-freedom for the massive spin 2 state. Indeed, assuming that the parameters $a$ and $b$ satisfy the condition \eq{UC}, one finds ghostly kinetic terms for the massive spin 2 multiplet \cite{Lu:2011mw}. For further properties of this model, see \cite{Lu:2011mw} where possible ways of evading the ghost issue by imposing certain boundary conditions \cite{Lu:2011ks} is also discussed.
%


\def\beps{{\bar\epsilon}}
\def\bpsi{{\bar\psi}}

\subsection{\texorpdfstring{$N=1,4D$}{N=1,4D} off-shell supergravity in the new minimal formulation and four-derivative invariants}

In the new minimal formulation, the auxiliary fields are an antisymmetric tensor gauge field $B_{\m\n}$ and a vector gauge field $V_\m$ which gauges the chiral U(1) symmetry in the supergravity multiplet \cite{Sohnius:1981tp}. The bosonic part of the off-shell supergravity Lagrangian, in conventions of  \cite{deRoo:1990zm}, is given by
\be
e^{-1} \cL_1 = -\frac12 R -\frac12 H^{\m\n\r} H_{\m\n\r} +\frac23 i \ve^{\m\n\r\s} V_\m \partial_\n B_{\r\s}\ ,
\label{NewMiminalN1D4}
\ee
where $H=dB$. The supersymmetry and gauge transformations are given by
\bea
\delta e_\m^a &=& \frac12 \beps \gamma^a \psi_\m \ ,\qquad \delta\psi_\m = D_\m (\Omega_+, V_+) \e + i\gamma_5 \Lambda \psi_\m\ ,
\nn\w2
\delta B_{\m\n} &=& \frac32 \beps \gamma_{\m} \psi_{\n]}\ ,\qquad \delta V_\m = \frac18 i \beps \gamma_5\gamma_\m \gamma^{ab} \psi_{ab} +\partial_\m \Lambda\ ,
\eea
where
\be
\Omega_{\m \pm}{}^{ab} = \omega_\m{}^{ab} (e,\psi) \pm H_\m{}^{ab}\ ,\qquad V_{\m +} = V_\mu +\frac16 i \ve_\m{}^{abc} H_{abc}\ .
\ee
Further definitions are
\bea
&& \psi_{\m\n} = 2 D_{[\m} (\Omega_+, V_+) \psi_{\n]}\ ,\qquad H_{\m\n\r} = \partial_{[\m} B_{\n\r]} \ ,
\nn\w2
&& D_\m (\Omega_+, V_+) \e = \left( \partial_\m -\frac14 \Omega_{\m +}{}^{ab} \gamma_{ab} -i\gamma_5 V_\m \right)\epsilon\ .
\eea
The new minimal set of auxiliary fields makes it possible to use the analogy between supergravity and Yang-Mills to construct higher derivative invariants. Indeed, three off-shell curvature-squared invariants are constructed in \cite{deRoo:1990zm}, with bosonic parts given by\footnote{For an on-shell construction of a curvature-squared invariant, we refer the reader to  \cite{deRoo:1990zm}.}
\bea
e^{-1} \cL_2 &=& -\frac14 R_{\m\n ab} (\Omega_-) R^{\m\n ab} (\Omega_-) -2 F_{\m\n} (V_+)F^{\m\n}(V_+)\ ,
\nn\w2
e^{-1}\cL_3 &=& -\frac14 F_{\m\n}(V) F^{\m\n}(V) -\frac18 \big( R(\omega)+H^2 \big)^2\ ,
\nn\w2
e^{-1}\cL_4 &=& \frac14 R_{\m\n ab} (\Omega_-) R^{\m\n ab} (\Omega_-) -\frac12 i \ve^{\m\n\r\s} D^\lambda H_{\lambda\m\n}\, F_{\r\s}(V_+)
\nn\w2
&& -F_{\m\n}(V_+) F^{\m\n}(V_+) -\frac16 D_\m H_{\n\r\s} D^\m H^{\n\r\s}\ .
		\label{L3Starobinsky}
\eea
Once again using the analogy between supergravity and Yang-Mills, the off-shell supersymmetrization of $ \tr (R\wedge R)$  has also been found in  \cite{deRoo:1990zm} and its bosonic part  takes the simple form
\be
e^{-1}\cL_5 = \frac14 i \partial_\m \big[   \varepsilon^{\m\n\r\s} \tr \big( \Omega_{\n -} \partial_\r \Omega_{\s -} -\frac23 \Omega_{\n -}\Omega_{\r -}\Omega_{\s -} \big) \big]\ ,
\ee
where $\tr (\Omega_\m \Omega_\n ) = \Omega_{\m a}{}^b \Omega_{\n b}{}^a$. Integrating this over a manifold $M$ with boundary $\partial M$, it yields action for supersymmetric Lorentz-Chern-Simons on $\partial M$.

The coupling of a scalar multiplet, $\{ A, B, \psi, F, G\}$, and Yang-Mills multiplet, $\{A_\m, \lambda, D)$, to new minimal supergravity has been given as well in  \cite{deRoo:1990zm}. The pseudoscalar appears everywhere in the action under a derivative as $\partial_\m B$, and therefore it can be dualized to a two-form potential $A_{\m\n}$. The resulting action, including the Lagrangians $\cL_1$ and $\cL_2$, is given by \cite{deRoo:1990zm}
\bea
 \cL &=&  \cL_1 + A \Big(\cL_2 + \cL_{YM} \Big) - e \left(\partial_\m A \partial^\m A+ F^2 + G^2\right) -\frac32 e F_{\m\n\r} F^{\m\n\r}\ ,
 \eea
where
\bea
F_{\m\n\r} &=& \partial_{[\m} A_{\n\r]} -\frac14 X_{\m\n\r}^L -\frac14 \beta X_{\m\n\r}^{YM} +\frac23 A \partial_{[\m} B_{\n\r]}\ ,
\nn\w2
X_{\m\n\r}^L &=& \tr \big( \Omega_{\m -} \partial_\n \Omega_{\r -} -\frac23 \Omega_{\m -}\Omega_{\n -}\Omega_{\r -} \big) \ ,
\nn\w2
X_{\m\n\r}^{YM} &=& \tr \big( A_{[\m} \partial_\n A_{\r]} -\frac23 A_{[\m} A_\n A_{\r]} \big)\ ,
\eea
and the bosonic part of the  supersymmetric Yang-Mills action is given by
\be
e^{-1} \cL_{YM} =- g^{-2}_{YM}{\rm tr}\big(\frac14 F_{\m\n} F^{\m\n}-\frac12 D^2\big)\ .
\ee
%

\subsection{Off-shell Gauss-Bonnet and its higher derivative scalar couplings in old minimal formulation}

The couplings of matter to $N=1, 4D$ higher derivative supergravity were initiated in
a series of papers \cite{Cecotti:1985nf,Cecotti:1985mf,Cecotti:1987mr}, motivated by
the low energy effective theory of heterotic string compactified on Calabi-Yau three-folds.
Let us first recall the coupling of off-shell scalar multiplets to supergravity in the old minimal formulation without higher derivative corrections. Denoting the fields of the old-minimal multiplet by $(e_\mu^a, u, b_\mu, \psi_\mu)$, the Lagrangian is given \cite{Cremmer:1982en}\footnote{Here we're using the notation and conventions of \cite{Cecotti:1987mr}.}
\bea
e^{-1}{\cal L}_{SG}&=&\frac1{12}\phi R  -\frac12\phi^{''}_{ij*} \partial_{\m}Z^i \partial^\m Z^{j*} +\frac{i}3 b_\m\phi^{'}_i\partial^\m Z^i  +\frac1{18}\phi(uu^*-b_\m b^\m)
\nn\\
&&+\frac12\phi^{''}_{ij*} h^i h^{j*}+\frac12W^{'}_i h^i+\frac13 u^*\big(\phi^{'}_ih^i+\frac32 W^* \big) + h.c.\ ,
\eea
where $\phi (Z, Z^\star) := -3 e^{-K(Z,Z^\star)/3}$ with $K(Z,Z^\star)$ representing the K\"ahler potential and $W(Z)$ is the holomorphic superpotential. The primes denote differentiation with respect to $Z^i$ or $Z^{\star i}$, and in particular  $\phi^{''}_{ij*}$ means the second derivative of $\phi$ with respect to $Z^i$ and $Z^{j*}$. The vector field $b_\m$, and the complex scalar $u=S-iP$ are the auxiliary fields in the old minimal supergravity multiplet, and $h^i$ is the auxiliary field of each the scalar multiplet.

Next, let us consider the coupling of the off-shell Gauss-Bonnet invariant to the off-shell scalar multiplets. This was achieved  in \cite{Cecotti:1987mr} with the Lagrangian given by
\bea
e^{-1}{\cal L}_{GB}&=&\frac98 f(Z)(\star R\star R-R\star R)+\Big\{ f(Z) D^\m L_\m + f^{'}_ih^i\big[\frac12 Ru^*-\frac13 u^*(uu^*+5b_\m b^\m)
\nn\\
&&+iu^*D_\m b^\m+2i b^\m\partial_\m u^*\big]\Big\}+ h.c.\ ,
\eea
where $f(Z)$ is an arbitrary function of the scalar fields $Z_i$ and
\bea
L_\m &=& u^*(\partial_\m -\frac{i}3) u+D_\n(b_\m b^\n) +2i b^\n B_{\n\m}-\frac23 i b_\m b_\n b^\n +\frac32\epsilon^{\m\n\r\s}b_\n D_\r b_\s\ ,
\nn\\
B_{\m\n}&=&\frac32(R_{\m\n}-\frac16 g_{\m\n} R)+\frac{i}2F_{\m\n} (b)-\frac16(uu^*+b_\r b^\r)g_{\m\n}+\frac13 b_\m b_\n\ ,
\eea
with $F_{\m\n}(b)= 2\partial_{[\m} b_{\n]}$.

A higher derivative coupling of the chiral multiplets to old minimal supergravity has also been constructed \cite{Koehn:2012ar,Farakos:2012qu}.
Its construction utilizes the superspace method and demands absence of propagating ghosts. In the notation of \cite{Koehn:2012ar}, the bosonic part of the Lagrangian is given by
\be
e^{-1}{\cal L}_{hds}=\left(\partial_\m Z^i\partial^\m Z^j \partial_\n Z^{k*}\partial^\n Z^{l*}-2h^i h^{k*} \partial_\m Z^j\partial^\m Z^{l*}+h^ih^j h^{k*} h^{l*}\right)T_{ijk*l*}|\ ,
\ee
where $T_{ijk*l*}|$ is the lowest component of the tensor superfield $T_{ijk*l*}$ which is chiral and it is required to be hermitian and symmetric in pairs of indices $i\,,j$
as well as $k*\,,l*$\footnote{Examples given in \cite{Koehn:2012ar} are $T_{ijk\star\ell\star}= g_{ik\star}g_{j\ell\star} +g_{jk\star}g_{i\ell\star}$, and $T_{ijk\star\ell\star} = R_{ik\star j\ell\star}$.} We can now consider a linear combination of all three off-shell super-Poincar\'e invariant Lagrangians,
\be
\cL = {\cal L}_{SG}+\a {\cal L}_{GB}+\b{\cal L}_{hds}\ ,
\ee
from which one can deduce that the auxiliary field $h^i$ obeys the cubic equation
\bea
0&=&\phi^{''}_{ij*}h^{j*}+\frac12 W^{'}_i +\frac13 u^*\phi^{'}_i+\a f^{'}_i\Big(\frac12 Ru^*-\frac13 u^*(uu^*+5 b_\m b^\m )+iu^*D_a A^a+2i D^\m\partial_\m u^* \Big)
\nn\w2
&&+2\b(-h^{k*}\partial_\m Z^j\partial^\m Z^{l*}+h^j h^{k*} h^{l*})T_{ijk*l*}|\ .
\eea
In the absence of Gauss-Bonnet coupling and restricting to a single chiral superfield,
the auxiliary field equation was analyzed in \cite{Koehn:2012ar}. It was found that
when a superpotential is present, the auxiliary field $h$ admits
three distinct solutions, which lead to three distinct theories. One of these solutions is related to the usual solution
for $h$ that one obtains in two-derivative chiral supergravity, while the other two solutions
correspond to new branches of the theory.

The Lagrangian $\cL= \cL_{SG} + \beta \cL_{hds}$ was generalized in \cite{Farakos:2012qu} to include the coupling of Yang-Mills multiplets such that the Yang-Mills fields gauge an isometry group of the K\"ahler sigma model parametrized by the scalars $Z_i$. The cubic field equation for the auxiliary field $h^i$, encountered in this case as well, was analyzed by considering a single scalar multiplet. We refer the reader to \cite{Farakos:2012qu} for further details.

\subsection{\texorpdfstring{Weyl$^4$}{Weyl**4} tensor invariant in old minimal formulation from superspace }

In the context of the old minimal formulation, an off-shell Weyl$^4$ invariant was introduced in superspace in \cite{Moura:2002ft}
\begin{eqnarray}
    \mathcal{L} = -\frac{1}{4\kappa^2} \int E \,\Big[ 3 \left({\bar\nabla}^2 +\frac13 {\bar R} \right)\left( \alpha W^2 \bar{W}^2 +1\right)\Big] d^2\theta  +h.c.\ ,
\end{eqnarray}
where $\alpha$ is an arbitrary constant. This Lagrangian has been worked out in components in \cite{Moura:2002ft}. While the terms involving the fermions are very complicated, the bosonic part turns out to be remarkably simple given by
\begin{eqnarray}
    e^{-1} \cL = - \frac12 R  -\frac13(M^2+N^2-A^\mu A_\mu) -\frac34 \alpha\, \left[ \big(C_{+\mu\nu\rho\sigma} C_+^{\mu\nu\rho\sigma}\big)\big(C_{-\mu\nu\rho\sigma} C_-^{\mu\nu\rho\sigma}\big) + \ldots \right] \ ,
\end{eqnarray}
where $C_{\pm\mu\nu\rho\sigma}$ are the (anti)self-dual part of the Weyl tensor, and ellipsis represents the terms involving the auxiliary fields $(M,N,A_\mu)$.


\subsection{\texorpdfstring{$N=1, 4D$}{N=1, 4D} supersymmetric extension of Starobinsky type \texorpdfstring{$R+R^2$}{R+R2} models}

The $R^2$ extension of general relativity, known as the Starobinsky model \cite{Starobinsky:1980te}, is one of the most studied and successful inflationary models. At the bosonic level, the Starobinsky model is given by
\be
e^{-1} \cL = R + \frac{1}{6M^2} R^2 \,.
\label{Starobinsky}
\ee
Its supersymmetric completion in the off-shell Type I multiplet formulation is given by the combination of $\cL_1$ and $\cL_{R^2}$ action from \eqref{StarobinskyLR2} \cite{LeDu:1997us}. In the new minimal setting, the off-shell Starobinsky model is given by the combination of \eqref{NewMiminalN1D4} and $\cL_3$ from \eqref{L3Starobinsky}. Note that in both the Type-I and new minimal settings, auxiliary fields can be consistently set to zero, leading to an identical on-shell result.

The $R + R^2$ Starobinsky model can also be expressed as a scalar-tensor theory, which is its most convenient form in cosmological applications, by considering the following Lagrangian \cite{DeFelice:2010aj}
\begin{eqnarray}
e^{-1} \cL = \Lambda + \frac{1}{6M^2} \Lambda^2 - \varphi (\Lambda - R) \,,
\label{FStarobinsky}
\end{eqnarray}
which, upon integrating out the $\varphi$, takes us back to \eqref{Starobinsky}. Varying this action with respect to $\Lambda$, imposing the resulting field equation, performing the Weyl rescaling $g_{\m\n} \to \vf g_{\m\n}$ and finally introducing a scalar field $\phi = \sqrt{3/2} \ln \vf$, the action reads
\begin{eqnarray}
    e^{-1} \cL &=& R - \partial_\m \f \partial^\m \f - \frac32 M^2 \left(1 - e^{-\sqrt{\frac{2}{3}}\phi} \right)^2 \,.
    \label{DualStarobinsky}
\end{eqnarray}
This action is known as the scalar-tensor form of the Starobinsky model. Its supersymmetric version has been obtained in \cite{Cecotti:1987sa}. Following the same logic that leads to \eqref{DualStarobinsky}, the following result has been obtained for the bosonic part of the supersymmetrized $R+R^2$ action in the old minimal formulation \cite{Farakos:2013cqa}:
\be
    e^{-1} \cL = \frac12 R - K_{I \bar J} \partial_\m Z^{(I)} \partial^\m \bar Z^{(\bar {J})} - V\ ,
\ee
where $I, J = 1,2$ and
\bea
V &=& e^K \left( K^{I \bar{J}} D_I W D_{\bar J} \bar W - 3 W \bar{W} \right) \ ,
\nn\w2
K&=& -3\log \left( Z^{(1)}+\bar{Z}^{(1)}-Z^{(2)}\bar{Z}^{(2)} \right)
\nn\w2
W &=& 3\sqrt{6} M Z^{(2)}\left(Z^{(1)}-\frac12 \right)\ ,
\eea
and $K_I := K_{,I} \  K_{I \bar{J}} := K_{,I\bar J}$ and $D_I W  = W_{,I} + K_I W$.
Parametrizing the complex scalar $Z^{(1)}$ as \cite{Cecotti:1987sa}
\be
Z^{(1)} = \frac12 e^{\sqrt{\frac{2}{3}}\f} + \rmi b \ ,
\label{DecompositionOfT}
\ee
the action is invariant under $b\to -b$ and  $Z^{(2)} \to -Z^{(2)}$, so that one can set both $b=0$ and $Z^{(2)}=0$ consistently. Doing so yields the scalar-tensor formulation of the Starobinsky model \eqref{DualStarobinsky}.

The Lagrangian \eq{FStarobinsky} can be generalized by replacing $\Lambda^2$ with $M^{2-2n} \Lambda^{2n}$. This gives  $R+R^n$, upon elimination of $\Lambda$, and the potential
\begin{eqnarray}
    V(\phi) &=& \frac{(n-1)}{2n^{n/(n-1)}} M^2 e^{-2\sqrt{\frac{2}{3}}\phi} \left( e^{\sqrt{\frac{2}{3}}\phi} - 1\right)^{\frac{n}{n-1}} \,.
    \label{RnScalarPotential}
\end{eqnarray}
The supersymmetric completion of $R+R^n$ model,  in the off-shell old-minimal formulation is given by \cite{Cecotti:1987sa,Ozkan:2014cua}\footnote{The first line is obtained from \eqref{StarobinskyLR2} by setting $g=0$ and letting
$b_\m\rightarrow 3A_\m,\ M\rightarrow {\sqrt 3} S,\ R\rightarrow -R$.},
\bea
e^{-1} \cL &=& \frac12 R - |S|^2  +3A_\mu A^\mu
\nn\w2
&& +\alpha  \text{Re} \Bigg[ \Big(R + 6 A^\m A_\m + 6 \rmi \nabla_\m A^\m + 2 |S|^2 \Big)^{n-2} \times\Big( \frac1{12} R^2 + R A^\m A_\m+ \frac{n - 1}{6} R |S|^2
\nn\\
&&\quad + 3 (A^\m A_\m)^2 + \frac{2n - 3}3|S|^4 + ( n - 1) A^\m A_\m S \bar{S}+ 3 (\nabla_\m A^\m)^2+(n-1) \bar{S} \Box S
\nn\\
&& \quad + (3n - 5) \rmi |S|^2 (\nabla_\m A^\m) + (2n - 2) \rmi A_\m \bar{S} \partial^\m S \Big) \Bigg] \ .
\label{RRn}
\eea
This was constructed in \cite{Cecotti:1987sa} in superspace, and in \cite{Ozkan:2014cua} by using superconformal tensor calculus in which the Weyl multiplet is coupled to a chiral multiplet compensator. In the latter approach, upon fixing the redundant symmetries, one obtains the off-shell Poincar\'e theory, in which the scalar $S$ of the chiral multiplet ends up being an auxiliary field. Note that in the Lagrangian above, the complex scalar $S$ has developed a kinetic term. If we do not treat the $R^n$ as a small perturbative extension of the Einstein term,  then the scalar $S$ is unstable during the inflationary phase \cite{Kallosh:2013lkr,Antoniadis:2014oya}. To avoid this problem, one can take the compensating chiral multiplet to be nilpotent \cite{Rocek:1978nb,Lindstrom:1979kq,Casalbuoni:1988xh}. In that case, the scalar $S$ becomes bilinear in the fermions, and the bosonic part of the action turns out to be \eqref{RRn} with $S$ set to zero. The dual scalar-tensor model can then be constructed by following the steps spelled out in action \cite{Ozkan:2014cua}, the resulting bosonic action being  $\cL= e (R -\partial_\mu \phi\partial^\mu \phi -V)$ where the potential $V$ is given by \eq{RnScalarPotential}.

In the case of the new minimal supergravity, one can again start with the supersymmetric completion of $R + R^2$ theory, which is now given by the lowest order in derivative Lagrangian \eqref{NewMiminalN1D4} and the Lagrangian $\cL_3$ in \eqref{L3Starobinsky} that contains the $R^2$ term. Next, one considers the analog of \eqref{FStarobinsky} in the new minimal formulation. This is done in \cite{Farakos:2013cqa} by introducing a linear multiplet playing the role of the Lagrange multiplier, and a vector multiplet,  which upon the use of the Lagrange multiplier equation of motions becomes equal to a composite vector multiplet whose highest component contains the Ricci scalar. In the conventions of \cite{Farakos:2013cqa}, integrating out the Lagrange multiplier gives a solution in which a chiral scalar multiplet with scalars $(\phi, a)$ arises. The resulting on-shell Lagrangian is given by \cite{Farakos:2013cqa}
\begin{eqnarray}
e^{-1} \cL &=& \frac12 R - \frac{1}{4 g^2} F^{\m\n}(V) F_{\m\n}(V) +  \frac{\rmi}{8 g^2} \e^{\m\n\r\s} F_{\m\n}(V) F_{\r\s}(V)   - 2 e^{4/\sqrt{6} \f} \left( \partial_\m a + V_\m \right)^2 \nonumber\\
&& - \frac12 \partial_\m \f \partial^\m \f - \frac{9}{2} g^2 \left(1 - e^{\sqrt{2/3} \f} \right)^2 \ .
\end{eqnarray}
Note that the vector $V_\mu$ eats up the scalar $a$, so that this Lagrangian describes the bosonic coupling of on-shell $N=1$ supergravity to a single massive vector multiplet. Analogous construction of $R+\sum_n\xi_n R^n$ can be found in \cite{Ferrara:2013kca}.

\subsection{ Off-shell Killing spinors in \texorpdfstring{$N=1, 4D$}{N=1, 4D} supergravity}

Motivated by constructing supersymmetric field theories on curved manifolds, off-shell Killing spinors in four-dimensional $N=1$ and $N=2$ supergravities with Lorentzian or Euclidean signature have been studied extensively. Here, we will focus on the off-shell $N=1$ supergravity with Lorentzian signature.  Readers interested in the Euclidean case are referred to \cite{Samtleben:2012gy,Dumitrescu:2012ha,Kehagias:1997cq, Klare:2012gn, Klare:2013dka,Butter:2015tra}. Supergravity backgrounds preserving certain amount of supercharges in off-shell $N=2$ (conformal) supergravity with Lorentzian and Euclidean signatures can be found in \cite{Klare:2013dka,Butter:2015tra,Gupta:2012cy}

\subsubsection*{ Off-shell Killing spinors in \texorpdfstring{$N=1, 4D$}{N=1, 4D} old minimal supergravity}

Off-shell Killing spinors in the $N=1, 4D$ old minimal supergravity were first studied by \cite{Festuccia:2011ws} in pursuit of supergravity backgrounds preserving maximal off-shell supersymmetry. Later on, more general supergravity backgrounds preserving less supersymmetry were investigated in \cite{Liu:2012bi}. Here our presentation will closely follow \cite{Liu:2012bi}. In the old minimal supergravity, the vanishing of supersymmetry variation of the gravitino gives the Killing spinor equation
\be
\left[\nabla_\m-\frac{i}6(\g_\m^{~\n}-2\d^\n_\m)\g_5 V_\n+\frac16\g_\m(S+i\g_5 P)\right]\epsilon=0\ .
\ee
From the integrability of the Killing spinor equation, one can deduce the following conditions for a supergravity background to preserve maximal supersymmetry \cite{Festuccia:2011ws, Liu:2012bi}
\bea
&&SV_\m=PV_\m=0,\quad \partial_\m S=\partial_\m P=0\ ,
\nn\\
&&\nabla_\m V_\n=0,\quad C_{\m\n\r\s}=0\ ,
\nn\\
&&R_{\m\n}=\frac29(V_\m V_\n-g_{\m\n}V^2)-\frac13 g_{\m\n}(S^2+P^2)\ .
\eea
When $V_\m=0$, and $S, P$ are non-vanishing constants, the background is AdS$_4$ with the radius $3/\sqrt{S^2+P^2}$. If instead, $V_\m$ is a non-vanishing covariantly constant vector, and $S=P=0$, the background turns out to be $R\times S^3$ or AdS$_3\times R$ \cite{Festuccia:2011ws, Liu:2012bi}.

In general, the background preserves only a fraction of maximal supersymmetry.
Assuming there exists a spinor satisfying the Killing spinor equation, one can show that its bilinear $
K_\m=\bar{\epsilon}\g_\m\epsilon$
is a null Killing vector. Assuming $K^\m V_\m=0$, one can show that $K_\m$ is hypersurface orthogonal. One can choose special coordinates $(u,\,v,\,y^m)$ so that $K^\m\partial_\m=\partial_v$,  and thus the metric admitting such a Killing vector can be parameterized as
\be
ds^2=H^{-1}\left({\cal F}du^2+2dudv+\widehat{g}_{mn}dy^mdy^n\right)\ ,
\label{4dn1bg}
\ee
where $H$, ${\cal F}$ and $\widehat{g}_{mn}$ depend only on coordinates $(u,\,y^m)$. To proceed, one introduces the vierbein basis
\be
e^+=H^{-1}du,\quad e^-=dv+\frac12{\cal F}du,\quad e^a=H^{-1/2}\widehat{e}^a_m dy^m\ .
\label{4dn1vb}
\ee
Exploring the consequences of Killing spinor equation leads to the solutions for auxiliary fields \cite{Liu:2012bi},
\bea
S&=&-H^2\widehat{\nabla}^m(H^{-3/2}X_m),\quad P=H^2\widehat{\epsilon}^{mn}\partial_m(H^{-3/2}X_n)\ ,\quad V_-=0\ ,
\nn\\
V_+&=&-H\widehat{\epsilon}_{mn}X^m\partial_u X^n,\quad V_m=H^{1/2}[-X_m(\widehat{\epsilon}^{np}\partial_n X_p)+\widehat{\epsilon}_{mn}X^n(\widehat{\nabla}^pX_p)]\ .
\eea
Thus the general supergravity background admitting at least one Killing spinor is characterized by $H(u,y^m)$, ${\cal F}(u,y^m)$, $\widehat{g}_{mn}(u,y^m)$ and a unit spacelike vector $X^m(u,y^m)$ obeying $X_m X^m=1$.  The Killing spinor satisfies the projection conditions
\be
\g^+\epsilon=0,\quad X_a\g^a\epsilon=\epsilon\ .
\ee
When $K^\m V_\m\neq 0$, the geometric meaning of the background is unclear and deserves further study. It should be noted that so far, one has not employed field equations that are model-dependent. In $N=1, 4D$ old minimal formulation of Einstein-Weyl and more general curvature squared supergravities, various supersymmetric solutions were studied in \cite{Lu:2012am,Lu:2012cz,Liu:2012mh}.

\subsubsection*{Off-shell Killing spinors in \texorpdfstring{$N=1, 4D$}{N=1, 4D} new minimal supergravity}

In the convention of \cite{Liu:2012bi}, vanishing of supersymmetry variation of gravitino
leads to the Killing spinor equation in $N=1, 4D$ new minimal supergravity
\be
\left[\nabla_\m + i\g_5 A_\m -\frac{i}2(\g_\m^{~\n}-2\d^\n_\m)\g_5V_{\n}\right]\epsilon=0\ ,
\ee
where $A_\m$, and $V_\m$ satisfying $\nabla^\m V_\m=0$, are auxiliary fields. Combining the integrability of the Killing spinor equation with the requirement of maximal supersymmetry gives rise to the conditions
\bea
&& \nabla_\m V_\n=0,\quad \partial_{[\m}A_{\n]}=0, \quad C_{\m\n\r\s}=0\ ,
\nn\\
&& R_{\m\n}=2(V_\m V_\n-g_{\m\n}V_\r V^\r)\ .
\eea
A solution to the equations above is given by $R\times S^3$ or AdS$_3\times R$ with
\be
V_i=A_i=0,\quad V_0=\frac1r,\quad A_0={\rm const}\ ,
\ee
where $r$ is the radius of $S^3$ (AdS$_3$), and subscript ``0" labels the component in the $R$-direction.

Supergravity backgrounds preserving less supersymmetry were also analyzed in \cite{Liu:2012bi}.
Using the same metric \eqref{4dn1bg} and vierbein \eqref{4dn1vb}, the existence of at least one Killing spinor is guaranteed by taking the vector fields to be
\bea
&&A_-=V_-=0,\quad V_m=\frac12 H^{-1/2}\widehat{\epsilon}_m^{~n}\partial_n H,\quad A_++\frac32V_+=-\frac12 H\widehat{\epsilon}_m^{~n}X^m\partial_uX^n\ ,
\nn\\
&&A_m=\frac12H^2\left[-X_m\widehat{\epsilon}^{np}\partial_n(H^{-3/2}X_p)+\widehat{\epsilon}_{mn}X^n\widehat{\nabla}^p(H^{-3/2}X_p)\right]\ ,
\eea
where $A_\pm$ and $V_\pm$ are lightcone projections of the vectors, and $H, X_m$ are described in the previous subsection. Different from the old-minimal case, here, the Killing spinor obeys one projection condition $\g^+\epsilon=0$. Thus a generic off-shell background in new minimal supergravity preserves two of the four supersymmetries. As noticed in \cite{Liu:2012bi}, this solution contains AdS$_4$ with non-trivial auxiliary vectors breaking the AdS isometry. The background discussed so far corresponds to the untwisting case, meaning that the Killing 1-form $K_\m dx^\m$ built from a bilinear of the Killing spinors satisfies $K\wedge dK=0$. More general backgrounds for which the untwisting condition is violated can be found in \cite{Cassani:2012ri}. There is also a close relation between $N=1$ conformal Killing spinors and Killing spinors in new minimal supergravity as described in \cite{Cassani:2012ri}.

\section{\texorpdfstring{$D=3$}{D=3}}
Ungauged $N=8$ and $N=16$ supergravities were  coupled to scalar multiplets in \cite{Marcus:1983hb}. Generalizations to other ungauged $N< 16$ cases were provided in \cite{deWit:1992psp}. The most general gaugings of these theories were achieved in \cite{deWit:2003ja}, which also provides references to earlier works. We shall review results for higher derivative invariants in the case of off-shell $N=1,2$ supergravities in $3D$. In the case of $N=8$ supergravity in $3D$, higher derivative extensions have been obtained in \cite{Eloy:2020dko} from the ordinary dimensional reduction of $N=1, 10D$ heterotic supergravity on torus $T^7$, and in \cite{Baron:2017dvb} from its double field theory formulation.

We shall first review the higher derivative superinvariants for the off-shell $N=1,2$ supergravities in $3D$, and their salient properties, including the issue of ghost freedom and the (non)unitarity of their holographic duals as certain $2D$ CFTs. In the case of $N=2$ supergravity in $3D$, there exist two distinct off-shell supergravities. They are also referred to as $N=(1,1)$ and $N=(2,0)$ supergravities, because they admit vacuum solutions with super AdS symmetry, $OSp(2,p) \oplus OSp(2,q)$, with $(p,q)=(1,1)$ and $(p,q)=(2,0)$, respectively. Since this terminology is associated only with the nature of the vacuum solution, we shall refer to the total amount of supersymmetry $N=p+q$ instead, in characterizing the supersymmetry of the actions in what follows.

Beyond $N=2$, we shall review off-shell $N=6$ conformal supergravity coupled to matter, in which the gravitational sector contains the Lorentz Chern-Simons term. Its consistent truncations easily yield such models for $N <6$. On-shell similar couplings will be reviewed for $N=8$. We shall also comment on superspace formulations for $4\le N\le 8$. Beyond four-derivative extensions, linearized results are available in \cite{Bergshoeff:2010ui}. Finally, we will also summarize results on exact solutions of higher derivative extensions of $3D$ (conformal) supergravities.

It is worthwhile to mention that three-dimensional higher-derivative supergravity only makes sense non-perturbatively. Indeed, if treated perturbatively, they can always be reduced to (cosmological) Einstein-Hilbert supergravity and a gravitational Chern-Simons supergravity by means of field redefinitions and truncations \cite{Gupta:2007th}.

\subsection{\texorpdfstring{$N=1, 3D$}{N=1, 3D} higher derivative  supergravities}

\subsubsection{Off-shell invariants from superconformal tensor calculus}

 The off-shell $N=1, 3D$ supergravity multiplet consists of a vielbein, a Majorana gravitino, and a real scalar auxiliary field. A general Lagrangian up to four-derivatives, has been constructed in \cite{Bergshoeff:2010mf} by using superconformal tensor calculus. Its bosonic part is given by
\begin{eqnarray}
 e^{-1} \cL_1 &=& \sigma \left( R - 2 S^2\right) + M S + \frac{1}{m^2} \left( R_{\mu\nu} R^{\mu\nu} - \frac38 R^2 - \frac12 R S^2  - \frac32 S^4\right)
 \nn\w2
 && + \frac{1}{8 \tilde m^2} \left( R^2 + 16 S \Box S + 12 R S^2 + 36 S^4 \right)  + \frac{1}{\check{m}^2} \left( S^4 + \frac{3}{10} R S^2 \right)
 \nn\w2
 &&  + \frac{1}{\check \mu} \left(S^3 +\frac{1}{2} RS\right) + \frac{1}{\mu} \varepsilon^{\mu\nu\l} \Gamma_{\lambda\sigma}^\rho \left( \partial_\mu \Gamma_{\rho\nu}^\sigma + \frac23 \Gamma^\sigma_{\mu\tau} \Gamma^\tau_{\nu\rho} \right) \ ,
 \label{3dN1CurvatureSquared}
\end{eqnarray}
where $(\sigma, M, m, \tilde m, \mu, \tilde \mu)$ are independent coupling constants. Thus, there are six independent off-shell invariants. The fermionic parts of $\cL_\sigma, \cL_M$ and $\cL_\mu$ can be found in \cite[Eqs. (2.8) and (2.10)]{Bergshoeff:2010mf}, and the fermionic part of the $\cL_{\tilde m^2}$ can be found in \cite[Eq. (2.39)]{Bergshoeff:2010mf}. The supersymmetric completion of a specific combination of $\cL_{m^2}$ and $\cL_{\tilde m^2}$, namely $-\cL_{m^2} + 1/8 \cL_{\tilde m^2}$ is given in  \cite[Eq. (2.24)]{Bergshoeff:2010mf}. The bosonic part of the most general six-derivative supersymmetric Lagrangian is given by \cite{Bergshoeff:2014ida}
\bea
&&e^{-1} \cL_2 = a_1 R_{\m\n} R^{\n\r} R_\r{}^\m  +a_2 R R_{\m\n} R^{\m\n} + a_3 R^3  \nn\\
&&\qquad\qquad  + \Big( - \frac{3}{8}a_1 + a_6 \Big) R_{\m\n} \partial^\m S \partial^\n S + \Big( \frac{31}{32}a_1 + 8 a_2 + 32 a_3 - \frac14 a_6 + a_7 \Big) R S \Box S  \nn\\
&& \qquad\qquad+ \Big( \frac{123}{32} a_1 + 4 a_2 - a_5 - \frac14 a_6 \Big) S^2 R_{\m\n} R^{\m\n} - \Big(  \frac{117}{64}a_1 + 2 a_2 + \frac58 a_6 - \frac 12 a_7  \Big) R \partial_\m S \partial^\m S  \nn\\
&&\qquad \qquad+ \Big( \frac{223}{512}a_1 + \frac72 a_2 + 14 a_3 + a_4 + \frac14 a_5 + \frac7{64} a_6 - \frac1{16} a_7 \Big) R^2 S^2 \nn\\
&&\qquad \qquad + \Big( - \frac{309}{16}a_1 - 104 a_2 - 384 a_3 - 16 a_4 + 2 a_5 + \frac32 a_6 - 10 a_7 \Big) S^2 \partial_\m S \partial^\m S \nn\\
&&\qquad \qquad + \Big( \frac{2357}{256}a_1 + 27 a_2 + 80 a_3 + \frac{19}{2} a_4 - a_5 + \frac5{32} a_6 - \frac{1}{8} a_7 + \frac{5}{22} a_8 \Big)  R S^4 \nn\\
&& \qquad \qquad + \Big( \frac{527}{32}a_1 + 52 a_2 + 160 a_3 + 25 a_4 - 3 a_5 + \frac14 a_6 + \frac12 a_7 + a_8 \Big) S^6 \,,
\label{GenCubAct}
\eea
where $a_i$ with $i = 1,\ldots,8$ are free parameters. The fermionic parts of these 8 independent invariants have not been provided in the literature. The following special cases of $\cL_1$ have been studied extensively in the literature:
\begin{align}
& N=1:\quad \mbox{Topological Massive Gravity (TMG):} & \sigma, M, \mu\ ,
\nn\w2
& N=1:\quad \mbox{New Massive Gravity (NMG):} & \sigma, {\check m}^2=3 m^2/5\ ,
\nn\w2
& N=1:\quad \mbox{Generalized Massive Gravity (GMG): } & \sigma, \mu, {\check m}^2=3 m^2/5\ ,
\label{3dMassiveGravityModels}
\end{align}
where the specified parameters are kept and $\check \mu = \infty$. The fermionic parts of the Lagrangians $\cL_{RS}$ and $\cL^{(6)}$ can be straightforwardly obtained from the ingredients provided in  \cite{Bergshoeff:2010mf,Bergshoeff:2014ida,Andringa:2009yc}. Off shell, the supertransformation of the sole fermionic field is
\bea
\d \p_\m &=&  \Big( \partial_\m + \frac14 \o_\m{}^{ab} \g_{ab} \Big) \e + \frac12 S \g_\m \e \ .
\label{OffShellTransformationRulesD3N1}
\eea
The supersymmetric completions of these massive gravity models admit a maximally supersymmetric AdS$_3$ vacuum. Thus they are holographically dual to supersymmetric $2D$ CFTs. With their higher derivative extensions, the left and right central charges for the CFT duals are given by \cite{Bergshoeff:2010mf}
\begin{eqnarray}
 c_{L/R}^{\textrm{GMG}} &=& \frac{3\ell}{2G} \left(\sigma + \frac{1}{2\ell^2 m^2} \mp \frac{1}{\mu\ell}\right)\ ,
 \end{eqnarray}
which reproduces those of NMG in the $\mu \to \infty$ limit while TMG is obtained in the $m \to \infty$ limit. The central charges for extended NMG (ENMG), corresponding to the following choice of free parameters in \eqref{GenCubAct}
\begin{align}
a_1 &= 1\ , &  a_2 & = - \frac{9}8, & a_3 &= \frac{17}{64}\ , & a_4 &= - \frac{3}{32}\ , \nonumber\\
a_5 &= - \frac{3}{4}\ , & a_6 &= \frac{3}8\ , & a_7 &= - \frac{3}8\ ,  & a_8 &= - \frac{33}{160} \ ,
\end{align}
are of the form
\begin{eqnarray}
  c_{L/R}^{\textrm{ENMG}} &=& \frac{3\ell}{2G} \left(\sigma + \frac{1}{2\ell^2 m^2} - \frac{a}{8 m^4\ell^2} \right) \ .
\end{eqnarray}
It is important to note that GR in three dimensions with or without a cosmological constant, has no propagating degrees of freedom and local dynamics may be generated by adding higher-curvature terms at the price of introducing potentially ghost-like excitations.  The special combinations listed in \eqref{3dMassiveGravityModels} are meticulously designed to avoid non-unitary perturbative degrees of freedom. Amongst them, TMG contains a single massive helicity-$2$ state where the mass is sourced by the gravitational Chern-Simons term, hence acquiring the name Topologically Massive Gravity. For NMG and GMG, their spectrum contains a pair of helicity-$\pm 2$ states.

\subsubsection{On shell minimal massive supergravity from third way consistency}

Finally, in three-dimensions, there exist bosonic higher-derivative gravity models based on the third way consistency \cite{Bergshoeff:2015zga, Ozkan:2018cxj}. The landmark of these models is that the integrability of the metric field equation  requires the metric field equation itself rather than being a geometric identity. Although the third way field equations cannot arise from the variation of a diffeomorphism invariant functional involving only the metric and its derivatives, a local Lagrangian formulation can be achieved by introducing a number of auxiliary fields whose elimination by field equations gives rise to the higher-derivative gravitational equations of motion.  So  far the only supersymmetric third way consistent model was given in \cite{Deger:2022gim,Deger:2023eah}
\begin{eqnarray}
\cL = \epsilon^{\mu\nu\rho} \left[ e_\m{}^a R_{\nu\rho\, a} +\lambda \e_{abc} e_\m{}^a e_\n{}^b e_\rho{}^c + \tau e_\m{}^a D_\nu (\bar\omega)e_\rho{}^a   + \kappa \big(\bar\omega_\mu{}^a \partial_\nu \bar\omega_{\rho\, a} + \frac13 \e_{abc} \bar\o_\m{}^a \bar\o_\n{}^b \bar\o_\r{}^c \big) \right].\nn\\
\end{eqnarray}
Here, $R^a = d\o^a + \ft12 \e^{abc} \o_b \o_c$ with $\o = \o(e)$ and $\bar\o$ is an independent (torsionful) spin connection.  $D(\bar\o)$ denotes the covariant derivative with respect to $\bar\o$ and $\{\tau,\kappa,\lambda\}$ are free parameters.  The bosonic part of this model is known as the Minimal Massive Gravity (MMG) \cite{Bergshoeff:2014pca}. In \cite{Deger:2022gim,Deger:2023eah}, the $N=1$ MMG model was derived up to and including quartic fermions. Up to cubic fermion terms, the supersymmetry transformation rules for the fermionic fields are given by
\be
\d \p_\m = D_\m(\omega) \e - \frac14 \left(\eta \tau + \frac{1}{\eta \kappa} \right)\gamma_\m \e ,\qquad
\d \P_\m = D_\m(\bar\o) \e - \frac12 \eta \tau \g_\m \e  \ ,
\ee
where the constant $\eta$ satisfies
\begin{eqnarray}
 \l = \frac{1}{12} \left(\eta\tau + \frac{1}{\eta \kappa}\right)^2 - \frac{\tau}{3} \left(\eta \tau - \frac{1}{\eta \kappa} \right) \,.
\end{eqnarray}
For the time being, the fully supersymmetric completion of the MMG is yet to be obtained and  possible ways of completing the model beyond the quartic-fermion level were suggested in \cite{Deger:2023eah}.

\subsection{Off-shell \texorpdfstring{$N=(1,1), 3D$}{N=(1,1), 3D} higher derivative  supergravities}

The field content of the off-shell $N=(1,1), 3D$ multiplet is
\be
\{ e_\mu{}^a, \psi_\mu, V_\mu, S \}\ ,
\ee
where the gravitino is a Dirac spinor, the auxiliary vector $V_\m$ is not associated with any gauge symmetry, and the auxiliary scalar $S$ is complex. Recall that the terminology of $N=(1,1)$ means that there exists a Lagrangian which admits an AdS vacuum with $OSp(2,1) \oplus OSp(2,1)$ symmetry. The bosonic part of off-shell $N=(1,1), 3D$ supergravity up to and including four-derivative terms is given by  \cite{Alkac:2014hwa}
\bea
e^{-1}\cL_{N=(1,1)} &=& \sigma \left(R + 2 V^\mu V_\mu - 2 |S|^2 \right) + M (S + S^\star) \nn\w2
&& + \frac{1}{\mu} \left[-\frac14\,\varepsilon^{\mu\nu\rho}\left( R_{\mu\nu}{}^{ab}(\omega)\, \omega_{\rho ab}
+ \frac23 \omega_\mu{}^{ab}\, \omega_{\nu b}{}^c\,\omega_{\rho ca} \right)
 +\varepsilon^{\mu\nu\rho} F_{\mu\nu} {V_\rho} \right]
\nn\w2
&& + \frac{1}{m^2} \Bigg[R_{\mu\nu}R^{\mu\nu} -R_{\mu\nu}V^\mu V^\nu + \frac74 RV^2 +\frac{17}{8} R\left|S\right|^2
+\frac{23}{4}|S|^4 - F_{\mu\nu} F^{\mu\nu} \nn
\\
&& +6 \left(\nabla_\mu V^\mu \right)^{2}
+\frac32 (V^2)^2 +\frac{11}{4} V^2 \left|S\right|^{2}  -6\partial_{\mu}S\partial^{\mu}S^*
-\frac{7}{2}\rmi V^\mu S^* \overleftrightarrow{\partial_\mu}S \Bigg] \nn\\
&& + \frac{1}{\tilde m^2} \Bigg[ R^{2}+16\left|S\right|^{4}+ 4(V^2)^2 +6 R\left|S\right|^{2}+4RV^2+12\left|S\right|^2 V^2
\nn\\
&&  -16\partial_\mu S\,\partial^\mu S^* -8\rmi V^\mu S^* \overleftrightarrow{\partial_\mu} S +16\left( \nabla_\mu V^\mu \right)^2 \Bigg] \nn\w2
&& + \frac{1}{\check m^2} \left[ R S^2+ \frac{10}{3} S^2 |S|^2
+2  S^2 V^2  -4i S^2 \nabla_{\mu} V^{\mu} + h.c. \right] \ .
\label{N11Action}
\eea
The fermionic part of the Lagrangian proportional to $\sigma, M$ and $1/\mu$ can be found in \cite{Alkac:2014hwa}. The fermionic terms of the four derivative action have not been given in component form explicitly, but it is straightforward to write them down using the ingredients provided in  \cite{Alkac:2014hwa}. The supersymmetry transformation rules for the gravitino is given by
\be
\delta\psi_\mu = (\partial_{\mu}+\frac14\omega_{\mu}{}^{ab}\,\gamma_{ab})\epsilon -\frac{1}{2} \rmi V_{\nu}\,\gamma^{\nu}\gamma_{\mu}\,\epsilon
-\frac12 S\gamma_\mu \left(B\epsilon\right)^{*}\ .
\label{3DN11KS}
\ee
Special cases of the $N=(1,1), 3D$ model arise as follows:
\begin{align}
& N=(1,1): \quad \mbox{Topological Massive Gravity (TMG):} & m={\tilde m}= {\check m} =\infty\ ,
\nn\w2
& N=(1,1): \quad \mbox{New Massive Gravity (NMG):} & \mu=\infty,\ {\tilde m}^2=\frac83 m^2,\ {\check m}^2= 8m^2\ ,
\nn\w2
& N=(1,1): \quad \mbox{Generalized Massive Gravity (GMG):} & {\tilde m}^2=\frac83 m^2,\ {\check m}^2= 8m^2\ .
\end{align}

\subsection{Off-shell \texorpdfstring{$N=(2,0), 3D$}{N=(2,0), 3D} higher derivative supergravities}

Next, we summarize the off-shell $N=(2,0), 3D$ higher derivative couplings. The field content is
\be
\{ e_\mu^a, \psi_\mu, C_\mu, V_\mu, D \}\ ,
\label{(2,0)}
\ee
where the gravitino is a Dirac spinor, $C_\mu$ is the gauge field, the auxiliary field $V_\mu$ is non-gauge and the auxiliary field $D$ is real. Note that the notation $N=(2,0)$ means that there exists an AdS vacuum solution with $OSp(2,2) \oplus O(2,1)$ symmetry. The bosonic part of off-shell $N=(2,0), 3D$ supergravity up to and including four-derivative terms is given by
\begin{align}
 e^{-1} \cL_{(2,0)} =& \ \sigma \left(R -2 G^2-8D^{2}-8\epsilon^{\mu\nu\rho}\,C_{\mu}\,\partial_{\nu}V_{\rho}\right) + M \left( 2D-\epsilon^{\mu\nu\rho}\,C_{\mu}\,G_{\nu\rho} \right)
 \nn\w2
 & + \frac{1}{\mu} \left[-\frac14\,\varepsilon^{\mu\nu\rho}\left( R_{\mu\nu}{}^{ab}(\omega)\, \omega_{\rho ab}
+ \frac23 \omega_\mu{}^{ab}\, \omega_{\nu b}{}^c\,\omega_{\rho ca} \right)
 +\varepsilon^{\mu\nu\rho} F_{\mu\nu} {V_\rho} \right]
\nn\w2
 & + \frac{1}{m^2} \Bigg[R_{\m\n}\, R^{\m\n} - \frac14 R^2 + 4 R D^2   +RG^2 -2R_{\mu\nu}\,G^\mu\, G^\nu + 48 D^4 + 8 D \Box D
\nn\\
&\qquad  +8D^2G^2 +(G^2)^2  - 2 ( F_{\mu\nu}  + \nabla_{[\mu} G_{\nu]}  )^2- \left(\nabla_\mu G_\nu + 4DG_{\mu\nu} \right)^2 \Bigg]
\nn\w2
& + \frac{1}{\tilde m^2} \left[ (R + 24 D^2 + 2G^2 )^2  -8 \left( F_{\mu\nu} + 2\nabla_{[\mu} G_{\nu]} +4 DG_{\mu\nu} \right)^2
+  64D\Box D \right]\ ,
\label{N20GMG}
\end{align}
where
\begin{eqnarray}
 G_\mu := \e_{\mu\nu\rho} G^{\nu\rho}\ , \qquad    G^2 := G_\mu G^\mu\,.
\end{eqnarray}
The fermionic part of $\cL_\sigma, \cL_M$ and $\cL_\mu$ can be found in  \cite{Alkac:2014hwa}. The fermionic terms of the four derivative action have not been given in component form explicitly, but it is straightforward to write them down using the ingredients provided in \cite{Alkac:2014hwa}. The supersymmetry transformation rules for the gravitino is given by
\bea \label{nmt}
\delta\psi_{\mu} =  \Big(\partial_{\mu}+\frac{1}{4}{\omega}_{\mu}{}^{ab}\,\gamma_{ab}
-\rmi V_{\mu}\Big)\epsilon-\frac{1}{2} \rmi\, \gamma_{\mu}\gamma^{\nu\lambda} {G}_{\nu\lambda} \epsilon-\gamma_{\mu}\,D\epsilon \ .
\eea

\subsection{Off-shell \texorpdfstring{$N = 4, 3D$}{N=4 D=3} higher derivative supergravities }

Off-shell $N=4$ topologically massive supergravity was constructed in \cite{Lauf:2016sac} in superspace by coupling the $N=4$ conformal supergravity \cite{Butter:2013rba} to a compensating hypermultiplet. Upon fixing the conformal symmetries, the resulting $20_B+20_F$ Poincar\'e supermultiplet consists of
\be
\{e_\mu^a, B_\mu^{IJ}, w, y\,; \psi_\mu^I, w^I \}\ ,
\ee
where $I=1,...,4$, the vector fields are in $(3,1)+(1,3)$ of $SO(4)$, $w,y$ are real scalars, the rest are Majorana spinors. The vector fields in $(3,1)$ are gauge fields, and the spinor $w^I$ is in $(2,1)$ of $SO(4)$. The degrees of freedom are $\{3,15,1,1; 16,4\}$. The off-shell topologically massive supergravity Lagrangian is given by \cite{Lauf:2016sac}

\bea
e^{-1}\cL &=& -\Big[ R+\frac12 B_{ij}^a B_a^{ij}+2y +2 w^2 \Big]
+\frac{1}{\mu} \Big[\ve^{abc}{\rm tr}\Big( R_{ab} \omega_c -\frac23\omega_a \omega_b\omega_c\Big)
\nn\w2
&& - \ve^{abc} {\rm tr}\Big(F_{ab}B_c - \frac23 B_a B_b B_c \Big) -2wy \Big]\ ,
\eea
where $B_a^{ij} = -\frac12 (\gamma^{IJ})_{ij} B_{a IJ}$ is the self-dual part of the vector fields. Note that going on shell gives $\omega=-\mu$ and $y=-2\mu^2$, thereby leading to
\be
\cL=-e(R+2\mu^2) +\frac{1}{\mu} L_{CS}\ ,
\label{CTMG}
\ee
which admits an AdS vacuum  with $\mu\ell=1$, where $\ell$ is the AdS radius.  This is the  critical point at which the bulk gravity mode disappears, and a single helicity 2 boundary graviton arises \cite{Li:2008dq}.
Aspects of the coupling of the model above to matter have been studied in superspace in \cite{Lauf:2017nfa}.

A higher derivative extension of $N=4, 3D$ supergravity  beyond the Lorentz CS term has been constructed, albeit at the linearized level, by using superconformal tensor calculus \cite{Bergshoeff:2010ui}. After fixing the conformal symmetries, the resulting off-shell supergravity multiplet consists of the following $24_B+24_F$ fields
\begin{eqnarray}
\{h_{\m\n}, V_\m{}^{ij}, E, D,  \Phi\,; \p_\m^i, \chi^i, \psi^i\,\}
\end{eqnarray}
where  $i = 1,...,4$, the scalars $D, E, \Phi$ are real, and the spinors are Majorana.  The degrees of freedom are $\{ 3,18,1,1,1; 16, 4 \}$. Up to and including five derivatives, the linearized level Lagrangian constructed in \cite{Bergshoeff:2010ui} has the following schematic form:
%
\bea
e^{-1}\cL &=&  \frac12\big( h^{\mu \nu} G^{(\mathrm{lin})}_{\mu \nu}+ \cdots \big) + \frac{1}{\mu} \big( h^{\mu\nu}C_{\mu\nu}^{\rm (lin)} +\cdots\big)
+\frac{1}{m^2} \big( -\frac{1}{2}\epsilon^{\mu\tau\rho}h_\mu{}^\nu\partial_\tau C_{\rho\nu}^{\rm (lin)} +\cdots \big)
\nn\w2
&& +\frac{1}{M^3} \big( R^{\mu\nu}_{\text{(lin)}} C_{\mu\nu}^{\rm (lin)} + \cdots \big)\ ,
\label{N4HDSugra}
\eea
where $G^{(\mathrm{lin})}_{\mu \nu}$ is the linearized Einstein tensor, and
\be
C_{\mu\nu}^{\rm (lin)}  =
\varepsilon_\mu{}^{\tau\rho} \partial_\tau S^{\rm (lin)}_{\rho\nu}\ , \qquad S^{\rm (lin)}_{\mu\nu} = R_{\mu\nu}^{\rm (lin)} - \frac14 \eta_{\mu\nu} R^{\rm (lin)}\ .
\label{defFphi}
\ee
For further details, including the supertransformations, see \cite{Bergshoeff:2010ui}.

The part of the Lagrangian proportional to $1/\mu$ in \eq{N4HDSugra} constitutes the bosonic part of conformal supergravity. Its superconformal coupling to the so called Chern-Simons matter is of considerable interest. In what follows we shall review such couplings for $N=6,8$ (denoted by $\mN$ below to save $N$ for flavor groups) from which  couplings with less supersymmetry can be obtained by consistent truncations. We shall comment briefly on their superspace formulations in section \ref{3dsuperspace}.

\subsection{\texorpdfstring{${\mathcal N}=6, 3D$}{N=6 D=3} higher derivative conformal supergravity coupled to matter}
\label{abjm}

The on-shell ${\mathcal N}=6$ conformal supergravity including the Lorentz-Chern-Simons term coupled to ABJM matter was constructed in \cite{Chu:2009gi} by employing Noether procedure. The complete off-shell version was constructed later in \cite{Nishimura:2013poa} also by means of Noether procedure. The off-shell field content of ${\mathcal N=6}, 3D$ conformal supergravity is \cite{Nishimura:2013poa, Kuzenko:2013vha}
\begin{eqnarray}
\{ e_\mu^r, B_\m{}^A{}_B, C_\mu, E^{ij}, D^{ij}\, ;\,  \psi_\mu^i, \chi^{ijk}, \chi^i  \}
\end{eqnarray}
where $i=1,...,6, A=1,...,4$ label the vector and spinor representations of the $R$-symmetry group. The $i,j$ indices are anti-symmetrized and the spinors are Majorana, $B_{\mu}{}^A{}_B$ and $C_\mu$  are the gauge fields of the $R$-symmetry group $SU(4)_R\times U(1)_R$. The fields $E_{ij}, D_{ij}, \chi_{ijk}, \chi_i$, which have conformal dimensions $-1,-2, -3/2$, respectively, thus satisfy algebraic field equations. This multiplet was coupled to on-shell $SU(N)$ gauge invariant ABJM model\footnote{The allowed gauge groups in the absence of coupling to supergravity were classified in \cite{Schnabl:2008wj}  (see also \cite{Bergshoeff:2008bh}). Upon coupling to supergravity, new possible groups arise, and their superspace formulation has been discussed in \cite{Lauf:2017nfa}. In the case of coupling to supergravity, there exists a model with gauge group $SU(N)\times U(1)$ and matter fields in the fundamental representation. The model discussed here corresponds to that model in which the $U(1)$ is decoupled by setting its charge to zero. } which has the field content
\be
\{\,Z_a^A, A_\mu{}^a{}_b; \psi_{aA} \}\ ,
\ee
where $A_\mu{}^a{}_b$ is the $SU(N)$ gauge field, $Z^A_a$ and $\psi_{aA}$ are the scalars and fermions. The bosonic part of the Lagrangian is given by \cite{Nishimura:2013poa}
\bea
\cL &=& \frac12 \e^{\mu\nu\rho} \tr\big(\omega_\mu \partial_\nu\omega_\rho +\frac23 \o_\mu \o_\nu\o_\rh \big)  -\e^{\mu\nu\rho} \tr\big( B_\m \partial_\n B_\rh +\frac23 B_\m B_\n B_\rh\big) -2\e^{\mu\nu\rho} C_\m\partial_\n C_\r
\nn\w2
&& -2e D_{ij}E^{ij}+\frac{1}{3\sqrt{2}}\e^{ijklmn}E_{ij}E_{kl}E_{mn}-eD_{\m}\bar{Z}^a_A D^{\m}Z^A_a-\frac23e|Y_{Aa}^{BC}|^2
-\frac18 R |Z|^2
\nn\w2
&&-\frac12\e^{\m\n\r}\big(f^{ab}{}_{cd}A_\m^c{}_b\partial_\n A_\r^d{}_a+\frac23 f^{ac}{}_{dh}f^{hg}{}_{eb}A_\m^b{}_{a}A_{\n}^d{}_c A_{\r}^{e}{}_{g}\big)
\nn\w2
&&+\frac1{\sqrt2}{\rm i}e E^{ij}f^{ab}{}_{cd} (Z\G^{ij} {\bar Z})_a{}^c (Z\bar Z)_b{}^d +\frac{1}{\sqrt2}{\rm i}e D^{ij} \tr (Z\Gamma^{ij} {\bar Z}) -\frac14 e E^{ij}E^{ij}|Z|^2\ ,
\label{Tanii6}
\eea
where
\begin{align}
Y^{BC}_{Aa} =& f^{cd}{}_{ba} \big(Z^B_c Z^C_d\bar{Z}^b_A+\delta_{A}^{[B}Z^{C]}_c\bar{Z}^b_D Z^D_d\big) \ , & (Z\Gamma^{ij} {\bar Z})_a{}^b & = Z_a^A (\Gamma^{ij})_A{}^B {\bar Z}_B{}^b\ ,
\nn\w2
 (Z\bar Z)_a{}^b =& Z_a^A {\bar Z}_A{}^b \ , &   |Z|^2 &= \bar{Z}^a_A Z^A_a\ .
\end{align}
Up to leading order in fermions, the model is invariant under the following superconformal transformations
of the fermionic fields
\bea
\delta \psi_\mu^i &=& D_\mu \e^i + \gamma_\mu \eta^i\ ,
\nn\w2
\delta \chi^{ijk} &=& - \frac{3}{4\sqrt{2}} \g^{\m\n} \e^{[i} G_{\m\n}^{jk]} +\frac12 \e^{ijklmn} \e^l D^{mn} + \frac14 \e^{ijklmn} \g^\m \e^l D_\m E^{mn} - \frac{3}{\sqrt{2}} \e^l E^{[ij} E^{kl]}
\nn\w2
&& - \frac12 \e^{ijklmn} \eta^l E^{mn} \,,\nn\w2
\delta \chi^i &=&  -\frac{1}{4\sqrt{2}} \g^{\m\n} \e^i G_{\m\n} + \e^j D^{ij} - \frac12 \gamma^\m \e^j D_\m E^{ij} + \frac{1}{8\sqrt{2}} \e^{ijklmn} \e^j E^{kl} E^{mn} + \eta^j E^{ij} \ ,
\nn\w2
\delta \psi_{Aa} &=&- \frac{1}{2\sqrt{2}} \gamma^\m \e^i (\Gamma^i)_{AB} D_\m Z_a^B + \frac{1}{2\sqrt{2}} \e^i (\Gamma^i)_{BC} Y^{BC}_{Aa} - \frac18 \rmi \e^{k} (\Gamma^{ij} \G^k)_{AB} Z_a^B E^{ij} \nn\\
&& +  \frac{1}{2\sqrt{2}} (\G^i)_{AB} \eta^i Z_a^B \ ,
\eea
where
\bea
G_{\m\n}^{ij} = \partial_\m B_\n^{ij} - \partial_\n B_\m^{ij} + B_\m^{ik} B_\n^{kj} -  B_\n^{ik} B_\m^{kj} \ ,\qquad   G_{\m\n} = 2\partial_{[\mu} C_{\nu]}\ .
\eea
Here, $B_\mu^{ij} = B_\mu^A{}_B (\Gamma^{ij})_A{}^B$, and the gauge covariant derivative $D_\m$ contains the $SO(6) \times U(1)$ gauge fields. The closure of the superconformal algebra on the matter fields requires their field equations and therefore while the conformal supergravity sector is off-shell, the total Lagrangian is invariant on-shell in the matter sector, as observed in \cite{Nishimura:2013poa}. The field equations of the auxiliary fields are algebraic and can be readily solved to express them in terms of the ABJM fields as
\begin{eqnarray}
D^{ij} &=& \frac{1}{8\sqrt{2}} \rmi \mu f^{ab}{}_{cd}  (Z\Gamma^{ij}\bar Z)_a{}^c (Z\bar Z)_b{}^d  + \frac{1}{64\sqrt{2}} \rmi \mu^2 (Z\Gamma^{ij}\bar Z)_b{}^a \big[ (Z\bar Z)_a{}^b - \delta_a^b |Z|^2 \big] \ ,
\nn\w2
E^{ij} &=& \frac{1}{8\sqrt{2}} \rmi \mu \tr (Z\Gamma^{ij} \bar Z) \ .
\end{eqnarray}
Using these expressions in \eq{Tanii6} yields the on-shell model obtained earlier in \cite{Chu:2009gi}, and the Lagrangian in convention of \cite{Chu:2009gi} takes the form\footnote{ According to \cite{Chu:2009gi}, the matter fields carry $U(1)_R$ charge $q=\pm 1/4$, as fixed by supersymmetry, while in \cite{Nishimura:2013poa} it is $\pm 1/2$.}
\bea
\cL &=&  \frac{1}{\mu} \Big[\, \frac12 \e^{\mu\nu\rho} \tr\left(\omega_\mu \partial_\nu\omega_\rho +\frac23 \o_\mu \o_\nu\o_\rh \right)  -\e^{\mu\nu\rho} \tr\left( B_\m \partial_\n B_\rh +\frac23 B_\m B_\n B_\rh\right)
\w2
&& +\frac12 \e^{\mu\nu\rho} C_\mu \partial_\nu C_\rho  \Big] -\frac18 e |Z^2|R   -|{\widetilde D}_\mu Z_a^A|^2  +\frac12 \e^{\m\n\rh} \tr \left( A_\m \partial_\n {\widetilde A}_\rh + \frac23 A_\m {\widetilde A}_\n {\widetilde A}_\rh \right) -V\ ,
\nn
\eea
where
\be
{\widetilde D}_\mu Z_a^A = \partial_\mu Z_\a^A + B_\mu{}^A{}_B Z_b^B + {\widetilde A}_\mu{}^b{}_a Z_b^A + q C_\mu Z_a^A\ ,
\ee
and
\be
{\widetilde A}_\mu{}^a{}_b = f^{ac}{}_{bd} A_\mu{}^d{}_c\ ,\qquad f^{ab}{}_{cd} = \lambda \, (t^\alpha)^{[a}{}_c (t^\alpha)^{b]}{}_d \ ,
\ee
with $\lambda$ an arbitrary constant, and $t^\a$ the generator of $SU(N)$ \cite{Bagger:2008se}.  The scalar potential takes the form
\bea
V &=&\frac23 \Big(f^{ab}{}_{c'd} f_{a'b'}{}^{cd}-\frac12 f^{ab}{}_{b'd} f_{a'c'}{}^{cd} \Big) \left(Z{\bar Z}\right)_a{}^{a'} \left(Z{\bar Z}\right)_b{}^{b'}\left(Z{\bar Z}\right)_c{}^{c'}
\nn\w2
&& +\frac12 \mu f^{ab}{}_{cd} \Big[\, -\left(Z{\bar Z}Z{\bar Z}\right)_a{}^c \left(Z{\bar Z}\right)_b{}^d   +\frac14|Z|^2 \left(Z{\bar Z}\right)_a{}^{c}\left(Z{\bar Z}\right)_b{}^d\Big]
\nn\w2
&& +\mu^2 \Big[ -\frac{1}{48} \tr \big((Z{\bar Z})^3\big)
+\frac{1}{32} |Z^2| \tr \big((Z{\bar Z})^2\big) - \frac{5}{12\cdot 64} (|Z|^2)^3 \Big]\ .
\label{pot1}
\eea
The on-shell supersymmetry transformations of the fermionic fields, in convention of \cite{Chu:2009gi}, are given by
\bea
\delta \psi_\mu^i &=& \pm \frac{1}{\sqrt 2} D_\mu (\omega,B)\e^i +\gamma_\mu \eta^i\ ,
\nn\w2
\delta \psi_{aA} &=& \gamma^\mu \Gamma^i_{AB} \e^i {\widetilde D}_\mu Z_a^B +f^{cd}{}_{ab} \Big[ -\left( {\bar Z}_c \Gamma^i Z_d \right){\bar Z}_A^b\e^i +\Gamma^i_{AB} Z^B_c \left(Z{\bar Z}\right)_d{}^b \e^i \Big]
\nn\w2
&& + \Gamma^i_{AB}\eta^i Z_a^B\ ,
\eea
where $\Gamma^i_{AB}$ are the chirally projected $SO(6)_R$ gamma matrices.

\subsection{\texorpdfstring{${\cal N}=8, 3D$}{N=8 D=3} higher derivative conformal supergravity coupled to matter}

The off-shell field content of $N=8, 3D$ conformal supergravity is
\begin{eqnarray}
\{ e_\mu^r, B_\m{}^{ij}, E^{ijkl}, D^{ijkl}\, ;\,  \psi_\mu^A, \chi^{ABC}  \}
\label{N=8,3D}
\end{eqnarray}
where $i=1,...,8$ and $A=1,...,8$ label the vector and spinor representations of the $R$-symmetry group $SO(8)$, $B_{\mu ij}$ are the gauge fields and anti-symmetrizations in $SO(8)$ indices are understood. The scalars $E^{ijkl}$ and $D^{ijkl}$ have opposite $SO(8)$ dualities. Writing $D_{ijkl} =\eta\e^{ijklmnpq} D^{mnpq}/4!$, it was shown in \cite{Nishimura:2012jh} that there are two distinct off-shell conformal supergravities for $\eta=1$ and $\eta=-1$. As far as the Lagrangians are concerned, in the case of $\eta=1$ the coupling of BLG matter with $SU(2)\times SU(2)$ gauge symmetry \cite{Bagger:2007jr,Gustavsson:2007vu} to off-shell conformal supergravity background was achieved in \cite{Nishimura:2012jh} but neither the Einstein-Hilbert Lagrangian  nor the Lorentz Chern-Simons conformal supergravity action exists in this off-shell setting. This is due to the fact that this would require a term of the form $E^{ijkl} D_{ijkl}$, which is not possible due to opposite duality properties \cite{Bergshoeff:2010ui}\footnote{ Decomposing the index $i=\{I,8\}$ with $I=1,...,7$ gives an ${\cal N}=7$ off-shell supermultiplet and a linearized off-shell action in which the Lorentz Chern-Simons term is present since the required term of the form $D_{IJKL} E^{IJKL}$ now exists. See \cite{Bergshoeff:2010ui} for further details.}.  Interestingly enough, it was shown in \cite{Nishimura:2012jh} that the coupling of $\mN=8$ conformal supergravity background to Chern-Simons matter was not possible for $\eta=-1$ unless one puts the conformal supergravity on-shell.

Focusing on the coupling of $\mN=8$ on-shell conformal supergravity to BLG matter in the presence of Einstein-Hilbert and Lorentz Chern-Simons terms, we turn to the construction of \cite{Gran:2008qx,Gran:2012mg}.
In the absence of coupling to supergravity the allowed gauge group is $SU(2)\times SU(2)$ \cite{Bagger:2007jr,Gustavsson:2007vu} but in the presence of supergravity more groups, and in particular $SO(N)$, are possible \cite{Gran:2012mg,Lauf:2017nfa}. Following \cite{Gran:2012mg}, we shall summarize the bosonic sector of the coupled system with $SU(2)\times SU(2)$ gauge symmetry, which brings in new couplings, just as in the $\mN=6$ case summarized in the previous subsection. It will be straightforward to deduce the case of $SO(N)$ gauge symmetry from these results as will be explained below. The fields of the on-shell supergravity and matter multiplets are
\be
\{ e_\mu{}^r, B_{\mu ij}, \psi_\mu^A \}\ ,\qquad \{ X_a^i, \psi_a^A , A_\mu^{ab} \}\ ,
\ee
where $a=1,...,4$ labels the vector representation of the gauge group $SO(4)\approx SU(2)\times SU(2)$. The bosonic part of the Lagrangian  constructed in \cite{Gran:2012mg} is of the form
\bea
\cL &=&  \frac{1}{\mu} \Big[  \frac12 \e^{\mu\nu\rho} \tr\big(\omega_\mu \partial_\nu\omega_\rho +\frac23 \o_\mu \o_\nu\o_\rh \big)  -\e^{\mu\nu\rho} \tr\big( B_\m \partial_\n B_\rh +\frac23 B_\m B_\n B_\rh\big)  \Big]
\nn\w2
&& -\frac{1}{16} e X^2 R   -{\widetilde D}_\mu X_a^i {\widetilde D}^\mu X_a^i   -\frac12 \e^{\m\n\rh} \tr \big( A_\m \partial_\n {\widetilde A}_\rh + \frac23 A_\m {\widetilde A}_\n {\widetilde A}_\rh \big) -V\ ,
\eea
where ${\widetilde D}$ is the supergravity covariant derivative (see \cite[Eq.(2.12)]{Gran:2012mg} for its definition) and
\be
{\widetilde A}_\mu^{ab} = \big(\lambda\, \e^{abcd}-\frac14 \mu\, \delta^{ab}_{cd}\big) A^{cd}_\mu\ ,
\ee
and the potential is given by
\bea
V &=& \frac{1}{12} e\lambda^2   \left(\e^{abcd} X_b^i X_c^j X_d^k\right)^2
+\frac{1}{32\cdot 64} e \mu^2 \left(X^2 X_a^i -4X_a^j X_b^j X_b^i\right)^2 \ ,
\eea
with  $X^2=X_a^i X_a^i$ and $\lambda$ a constant. The supertransformations of the fermionic fields,  up to cubic fermions, are \cite{Gran:2012mg}
\bea
\delta \psi_\mu &=&\pm \frac{1}{\sqrt 2} D_\mu (\omega,B) \e +\gamma_\mu \eta\ ,
\nn\w2
\delta \psi_a &=& \gamma^\mu \Gamma^i \e {\widetilde D}_\mu X_a^i -\frac16\lambda \e_{abcd}\Gamma^{ijk} \e X_b^i X_c^j X_d^k  +\frac18 \mu\Gamma^i \e \Big(X_b^i X_b^j X_a^j -\frac14 X_a^i X^2 \Big)
\nn\w2
&& +X_a^i\Gamma^i\eta\ .
\eea
It was observed in \cite{Gran:2012mg} that setting $\lambda=0$ gives a result which can be readily extended to an $SO(N)$ invariant one by simply declaring the range of the index $a$ to be 1 to $N$.

Models summarized above accommodate topologically massive gravity. A massive gravity version is known at the linearized level. The linearized ${\cal N}=8$ off-shell supergravity multiplet consists of the following fields
\begin{eqnarray}
\{h_{\m\n}, B_\m{}^{ij}, E^{ijkl}, D^{ijkl}, \Phi^{ijkl}\, ;\, \psi_\m^i, \chi^{ijk}, \p^{ijk} \}
\end{eqnarray}
where $i, j =1 \, \ldots , 8$. All but $\p^{ijk}$ and $\Phi^{ijkl}$ constitute the Weyl multiplet, with triality used to replace the $SO(8)$ spinor indices on the fermions with the bosonic ones. The extra fields arise from taking eight copies of ${\cal N}=8$ scalar multiplets, each containing $8_B+8_F$ degrees of freedom, and imposing certain constraints on them in order to fix the redundant local symmetries, as detailed in \cite{Bergshoeff:2010ui}.
At the linearized level, the bosonic part of the Lagrangian is given by \cite{Bergshoeff:2010ui}. Schematically, it takes the form
\be
e^{-1}\cL = \frac12\Big( h^{\mu \nu} G^{(\mathrm{lin})}_{\mu \nu}+ \cdots \Big)  +\frac{1}{m^2} \Big( -\frac{1}{2}\epsilon^{\mu\tau\rho}h_\mu{}^\nu\partial_\tau C_{\rho\nu}^{\rm (lin)} +\cdots \Big)  +\frac{1}{M^4} \Big( C_{\rm (lin)}^{\mu\nu}C_{\mu\nu}^{\rm (lin)} +\cdots\Big)   \ .
\label{action4}
\ee
The linearized supertransformations of the fermionic fields are given by
\begin{eqnarray}
\delta\psi_\mu^i &=& -\frac{1}{4}\gamma^{\rho\sigma}\partial_\rho
h_{\mu\sigma}\epsilon^i
- B_\mu^{ij}\epsilon^j + \frac{1}{2} \gamma_\mu \gamma^\rho B_{\rho}^{ij} \epsilon^{j}\ \ ,
\nn\w2
\delta\chi^{ijk} &=& -\frac{3}{4} \gamma^\mu F_\mu^{[ij}\epsilon^{k]} +  \gamma^\mu(\partial_\mu E^{ijkl})\epsilon^l  + D^{ijkl}\epsilon^l\ ,
\nn\w2
\delta\psi^{ijk} &=& -\frac{3}{4} \gamma^\mu
B_\mu^{[ij}\epsilon^{k]} +   E^{ijkl}\epsilon^l +
\gamma^\mu(\partial_\mu \phi^{ijkl})\epsilon^l\ ,
\label{N=8NMG2dversion}
\end{eqnarray}
where $F^{\mu ij} = \ve^{\mu\nu\rho} \partial_\nu B_\rh^{ij}$.

\subsection{ Comments on superspace formulation for matter coupled \texorpdfstring{$4\le {\cal N} \le 8$}{ {4<= N <=8}} models}
\label{3dsuperspace}

So far we have focused on $\mN=1,2,4,6,8$. On-shell $\mN=3,5,7$ models can be obtained by suitable truncations of the $\mN=8$ model summarized above. Off-shell, on the other hand,  one can start from the $\mN=6$ model that describes the coupling of conformal supergravity to Chern-Simons matter and perform consistent truncation as follows \cite{Nishimura:2013poa}
\begin{eqnarray}
\mathcal{N} = 5 &:& e_\m{}^r\,,  \p_\m^i\,,  B_\m^{ij} \,, \chi^{ijk} \,, \chi^6\,, E^{i6}\,, D^{i6} \ ,\nn\w2
\mathcal{N} = 3 &:& e_\m{}^r\,,  \p_\m^i\,,  B_\m^{ij} \,, \chi^{123} \ .
\end{eqnarray}
There exists a large body of literature on superspace formulations of these theories which go beyond the scope of this review. In particular, Chern-Simons matter coupled $\mN=6,8$ models were treated in \cite{Gran:2012mg}, while more extensive studies covering the range $4 \le \mathcal{N} \le 8$ were conducted in \cite{Lauf:2017nfa}\footnote{For purely conformal supergravity sector, earlier work existed; see, for example, \cite{Butter:2013rba} where new results for off-shell ${\cal N}=3,4,5$ supergravity actions were given.}.
In terms of on-shell superfields, the spinor derivative of the scalar is
\be
{\cal D}_{\a i} Q={\rm i}\G_i \L_\a\ ,\qquad \a=1,2,\quad i=1,..., \mN\ .
\ee
The lowest component of $Q$ represents the matter scalars, which in general carry the spinor index of the $R$ symmetry group and flavor indices for fundamental or bifundamental representations.
The derivative of $\L$ takes the form
\be
{\cal D}^i_\a\L_\b=\g^\m_{\a\b}\G^i{\cal D}_\m Q+\frac12\epsilon_{\a\b}H^i\ ,
\ee
where $H^i=H^i_{SG}+H^i_{CS}$ with constraints
\bea
\G^{[j} H^{i]}_{SG}&=&-\frac12\big(W^{ijkl}\G_{kl}+4K\G^{ij}\big)Q\ ,\nn\w2
\G^{[j} H^{i]}_{CS}&=&F^{ij}Q+QG^{kl}\ ,
\eea
where $F^{ij}$ and $G^{ij}$ correspond to the field strengths of gauge fields associated with gauging a flavour group of the form $F \times G$. The constraints on $H^i$ are solved cases by case for $4\le \mN \le 8$, and the field equations for the matter fields are presented in superspace in \cite{Lauf:2017nfa}. It is also found that on-shell, the super Cotton tensor satisfies
\be
W^{ijkl}=-\frac{\m}{16}\bar{Q}\G^{ijkl}Q\ .
\label{CT}
\ee
Imposing $Q$ to be constant, supersymmetry requires that
\be
4K\G^{ij}Q=-W^{ijkl}\G_{kl}Q\ .
\ee
Using \eq{CT} in this equation, $K$ can be solved for, and it turns out to be the cosmological constant, as can be deduced from the commutator $[{\cal D}_\m,\, {\cal D}_\n]=4K^2 M_{\m\n}$.

\subsection{Higher derivative \texorpdfstring{$N=8, 3D$}{N=8, 3D} supergravity from higher dimensions}

Ungauged $N=8, 3D$ supergravity coupled to scalar multiplet whose scalars parametrize the coset $SO(n,8)/SO(n)\times SO(8)$ was constructed in \cite{Marcus:1983hb}.  We are not aware of a direct construction of its higher derivative extension. However, such an extension can be obtained from ordinary dimensional reduction of heterotic supergravity extended by Riemann-squared term on torus $T^7$. In the resulting action, the scalar fields parametrize the coset $SO(n+7,7)/(SO(n+7)\times SO(7))$ \cite{Baron:2017dvb, Eloy:2022vsq}. A detailed construction  in which the $SO(7,7)/(SO(7)\times SO(7))$ coset is enlarged to $SO(8,8)/(SO(8)\times SO(8)$ upon dualization of the $7+7$ vectors coming from the metric and Kalb-Ramond field, was achieved in \cite{Eloy:2022vsq} at least in the bosonic sector. This construction also incorporates the dilaton into the parametrization of the enlarged coset. Here we shall summarize this result, without specifying the embedding into the heterotic supergravity, the details of which can be found in \cite{Eloy:2022vsq}.

An essential ingredient in describing the action in $3D$ is the scalar current defined in terms of the $O(8,8)$ scalar matrix $M$, as follows
\be
J_\mu = \partial_\mu M\, M^{-1}\ .
\ee
The  $O(7,7)$ invariant metric $\eta_{MN}$ and projector $P_{ MN}$ are defined as
\begin{equation} \label{eq:etad+1}
    \eta_{MN} =\begin{pmatrix}
                      0  & \delta_m{}^n  & 0 &0 \\
                      \delta^m{}_n & 0 & 0 &0  \\
                      0 & 0&0 & 1\\
                      0&0&1&0
                    \end{pmatrix}\ ,\qquad
                       P_{MN} = \frac{1}{2}\,\left(\eta_{ MN}-M_{MN}\right)\ , \quad m,\,n=1,\cdots,7\ .
\end{equation}
It is also useful to define the $O(8,8)$ compensating vector
\be
\veccomp = \{ {\bf 0}, 1,0\}\ .
\ee
Given that the field equation for the $3$-form field strength implies that it is a constant, and choosing that constant to be zero, the bosonic sector of the Riemann-squared extended heterotic supergravity on $T^7$ gives the action \cite{Eloy:2022vsq}
\bea
I &=& \int d^{3}x\,\sqrt{-g}\left(R+\frac{1}{8}\,\Tr{\partial_{\mu}M\partial^{\mu}M^{-1}}\right)
\nn\\
&&  -\frac{\alpha'}{4}e^{-2\Phi}\Big[-\frac{1}{32}\Tr\left(J_{\mu}J_{\nu}J^{\mu}J^{\nu}\right)-\frac{1}{16}\,{\rm Tr}\,\left(J_{\mu}J^{\mu}J_{\nu}J^{\nu}M\eta \right)
\nn\\
&& -\frac{1}{64}\,\Tr \left(J_{\mu}J_{\nu}\right)\Tr \left(J^{\mu}J^{\nu}\right)+\frac{1}{128}\,\Tr \left(J_{\mu}J^{\mu}\right)\Tr \left(J_{\nu}J^{\nu} \right)
\nn\\
&& +e^{-2\Phi}\bigg(-\frac{1}{2}\,\left(\veccomp P\eta J_{\mu}J_{\nu}J^{\mu}J^{\nu}P\veccomp\right)+\frac{1}{2}\,\left(\veccomp P\eta J_{\mu}J^{\mu}J_{\nu}J^{\nu}P\veccomp\right)-\frac{1}{2}\,\left(\veccomp P\eta J_{\mu}J_{\nu}J^{\nu}J^{\mu}P\veccomp\right)
\nn\\
&&-\frac{1}{4}\,\Tr \left(J_{\mu}J_{\nu}\right)\left(\veccomp P\eta J^{\mu}J^{\nu}P\veccomp\right)+\frac{1}{8}\,\Tr \left(J_{\mu}J^{\mu}\right)\left(\veccomp P\eta J_{\nu}J^{\nu}P\veccomp\right)\bigg)
\nn\\
&& +e^{-4\Phi}\bigg(-2\,\left(\veccomp P\eta J_{\mu}J_{\nu}P \veccomp\right)\left(\veccomp P\eta J^{\mu}J^{\nu}P\veccomp\right)+\left(\veccomp P\eta J_{\mu}J^{\mu}P\veccomp\right)\left(\veccomp P\eta J_{\nu}J^{\nu}P\veccomp\right)\bigg)\Big]\ ,
\eea
where $\Phi$ is the dilaton field. While the $O(8,8)$ symmetry is manifest in the two-derivative action, it is clear that the four-derivative part breaks that symmetry due to the presence of the compensating vector $\veccomp$, and the appearance of the dilaton. In describing the lift of this theory to $10D$, the required formulae that give the expression for the coset scalar matrix $M$ in terms of the $10D$ heterotic supergravity fields are provided in \cite{Eloy:2022vsq}.

A massive deformation of the theory above is obtained by switching on the three-form field strength \cite{Eloy:2022vsq}. The mass parameter arises from the dualization of the two-form field $B_{\mu\nu}$. At the two-derivative level,
the massive deformation results in a topological mass for the vectors and a potential for the dilaton which breaks $O(8,8)$ to $O(7,7)$ \cite{Kaloper:1993fg}. A four-derivative extension of this theory is worked out in \cite{Eloy:2022vsq} where it is also shown that a novel Chern-Simons term based on composite connections arises and that remarkably it is O(8,8) invariant to leading order in the deformation parameter.

\subsection{Killing spinors and exact solutions}

In what follows, we shall review the solutions to the Killing spinor equations for $N=1, 2, 3D$ supergravities. While there is an extensive literature on the solutions for these theories, we shall focus on the review of those of topologically massive supergravities. Exact solutions of $N=(1,1)$ and NMG theory theory can be found in \cite{Alkac:2015lma}, and those of $N=(1,1)$ and GMG theory in \cite{Deger:2018kur, Sarioglu:2011vz}.

\subsubsection{On-shell Killing spinors in \texorpdfstring{$N=1$}{N=1} TMG and exact solutions}

In $N=1, 3D$ supergravity, all fermions are two-component Majorana spinors. This severely restricts the structure of supersymmetric background configurations, that is, only planar-wave type solutions with a null Killing vector as well as maximally supersymmetric AdS$_3$ and Minkowski background solutions are possible. For TMG, the Killing spinor equation is given by setting $S=m$ and $\delta \psi_\m = 0$ in \eqref{OffShellTransformationRulesD3N1}.
%
%
In this case, $K^\m = \bar\e \g^\m \e$ is a null Killing vector for commuting Killing spinor $\e$, i.e.
\be
K^\m K_\m  = 0 \,, \qquad \nabla_\m K_\n + \nabla_\n K_\m  = 0 \,.
\ee
The integrability condition for the Killing spinor
\begin{eqnarray}
\left(G^{\m\n} - m^2 g^{\m\n} \right) \g_\n \e = 0 \,,
\end{eqnarray}
implies that the only maximally supersymmetric configurations are the Minkowski space with $m = 0$ and the anti-de Sitter space with $G_{\m\n} = m^2 g_{\m\n}$. The most general local forms of the supersymmetric solutions are classified depending on the value of gravitational Chern-Simons coupling $\mu$ \cite{Gibbons:2008vi}
\bea
\mu  \neq -1  &:& ds^2  = d\rho^2 + 2 e^{2\rho} du dv + e^{(1-\mu) \rho} f(u) du^2 \,,
\nn\\
\mu  = 1  &:& ds^2  = d\rho^2 + 2 e^{2\rho} du dv + \rho f(u) du^2 \,,
\nn\\
\mu  = -1  &:& ds^2  = d\rho^2 + 2 e^{2\rho} du dv + \rho e^{2\rho} f(u) du^2 \,,
\eea
where $f(u)$ is an arbitrary function of $u$, and we have set $m=1$. The Killing spinor equation is solved by a single, $v$-independent Killing spinor for all these plane-wave solutions \cite{Gibbons:2008vi}. If $f(u) = 0$, however, there is a supersymmetry enhancement with two Killing spinors and the solutions become the AdS$_3$ in the Poincar\'e patch. There is also an extremal BTZ black hole solution with a single, globally defined, Killing spinor.

\subsubsection{Off-shell Killing spinors in \texorpdfstring{$N=(1,1), 3D$}{N=(1,1), 3D} theory and \texorpdfstring{$1/4$}{1/4} exact supersymmetric solutions of TMG}

In the case of ${N} = (1,1)$ supersymmetry, the off-shell Killing spinor equation is given by setting $B=1$ and $\delta \psi_\mu = 0$ in \eqref{3DN11KS}.
The consequences of this equation are described in detail in \cite{Deger:2013yla}. Here we shall summarize the exact and supersymmetric solutions that use the properties of the Killing spinors.

For the time-like Killing vector, and focusing on the solutions with
\be
S = m\,, \qquad   V_0, V_1\ \ \mbox{constants}\ , \qquad  V_2 = 0 \ ,
\ee
where the indicated components of the vector are in the tangent space, the solutions can be summarized as follows \cite{Deger:2013yla}.
\begin{itemize}
\item[$\bullet$]{\textbf{Round AdS$_3$:} The solution is given by
\be
ds^2= \frac{\ell^2}{y^2} \left(-d\tau^2+dx^2+dy^2\right)\ , \qquad \ell \equiv m^{-1}\ .
\label{Poincare}
\ee
For this solution, the components of the vector field are $V_\m=0$.
}
\item[$\bullet$]{\textbf{AdS$_3$ pp-wave:} The solution is given by
\bea
ds^2 &=& \ell^2 \left[ \frac{du^2+dx^+ dx^-}{u^2} - u^{2 (\mu \ell +2)}  \left(\frac{dx^-}{u^2}\right)^2 \right] \,,
\label{pp}
\eea
and the vector field $V$ in this coordinates takes the form
\bea
V &=& (\mu\ell+1)\,u^{\mu\ell}\,dx^-\;.
\eea
The limit $\mu\ell \to -2$ leads to the ``minus'' null warped AdS$_3$ metric.
}
\item[$\bullet$]{\textbf{Null warped AdS$_3$:} If the non-vanishing components of the vector field are set to
\begin{eqnarray}
V_0 = \frac{\mu}{2}\ ,\quad V_1=m= -\ell^{-1} \,,
\label{VectSet}\;
\end{eqnarray}
and $|\mu \ell| = 2$, indicating that $V_2 = 0$, then the following metric is a solution to the equations of motion
\bea
ds^2 &=& -e^{-4my} dt^2 \mp 2 e^{-2my}dtdx +dy^2 \ .
\eea
Upon change of coordinates
\bea
y&=&\ell \,{\rm log}\,u\;,\qquad t~=~\ell \,x^-\;,\qquad x~=~\mp\ell \,x^+ \;,
\eea
one recovers the ``minus'' null warped AdS$_3$ metric, which is the $\mu \ell \to -2$ limit of \eqref{pp}.
}
\item[$\bullet$]{\textbf{Spacelike squashed AdS$_3$:} Imposing the components of the vector fields to be \eqref{VectSet} and setting $|\mu \ell| < 2$, the following metric is a solution to the equations of motion
\be
ds^2 = \frac{\ell^2}{4} \left[ \frac{-dt'^2+dz^2}{z^2}+ \nu^2 \left(dx'+ \frac{dt'}{z}\right)^2 \right]\ .
\label{ss1}
\ee
This is a spacelike squashed AdS$_3$ with squashing parameter $\nu^2$ given by $\nu^2 = 1/4 (\mu \ell)^2$. The terminology of ``squashed'' is due to $\nu^2 <1$.
}
\item[$\bullet$]{\textbf{Timelike stretched AdS$_3$:} Imposing the components of the vector fields to be \eqref{VectSet} and setting $|\mu \ell| > 2$, the following metric is a solution to the equations of motion
\be
ds^2 = \frac{\ell^2}{4} \left[ - \nu^2 \left(dx'+ \frac{dt'}{z}\right)^2+\frac{dt'^2+dz^2}{z^2} \right]\ .
\label{TLStr11}
\ee
This is a timelike stretched AdS$_3$ with squashing parameter $\nu^2$, again defined as $\nu^2 = 1/4 (\mu \ell)^2$. The terminology of ``stretched'' is due to $\nu^2 >1$.
}
\end{itemize}
All these background solutions, except for the AdS$_3$ metric, preserve $1/4$ of the supersymmetries.

\subsubsection{Killing spinors in  \texorpdfstring{$N=(2,0), 3D$}{N=(2,0), 3D} theory and exact solutions in TMG}

In the case of ${N}=(2,0)$ supersymmetry, the Killing spinor equation is given by setting $\delta \psi_\m = 0$ in \eqref{nmt}.
In ${N} = (2,0)$ four-derivative gravity \eqref{N20GMG}, the presence of the $RD^2$ term is problematic for ghost-freedom on AdS background, and the combination that cancels $RD^2$ term is not the NMG combination, again leading to ghost-like fluctuations around AdS vacua. This problem does not exist in ${N} = (1,1)$ theory owing to the existence of an off-diagonal $RS^2$ action that cancels out the $RS^2$ term in the supersymmetric NMG action \cite{Alkac:2014hwa}. Consequently, the existing literature on the supersymmetric backgrounds and black hole solutions only focuses on ${N} = (2,0)$ topological massive gravity, which is equivalent to setting $m = \tilde m = \infty$ in \eqref{N20GMG}.
					
In the case of a null Killing vector, the analysis is identical to the ${N}=1$ theory \cite{Gibbons:2008vi}. For a timelike Killing vector, it is possible to make a weaker ansatz compared to ${N}=(1,1)$ theory, i.e. the components of the vector field $V_\mu$ can be non-constant in the flat basis. In fact, if the metric is assumed to be of the form \cite{Deger:2016vrn}
\begin{eqnarray}
ds^2 = -f^2 \left(dt + A \right)^2 + e^{2\sigma} \left(dx^2 + dy^2 \right) \,,
\end{eqnarray}
in the adapted coordinates $\left(t,x^1 \equiv x, x^2 \equiv y\right)$, then the solutions can be classified by the following ansatz \cite{Deger:2016vrn}
\begin{eqnarray}
f \rho = \text{constant}\,,\quad D=\frac{M}8\ ,
\label{ConsistencyAnsatz}
\end{eqnarray}
where $\rho = e^{-2\sigma} \partial_{[1} A_{2]}$ and $M$ is constant. Based on the ansatz for the metric and the equations of motion for the vector fields $C_\mu$ and $V_\mu$, the ansatz \eqref{ConsistencyAnsatz} can be further split into two sub-cases: (i.) $f$ and $\rho$ are separately constant and (ii.) $f$ and $\rho$ are not necessarily constant.
	
Taking $\rho$ and $f$ to be constants, all solutions are characterized by the constant values of two parameters; $\rho = \rho_0 =\text{const.}$ and $\kappa$ which is defined as
\begin{equation}
\kappa \equiv 2  \left( -\mu + 2D + \rho f \right) \left( \frac{\rho f}{4} - D \right)- f^2\rho^2~.
\label{kappa}
\end{equation}
If $f$ is not constant, choosing a special form $f \rho = -2\mu$ for the ansatz \eqref{ConsistencyAnsatz}, the solutions are classified in terms of four parameters, $\{\k_1,\k_2,c_1, d_1\}$. In this case, $f$ and $\rho$ are given by
\be
f = e^h x^{-\k_2/\k_1}\,, \qquad \rho = -2 \mu e^{-h} x^{\k_2/\k_1} \,.
\ee
where $h = c_1 x + d_1$.

The consequences of the Killing spinor equation are described in detail in \cite{Deger:2016vrn}, where several supersymmetric solutions of TMG can be found, and which we list below.

\begin{itemize}

\item[$\bullet$] \textbf{Fully supersymmetric solutions:}

In this category, only the first two solutions have $f=\text{constant}$.
\item[\null]\textbf{Round AdS:} For $\kappa < 0$ and $\rho_0^2 = \left|\kappa\right| $, the solution is given by a round AdS metric \eqref{Poincare} as long as $M \neq 0$.
In this $f=$ constant and $\kappa$ is defined in \eq{kappa}.

\item[\null] \textbf{Warped timelike flat:} For $\kappa = 0$ and $\rho_0 \neq 0$, the solution is given by the warped timelike flat metric
\begin{eqnarray}
ds^2 = - \left( d t + \rho_0 \, x \,d y \right)^2 + d x^2 + d y^2\ .
\label{eq:eucwarp}
\end{eqnarray}

\item[\null] \textbf{$z$-warped null flat}: For $\kappa_1\neq0$ and ${ c_1 }=0$ along with $\kappa_1 = - \kappa_2$, the solution is given by
\begin{equation}\label{eq:zflat}
ds^2= - e^{2w/\mu} d t^2 - 2 d t d y + d w^2~.
\end{equation}

\item[\null] \textbf{Spacelike squashed AdS}: For $\kappa_1 = \kappa_2$, $c_1 \neq 0$ and $c_1 > 2\mu$, the solution is identical to \eqref{ss1} with the squashing parameter $\nu = 4\mu^2/{ c_1 }^2<1$.

\item[\null] \textbf{Timelike warped AdS}: For $\kappa_1 = \kappa_2$, $c_1 \neq 0$ and $c_1 < 2\mu$, the solution is identical to \eqref{TLStr11}.
In this case, the deformation from AdS$_3$ is stretched, i.e. $\nu = 4\mu^2/{c_1 }^2>1$.

\item[\null] \textbf{Null warped AdS}: For $\kappa_1 = \kappa_2$, $c_1 \neq 0$ and $c_1 = 2\mu$, the metric is null warped AdS given by the $\mu \ell \to -2$ limit of the AdS$_3$ pp-wave metric \eqref{pp}.
\item[$\bullet$] \textbf{The half-supersymmetric solutions:}

In this category, only the first two solutions have $f=\text{constant}$.

\item[\null] \textbf{Warped timelike AdS:} For $\kappa < 0$, $\rho_0^2 \neq  \left|\kappa\right| $ and $\rho_0 \neq 0$, the solution is a warped timelike AdS.
\begin{eqnarray}
ds^2= - \left( d t - \frac{\rho_0}{|\kappa|} \frac{1}{x} d y \right)^2 + \frac{1}{|\kappa| x^2} \left( d x^2 + d y^2 \right)~. \label{eq:wadst}
\end{eqnarray}
 For $\rho_0 = 0$ and $\mu = - M/4$, the solution is given by $R_t \times H_2$.

\item[\null] \textbf{Lorentzian Sphere}: For $\kappa >0$ and $\rho_0 = -\mu - 2D$, the solution is given by the Lorentzian sphere
\begin{equation}
ds^2 = - \left( d t + A \right)^2 + \frac{1}{\kappa} \left( d \theta^2 + \sin^2\theta d \phi^2 \right)\ ,
\label{eq:lhopf}
\end{equation}
where
\begin{eqnarray}
    A= - \frac{\rho_0}{\kappa}  \cos \theta \, d \phi \,.
\end{eqnarray}
\end{itemize}

\begin{itemize}
\item[\null] \textbf{$\Gamma$-Metric:} For $\k_1 \neq 0$ and $c_1 > 0$, the solution is given by
\begin{equation}\label{eq:gammametric}
ds^2 = - e^{2x} x^{-2 \frac{\kappa_2}{\kappa_1}} \left( dt +
\frac{2\mu}{\kappa_1}  \Gamma\left(\frac{\kappa_2}{\kappa_1}-1, x \right) d y
\right)^2 + \frac{d x^2 + d y^2}{\kappa_1 \, x^2}~.
\end{equation}
which is referred to as the $\Gamma$-metric \cite{Deger:2016vrn} due to the appearance of the gamma function.

\item[\null] \textbf{$z$-warped null AdS}: For $\kappa_1\neq0$ and ${ c_1 }=0$ along with $\k_1 \neq \k_2$ and $\kappa_2 \neq 0$, the solution is given by
\begin{equation}
ds^2 = \frac{16D^2}{\left(2D-\mu\right)^4}\left(- w^{2z}  d u^2  + \frac{d w^2 + 2 d u d v}{w^2}\right)~,  \label{eq:zwarped}
\end{equation}
which is an AdS pp-wave metric. Nevertheless, the solution has the following non-relativistic rescaling symmetry
\begin{equation}
u \mapsto \Lambda^{-z} u , \quad w \mapsto \Lambda w, \quad v \mapsto \Lambda^{2+z} v~.
\end{equation}
Hence, \eqref{eq:zwarped} is also referred to as null $z$-warped metric \cite{Deger:2016vrn}.
\end{itemize}

\subsubsection{Exact solutions of higher derivative conformal supergravity coupled to matter}

For models describing ${\cal N}=4,5,7$ conformal supergravity coupled to matter, when the scalar compensator acquires a vev of form
\be
Q = {\rm diag}\, (v,0,...,0)\ ,
\ee
and choosing $v$ that gives the canonical Einstein-Hilbert term, the potential gives rise to a cosmological constant whose value is listed below \cite{Lauf:2016sac,Lauf:2017nfa}
\be
\Lambda =
\begin{cases} \m^2 & \text{for} \ \ {\cal N}=4
\\
\frac{9}{25}\m^2 & \text{for}\ \  {\cal N}=5
\\
4\m^2 &  \text{for} \ \ {\cal N}=7 \ .
\end{cases}
\ee
Given that the gravitational part of the Lagrangian relevant for determining the propagating degrees of freedom is given by
\be
\cL=e(R+2\Lambda) + \frac{1}{\mu} L_{CS}\ ,
\label{GTMG}
\ee
it follows that the case of ${\cal N}=4$ has chiral gravity sector with Lagrangian \eq{CTMG} discussed earlier.

\subsubsection*{The case of ${\mathcal N}=6$:}

In the model reviewed in section \ref{abjm}, the $\mu$ dependent terms in the potential are due to coupling of supergravity to the ABJM model, and they play a role in finding an AdS vacuum solution.  It has been noted in \cite{Nilsson:2013fya} that in this model, upon setting
\be
Z_a^A = {\rm diag}\, (\underbrace{v,...,v}_p,0,...,0)\ , \qquad p=1,...,4\ ,
\label{zvev}
\ee
and choosing $v$ that gives the canonical Einstein-Hilbert term, the potential gives rise to a cosmological constant
\be
\Lambda = \mu^2 \Big| \frac{5p^2-24p+16}{3p^2} \Big| \ , \qquad p=1,...,4\ .
\ee
In particular for $p=1$, in the gravitational sector, one gets the so called chiral gravity Lagrangian, displayed in \eq{CTMG},  as observed in \cite{Chu:2009gi}. As was mentioned in the previous footnote, there are other possible gauge groups. The couplings to supergravity have not been spelled out in components in those cases but they have been formulated in superspace in \cite{Gran:2012mg} for  $4\le{\cal N}\le 8$, following the framework laid out in \cite{Lauf:2017nfa}. In particular, $SU(N)\times U(1)$ is among the possible gauge groups, where the  $U(1)$ is not to be confused with $U(1)_R$, and the coupling constants depend on a single parameter. For a particular choice of this parameter, taking $Z_a^A$ as in \eq{zvev}, and choosing $v$ that gives the canonical Einstein-Hilbert term, the potential gives rise to a cosmological constant \cite{Lauf:2017nfa}
\be
\Lambda = \mu^2 \Big|\frac{2}{p} -1\Big|^2\ ,\qquad p=1.,...,4\ .
\ee
In this case, $p=1$ gives chiral gravity, as discussed above.

\subsubsection*{The case of ${\mathcal N}=8$:}

The vacuum solutions of the $SO(N)$ model were considered in \cite{Gran:2012mg,Nilsson:2013fya}. Taking the scalars to have the form \eq{zvev}, and choosing $v^2=16/p$ to get the canonical Einstein-Hilbert term, the potential gives rise to the cosmological constant \cite{Nilsson:2013fya}
\be
\Lambda = \mu^2 \Big(\frac{4}{p} -1\Big)^2\ ,\qquad p=1.,...,8\ .
\ee
In this case $p=2$ gives the chiral gravity. Furthermore, it has been noted in \cite{Nilsson:2013fya} that in the case of $p=3$ and $p=6$, the model admits null-warped AdS$_3$ solution.
\end{itemize}	
\section{Lower dimensions }

\subsection{\texorpdfstring{$D=2$}{D=2}}

In $2D$, there exists $(p,q)$ type of supersymmetry where $p$ and $q$ refer to the number of left- and right-handed supersymmetry generators. There are subtleties in the characterization of supermultiplet structures in this case, as explained, for example, in the appendix of \cite{Strathdee:1986jr}. We shall focus on $(p,p)$ supergravities below.

Supergravity in $2D$ is topological. In the case of $N=(1,1), 2D$ supergravity, its particular coupling to a single scalar multiplet gives rise to the Jackiw-Teitelboim (JT) supergravity, see, for example, \cite{Mertens:2022irh} for a review. This model is much studied due to the fact that it is a soluble quantum gravity model, and holographically dual to the so-called SYK quantum mechanical model. To begin with, we shall briefly recall the $N=(1,1)$ super JT model. The off-shell $N=(1,1)$ supergravity multiplet  consists of a graviton $e_\mu{}^a$, a gravitino $\psi_\m$ and an auxiliary scalar field $A$. The $N=(1,1)$ super JT model was constructed in \cite{Chamseddine:1991fg}, by using a curvature multiplet superfield $S$, a scalar superfield $\Phi$, and the determinant of the supervielbein $E$. In the convention of \cite{Hindawi:1995fy}, their expansions are given by
\bea
S &=& A+ i\theta^\alpha \Sigma_\alpha + \theta^2 C\ ,
\nn\w2
\Phi &=& \phi +i\theta^\alpha \pi_\alpha + \theta^2 F\ ,
\nn\w2
E &=& e(1-\frac12 i\theta^\alpha\theta_\alpha A^2 +{\rm ferms})\ ,
\eea
where $\theta^\alpha$ is a two-component Majorana spinor, $\theta^2=\theta^\alpha\theta_\alpha$, $C=-R-\frac12 A^2+ {\rm ferms}$, and $\Sigma_\alpha = -2\gamma^5_\alpha{}^\beta \epsilon^{ab} D_a \psi_{b\b} -\frac12 \gamma^a_\alpha{}^\beta \psi_{a\beta}\,A$. The action is given by \cite{Chamseddine:1991fg}
\be
I = \int d^2 x d^2\theta\, E \Phi (S-K)\ ,
\label{JT1}
\ee
where $K$ is a constant. In components, this readily gives
\be
I= \int d^2 x\, e \big[ \phi\ ( R-\frac12 K A)  + F(A+K) +{\rm ferms} \big]\ .
\ee
Substituting the algebraic field equation of $F$ into the action gives the well-known JT action $I = d^2 x\, e\, \phi (R +\Lambda)$ where $\Lambda =\frac12 K^2$. The off-shell action \eqref{JT1} can be extended by elevating $K$ to be a scalar superfield, and introducing two new off-shell invariants involving two arbitrary functions of the superfield $K$, namely $f(K)$ and $g(K)$, as follows \cite{Hindawi:1995fy},
\be
I = \int d^2 x d^2\theta\, E \big[ \Phi (S-K) + f(K) + i g(K) D^\alpha K D_\alpha K \big] \ .
\label{JT2}
\ee
Integrating out $\Phi$ this time gives
\be
I = \int d^2 x d^2\theta\, E \big[ f(S) + i g(S) D^\alpha S D_\alpha S \big] \ .
\label{JT3}
\ee
In components, the bosonic part of this action takes the form \cite{Hindawi:1995fy}
\begin{eqnarray}
  e^{-1}  \cL &=& - f^\prime(A) \left( R  +\frac12 A^2 \right) - g(A) \left( 2A^2 R + 2R^2 - 2(\nabla A)^2  -\frac12  A^4 \right) \ ,
  \label{ts}
\end{eqnarray}
and the supertransformation of the gravitino is given by
\be
\delta \psi_\mu = 2 D_\mu \epsilon +\frac12 \gamma_\mu\epsilon A\ .
\ee
Despite the presence of $R^2$ in the model \eqref{ts}, this is JT supergravity coupled to an extra scalar multiplet in disguise. The vacuum solution of the model \eqref{ts} has been analyzed in \cite{Hindawi:1995qa}, where it was shown that by choosing $f$ and $g$ suitably, one can have nontrivial extremum of the potential breaking supersymmetry spontaneously.

The $N=(2,2)$ and $N=(0,4)$ dilaton supergravities  and their matter couplings are known \cite{Gates:2000fj,Haack:2000di}. Here $N=(p,q)$ refers to $p$ left-handed and $q$ right-handed Majorana-Weyl spinors. The case of $N=16, 2D$ supergravity, where $N$ counts the number of Majorana spinors, can be obtained from circle reduction of $N=16, 3D$ supergravity, and it has been a subject of several studies, in part owing to its infinite-dimensional symmetries \cite{Julia:1980gr, Nicolai:1998gi}  and integrability \cite{Nicolai:1987kz,Nicolai:1988jb}.     We are not aware of higher derivative extensions of the dilaton supergravities discussed so far, apart from the fact that a dimensional reduction of Bergshoeff-de Roo heterotic supergravity on torus $T^8$ which gives higher derivative extension of $N=8, 2D$ supergravity. So far, the bosonic sector of the general result obtained from the reduction on torus $T^d$ has been worked out in \cite{ Baron:2017dvb, Eloy:2020dko}, as discussed in section \ref{sec987}. To express the action in $2D$, it suffices to set $d=8$ in \eq{na}, taken from \cite{Baron:2017dvb}.  In the resulting four-derivative extended $N=8, 2D$ theory, the scalars parametrize the coset $SO(8,8+n_V)/(SO(8)\times SO(8+n_V))$, where $n_V$ is the dimension of the Yang-Mills group in  $10D$ heterotic supergravity.

\subsection{\texorpdfstring{$D=1$}{D=1}}
Reduction of heterotic string on 9-torus leads to a half-maximal supersymmetric mechanics model with higher time derivatives.
The heterotic string tree level effective action up to and including order $\a'^3$ terms, as discussed in section \ref{hetsugra}, is given by
\be
I_{\rm het}= \int d^{10}x\sqrt{-g} \Big[ \cL^{\rm het}_{BdR}+ e^{-2\phi}\frac{\zeta(3)\a'^3}{3\cdot 2^{14}} \left( t_8t_8 R^4+\frac18\epsilon_{10}\epsilon_{10}R^4\right)+{\cal O}(\a'^4)\Big]\ ,
\label{hetagain}
\ee
where $\cL^{\rm het}_{BdR}$ is from \eq{BdR1} in which both $\alpha$ and $\beta$ are understood to be proportional to $\alpha'$. It has been shown that setting $B_{ij}$ to zero, where $i,j=1,...,9$, and parameterizing the $10D$ metric $G_{\m\n}$ and dilaton $\phi$ as
\be
\phi=\frac12\Phi+\frac12\log\sqrt{g},\quad G_{\m\n}={\rm diag}(-n^2,\, g_{ij})
\ ,
\ee
the action \eq{hetagain} reduces to (up to field redefinitions) \cite{Codina:2021cxh}
\bea
I &=&\int dt\, n e^{-\Phi}\Big\{-{\dot\Phi}^2-\frac18 {\rm Tr}(\dot{S}^2)+\frac1{128} \a'{\rm Tr}(\dot{S}^4)
\nn\\
&&- \frac{1}{2^{19}} \a'^3 \Big[ 15{\rm Tr}(\dot{S}^4)^2+  128 \zeta(3) \left(3{\rm Tr}(\dot{S}^8)-{\rm Tr}(\dot{S}^4)^2 \right) \Big] \Big\}\ ,
\label{o99}
\eea
where $S$ is given by
\be
S=\begin{pmatrix}0& g\\
g^{-1}&0\end{pmatrix}\ .
\ee
The terms in the first line are straightforwardly obtained from the dimensional reduction of \eq{PreD2}, followed by field redefinitions of $n, \Phi$ and $g$ along the lines described in \cite[Eq. (2.31)]{Codina:2021cxh}. Restoring the $B_{ij}$ dependence by duality transformation rules, the action becomes manifestly $O(9,9)$ invariant, as explained in \cite{Codina:2021cxh}. The eight-derivative terms appearing in \eqref{o99} also arise in type IIA string on torus $T^9$, and they are given in \cite{Garousi:2021ikb} with $B_{ij}$ dependence retained as well (see also \cite{David:2021jqn}).

\section{\texorpdfstring{$R^4, D^4 R^4$}{R4, D4 R4} and \texorpdfstring{$D^6 R^4$}{D6 R4} invariants and duality symmetry in diverse dimensions}
\label{lastsection}

So far we have mostly discussed the four-derivative extensions of supergravities, with few exceptions. Here we shall turn to eight and higher derivative extensions.  Even though they are very difficult to construct explicitly, if one assumes the existence of a hidden symmetry, such as those listed in Table \ref{dualgr} for maximal supergravities, the terms of the form $R^4, D^4 R^4, D^6 R^4$ multiplied by the so-called modular functions of the moduli have been studied, in which the focus is on the construction of these moduli dependent functions. These functions carry information about duality symmetries, as well as nonperturbative contributions of branes to the effective action. In the next section we shall review briefly the modular functions in dimensions $3\le D\le 10$, mostly obtained from the analysis of the four-point supergraviton amplitudes, and their various limits\footnote{In the case of maximal supergravity in $1D$, the expected duality symmetry $E_{10}$ has been utilized in \cite{Damour:2005zb} (see also \cite{Damour:2005zs,Damour:2006ez,Fleig:2012xa,Fleig:2013psa}) to find restrictions on the higher derivative corrections to $11D$ supergravity. }. We shall then recall the relevance of these results to the UV divergences and counterterms in supergravities. Finally, we shall review, again briefly, the construction of the higher derivative actions as integrals in ordinary or harmonic superspace, and in the ectoplasm approach.

\subsection{Eisenstein series in leading gravitational part of the invariants}
\label{modular functions}

A great deal of information can be obtained on string theory effective actions, and thus higher derivative extensions of supergravities that describe their low energy limits, by studying the four-point supergraviton amplitude. This amplitude has analytic and non-analytic parts. The analytic part has a low energy expansion in Mandelstam variables, and it can be expressed in terms of a local effective action, which schematically takes the form (in Einstein frame)
\be
S =  \sum_{p\ge 0, q\ge -1} \ell_D^{4p+6q+8-D} \int d^{D}x\, \sqrt{-g}\, \vE^{(D)}_{(p,q)} \partial^{4p+6q} {\cal R}^4\ ,
\label{defE}
\ee
where $\ell_D$ is the Planck length, $p$ and $q$ denote powers of $\s_2$ and $\s_3$,  where $\sigma_k := \left(\frac{\ell_D}{4}\right)^k (s^k+t^k+u^k)$, with $s,t,u$ representing the standard Mandelstam variables built out of the external momenta in the corresponding amplitude, and ${\cal R}^4$ denotes the fourth order polynomial $t_8 t_8 R^4$ in the Riemann tensor. The $p=0,\, q=-1$ term is the Hilbert-Einstein term $ \ell_D^2\int d^Dx \sqrt{-g} R$. The functions $\vE^{(D)}_{(p,q)}$ depend on the coordinates on the moduli space ${\cal M}_D = E_{11-D}/K_{11-D}$, where $E_{11-D}$ is the duality group in $D$ dimensions and $K_{11-D}$ is its maximal compact subgroup. These functions must be invariant under the left-action of the discrete subgroup $E_{11-D}(\mathbb{Z}) \subset E_{11-D}(\mathbb{R})$ on ${\cal M}_D$ \cite{Hull:1994ys,Green:1997tv}. Moreover, they must also satisfy differential equations from supersymmetry for $4p+6q <8$ \cite{Pioline:1998mn,Green:1998by,Green:2005ba,Basu:2006cs,Green:2010wi,Bossard:2014lra,Basu:2014hsa,Bossard:2014aea,Wang:2015jna,Wang:2015aua,Bossard:2015uga} that can also be understood representation-theoretically \cite{Green:2010kv,Pioline:2010kb,Green:2011vz}. These are Poisson-type equations of the form \cite{Green:2010wi,Pioline:2015yea}
\bea
\Big(\Delta-\frac{3(11-D)(D-8)}{D-2}\Big)\vE^{(D)}_{(0,0)}&=&6\pi\d_{D,8}\ .
\label{E1}\w2
\Big(\Delta -\frac{5(12-D)(D-7)}{D-2}\Big)\vE^{(D)}_{(1,0)} &=&40\zeta(2)\d_{D,7}+7\vE^{(4)}_{(0,0)}\d_{D,4}\ ,
\label{E2}\w2
\Big(\Delta -\frac{6(14-D)(D-6)}{D-2}\Big)\vE^{(D)}_{(0,1)} &=& -(\vE^{(D)}_{(0,0)})^2+40\zeta(3)\d_{D,6}+\frac{55}3\vE^{(5)}_{(0,0)}\d_{D,5}
\nn\w2
&& +\frac{85}{2\pi}\vE^{(4)}_{(1,0)}\d_{D,4}\ .
\label{E3}
\eea
The scalar Laplace operator $\Delta$ is defined on ${\cal M}_D$, and in a convenient parametrization of ${\cal M}_D$, it can be found, for example, in \cite[p. 513]{Fleig:2015vky}.

\begin{table}[ht!]
\centering
\begin{tabular}{|ccccc|}
\hline
d&  $G_d(\mathbb{R})$ & $K$ & $G_d(\mathbb{Z})$  & $D$ \\ \hline
0   & $SL(2,\mathbb{R})$    & $SO(2)$&$SL(2,\mathbb{Z})$ &10       \\
1 &$SL(2,\mathbb{R})\times \mathbb{R}^+$    & $SO(2)$& {$SL(2,\mathbb{Z})\times\mathbb{Z}_2$} & 9    \\
2 &$SL(3,\mathbb{R})\times SL(2,\mathbb{R})$   &   $SO(3)\times SO(2)$& $SL(3,\mathbb{Z})\times SL(2,\mathbb{Z})$ & 8 \\
3 &$SL(5,\mathbb{R})$   &  $SO(5)$ &$SL(5,\mathbb{Z})$ & 7 \\
4 &$Spin(5,5,\mathbb{R})$    & $(Spin(5)\times Spin(5))/\mathbb{Z}_2$ & $Spin(5,5,\mathbb{Z})$ &6  \\
5 &$E_{6(6)}(\mathbb{R}) $ & {$USp(8)/\mathbb{Z}_2$} &$E_{6(6)}(\mathbb{Z}) $ & 5  \\
6 &$E_{7(7)}(\mathbb{R})$   & $SU(8)/\mathbb{Z}_2$ &$E_{7(7)}(\mathbb{Z})$& 4\\
 7 &$E_{8(8)}(\mathbb{R})$  &  {$SO(16)/\mathbb{Z}_2$} &$E_{8(8)}(\mathbb{Z})$  & 3  \\
\hline
\end{tabular}
\caption{ $U$ duality groups $G_d(\mR)=E_{d+1(d+1)}(\mR)$ and their maximal compact subgroups. }
\label{dualgr}
\end{table}
 We recall that the string amplitudes contain analytical part and non-analytical part. The latter is due to massless thresholds meaning that the internal lines of massless particles are on mass-shell. The Kronecker delta terms in $D=6,7,8$ arise from the non-analytic part of the string amplitude roughly as follows \cite{Pioline:2015yea, Green:2010sp, Bossard:2015oxa}. The massless threshold contributions to string amplitudes contain logarithmic terms of the form $\log (-\ell_s^2 s f(x))$ where $f(x)$ is a complicated function of $x=-t/s$. In going to Einstein frame, this term gives rise to an additional term which is analytic and it involves a proportionality factor $\log y_{D_L}$, where $D_L$ is the lowest dimensions in which $L$-loop maximal supergravity has ultraviolet divergences, and $y_D=\ell_s^d g_s^2/V_{(d)}= (\ell_D/\ell_s)^{D-2}$ where $V_{(d)}$ is the volume of $d$-torus. The delta terms in $D=6,7,8$ in \eqref{E1}-\eqref{E3} arise from the action of the Laplace operator on these $\log y_D$ terms \cite{Green:2010sp}.
Turning to the $\log s$ dependent terms in maximal supergravity, they arise after the subtraction of the $\e$ pole in the amplitude evaluated at $D=D_L+2\e$ in a dimensional regularization scheme. More specifically, the single pole in $\e$ arises as $(-s/\m)^{2\e}/\e$, where $\m$ is an arbitrary scale introduced in dimensional regularization. It can be removed by adding a counterterm so as to give $\big( (-s/\m)^{2\epsilon} -1\big)/\epsilon$, which gives the finite result $\log(-s/\mu)$ in the limit $\e\to 0$\footnote{If one uses UV momentum cut-off $\Lambda$ instead, it will manifest itself as the divergence $\log(-s/\L^2)$. Thus, the terminology of ``logarithmic divergence" refers to the $\Lambda$ cut-off scheme.}. In $D=6,7,8$, the counterterms are of the form $D^6 R^4, D^4 R^4, R^4$, respectively.  In general, in $D_L$ dimensions, logarithmic divergence can appear at $L$ loop, associated with $\partial^n R^m$,
\be
\partial^n R^m: \qquad n+2m=(D_L-2)L+2\ .
\ee
Turning to Langlands-Eisenstein series (often referred to as Eisenstein series), which are associated with the maximal parabolic subgroup $P_\beta\in G$, they are defined as \cite{Green:2010kv}
\be
\vE^G_{\b,s}(g):=\sum_{\g\in P_\beta(\mZ)\backslash G(\mZ)}
e^{2s\langle\o_\b, H(\gamma g)\rangle}\ ,
\ee
where $\b$ is the simple root labeling the maximal parabolic subgroup $P_\beta$ and $\o_\b$ is a basis vector in the space dual to the root space obeying $\langle\o_\b,\,\b\rangle$=1. $H(g)$ resides in the Cartan subalgebra of $G(\mR)$ and is defined via the Iwasawa decomposition of an arbitrary group element $g\in G$ according to
\be
g=n e^{H(g)}k,\quad {\rm with }\quad n \in N(\mR)\ ,\quad  k\in K(\mR)\ ,
\ee
where $N$ is the unipotent subgroup of $G$, and $K$ is the maximal compact subgroup of $G$. The convergence of $\vE^G_{\b,s}(g)$ requires that the complex parameter $s$ satisfies the condition
$ \langle s \omega_\b -\r, \omega_\b\rangle > 0$; see, for example, \cite{Fleig:2015vky}. In the normalization chosen above, $\vE^G_{\b,0}(g)=1.$
\begin{figure}
\centering
\includegraphics[scale=0.36]{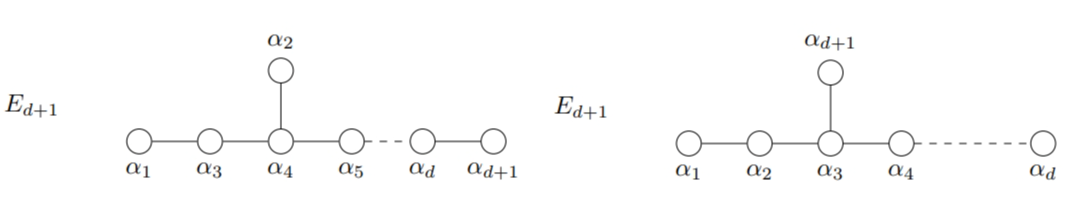}
\caption{Bourbaki  (on the left) and standard Dynkin diagram labeling conventions for the group $E_{d+1}$.}
\label{fig1}
\end{figure}

We shall now summarize the existing results for $\vE^G_{\b,s}(g)$ for the $U$ duality groups displayed in Table \ref{dualgr}. To begin with, it is convenient to define
\be
E^{E_{d+1}}_{[0^\a 1 0^{d-\a}];s} := 2\zeta(2s) \vE_{\beta;s}^{E_{d+1}}\ ,
\ee
where $[0^\a 1 0^{d-\a}]$ is the Dynkin label associated with the simple root $\beta$ in Bourbaki conventions; see Figure \ref{fig1}. In the case of $SL(d)$ with $\beta = [0^\a 1 0^{d-\a}]$, it can be expressed as (see, for example, \cite{Green:2010sp,Obers:1999um})
\be
E^{SL(d)}_{[0^{\a-1} 1 0^{d-\a-1}];s} = \sum_{ \{m^i\} \in \mZ^d/\{0\}}
\frac{1}{\big(c^{i_1...i_\a} g_{i_1j_1}\cdots g_{i_\a j_\a} c^{j_1...j_\a}\big)^s} \ ,
\ee
where the sum is over all values of $m^i$ with the values $m^1 = m^2 = \cdots = 0$ omitted, $g_{ij}$ is the metric on $SO(d)\backslash SL(d)$, and $c^{i_1...i_\a}= m_1^{[i_1}\cdots m_\a^{i_\a]}$. The ones for $E_{d+1}$ with $d\ge 4$ are more subtle; see, for example, \cite{Green:2010sp,Fleig:2015vky}. The results for ${\cal E}^{D}_{(0,0)}$ are \cite{Green:1997as, Kiritsis:1997em, Green:1997di, Obers:1999um, Green:2010wi,Pioline:2010kb, Green:2010kv}
\begin{align}
3\le D \le 7 : &\qquad  \vE_{(0,0)}^{(D)}= E^{E_{d+1}}_{[10^d];\frac32}\ ,
\label{Ed00}\w2
D=8: & \qquad\vE^{(8)}_{(0,0)} = {\widehat E}^{SL(3)}_{[10];\frac32} +2 {\widehat E}^{SL(2)}_{[1];1} (U)\ ,
\label{8d1}\w2
D=9: & \qquad  \vE_{(0,0)}^{(9)} = \nu^{-3/7} E^{SL(2)}_{[1];\frac32} (\tau) + 4\zeta(2) \nu^{4/7}\ ,
\w2
D=10: & \qquad \vE^{(10)}_{(0,0)} = E^{SL(2)}_{1;\frac32} (\Omega)\ ,\qquad \mbox{for IIB supergravity}\ ,
\label{2B1}
\end{align}
where $\nu$ parametrizes $\mR^+$ and $(\tau, U, \Omega)$ parametrize the $SL(2,\mR)/SO(2)$ coset.
The results for $\vE^{(D)}_{(1,0)}$ are \cite{Green:1999pu, Green:2010wi,Basu:2007ru,Green:2010kv}
\begin{align}
3\le D\le 5: & \qquad  \vE^{(D)}_{(1,0)} = \frac12 E^{E_{d+1}}_{[10^d];\frac52}\ ,
\w2
D=6: & \qquad \vE^{(6)}_{(1,0)} = \frac12 {\widehat E}^{SO(5,5)}_{[10000];\frac52} +\frac{4}{45} {\widehat E}^{SO(5,5)}_{[00001];3}\ ,
\w2
D=7: &\qquad \vE^{(7)}_{(1,0)}= \frac12 {\widehat E}^{SL(5)}_{[1000];\frac52} +\frac{\pi}{30} {\widehat E}^{SL(5)}_{[0001];3}\ ,
\w2
D=8: & \qquad \vE^{(8)}_{(1,0)}= \frac12 E^{SL(3)}_{[10];\frac52} -4E^{SL(2)}_{[1];2} E^{SL(3)}_{[10],-1/2}\ ,
\label{8d2}\w2
D=9: &\qquad \vE^{(9)}_{(1,0)}= \frac12 \nu^{-5/7} E^{SL(2)}_{[1];\frac52} +\frac{2}{15}\zeta(2) \nu^{9/7} E^{SL(2)}_{[1];\frac32} +\frac{4\zeta(2)\zeta(3)}{15} \nu^{-12/7}\ ,
\w2
D=10: &\qquad \vE^{(10)}_{(1,0)}=E^{SL(2)}_{[1];\frac52}\qquad \mbox{for IIB supergravity}\ ,
\label{2B2}
\end{align}
where the hats indicate the finite part of the series after subtraction of an $\e$ pole as in \cite{Green:2010wi}, and $\nu$ is an element of $GL(1) \subset GL(2,\mR)$. Note that the  perturbative contributions to $R^4$, $D^4R^4$ and $D^6R^4$
are identical in type IIA and type IIB \cite{Pioline:2015yea}. Since there are no $D$-instantons in type IIA, there are no non-perturbative contributions in that case.
In general, the expression for $\vE^{(D)}_{(0,0)}$ has perturbative part in string coupling constant expansion in $D$ dimensions, and a nonperturbative part. The perturbative part consists of a tree level and one-loop level term.  In the case of $\vE^{(D)}_{(1,0)}$, the perturbative contributions consist of tree-level, one-loop  (vanishing in $D=10$), and two-loop contributions.

In obtaining many of the results above, the following three limits have been used in \cite{Green:2010wi}:

\begin{itemize}

\item \mbox{\it The decompactification limit from $D$ to $D+1$ dimensions:} This is the limit in which the radius $r_d$ of one compact dimension becomes large. In this limit  $\vE^{D)}_{(p,q)}$ leads to a finite term which is required to produce $\vE^{(D+1)}_{(p,q)}$.

\item \mbox{\it Perturbative string theory limit:} This is the limit in which the $D$-dimensional
string coupling constant becomes small, and in this limit, the expansion of $\vE^{D)}_{(p,q)}$ in powers of the $D$-dimensional string coupling is required to reproduce the known perturbative string theory results.

\item \mbox{\it The semiclassical M-theory limit:} This is the limit in which the effects of wrapped p-branes are suppressed and the Feynman diagrams of compactified eleven-dimensional quantum supergravity should give a valid expansion in powers of the inverse volume of the torus.

\end{itemize}
For more details, see \cite{Green:2010wi}. Note also that the explicit forms of these couplings can also be obtained by direct calculation from exceptional field theory loops \cite{Bossard:2015foa}. In the case of type II string theory, the analytic contribution to the low energy expansion of the amplitudes at genus-one has been treated as a power series in space-time derivatives with coefficients that are determined by integrals of modular functions over the complex structure modulus of the world-sheet torus \cite{Green:2008uj, DHoker:2015gmr}.

As to $\vE^{(D)}_{(0,1)}$ which cannot be represented as an
Eisenstein series, apparently no closed form expression is known for it as yet. Nonetheless, an implicit expression for it as a two-loop Schwinger integral can be found in \cite{Bossard:2020xod}. See also \cite{Pioline:2015yea}, where its weak coupling and decompactification limits have been explored.

\subsection{Relevance to UV divergences and counterterms in supergravity}
\label{divergences}

From the point of view of supergravity divergences appearing in explicit loop computations in diverse dimensions, it is also useful to note the following points, which are essentially summarized in Table \ref{sugradiv}.
\begin{itemize}

\item In $D=9$, $\vE^{(D)}_{(0,0)}$ and $\vE^{(D)}_{(1,0)}$ cannot be deduced from supergravity divergences in maximal theory using dimensional regularization. This is due to the fact that these two functions receive contributions from tree and one-loop levels. However, divergences in maximal $9D$ supergravity start showing up at two-loop (one-loop divergences are absent in odd dimensions when using dimensional
regularisation) which would contain information on the $D^8 R^4$ term \cite{Bern:2000fm}.

\item $D=8$ is the lowest dimension where the $R^4$ term first appears
in one loop UV divergences, both in the maximal and half-maximal cases. In addition, there is a two-loop divergent $D^6 R^4$ counterterm in the maximal theory.

\item $D=7$ is the lowest dimension where the $D^4R^4$ term first appears
in two-loop UV divergences in the maximal theory.

\item In $D=6$, maximal supergravity admits a logarithmic divergence at three-loop level, which is related to the logarithmic term in the weak coupling expansion of $\vE^{(6)}_{(0,1)}$ \cite{Green:2010sp}, namely the function in front of the $D^6R^4$ term which we have not discussed so far. In the $N=(1,1)$ theory, there is also a $D^2R^4$ divergent term appearing at two-loop level.

\item In $D=5$, maximal supergravity is finite at four-loop level \cite{Bern:2000fm,Bern:2014sna}. It is likely that the first divergence in 4-pt supergraviton amplitude appears at six-loop level and is of the form $D^{12}R^4$.

\item In $D=4$, pure $N=4$ supergravity is divergent at four loop \cite{Bern:2013uka}. The recent five-loop computations show that maximal supergravity possesses a UV divergence at $D=\frac{24}5$ \cite{Bern:2018jmv}, corresponding to the operator $D^8 R^4$ which is the same operator that may appear in the seven-loop divergence of $D=4$ maximal supergravity. $D=4, N\ge 5$ supergravities are finite up to and including four loops \cite{Bern:2014sna}, and $D=4, N=1,2$ supergravities are finite up to and including two loops \cite{Deser:1977yyz}.

\end{itemize}
As discussed earlier in more detail, it is also worth recalling that the logarithmic divergences are related to the dimension dependent source terms in \eq{E1}-\eq{E3}\cite{Green:2010sp, Pioline:2015yea,Bossard:2015oxa}.

The UV divergences in supergravities summarized above are based on the loop calculations that have been carried out explicitly. These results  suggest that there exist fully nonlinear supersymmetric extensions of the $D^{2k}R^4$ terms that arise in the local counterterms, assuming that the quantization schemes employed respect supersymmetry. Turning this argument around, in cases where loop calculations are not available as yet, if one can construct the (nonlinear) supersymmetric completion  $D^{2k}R^4$, it would potentially imply the existence of divergences at appropriate loop order. For example, in $D=4$ the maximal theory seems to allow a $D^8 R^4$ invariant, which means that potentially a seven-loop divergence may arise \cite{Howe:1980th}. As to duality symmetries, their fate at the quantum level deserves scrutiny separately.

\begin{table}[H]
\centering

\begin{tabular}{|c|c|c|c|c|}
\hline
Local operator               &  $D$       & $L$      & $N$  & refs                           %
\\ \hline \hline
\multirow{2}{*}{$R^4$} &   8              & 1   & Max    &   \cite{Bern:2000fm}            \\ \cline{2-5 }
                   & 8 & 1   &      Half max      &  \cite{Dunbar:1999nj}
                   \\ \hline
\multirow{3}{*}{$D^2R^4$} &  10           & 1   &   Half max   & \cite{Dunbar:1999nj}         \\ \cline{2-5}
                   & 6 & 2   &    (1,1)  & \cite{Bern:2012gh}                        \\ \cline{2-5}
                   & 4 & 4   &  Half max  & \cite{Bern:2013uka} \\ \cline{2-5}
                    \hline
       $D^4R^4$      &  7 & 2  & Max &   \cite{Bern:2000fm}        \\
                    \hline
\multirow{2}{*}{$D^6R^4$}      &  8 & 2  & Max &   \cite{Bern:2000fm}        \\ \cline{2-5}
                               &   6 & 3  &(2,2)&  \cite{Bern:2008pv}  \\ \cline{2-5}
                    \hline
\multirow{3}{*}{$D^8R^4$}      &  9 & 2  & Max  &  \cite{Bern:2000fm}         \\ \cline{2-5}
                              & 4 & 7 & Max & \cite{Bossard:2010bd}\\ \cline{2-5}
                              &$\frac{24}5$& 5 &Max & \cite{Bern:2018jmv} \\ \cline{2-5}
                    \hline
$D^{9}R^4$      &  7 & 3  & Max &   \cite{Bern:2008pv}         \\
                    \hline
    $D^{10}R^4$      &  10 & 2  & Max &   \cite{Bern:2000fm}         \\
                    \hline
    $D^{12}R^4$      &  11 & 2  & Max  &   \cite{Deser:1998jz,Bern:2000fm}        \\
                    \hline
     $D^{15}R^4$      &  9 & 3  & Max  &   \cite{Bern:2008pv}        \\
                    \hline
       $D^{21}R^4$      &  11 & 3  & Max  &   \cite{Bern:2008pv}        \\
                    \hline
\end{tabular}

\caption{This table exhibits supergravity divergences in four-point supergraviton amplitude obtained by explicit computations in $D$ dimensions using dimensional regularization, with the exception of $D=4,N=8$ at $7$ loops, in which case an acceptable invariant has been put forward. See next section for further remarks. }
\label{sugradiv}
\end{table}

\subsection{Invariants in \texorpdfstring{$D=4$}{D=4} supergravities in ordinary and harmonic superspace}
\label{ordinary invariants}

Here we shall summarize known constructions of invariants in $4D$ maximal, as well as $N=4,5,6$ supergravities, in ordinary and mostly harmonic superspace approaches. The leading curvature terms will be of the form $D^{2k} R^4$ for various values of $k$.
We shall first recall the two maximally supersymmetric nonlinear invariants in ordinary superspace. Next, we shall summarize the number of invariants with maximal as well as $N=4,5,6$ supersymmetry in the harmonic superspace approach. Some of them will be available only in linearized harmonic superspace, and those which are known at the nonlinear level may not exhibit the full duality symmetry, as shall be summarized below.

In ordinary curved superspace, $R^4$ and $D^4 R^4$ invariants can only be realized at the linearized level. On the other hand, there exist nonlinear $D^{10}R^4$  and $D^8 R^5$ actions in ordinary superspace, as given in \eqref{e:chi4} and \eqref{e:ccF2}, which also preserve the full duality symmetry.  Finally, there exist also nonlinear invariants in harmonic superspace with the leading term $f(\phi)\nabla^k R^4$ in $4D$, $N=4,5,6,8$ supergravities preserving only the maximal compact subgroup of the full duality group.

In \cite{Howe:1980th,Kallosh:1980fi}, the following two nonlinear invariants have been provided:
\begin{align}
I_1  &= \int d^4x\, d^{32}\theta\,E(x,\theta)\,  \varepsilon^{\alpha\beta}\varepsilon^{\ad\bd}
\chi_{\alpha\, ijk}\bar\chi_\ad^{ijk}\chi_{\beta\,mnp}\bar\chi_\bd^{mnp}   \sim \int d^4x\,e \bigl( (\nabla^5 R^2 )^2 + \cdots  \bigr)\ ,
\label{e:chi4}\w2
I_2 &=\int d^4x\, d^{32}\theta\,E(x,\theta)\,\varepsilon^{\alpha\beta} \varepsilon^{\ad\bd} \chi_{\alpha\, i[jk} \chi_{\beta\, lmn]} \bar \chi_\ad^{i[jk} \bar \chi_\bd^{lmn]} \sim \int d^4x\,e ( \nabla^8 R^5 +\cdots) \ ,
\label{e:ccF2}
\end{align}
where $\chi_{\alpha ijk}$ is a spinor superfield whose lowest component represents the spin 1/2 fields in the $56$-plet of $SU(8)$. These invariants are fully $E_{7(7)}$ invariant because they are constructed from a full superspace integral of a superfield entering the superspace torsion. It is also worth mentioning that a dimension 16 invariant starting with $D^8 R^4$ as an ordinary superspace action integral does not exist because it may come from the integral  $\int E $, which, however, vanishes \cite{Bossard:2011tq}.

We now turn to the summary of invariants that have been constructed in harmonic superspace in $4D$. There exist linearized supersymmetric action integrals for $N$-extended supergravities in $4D$ which can be written as integrals in harmonic superspace, some of which admit a nonlinear extension \cite{Galperin:1984av}. Denoting the $N$-extended superspace by $M_{N}$, the $(N,p,q)$ harmonic superspace is defined as the direct product $M_N \times K_{p,q}$ where \cite{Hartwell:1994rp}
\be
K_{p,q}= S(U(p)\times U(N-p-q) \times U(q))\backslash SU(N)\ .
\ee
It is convenient to define the following projections of the spinorial covariant derivatives
\be
D_{\a I} := u_I{}^i D_{\a i}\ ,\qquad \bD^I_\ad := (u^{-1})_i{}^I \bD^i_\ad\ ,
\ee
where $u_I{}^i =(u_r{}^i, u_R{}^i, u_{r'}{}^i)$ with the indices $(r,R,r')$ labeling the fundamental representations of $SU(p), SU(N-p-q)$ and $SU(q)$, respectively. In terms of these derivatives, analytic fields in harmonic superspace are defined to be those which are annihilated by $(D_{\a r}, \bD_{\ad}^{r'})$. They can also be harmonic analytic, which means they are holomorphic with respect to the ${\bar\partial}$ operator on $K_{p,q}$, see, for example, \cite[Eqs. (3.1) and (3.2)]{Bossard:2010bd} and \cite{Drummond:2003ex} for further details.  Invariant action integrals in harmonic superspace are constructed by integrating analytic fields with respect to an appropriate measure. For $(N,p,q)$ harmonic superspace, such a measure at the linearized level in $4D$ is given by \cite{Hartwell:1994rp, Drummond:2003ex}
\be
d\mu_{p,q}^N := d^4 x du [D_{p+1} \cdots D_N \bD^1 \cdots \bD^{N-q} ]^2\ ,
\label{lmea}
\ee
where $du$ is is the standard Haar measure on $K_{p,q}$.

\subsubsection*{ \boldmath $D=4, N=8$}

To construct the linearized $N=8$ supersymmetric action integrals, we need the superfield $W_{ijkl}$ which is in the $70$-plet of $SU(8)$ with the appropriate reality condition, and satisfying the constraint
\be
D_{\a i} W_{jklm} = D_{\a[i} W_{jklm]}\ .
\ee
With the above ingredients, the following three linearized supersymmetric $SU(8)$ invariant action integrals have been constructed \cite{Drummond:2003ex}
\bea
I_{\frac12} &=& \int d\mu^8_{4,4}\, W^4  \sim  \int d^4 x\, t_8 t_8 R^4\ ,
\label{half} \w2
I_{\frac14} &=& \int d\mu^8_{2,2} \left( \e^{RSTU} W_{RS} W_{TU} \right)^2  \sim  \int d^4 x\,D^4 R^4\ ,
\label{quater}\w2
I_{\frac18} &=& \int d\mu^8_{1,1}\, \e^{R_1...R_6} \e^{S_1...S_6} W_{R_1R_2R_3} W_{R_4 R_5 S_1 }W_{R_6 S_2 S_3} W_{S_4 S_5 S_6} \sim  \int d^4 x\, D^6 R^4\ ,
\label{8th}
\eea
where
\bea
W: &=& u_1{}^i u_2{}^j u_3{}^k u_4{}^l\, W_{ijkl}\ ,
\label{12}\w2
W_{RS} &:=& u_1{}^i u_2{}^j u_R{}^k u_S{}^l\, W_{ijkl}\ ,
\label{14}\w2
W_{RST} &:=& u_1{}^i u_R{}^j u_S{}^k u_T{}^l\, W_{ijkl}\ .
\label{18}
\eea
These are the only ones that can be written down. However, it has been shown that the first two do not have a generalization to curved superspace. A nonlinear version of the first invariant in the superform (ectoplasm) approach has been considered in \cite{Bossard:2010bd} where it has been shown that it does not have $E_{7(7)}(\mathbb{R})$ symmetry. It has also been shown \cite{Bossard:2010bd} that the dimensional reduction of the $R^4$ term in $11D$ on seven-torus gives the leading term $f R^4$  in $4D$ where $f$ is a scalar-dependent function which satisfies the Laplace equation $(\Delta +42) f=0$, in agreement with \eq{E1}, and therefore it is consistent with $E_{7(7)}(\mathbb{Z})$. Note that in terms of the notation introduced in \eq{defE}, $f={\cal E}^{(4)}_{(0,0)}$, which is given in \eq{Ed00}.

The invariant given in \eq{8th} admits a nonlinear extension with the measure Eq. \eqref{lmea}, denoted by $d\mu_{(N,p,q)}$, given in \cite{Bossard:2011tq} for $d\mu_{(8,1,1)}$\footnote{Note that the linearized measure is denoted by $d\mu^N_{p,q}$, while its nonlinear extension is denoted by $d\mu_{(N,p,q)}$.}. However, it turns out that this invariant does not have $E_{7(7)}$ symmetry. Another invariant that has $SU(8)$ but not $E_{7(7)}$ symmetry is given by
\be
 I^\prime_{1/8}=\int  d\mu_{\scriptscriptstyle  (8,1,1)}  \; {\cal  F}^{11}_{88}({\cal V})
 \sim \int  d^4x \,  e \biggl( f^8_{6}(\phi) \nabla^3 R^2 \cdot  \nabla^3 R^2   + \cdots
 \biggr) \ ,
\ee
where $f^8_6(\phi)$ is the (appropriately normalised) $SU(8)$ invariant function of the 70 scalar fields discussed in \cite{Beisert:2010jx,Bossard:2010bd} and
\be  \label{e:F1188}
 {\cal F}^{11}_{88}({\mathcal V}) :=  u^1{}_i u^1{}_j  u^k{}_8 u^l{}_8  \, \bar
V^{im{\cal IJ}} \bar V^{jn{\cal K}{\cal L}}   V_{km{\cal K}{\cal L}}
V_{ln{\cal IJ}} \ ,
\ee
where $V^{im{\cal IJ}}   $ is a superfield whose lowest order component is the representative of the coset $E_{7(7)}(\mathbb{R})/SU(8)$. Note that this invariant differs from \eq{8th} by having a scalar dependent function in front of the leading gravitational term.

Finally, it is worth noting that an $E_{7(7)}(\mathbb{R})$ invariant and nonlinear action integral in full superspace which integrates to $D^8 R^4$ is known to exist and it is given by \cite{Howe:1980th}
\be
I = \int  d\mu_{(8,1,1)}\,B_{\alpha\bd}\,B^{\a\bd} \ ,
\label{BB1}
\ee
where $B_{\alpha\bd}= \bar\chi_{\dot\beta}^{1ij}\chi_{\alpha\,8 ij}$. It can be shown that this invariant reduces to $\int  d^4 x d^{32} \theta\,( W_{ijkl} \bar W^{ijkl} )^2$ $\sim \int d^4 x \big(\nabla^8 R^4 +\cdots\big) $ in the linearized approximation.

\subsubsection*{\boldmath $D=4, N=4,5,6$}

 All results summarized below for the $D=4, N=4,5,6$ superinvariants are taken from \cite{Bossard:2011tq}. An analog of the invariant \eqref{BB1} exists also for $D=4, N=4,5,6$ and it takes the form
\be
  \label{e:In}
  I^{{ N}}_1:= \kappa^{2({ N}-2)}\, \int    d\mu_{\scriptscriptstyle ({ N},1,1)}                \,
    B_{\alpha\bd}\,B^{\a\bd} \ ,
\ee
where
\be
  \label{e:Bdef}
  B_{\alpha\bd}=
  \begin{cases}
\bar\chi_{\dot\beta}^{1ij}\chi_{\alpha\, { N} ij}
& \mbox{for}~{ N}=4,5\cr
\bar\chi_{\dot\beta}^{1ij}\chi_{\alpha\, 6 ij} +\frac{1}{3} \chi_{\alpha}^{
1ijkl}\bar\chi_{\dot\beta\, 6 ijkl}
&
 \mbox{for}~{ N}= 6 \ ,\cr
  \end{cases}
\ee
where $\chi_{\alpha}^{
1ijkl}, \bar\chi_{\dot\beta\, 6 ijkl}$ are defined in \cite[Eq.(4.14)]{Bossard:2011tq}. Schematically, these invariants contain $D^{2(N-4)}R^4$. It is stated in \cite{Bossard:2011tq} that these have the full duality symmetries, namely $SO^\star(12), SU(5,1), SU(4)\times SU(1,1)$ for $N=6,5,4$ supergravities, respectively. For $N=6$ there are two other invariants given by
\begin{align}
I^{6}_2  &:= \int  d\mu_{\scriptscriptstyle (6,1,1)}                \,
  \varepsilon^{\alpha\beta}\varepsilon^{\ad\bd}    \biggl(J_{\alpha\bd}{}^{1i}_{6  i} J_{\beta\ad}{}^{1j}_{6  \,j} +\frac{4}{3}\,
J_{\alpha\bd}{}^{1i}_{6\,j} J_{\beta\ad}{}^{1j}_{6 \,i}\biggr)\ ,
\nn\w2
 I^6_3 & =\int d\mu_{\scriptscriptstyle (6,1,1)} \; {\cal F}^{11}_{66}({\cal V})  \sim   \int d^4x \,e \biggl(f^6_4(\phi)  \nabla  R^2 \cdot \nabla  R^2 + \cdots \biggr)\ ,
\end{align}
where
\begin{align}
J_{\alpha\dot\beta}{}^{ij}_{kl} &=\bar\chi_{\dot\beta}^{ijm}\chi_{\alpha\,klm}\ ,
\nn\w2
{\cal  F}^{11}_{66}({\cal V}) &:=  u^1{}_i u^1{}_j  u^k{}_6  u^l{}_6 \,
 \bar V^{i{\cal I}} \bar V^{j{\cal J}} V_{k{\cal I}} V_{l{\cal J}} \ ,
\end{align}
and $V_{i \cal I}$ is a superfield whose lowest component is the representative of the coset  $SO^\star(12)/U(6)$, and  $f_4^6(\phi)$ is a function on that coset. It is stated in \cite{Bossard:2011tq} that both of these invariants have only the $U(6)$ symmetry. Similarly, there is an additional $D=4,N=5$ invariant given by
\be
I^5_2= \int d\mu_{\scriptscriptstyle (5,1,1)} \; {\cal F}^{11}_{55}({\cal V}) \sim \int d^4x \, e \bigl( f^5_3(\phi) R^2 \cdot R^2  + \cdots \bigr) \  ,
\ee
where
\be
 {\cal F}^{11}_{55}({\cal V})  := u^1{}_i  u^1{}_j u^k{}_5 u^l{}_5  \,
 V^{i}  V^{j} \bar V_{k} \bar V_{l} \ .
 \ee
Here $V_i$ is a superfield whose lowest component is the representative of the coset  $SU(5,1)/U(5)$, and  $f_3^5(\phi)$ is a function on that coset. It is stated in \cite{Bossard:2011tq} that this invariant has only $U(5)$ symmetry\footnote{In a recent paper \cite{Kallosh:2023pkr}, it has been asserted that in the $D=4, N\ge 4$ actions discussed above, the Grassmann analyticity constraint on the harmonic superspace fields breaks the composite local $H$ symmetry, where $H$ refers to the stability subgroup in the $G/H$ cosets, with $G$ representing the duality symmetry group.}.

\subsection{Invariants in ectoplasm approach in \texorpdfstring{$4\le D\le 8$}{4<= D <=8}}
\label{ectoplasm}

Here we shall summarize briefly the results of \cite{Bossard:2014aea} for the action integrals in  ectoplasm approach in harmonic superspace and in the linearized approximation.  The formulae below refer to the integrands for the action \eq{ep}, in which the closed superform is defined in $(D|32)$ superspace extended by harmonic variables as described in the previous subsection, and its pull back to the $D$ dimensional spacetime is evaluated at $\theta=0$.
\begin{itemize}
\item {$D$=8, {\rm maximal}}
\bea\label{7Dresume}
 R^4&:& \qquad \sum_{n = 0}^{12}  \bar{U}^{- 2 n}\left(\bar{\mathcal{D}}^n \vE_\grad{2}{2}{0}\right) \cL^\ord{4n}   \ , \qquad  \sum_{n = 0}^{12} \left( \mathcal{D}_{[4n]}^n \vE\grad{2}{1}{1}\right) \cL^{[4n]} \ , \\
 \nabla^4 R^4&:& \qquad \sum_{n=0}^{14} \left( \  \sum_{ k = 0}^{20\mbox{-}n} \bar{U}^{- 2 k} \left(\bar{\mathcal{D}}^k \mathcal{D}^{n}_{[4n]} \vE_\grad{2}{1}{0} \right)\, \cL^{\ord{4k} [4n]}  + {U}^{- 2 } \left({\mathcal{D}} \mathcal{D}^{n}_{[4n]} \vE_\grad{2}{1}{0}\right) \, \cL^{\ord{-4} [4n]} \right) \ , \nn\\
&&\qquad  \sum_{n = 0}^{14} \left(\mathcal{D}_{[4n]}^n\vE'_{\scriptscriptstyle \frac{1}{4}}\right)   \, \cL^{[4n]} \ ,
\eea
In $8D$, maximal supergravity has the duality group $SL(2)\times SL(3)$, and the
scalar fields parametrize the symmetric space $SL(2)/SO(2)\times SL(3)/SO(3)$. The K\"ahler derivative on $SL(2)/SO(2)$, parametrized by the complex scalar $U$, is denoted by $\mathcal{D}$, while the $SU(2)$ isospin 2 tangent derivatives on $SL(3)/SO(3)$ are defined as ${\mathcal D}_{ijkl}$, with $i, j, k, l$ running from 1 to 2 of $SU(2)$. The $\cL^{(4k)[4n]}$ are $SL(2)\times SL(3)$ invariant eight-superforms in the isospin $2n$ representation of $SU(2)$ with $U(1)$ weight $4k$. In particular \cite{Bossard:2014aea}
\be
\cL^{(0)[0]} \propto \left( t^8 t^8 \partial_a\partial_b R \partial^a\partial^b RRR\right) +\cdots
\ee

The indices of the function $\mathcal{E}(n,p,q)$ refers to the harmonic superspace
construction of the associated invariant in the linearized approximation, whereas the notation ${\mathcal E}'_{1/4}$ indicates that the corresponding invariant cannot be written as a Lorentz invariant harmonic superspace integral in the linearized approximation. A combination of the functions $\vE_{(2,2,0)}$ and its complex conjugate $\vE_{(2,0,2)}$ is related to  ${\widehat E}^{SL(2)}_{[1];1}$, the function $\vE_{(2,1,1)}$ is proportional to ${\widehat E}^{SL(3)}_{[10];\frac32}$, a combination of $\vE_{(2,1,0)}$ and its complex conjugate $\vE_{(2,0,1)}$ is proportional to $E^{SL(2)}_{[1];2} E^{SL(3)}_{[10],-1/2}$ and $\vE'_{1/4}$ is proportional to $E^{SL(3)}_{[10];\frac52}$ as specified in \cite{Bossard:2014aea}.

    \item{$D=7, {\rm maximal} $
    \bea
 R^4&:& \qquad \sum_{n = 0}^{12} \left(\mathcal{D}_{[0,2n]}^n \vE_\gra{4}{2}\right)\,   \cL^{[0,2n]} \ ,  \\
 \nabla^4 R^4&:& \qquad \sum_{n,k = 0}^{n + 2 k \leq 20} \left(\mathcal{D}_{[4k,2n]}^{n + 2k}\vE_\gra{4}{1} \right) \, \cL^{[4k,2n]} \ , \qquad \sum_{n = 0}^{20} \left(\mathcal{D}_{[0,2n]}^n \vE'_{\scriptscriptstyle \frac{1}{4}}\right)  \, \cL^{[0,2n]} \ , \label{D4R48D}
\eea
}

In $7D$, maximal supergravity admits the duality group $SL(5)$, and the
scalar fields parametrize the coset $SL(5)/SO(5)$.
The covariant derivative ${\cal D}^{n+2k}_{[4k,2n]}$ acts on this coset and it denotes $(n+2k)$-fold product of the $[0,2]$ representation of $Sp(2)\approx SO(5)$ projected to the $[4k,2n]$ irreps.
$\cL^{[4k,2n]}$ denotes $SL(5)$ invariant superform in the $[4k,2n]$ of $Sp(2)$, i.e. traceless tensors of $SO(5)$ with $2k$ pairs of antisymmetric indices and $2n$ additional symmetrized indices.
The indices of the function $\mathcal{E}(n,p)$ refer to the harmonic superspace
construction of the associated invariant in the linearized approximation.
The function $\vE_{(4,2)}$ is related to  $E^{SL(5)}_{[1000];\frac32}$, and $\vE_{(4,1)}$ and  $\vE'_{1/4}$ are proportional to ${\widehat E}^{SL(5)}_{[0001];3}$ and ${\widehat E}^{SL(5)}_{[1000];\frac52}$, respectively. A combination of the last two functions defines
${\cal E}^{(7)}_{(1,0)}$. The notation  $\vE'_{1/4}$ denotes that the corresponding invariant cannot be written as a Lorentz invariant harmonic superspace integral in the linearized approximation. It was noted in \cite{Bossard:2015uga} that in $D\ge 7$ the $D^6R^4$ type invariants cannot be defined in the linearized approximation as harmonic superspace integrals.

\item {$D=6, {\rm maximal}$}

\bea
 R^4 &:& \qquad \sum_{n = 0}^{12} \left(\mathcal{D}_{[0,n],[0,n]}^n \vE_\grad{4}{2}{2}\right) \, \cL^{[0,n],[0,n]}  \ , \\
\nabla^4 R^4 &:& \qquad \sum_{n, k = 0}^{n + 2k \leq 20} \left(\mathcal{D}^{n + 2k}_{[0,n],[0,n+2k]} \vE_\grad{4}{2}{0}\right) \, \cL^{[0,n],[0,n + 2k]} \ , \nn\\
&& \qquad \sum_{n, k = 0}^{n + 2k \leq 20} \left(\mathcal{D}^{n + 2k}_{[2k,n],[2k,n]} \vE_\grad{4}{1}{1} \right)\, \cL^{[2k,n],[2k,n]} \ ,
\eea

In $6D$, maximal supergravity admits the duality group $SO(5,5)$, and the
scalar fields parameterize the coset $SO(5,5)/(SO(5)\times SO(5))$.
The covariant derivative ${\cal D}^{n+2k}_{[2k,n],[2k,n]}$ acts on this coset and it denotes $(n+2k)$-fold product of the $[0,1]\times [0,1]$ representation of $Sp(2)\times Sp(2)\approx SO(5)\times SO(5)$ projected to the $[2k,n]\times[2k,n]$ irreps.
$\cL^{[2k,n],[2k,n]}$ denotes the $SO(5,5)$ invariant superform in the $([2k,n],[2k,n])$ of $Sp(2)\times Sp(2)$.
The indices of the function $\mathcal{E}(n,p,q)$ refer to the harmonic superspace
construction of the associated invariant in the linearized approximation. The function $\vE_{(4,2,2)}$ is proportional to $ E^{SO(5,5)}_{[10000];\frac32}$. A combination of the functions $\vE_{(4,2,0)}$ and $\vE_{(4,0,2)}$ is related to ${\widehat E}^{SO(5,5)}_{[10000];\frac52}$. The function $\vE_{(4,1,1)}$ is proportional to $ {\widehat E}^{SO(5,5)}_{[00001];3}$. At the linearized level, the $D^6 R^4$ invariant was given in \cite[Eq. (3.27)]{Bossard:2015uga}. Its nonlinear version was suggested in \cite[Eq. (3.28)]{Bossard:2015uga}; see, however, \cite{Kallosh:2023pkr} for a recent discussion on this problem.

\item{$D=5, {\rm maximal} $}

\bea
R^4&:& \qquad \sum_{n = 0}^{12} \left(\mathcal{D}_{[0,0,0,n]}^{n} \vE_\gra{8}{4}\right)\,  \cL^{[0,0,0,n]} \ , \\
\nabla^4 R^4&:& \qquad \sum_{n,k = 0}^{n + 2k \leq 20} \left(\mathcal{D}_{[0,2k,0,n]}^{n+2k} \vE_\gra{8}{2}\right) \, \cL^{[0,2k,0,n]} \ .
\eea

In $5D$, maximal supergravity admits the duality group $E_{6(6)}$, with the maximal compact subgroup
$USp(8)\approx Sp(4)/\mathbb{Z}_2$. The
scalar fields parametrize the coset $E_{6(6)}/USp(8)$.
The covariant derivative ${\cal D}^{n+2k}_{[0,2k,0,n]}$ acts on this coset and it denotes $(n+2k)$-fold product of the $[0,0,0,1]$ representation of $Sp(4)$ projected to the $[0,2k,0,n]$ irreps.
$\cL^{[0,2k,0,n]}$ denotes $E_{6(6)}$ invariant superform in the $[0,2k,0,n]$ of $Sp(4)$. The function $\vE_{(8,4)}$ is proportional to $ E^{E_{6(6)}}_{[100000];\frac32}$, and the function $\vE_{(8,2)}$ is proportional to $ E^{E_{6(6)}}_{[100000];\frac52}$. At the linearized level, the $D^6 R^4$ invariant was given in \cite[Eq. (3.7)]{Bossard:2015uga}. Its nonlinear version was suggested in \cite[Eq. (3.8)]{Bossard:2015uga}.

\item{ $D=4, {\rm maximal}$ }
\bea
R^4&:& \qquad \sum_{n = 0}^{12} \left(\mathcal{D}_{[0,0,0,n, 0,0,0]}^{n} \vE_\gra{8}{4,4}\right) \,  \cL^{[0,0,0,n,0,0,0]} \ ,
\\
\nabla^4 R^4&:& \qquad \sum_{n,k = 0}^{n + 2k \leq 20} \left(\mathcal{D}_{[0,k,0,n,0,k,0]}^{n+2k} \vE_\gra{8}{2,2}\right) \, \cL^{[0,k,0,n,0,k,0]} \ .
\eea

In $4D$, maximal supergravity admits the duality group $E_{7(7)}$, with the maximal compact subgroup
$SU(8)/\mathbb{Z}_2$. The scalar fields parametrize the coset $E_{7(7)}/SU(8)$.
The covariant derivative ${\cal D}^{n+2k}_{[0,k,0,n,0,k,0]}$ acts on this coset and it denotes $(n+2k)$-fold product of the $[0,0,0,1,0,0,0]$ representation of $SU(8)$ projected to the $[0,k,0,n,0,k,0]$ irreps.
$\cL^{[0,k,0,n,0,k,0]}$ denotes $E_{7(7)}$ invariant superform in the $[0,k,0,n,0,k,0]$ of $SU(8)$. The function $\vE_{(8,4,4)}$ is proportional to $ E^{E_{7(7)}}_{[1000000];\frac32}$, and the function $\vE_{(8,2,2)}$ is proportional to $ E^{E_{7(7)}}_{[1000000];\frac52}$. At the linearized level, the $D^6 R^4$ invariant was given in \cite[Eq. (2.17)]{Bossard:2015uga}. Its nonlinear version was suggested in \cite[Eq. (2.18)]{Bossard:2015uga}.

\item{ $D=4, {\rm non-maximal}$ }

$N=5,6$ supergravities in $4D$ have also been investigated in \cite{Bossard:2010bd}. Explicit expressions for the attendant closed super four-forms have not been constructed but their analyses at the linearized level are sufficient to show that the $R^4$ invariant with $N=6,5$, and $D^2 R^4$  invariant with $N=6$ are not invariant under linearized duality transformations of $SO^\star(12)$ and $SU(5,1)$, respectively, which involve constant shifts in scalars. This is not surprising since there are no appropriate functions of the scalars in front of the leading terms in these invariants, which preserve the continuous duality symmetry. See \cite{Bossard:2010bd} for further details.

\end{itemize}

\section{Concluding remarks}

There are several open problems in the construction and analysis of higher derivative supergravities. We have summarized a number of approaches to their construction in section \ref{sec2}, but we also saw, especially in the example of $11D$ supergravity, the enormity of the complications in the construction of even the leading higher derivative terms. The situation is more manageable for four derivative extensions when they exist. But beyond that, for example in the case of much studied eight derivative extensions, complete results even in the bosonic sector are very rare, let alone the fermionic terms. Taking into account duality symmetries turns out to have profound consequences but its implementation has challenges as the number of derivatives grows, and even when the remarkable functions of the moduli needed for the duality symmetries are known, it is always a challenge to go beyond the leading terms in curvature. In filling the several existing gaps in the landscape of higher derivative supergravities, it is not entirely clear which methods among those surveyed in section \ref{sec2} are the most promising ones, and if they are even feasible, say beyond eight derivatives. Ultimately, it may be necessary to develop new and extremely powerful computer-based techniques.

Putting aside the problem of finding more extensive results on higher derivative extensions of supergravities, one may focus on the analysis of those which have been already constructed, and investigate further their applications in black holes, cosmology, holography and the Swampland program, as mentioned in the introduction.
In studying higher derivative extensions of matter coupled gravity, it is of course essential to ensure that unitarity and causality principles hold. Here we shall summarize very briefly some of the key conclusions drawn so far from the criteria of unitarity and causality, and we shall do so in chronological order.

First, in \cite{Hofman:2008ar} it was shown that in the context of AdS$_5/$CFT$_4$ correspondence, the requirement of the positivity of the energy on the CFT side puts restrictions on the conformal anomaly coefficients, which are related to the coefficients $a$ and $c$ of the Euler and Weyl-squared invariants in CFT$_4$, respectively. This, in turn, implies a restriction on the coefficient of quadratic curvature corrections to the bulk action, the relevant one being equivalently the Riemann-squared, or Weyl-squared or Gauss-Bonnet term \cite{Buchel:2009tt}. These results were generalized to AdS$_7/$CFT$_6$ in \cite{deBoer:2009pn}. Subsequently, using a relationship between the positivity of the energy flux in CFT and the causality in the bulk theory, and under certain assumptions, a range in which $\lambda$ is constrained to lie was found in any dimensions in \cite{Camanho:2009vw}.
Next, a different setting for causality considerations involving the scattering of gravitons was considered in \cite{Camanho:2014apa}. In the context of AdS$_5/$CFT$_4$, as well as bulk gravity that admits Minkowski vacuum, it was shown that an infinite tower of states with $J>2$ are needed to restore causality. Furthermore, the unitarity and analyticity of graviton amplitudes were shown to constrain the coefficients of quartic in Riemann curvature terms \cite{Bellazzini:2015cra}.
A constraint on the coefficient of the Riemann-squared term was found later in \cite{Cheung:2016wjt} which simply requires that it is positive.
Further advances were made in \cite{Arkani-Hamed:2020blm} where constraints imposed by causality and unitarity on the low-energy effective field theory expansion of four-particle scattering amplitudes were studied in flat space. The constraints found on the amplitudes can be translated to restrictions on the coefficients of $R^4, D^4R^4$ and $D^8 R^4$ terms.
In \cite{Caron-Huot:2021rmr}, again in flat space, using $S$-matrix and dispersion relations, it was derived that in $10D$ maximal supergravity, the coefficient in front of $R^4$ term, denoted by $g_0$,  resides in the region $0\le g_0\le 3(8\pi G/M^6)$ where $M$ is the scale beyond which the EFT breaks down. In string theory, this constraint is satisfied as $g_0 M^6/(8\pi G)=2\zeta(3)\approx2.4$. These computations were generalized to maximal AdS supergravity in $5D$ in \cite{Caron-Huot:2021enk}, where similar bounds were found.  Finally, the consequences of the causality, analyticity and IR divergence obstructions to UV completion were sharpened further in \cite{Caron-Huot:2022jli}, where more comprehensive references are provided as well.

In conclusion, it is well motivated to pin down the role of supersymmetry, duality symmetries, and the physical requirements such as unitarity and causality in determining the structure of the higher derivative extensions of matter coupled quantum gravity, and much remains to be done. These considerations may provide a sound framework for effective field theory approach to quantum gravity up to an appropriate cut-off energy scale. More ambitiously, contemplating a UV completion, given that the need for introduction of an infinite tower of massive higher spin states is widely appreciated, it would be interesting to understand the nature of such states in comparison with those arising in string theory. In the context of studying the constraints on the coefficient of the Riemann-squared term,  a general UV completion of the Gauss-Bonnet term, which involves the coupling of massive higher spin states, were considered, for example, in \cite{Cheung:2016wjt}. In a more general setting, the important question of whether UV complete theory of quantum gravity is uniquely determined by string theory has been addressed in many works, see, for example \cite{Arkani-Hamed:2023jwn} and references therein. If the bottom to top approach turns out to yield results that differ from those in string theory, it will be natural to study if such results can offer a progress in addressing some of the challenging problems in matter coupled quantum gravity, such as the very early universe and black hole physics.

\section*{Acknowledgement}

We thank Arash Ardehali, Guillaume Bossard, Daniel Butter, Hao-Yuan Chang, Aidan Herderschee, Kiril Hristov, Yoshifumi Hyakutake, Axel Kleinschmidt, Renata Kallosh, Yue-Zhou Li, Hong L\"u, Liang Ma, Bengt Nilsson, Kelly Stelle, Yoshiaki Tanii and Gabriele Tartaglino-Mazzucchelli for useful discussions. Y.P. would also like to thank Jenny Zhao for her support. M.O. is supported in part by TUBITAK grant 121F064, the Distinguished Young Scientist Award BAGEP of the Science Academy, and the Outstanding Young Scientist Award of the Turkish Academy of Sciences (TUBA-GEBIP). The work of Y.P. is supported by the National Key Research and Development Program under grant No. 2022YFE0134300 and the National Natural Science Foundation of China (NSFC) under grant No. 12175164. This work is also partially supported by Peng Huanwu Center for Fundamental Theory, under grant No.12247103. The work of E.S. is supported in part by NSF grant PHYS-2112859. We also thank Istanbul Technical University, where this work began, for hospitality.

\begin{appendices}

\section{Notations}


Given matrices $M_i, i=1,...,4$, the $t_8$ symbol is defined as
\bea
t^{a_1...a_8}M_{1 a_1a_2} \cdots M_{4 a_7a_8} &=& -2\big(\tr M_1M_2 \tr M_3 M_4 +\tr M_1 M_3 \tr M_2 M_4 +\tr M_1 M_4 \tr M_2 M_3\big)
\nn\w2
&& + 8\tr \big( M_1 M_2 M_3 M_4 + M_1 M_3 M_2 M_4 + M_1 M_4 M_2 M_3 \big)\ .
\eea
Definitions specific to $11D$ and $10D$ used in sections 3 and 4 are as follows:

\subsubsection*{Definitions in $D=11$}
%
\bea
\epsilon_{11}\epsilon_{11} R^4 &=& \e^{\m_1...\m_8 abc} \e_{\n_1...\n_8 abc} R_{\m_1\m_2}{}^{\n_1\n_2}\cdots R_{\m_7\m_8}{}^{\n_7\n_8}\ ,
\w2
t_8t_8 \left(ABCD\right) &=& t^{a_1...a_8} t^{b_1...b_8} A_{a_1a_2 b_1b_2} B_{a_3a_4 b_3b_4} C_{a_5a_6 b_5b_6} D_{a_7a_8 b_7b_8}\ ,
\w2
t_8t_8 \left(F^2 R^3\right) &=& t^{a_1...a_8} t^{b_1...b_8} F_{a_1b_1 cd}{}F_{a_2b_2}{}^{cd} R_{a_3a_4b_3b_4}\dots R_{a_7a_8 b_7b_8}
\w2
\epsilon_{11}\epsilon_{11} F^2 R^3 &=& \epsilon^{a\mu_1\cdots\mu_{10}} \epsilon_{a\nu_1\cdots\nu_{10}} F_{\mu_1\mu_2}{}^{\nu_1\nu_2} F_{\mu_3\mu_4}{}^{\nu_3\nu_4}\cdots R_{\mu_9\mu_{10}}{}^{\nu_9\nu_{10}}\ .
\label{e11e11}
\eea

\subsubsection*{Definitions in $D=10$}

%
\bea
\epsilon_{10}\epsilon_{10} R^4 &=& \e^{\m_1...\m_8 ab} \e_{\n_1...\n_8 ab} R_{\m_1\m_2}{}^{\n_1\n_2}\cdots R_{\m_7\m_8}{}^{\n_7\n_8}\ ,
\label{1010}\w2
\epsilon_9\epsilon_9 |G_3|^2 R^3 &=& \epsilon_{ab\mu_1...\mu_8} \epsilon^{ac\nu_1...\nu_8} G^{\mu_1\mu_2}{}_{c} {\bar G}_{\nu_1\nu_2}{}^{b}R^{\mu_3\mu_4}{}_{\nu_3\nu_4}\dots R^{\mu_7\mu_8}{}_{\nu_7\nu_8}\ ,
\label{99}\w2
t_8 t_8 |G_3|^2 R^3 &=& t_{\mu_1...\mu_8} t^{\nu_1...\nu_8} G^{[\mu_1}{}_{\nu_1 a} {\bar G}^{\mu_2]a}{}_{\nu_2} R^{\mu_3\mu_4}{}_{\nu_3\nu_4}\dots R^{\mu_7\mu_8}{}_{\nu_7\nu_8}\ ,
\label{tt8}\w2
\epsilon_8\epsilon_8 |G_3|^2 R^3 &=& -\frac12 \epsilon^{ab\mu_1...\mu_8} \epsilon_{ab}{}^{\nu_1...\nu_8} G_{[\mu_1|\nu_1 a} {\bar G}_{\mu_2]}{}^a{}_{\nu_2} R_{\m_3\m_4\n_3\n_4} \cdots R_{\m_7\m_8\n_7\n_8}\ ,
\label{88}
\w2
\epsilon_9 \epsilon_9 H^2 R^3 &=& -\left(\epsilon_{10} \epsilon_{10}\right)_{\mu_0...\mu_8}^{\nu_0...\nu_8} H^{\mu_1\mu_2}{}_{\nu_0} H_{\nu_1\nu_2}{}^{\mu_0} R^{\mu_3\mu_4}{}_{\nu_3\nu_4}\cdots R^{\mu_7\mu_8}{}_{\nu_7\nu_8}\ ,
\w2
\epsilon_9 \epsilon_9 H^2 (\nabla H)^2 R &=& -\left(\epsilon_{10} \epsilon_{10}\right)_{\mu_0...\mu_8}^{\nu_0...\nu_8} H^{\mu_0\mu_1\mu_2} H_{\nu_0\nu_1\nu_2} \nabla^{\mu_3} H^{\mu_4}{}_{\nu_3\nu_4} \cdots R^{\mu_7\mu_8}{}_{\nu_7\nu_8}\ .
\eea

\end{appendices}



\section*{References}

\begin{thebibliography}{100}
\expandafter\ifx\csname url\endcsname\relax
  \def\url#1{\texttt{#1}}\fi
\expandafter\ifx\csname urlprefix\endcsname\relax\def\urlprefix{URL }\fi
\expandafter\ifx\csname href\endcsname\relax
  \def\href#1#2{#2} \def\path#1{#1}\fi

\bibitem{Deser:1977yyz}
S.~Deser, J.~H. Kay, K.~S. Stelle, {Renormalizability Properties of
  Supergravity}, Phys. Rev. Lett. 38 (1977) 527.
\newblock \href {http://arxiv.org/abs/1506.03757} {\path{arXiv:1506.03757}},
  \href {https://doi.org/10.1103/PhysRevLett.38.527}
  {\path{doi:10.1103/PhysRevLett.38.527}}.

\bibitem{Deser:1978br}
S.~Deser, J.~H. Kay, {Three Loop Counterterms for Extended Supergravity}, Phys.
  Lett. B 76 (1978) 400--403.
\newblock \href {https://doi.org/10.1016/0370-2693(78)90892-4}
  {\path{doi:10.1016/0370-2693(78)90892-4}}.

\bibitem{Ferrara:1978wj}
S.~Ferrara, P.~Van~Nieuwenhuizen, {Structure of Supergravity}, Phys. Lett. B 78
  (1978) 573--576.
\newblock \href {https://doi.org/10.1016/0370-2693(78)90642-1}
  {\path{doi:10.1016/0370-2693(78)90642-1}}.

\bibitem{Howe:1980th}
P.~S. Howe, U.~Lindstrom, {Higher Order Invariants in Extended Supergravity},
  Nucl. Phys. B 181 (1981) 487--501.
\newblock \href {https://doi.org/10.1016/0550-3213(81)90537-X}
  {\path{doi:10.1016/0550-3213(81)90537-X}}.

\bibitem{Donoghue_2012}
J.~F. Donoghue, \href{https://doi.org/10.1063%2F1.4756964}{The effective field
  theory treatment of quantum gravity}, in: {AIP} Conference Proceedings,
  {AIP}, 2012.
\newblock \href {https://doi.org/10.1063/1.4756964}
  {\path{doi:10.1063/1.4756964}}.
\newline\urlprefix\url{https://doi.org/10.1063%2F1.4756964}

\bibitem{Vafa:2005ui}
C.~Vafa, {The String landscape and the swampland} (9 2005).
\newblock \href {http://arxiv.org/abs/hep-th/0509212}
  {\path{arXiv:hep-th/0509212}}.

\bibitem{Palti:2019pca}
E.~Palti, {The Swampland: Introduction and Review}, Fortsch. Phys. 67~(6)
  (2019) 1900037.
\newblock \href {http://arxiv.org/abs/1903.06239} {\path{arXiv:1903.06239}},
  \href {https://doi.org/10.1002/prop.201900037}
  {\path{doi:10.1002/prop.201900037}}.

\bibitem{Harlow:2022ich}
D.~Harlow, B.~Heidenreich, M.~Reece, T.~Rudelius, {Weak gravity conjecture},
  Rev. Mod. Phys. 95~(3) (2023) 035003.
\newblock \href {http://arxiv.org/abs/2201.08380} {\path{arXiv:2201.08380}},
  \href {https://doi.org/10.1103/RevModPhys.95.035003}
  {\path{doi:10.1103/RevModPhys.95.035003}}.


\bibitem{Mohaupt:2000mj}
T.~Mohaupt, {Black hole entropy, special geometry and strings}, Fortsch. Phys.
  49 (2001) 3--161.
\newblock \href {http://arxiv.org/abs/hep-th/0007195}
  {\path{arXiv:hep-th/0007195}}.



\bibitem{deWit:2007maa}
B.~de~Wit, {BPS black holes}, Nucl. Phys. B Proc. Suppl. 171 (2007) 16--38.
\newblock \href {http://arxiv.org/abs/0704.1452} {\path{arXiv:0704.1452}},
  \href {https://doi.org/10.1016/j.nuclphysbps.2007.06.004}
  {\path{doi:10.1016/j.nuclphysbps.2007.06.004}}.

\bibitem{Castro:2008ne}
A.~Castro, J.~L. Davis, P.~Kraus, F.~Larsen, {String Theory Effects on
  Five-Dimensional Black Hole Physics}, Int. J. Mod. Phys. A 23 (2008)
  613--691.
\newblock \href {http://arxiv.org/abs/0801.1863} {\path{arXiv:0801.1863}},
  \href {https://doi.org/10.1142/S0217751X08039724}
  {\path{doi:10.1142/S0217751X08039724}}.

\bibitem{Blau:1999vz}
M.~Blau, K.~S. Narain, E.~Gava, {On subleading contributions to the AdS / CFT
  trace anomaly}, JHEP 09 (1999) 018.
\newblock \href {http://arxiv.org/abs/hep-th/9904179}
  {\path{arXiv:hep-th/9904179}}, \href
  {https://doi.org/10.1088/1126-6708/1999/09/018}
  {\path{doi:10.1088/1126-6708/1999/09/018}}.

\bibitem{Bobev:2021qxx}
N.~Bobev, K.~Hristov, V.~Reys, {AdS$_{5}$ holography and higher-derivative
  supergravity}, JHEP 04 (2022) 088.
\newblock \href {http://arxiv.org/abs/2112.06961} {\path{arXiv:2112.06961}},
  \href {https://doi.org/10.1007/JHEP04(2022)088}
  {\path{doi:10.1007/JHEP04(2022)088}}.

\bibitem{Maldacena:2011nz}
J.~M. Maldacena, G.~L. Pimentel, {On graviton non-Gaussianities during
  inflation}, JHEP 09 (2011) 045.
\newblock \href {http://arxiv.org/abs/1104.2846} {\path{arXiv:1104.2846}},
  \href {https://doi.org/10.1007/JHEP09(2011)045}
  {\path{doi:10.1007/JHEP09(2011)045}}.

\bibitem{Nojiri:2017ncd}
S.~Nojiri, S.~D. Odintsov, V.~K. Oikonomou, {Modified Gravity Theories on a
  Nutshell: Inflation, Bounce and Late-time Evolution}, Phys. Rept. 692 (2017)
  1--104.
\newblock \href {http://arxiv.org/abs/1705.11098} {\path{arXiv:1705.11098}},
  \href {https://doi.org/10.1016/j.physrep.2017.06.001}
  {\path{doi:10.1016/j.physrep.2017.06.001}}.

\bibitem{VanNieuwenhuizen:1981ae}
P.~Van~Nieuwenhuizen, {Supergravity}, Phys. Rept. 68 (1981) 189--398.
\newblock \href {https://doi.org/10.1016/0370-1573(81)90157-5}
  {\path{doi:10.1016/0370-1573(81)90157-5}}.

\bibitem{Freedman:2012zz}
D.~Z. Freedman, A.~Van~Proeyen, {Supergravity}, Cambridge Univ. Press,
  Cambridge, UK, 2012.

\bibitem{Bergshoeff:1985mz}
E.~Bergshoeff, E.~Sezgin, A.~Van~Proeyen, {Superconformal Tensor Calculus and
  Matter Couplings in Six-dimensions}, Nucl. Phys. B 264 (1986) 653, [Erratum:
  Nucl.Phys.B 598, 667 (2001)].
\newblock \href {https://doi.org/10.1016/0550-3213(86)90503-1}
  {\path{doi:10.1016/0550-3213(86)90503-1}}.

\bibitem{deWit:1984rvr}
B.~de~Wit, P.~G. Lauwers, A.~Van~Proeyen, {Lagrangians of N=2 Supergravity -
  Matter Systems}, Nucl. Phys. B 255 (1985) 569--608.
\newblock \href {https://doi.org/10.1016/0550-3213(85)90154-3}
  {\path{doi:10.1016/0550-3213(85)90154-3}}.

\bibitem{Lauria:2020rhc}
E.~Lauria, A.~Van~Proeyen, {${\cal N}=2$ Supergravity in $D=4,5,6$ Dimensions},
  Vol. 966, 2020.
\newblock \href {http://arxiv.org/abs/2004.11433} {\path{arXiv:2004.11433}},
  \href {https://doi.org/10.1007/978-3-030-33757-5}
  {\path{doi:10.1007/978-3-030-33757-5}}.

\bibitem{Bergshoeff:1980is}
E.~Bergshoeff, M.~de~Roo, B.~de~Wit, {Extended Conformal Supergravity}, Nucl.
  Phys. B 182 (1981) 173--204.
\newblock \href {https://doi.org/10.1016/0550-3213(81)90465-X}
  {\path{doi:10.1016/0550-3213(81)90465-X}}.

\bibitem{Bergshoeff:2001hc}
E.~Bergshoeff, T.~de~Wit, R.~Halbersma, S.~Cucu, M.~Derix, A.~Van~Proeyen,
  {Weyl multiplets of N=2 conformal supergravity in five-dimensions}, JHEP 06
  (2001) 051.
\newblock \href {http://arxiv.org/abs/hep-th/0104113}
  {\path{arXiv:hep-th/0104113}}, \href
  {https://doi.org/10.1088/1126-6708/2001/06/051}
  {\path{doi:10.1088/1126-6708/2001/06/051}}.

\bibitem{Fujita:2001kv}
T.~Fujita, K.~Ohashi, {Superconformal tensor calculus in five-dimensions},
  Prog. Theor. Phys. 106 (2001) 221--247.
\newblock \href {http://arxiv.org/abs/hep-th/0104130}
  {\path{arXiv:hep-th/0104130}}, \href {https://doi.org/10.1143/PTP.106.221}
  {\path{doi:10.1143/PTP.106.221}}.

\bibitem{Hutomo:2022hdi}
J.~Hutomo, S.~Khandelwal, G.~Tartaglino-Mazzucchelli, J.~Woods, {Hyperdilaton
  Weyl multiplets of 5D and 6D minimal conformal supergravity}, Phys. Rev. D
  107~(4) (2023) 046009.
\newblock \href {http://arxiv.org/abs/2209.05748} {\path{arXiv:2209.05748}},
  \href {https://doi.org/10.1103/PhysRevD.107.046009}
  {\path{doi:10.1103/PhysRevD.107.046009}}.

\bibitem{Coomans:2012cf}
F.~Coomans, M.~Ozkan, {An off-shell formulation for internally gauged D=5, N=2
  supergravity from superconformal methods}, JHEP 01 (2013) 099.
\newblock \href {http://arxiv.org/abs/1210.4704} {\path{arXiv:1210.4704}},
  \href {https://doi.org/10.1007/JHEP01(2013)099}
  {\path{doi:10.1007/JHEP01(2013)099}}.

\bibitem{Ozkan:2013nwa}
M.~Ozkan, Y.~Pang, {All off-shell $R^{2}$ invariants in five dimensional
  $\mathcal{N} =$ 2 supergravity}, JHEP 08 (2013) 042.
\newblock \href {http://arxiv.org/abs/1306.1540} {\path{arXiv:1306.1540}},
  \href {https://doi.org/10.1007/JHEP08(2013)042}
  {\path{doi:10.1007/JHEP08(2013)042}}.

\bibitem{Bergshoeff:2004kh}
E.~Bergshoeff, S.~Cucu, T.~de~Wit, J.~Gheerardyn, S.~Vandoren, A.~Van~Proeyen,
  {N = 2 supergravity in five-dimensions revisited}, Class. Quant. Grav. 21
  (2004) 3015--3042.
\newblock \href {http://arxiv.org/abs/hep-th/0403045}
  {\path{arXiv:hep-th/0403045}}, \href
  {https://doi.org/10.1088/0264-9381/23/23/C01}
  {\path{doi:10.1088/0264-9381/23/23/C01}}.

\bibitem{Coomans:2011ih}
F.~Coomans, A.~Van~Proeyen, {Off-shell N=(1,0), D=6 supergravity from
  superconformal methods}, JHEP 02 (2011) 049, [Erratum: JHEP 01, 119 (2012)].
\newblock \href {http://arxiv.org/abs/1101.2403} {\path{arXiv:1101.2403}},
  \href {https://doi.org/10.1007/JHEP02(2011)049}
  {\path{doi:10.1007/JHEP02(2011)049}}.

\bibitem{Howe:1981gz}
P.~S. Howe, {Supergravity in Superspace}, Nucl. Phys. B 199 (1982) 309--364.
\newblock \href {https://doi.org/10.1016/0550-3213(82)90349-2}
  {\path{doi:10.1016/0550-3213(82)90349-2}}.

\bibitem{Gates:1983nr}
S.~J. Gates, M.~T. Grisaru, M.~Rocek, W.~Siegel, {Superspace Or One Thousand
  and One Lessons in Supersymmetry}, Vol.~58 of Frontiers in Physics, 1983.
\newblock \href {http://arxiv.org/abs/hep-th/0108200}
  {\path{arXiv:hep-th/0108200}}.

\bibitem{Buchbinder:1998qv}
I.~L. Buchbinder, S.~M. Kuzenko, {Ideas and methods of supersymmetry and
  supergravity: Or a walk through superspace}, 1998.

\bibitem{Wess:1977fn}
J.~Wess, B.~Zumino, {Superspace Formulation of Supergravity}, Phys. Lett. B 66
  (1977) 361--364.
\newblock \href {https://doi.org/10.1016/0370-2693(77)90015-6}
  {\path{doi:10.1016/0370-2693(77)90015-6}}.

\bibitem{Grimm:1978ch}
R.~Grimm, J.~Wess, B.~Zumino, {A Complete Solution of the Bianchi Identities in
  Superspace}, Nucl. Phys. B 152 (1979) 255--265.
\newblock \href {https://doi.org/10.1016/0550-3213(79)90102-0}
  {\path{doi:10.1016/0550-3213(79)90102-0}}.

\bibitem{West:1990tg}
P.~C. West, {Introduction to supersymmetry and supergravity}, 1990.

\bibitem{Howe:1980sy}
P.~S. Howe, {A SUPERSPACE APPROACH TO EXTENDED CONFORMAL SUPERGRAVITY}, Phys.
  Lett. B 100 (1981) 389--392.
\newblock \href {https://doi.org/10.1016/0370-2693(81)90143-X}
  {\path{doi:10.1016/0370-2693(81)90143-X}}.

\bibitem{Cederwall:2004cg}
M.~Cederwall, U.~Gran, B.~E.~W. Nilsson, D.~Tsimpis, {Supersymmetric
  corrections to eleven-dimensional supergravity}, JHEP 05 (2005) 052.
\newblock \href {http://arxiv.org/abs/hep-th/0409107}
  {\path{arXiv:hep-th/0409107}}, \href
  {https://doi.org/10.1088/1126-6708/2005/05/052}
  {\path{doi:10.1088/1126-6708/2005/05/052}}.

\bibitem{Galperin:1984av}
A.~Galperin, E.~Ivanov, S.~Kalitsyn, V.~Ogievetsky, E.~Sokatchev,
  {Unconstrained N=2 Matter, Yang-Mills and Supergravity Theories in Harmonic
  Superspace}, Class. Quant. Grav. 1 (1984) 469--498, [Erratum:
  Class.Quant.Grav. 2, 127 (1985)].
\newblock \href {https://doi.org/10.1088/0264-9381/1/5/004}
  {\path{doi:10.1088/0264-9381/1/5/004}}.

\bibitem{Galperin:2001seg}
A.~S. Galperin, E.~A. Ivanov, V.~I. Ogievetsky, E.~S. Sokatchev, {Harmonic
  superspace}, Cambridge Monographs on Mathematical Physics, Cambridge
  University Press, 2007.
\newblock \href {https://doi.org/10.1017/CBO9780511535109}
  {\path{doi:10.1017/CBO9780511535109}}.

\bibitem{deHaro:2002vk}
S.~de~Haro, A.~Sinkovics, K.~Skenderis, {On a supersymmetric completion of the
  R4 term in 2B supergravity}, Phys. Rev. D 67 (2003) 084010.
\newblock \href {http://arxiv.org/abs/hep-th/0210080}
  {\path{arXiv:hep-th/0210080}}, \href
  {https://doi.org/10.1103/PhysRevD.67.084010}
  {\path{doi:10.1103/PhysRevD.67.084010}}.

\bibitem{Voronov:1992}
T.~Voronov, {Geometric integration theory on supermanifolds},
Soviet Scientific Review, Section C: Mathematical Physics, 9, Part 1. Harwood Academic Publisher, Chur, 1991.

\bibitem{Gates:1997kr}
S.~J. Gates, Jr., {Ectoplasm has no topology: The Prelude}, in: {2nd
  International Seminar on Supersymmetries and Quantum Symmetries}: {Dedicated
  to the Memory of Victor I. Ogievetsky}, 1997, pp. 46--57.
\newblock \href {http://arxiv.org/abs/hep-th/9709104}
  {\path{arXiv:hep-th/9709104}}.

\bibitem{Gates:1998hy}
S.~J. Gates, Jr., {Ectoplasm has no topology}, Nucl. Phys. B 541 (1999)
  615--650.
\newblock \href {http://arxiv.org/abs/hep-th/9809056}
  {\path{arXiv:hep-th/9809056}}, \href
  {https://doi.org/10.1016/S0550-3213(98)00819-0}
  {\path{doi:10.1016/S0550-3213(98)00819-0}}.

\bibitem{Berkovits:2008qw}
N.~Berkovits, P.~S. Howe, {The Cohomology of superspace, pure spinors and
  invariant integrals}, JHEP 06 (2008) 046.
\newblock \href {http://arxiv.org/abs/0803.3024} {\path{arXiv:0803.3024}},
  \href {https://doi.org/10.1088/1126-6708/2008/06/046}
  {\path{doi:10.1088/1126-6708/2008/06/046}}.

\bibitem{Gates:1997ag}
S.~J. Gates, Jr., M.~T. Grisaru, M.~E. Knutt-Wehlau, W.~Siegel, {Component
  actions from curved superspace: Normal coordinates and ectoplasm}, Phys.
  Lett. B 421 (1998) 203--210.
\newblock \href {http://arxiv.org/abs/hep-th/9711151}
  {\path{arXiv:hep-th/9711151}}, \href
  {https://doi.org/10.1016/S0370-2693(97)01557-8}
  {\path{doi:10.1016/S0370-2693(97)01557-8}}.

\bibitem{Aldazabal:2013sca}
G.~Aldazabal, D.~Marques, C.~Nunez, {Double Field Theory: A Pedagogical
  Review}, Class. Quant. Grav. 30 (2013) 163001.
\newblock \href {http://arxiv.org/abs/1305.1907} {\path{arXiv:1305.1907}},
  \href {https://doi.org/10.1088/0264-9381/30/16/163001}
  {\path{doi:10.1088/0264-9381/30/16/163001}}.

\bibitem{Hohm:2013bwa}
O.~Hohm, D.~L\"ust, B.~Zwiebach, {The Spacetime of Double Field Theory: Review,
  Remarks, and Outlook}, Fortsch. Phys. 61 (2013) 926--966.
\newblock \href {http://arxiv.org/abs/1309.2977} {\path{arXiv:1309.2977}},
  \href {https://doi.org/10.1002/prop.201300024}
  {\path{doi:10.1002/prop.201300024}}.

\bibitem{Berman:2013eva}
D.~S. Berman, D.~C. Thompson, {Duality Symmetric String and M-Theory}, Phys.
  Rept. 566 (2014) 1--60.
\newblock \href {http://arxiv.org/abs/1306.2643} {\path{arXiv:1306.2643}},
  \href {https://doi.org/10.1016/j.physrep.2014.11.007}
  {\path{doi:10.1016/j.physrep.2014.11.007}}.

\bibitem{Baron:2017dvb}
W.~H. Baron, J.~J. Fernandez-Melgarejo, D.~Marques, C.~Nunez, {The Odd story of
  \ensuremath{\alpha}'-corrections}, JHEP 04 (2017) 078.
\newblock \href {http://arxiv.org/abs/1702.05489} {\path{arXiv:1702.05489}},
  \href {https://doi.org/10.1007/JHEP04(2017)078}
  {\path{doi:10.1007/JHEP04(2017)078}}.

\bibitem{Elvang:2015rqa}
H.~Elvang, Y.-t. Huang, {Scattering Amplitudes in Gauge Theory and Gravity},
  Cambridge University Press, 2015.

\bibitem{Peeters:2005tb}
K.~Peeters, J.~Plefka, S.~Stern, {Higher-derivative gauge field terms in the
  M-theory action}, JHEP 08 (2005) 095.
\newblock \href {http://arxiv.org/abs/hep-th/0507178}
  {\path{arXiv:hep-th/0507178}}, \href
  {https://doi.org/10.1088/1126-6708/2005/08/095}
  {\path{doi:10.1088/1126-6708/2005/08/095}}.

\bibitem{Baron:2018lve}
W.~H. Baron, E.~Lescano, D.~Marqu\'es, {The generalized Bergshoeff-de Roo
  identification}, JHEP 11 (2018) 160.
\newblock \href {http://arxiv.org/abs/1810.01427} {\path{arXiv:1810.01427}},
  \href {https://doi.org/10.1007/JHEP11(2018)160}
  {\path{doi:10.1007/JHEP11(2018)160}}.

\bibitem{Lescano:2021guc}
E.~Lescano, C.~A. N\'u\~nez, J.~A. Rodr\'\i{}guez, {Supersymmetry, T-duality
  and heterotic \ensuremath{\alpha}'-corrections}, JHEP 07 (2021) 092.
\newblock \href {http://arxiv.org/abs/2104.09545} {\path{arXiv:2104.09545}},
  \href {https://doi.org/10.1007/JHEP07(2021)092}
  {\path{doi:10.1007/JHEP07(2021)092}}.

\bibitem{Hronek:2022dyr}
S.~Hronek, L.~Wulff, S.~Zacarias, {The \ensuremath{\alpha}'$^{2}$ correction
  from double field theory}, JHEP 11 (2022) 090.
\newblock \href {http://arxiv.org/abs/2206.10640} {\path{arXiv:2206.10640}},
  \href {https://doi.org/10.1007/JHEP11(2022)090}
  {\path{doi:10.1007/JHEP11(2022)090}}.

\bibitem{Hronek:2020xxi}
S.~Hronek, L.~Wulff, {$O(D,D)$ and the string $\alpha'$ expansion: an
  obstruction}, JHEP 04 (2021) 013.
\newblock \href {http://arxiv.org/abs/2012.13410} {\path{arXiv:2012.13410}},
  \href {https://doi.org/10.1007/JHEP04(2021)013}
  {\path{doi:10.1007/JHEP04(2021)013}}.

\bibitem{Aharony:2008ug}
O.~Aharony, O.~Bergman, D.~L. Jafferis, J.~Maldacena, {N=6 superconformal
  Chern-Simons-matter theories, M2-branes and their gravity duals}, JHEP 10
  (2008) 091.
\newblock \href {http://arxiv.org/abs/0806.1218} {\path{arXiv:0806.1218}},
  \href {https://doi.org/10.1088/1126-6708/2008/10/091}
  {\path{doi:10.1088/1126-6708/2008/10/091}}.

\bibitem{Chester:2018aca}
S.~M. Chester, S.~S. Pufu, X.~Yin, {The M-Theory S-Matrix From ABJM: Beyond 11D
  Supergravity}, JHEP 08 (2018) 115.
\newblock \href {http://arxiv.org/abs/1804.00949} {\path{arXiv:1804.00949}},
  \href {https://doi.org/10.1007/JHEP08(2018)115}
  {\path{doi:10.1007/JHEP08(2018)115}}.

\bibitem{Binder:2018yvd}
D.~J. Binder, S.~M. Chester, S.~S. Pufu, {Absence of $D^4 R^4$ in M-Theory From
  ABJM}, JHEP 04 (2020) 052.
\newblock \href {http://arxiv.org/abs/1808.10554} {\path{arXiv:1808.10554}},
  \href {https://doi.org/10.1007/JHEP04(2020)052}
  {\path{doi:10.1007/JHEP04(2020)052}}.

\bibitem{Dorigoni:2022iem}
D.~Dorigoni, M.~B. Green, C.~Wen, {The SAGEX review on scattering amplitudes
  Chapter 10: Selected topics on modular covariance of type IIB string
  amplitudes and their~~supersymmetric Yang\textendash{}Mills duals}, J. Phys.
  A 55~(44) (2022) 443011.
\newblock \href {http://arxiv.org/abs/2203.13021} {\path{arXiv:2203.13021}},
  \href {https://doi.org/10.1088/1751-8121/ac9263}
  {\path{doi:10.1088/1751-8121/ac9263}}.

\bibitem{Cremmer:1978km}
E.~Cremmer, B.~Julia, J.~Scherk, {Supergravity Theory in Eleven-Dimensions},
  Phys. Lett. B 76 (1978) 409--412.
\newblock \href {https://doi.org/10.1016/0370-2693(78)90894-8}
  {\path{doi:10.1016/0370-2693(78)90894-8}}.

\bibitem{Tsimpis:2004rs}
D.~Tsimpis, {11D supergravity at O (l**3)}, JHEP 10 (2004) 046.
\newblock \href {http://arxiv.org/abs/hep-th/0407271}
  {\path{arXiv:hep-th/0407271}}, \href
  {https://doi.org/10.1088/1126-6708/2004/10/046}
  {\path{doi:10.1088/1126-6708/2004/10/046}}.

\bibitem{Metsaev:1986yb}
R.~R. Metsaev, A.~A. Tseytlin, {Curvature Cubed Terms in String Theory
  Effective Actions}, Phys. Lett. B 185 (1987) 52--58.
\newblock \href {https://doi.org/10.1016/0370-2693(87)91527-9}
  {\path{doi:10.1016/0370-2693(87)91527-9}}.

\bibitem{deWit:1978pd}
B.~de~Wit, S.~Ferrara, {On Higher Order Invariants in Extended Supergravity},
  Phys. Lett. B 81 (1979) 317--320.
\newblock \href {https://doi.org/10.1016/0370-2693(79)90343-5}
  {\path{doi:10.1016/0370-2693(79)90343-5}}.

\bibitem{Hyakutake:2006aq}
Y.~Hyakutake, S.~Ogushi, {Higher derivative corrections to eleven dimensional
  supergravity via local supersymmetry}, JHEP 02 (2006) 068.
\newblock \href {http://arxiv.org/abs/hep-th/0601092}
  {\path{arXiv:hep-th/0601092}}, \href
  {https://doi.org/10.1088/1126-6708/2006/02/068}
  {\path{doi:10.1088/1126-6708/2006/02/068}}.

\bibitem{Hyakutake:2007vc}
Y.~Hyakutake, {Higher derivative corrections in M-theory via local
  supersymmetry}, in: {15th International Conference on Supersymmetry and the
  Unification of Fundamental Interactions (SUSY07)}, 2007, pp. 606--609.
\newblock \href {http://arxiv.org/abs/0710.2673} {\path{arXiv:0710.2673}}.

\bibitem{Hyakutake:2007sm}
Y.~Hyakutake, {Toward the determination of $R^3 F^2$ terms in M-theory}, Prog.
  Theor. Phys. 118 (2007) 109.
\newblock \href {http://arxiv.org/abs/hep-th/0703154}
  {\path{arXiv:hep-th/0703154}}, \href {https://doi.org/10.1143/PTP.118.109}
  {\path{doi:10.1143/PTP.118.109}}.

\bibitem{Peeters:2000qj}
K.~Peeters, P.~Vanhove, A.~Westerberg, {Supersymmetric higher derivative
  actions in ten-dimensions and eleven-dimensions, the associated superalgebras
  and their formulation in superspace}, Class. Quant. Grav. 18 (2001) 843--890.
\newblock \href {http://arxiv.org/abs/hep-th/0010167}
  {\path{arXiv:hep-th/0010167}}, \href
  {https://doi.org/10.1088/0264-9381/18/5/307}
  {\path{doi:10.1088/0264-9381/18/5/307}}.

\bibitem{Hyakutake:2008}
Y.~Hyakutake, {Determination of $R^3 F^2$ and $R^2 (DF)^2$ in M-theory},
  unpublished (2008).

\bibitem{Gross:1986iv}
D.~J. Gross, E.~Witten, {Superstring Modifications of Einstein's Equations},
  Nucl. Phys. B 277 (1986) 1.
\newblock \href {https://doi.org/10.1016/0550-3213(86)90429-3}
  {\path{doi:10.1016/0550-3213(86)90429-3}}.

\bibitem{Duff:1995wd}
M.~J. Duff, J.~T. Liu, R.~Minasian, {Eleven-dimensional origin of string-string
  duality: A One loop test}, Nucl. Phys. B 452 (1995) 261--282.
\newblock \href {http://arxiv.org/abs/hep-th/9506126}
  {\path{arXiv:hep-th/9506126}}, \href
  {https://doi.org/10.1016/0550-3213(95)00368-3}
  {\path{doi:10.1016/0550-3213(95)00368-3}}.

\bibitem{deAlwis:1997gq}
S.~P. de~Alwis, {Coupling of branes and normalization of effective actions in
  string / M theory}, Phys. Rev. D 56 (1997) 7963--7977.
\newblock \href {http://arxiv.org/abs/hep-th/9705139}
  {\path{arXiv:hep-th/9705139}}, \href
  {https://doi.org/10.1103/PhysRevD.56.7963}
  {\path{doi:10.1103/PhysRevD.56.7963}}.

\bibitem{Freed:1998tg}
D.~Freed, J.~A. Harvey, R.~Minasian, G.~W. Moore, {Gravitational anomaly
  cancellation for M theory five-branes}, Adv. Theor. Math. Phys. 2 (1998)
  601--618.
\newblock \href {http://arxiv.org/abs/hep-th/9803205}
  {\path{arXiv:hep-th/9803205}}, \href
  {https://doi.org/10.4310/ATMP.1998.v2.n3.a8}
  {\path{doi:10.4310/ATMP.1998.v2.n3.a8}}.

\bibitem{Henningson:1998gx}
M.~Henningson, K.~Skenderis, {The Holographic Weyl anomaly}, JHEP 07 (1998)
  023.
\newblock \href {http://arxiv.org/abs/hep-th/9806087}
  {\path{arXiv:hep-th/9806087}}, \href
  {https://doi.org/10.1088/1126-6708/1998/07/023}
  {\path{doi:10.1088/1126-6708/1998/07/023}}.

\bibitem{Tseytlin:2000sf}
A.~A. Tseytlin, {$R^4$ terms in 11 dimensions and conformal anomaly of (2,0)
  theory}, Nucl. Phys. B 584 (2000) 233--250.
\newblock \href {http://arxiv.org/abs/hep-th/0005072}
  {\path{arXiv:hep-th/0005072}}, \href
  {https://doi.org/10.1016/S0550-3213(00)00380-1}
  {\path{doi:10.1016/S0550-3213(00)00380-1}}.

\bibitem{Green:1999by}
M.~B. Green, M.~Gutperle, H.~H. Kwon, {Light cone quantum mechanics of the
  eleven-dimensional superparticle}, JHEP 08 (1999) 012.
\newblock \href {http://arxiv.org/abs/hep-th/9907155}
  {\path{arXiv:hep-th/9907155}}, \href
  {https://doi.org/10.1088/1126-6708/1999/08/012}
  {\path{doi:10.1088/1126-6708/1999/08/012}}.

\bibitem{Green:2006gt}
M.~B. Green, J.~G. Russo, P.~Vanhove, {Non-renormalisation conditions in type
  II string theory and maximal supergravity}, JHEP 02 (2007) 099.
\newblock \href {http://arxiv.org/abs/hep-th/0610299}
  {\path{arXiv:hep-th/0610299}}, \href
  {https://doi.org/10.1088/1126-6708/2007/02/099}
  {\path{doi:10.1088/1126-6708/2007/02/099}}.

\bibitem{Green:2008bf}
M.~B. Green, J.~G. Russo, P.~Vanhove, {Modular properties of two-loop maximal
  supergravity and connections with string theory}, JHEP 07 (2008) 126.
\newblock \href {http://arxiv.org/abs/0807.0389} {\path{arXiv:0807.0389}},
  \href {https://doi.org/10.1088/1126-6708/2008/07/126}
  {\path{doi:10.1088/1126-6708/2008/07/126}}.

\bibitem{Bern:1998ug}
Z.~Bern, L.~J. Dixon, D.~C. Dunbar, M.~Perelstein, J.~S. Rozowsky, {On the
  relationship between Yang-Mills theory and gravity and its implication for
  ultraviolet divergences}, Nucl. Phys. B 530 (1998) 401--456.
\newblock \href {http://arxiv.org/abs/hep-th/9802162}
  {\path{arXiv:hep-th/9802162}}, \href
  {https://doi.org/10.1016/S0550-3213(98)00420-9}
  {\path{doi:10.1016/S0550-3213(98)00420-9}}.

\bibitem{Green:2005ba}
M.~B. Green, P.~Vanhove, {Duality and higher derivative terms in M theory},
  JHEP 01 (2006) 093.
\newblock \href {http://arxiv.org/abs/hep-th/0510027}
  {\path{arXiv:hep-th/0510027}}, \href
  {https://doi.org/10.1088/1126-6708/2006/01/093}
  {\path{doi:10.1088/1126-6708/2006/01/093}}.

\bibitem{Cremmer:1980ru}
E.~Cremmer, S.~Ferrara, {Formulation of Eleven-Dimensional Supergravity in
  Superspace}, Phys. Lett. B 91 (1980) 61--66.
\newblock \href {https://doi.org/10.1016/0370-2693(80)90662-0}
  {\path{doi:10.1016/0370-2693(80)90662-0}}.

\bibitem{Brink:1980az}
L.~Brink, P.~S. Howe, {Eleven-Dimensional Supergravity on the Mass-Shell in
  Superspace}, Phys. Lett. B 91 (1980) 384--386.
\newblock \href {https://doi.org/10.1016/0370-2693(80)91002-3}
  {\path{doi:10.1016/0370-2693(80)91002-3}}.

\bibitem{Howe:1997he}
P.~S. Howe, {Weyl superspace}, Phys. Lett. B 415 (1997) 149--155.
\newblock \href {http://arxiv.org/abs/hep-th/9707184}
  {\path{arXiv:hep-th/9707184}}, \href
  {https://doi.org/10.1016/S0370-2693(97)01261-6}
  {\path{doi:10.1016/S0370-2693(97)01261-6}}.

\bibitem{Howe:2003cy}
P.~S. Howe, D.~Tsimpis, {On higher order corrections in M theory}, JHEP 09
  (2003) 038.
\newblock \href {http://arxiv.org/abs/hep-th/0305129}
  {\path{arXiv:hep-th/0305129}}, \href
  {https://doi.org/10.1088/1126-6708/2003/09/038}
  {\path{doi:10.1088/1126-6708/2003/09/038}}.

\bibitem{Nilsson:1998ca}
B.~E.~W. Nilsson, {A Superspace approach to branes and supergravity}, in: {31st
  International Ahrenshoop Symposium on the Theory of Elementary Particles},
  1998, pp. 52--57.
\newblock \href {http://arxiv.org/abs/hep-th/0007017}
  {\path{arXiv:hep-th/0007017}}.

\bibitem{Cederwall:2000ye}
M.~Cederwall, U.~Gran, M.~Nielsen, B.~E.~W. Nilsson, {Manifestly supersymmetric
  M theory}, JHEP 10 (2000) 041.
\newblock \href {http://arxiv.org/abs/hep-th/0007035}
  {\path{arXiv:hep-th/0007035}}, \href
  {https://doi.org/10.1088/1126-6708/2000/10/041}
  {\path{doi:10.1088/1126-6708/2000/10/041}}.

\bibitem{Cederwall:2001bt}
M.~Cederwall, B.~E.~W. Nilsson, D.~Tsimpis, {The Structure of maximally
  supersymmetric Yang-Mills theory: Constraining higher order corrections},
  JHEP 06 (2001) 034.
\newblock \href {http://arxiv.org/abs/hep-th/0102009}
  {\path{arXiv:hep-th/0102009}}, \href
  {https://doi.org/10.1088/1126-6708/2001/06/034}
  {\path{doi:10.1088/1126-6708/2001/06/034}}.

\bibitem{Cederwall:2001dx}
M.~Cederwall, B.~E.~W. Nilsson, D.~Tsimpis, {Spinorial cohomology and maximally
  supersymmetric theories}, JHEP 02 (2002) 009.
\newblock \href {http://arxiv.org/abs/hep-th/0110069}
  {\path{arXiv:hep-th/0110069}}, \href
  {https://doi.org/10.1088/1126-6708/2002/02/009}
  {\path{doi:10.1088/1126-6708/2002/02/009}}.

\bibitem{Berkovits:2000fe}
N.~Berkovits, {Super Poincare covariant quantization of the superstring}, JHEP
  04 (2000) 018.
\newblock \href {http://arxiv.org/abs/hep-th/0001035}
  {\path{arXiv:hep-th/0001035}}, \href
  {https://doi.org/10.1088/1126-6708/2000/04/018}
  {\path{doi:10.1088/1126-6708/2000/04/018}}.

\bibitem{Berkovits:2002zk}
N.~Berkovits, {ICTP lectures on covariant quantization of the superstring},
  ICTP Lect. Notes Ser. 13 (2003) 57--107.
\newblock \href {http://arxiv.org/abs/hep-th/0209059}
  {\path{arXiv:hep-th/0209059}}.

\bibitem{Liu:2013dna}
J.~T. Liu, R.~Minasian, {Higher-derivative couplings in string theory:
  dualities and the B-field}, Nucl. Phys. B 874 (2013) 413--470.
\newblock \href {http://arxiv.org/abs/1304.3137} {\path{arXiv:1304.3137}},
  \href {https://doi.org/10.1016/j.nuclphysb.2013.06.002}
  {\path{doi:10.1016/j.nuclphysb.2013.06.002}}.

\bibitem{Frolov:2001jh}
S.~Frolov, A.~A. Tseytlin, {R**4 corrections to conifolds and G(2) holonomy
  spaces}, Nucl. Phys. B 632 (2002) 69--100.
\newblock \href {http://arxiv.org/abs/hep-th/0111128}
  {\path{arXiv:hep-th/0111128}}, \href
  {https://doi.org/10.1016/S0550-3213(02)00241-9}
  {\path{doi:10.1016/S0550-3213(02)00241-9}}.

\bibitem{Russo:1997mk}
J.~G. Russo, A.~A. Tseytlin, {One loop four graviton amplitude in
  eleven-dimensional supergravity}, Nucl. Phys. B 508 (1997) 245--259.
\newblock \href {http://arxiv.org/abs/hep-th/9707134}
  {\path{arXiv:hep-th/9707134}}, \href
  {https://doi.org/10.1016/S0550-3213(97)00631-7}
  {\path{doi:10.1016/S0550-3213(97)00631-7}}.

\bibitem{Deser:1998jz}
S.~Deser, D.~Seminara, {Counterterms / M theory corrections to D = 11
  supergravity}, Phys. Rev. Lett. 82 (1999) 2435--2438.
\newblock \href {http://arxiv.org/abs/hep-th/9812136}
  {\path{arXiv:hep-th/9812136}}, \href
  {https://doi.org/10.1103/PhysRevLett.82.2435}
  {\path{doi:10.1103/PhysRevLett.82.2435}}.

\bibitem{Deser:2000xz}
S.~Deser, D.~Seminara, {Tree amplitudes and two loop counterterms in D = 11
  supergravity}, Phys. Rev. D 62 (2000) 084010.
\newblock \href {http://arxiv.org/abs/hep-th/0002241}
  {\path{arXiv:hep-th/0002241}}, \href
  {https://doi.org/10.1103/PhysRevD.62.084010}
  {\path{doi:10.1103/PhysRevD.62.084010}}.

\bibitem{Deser:2005kb}
S.~Deser, D.~Seminara, {Graviton-form invariants in D=11 supergravity}, Phys.
  Rev. D 72 (2005) 027701.
\newblock \href {http://arxiv.org/abs/hep-th/0506073}
  {\path{arXiv:hep-th/0506073}}, \href
  {https://doi.org/10.1103/PhysRevD.72.027701}
  {\path{doi:10.1103/PhysRevD.72.027701}}.

\bibitem{Green:1997as}
M.~B. Green, M.~Gutperle, P.~Vanhove, {One loop in eleven-dimensions}, Phys.
  Lett. B 409 (1997) 177--184.
\newblock \href {http://arxiv.org/abs/hep-th/9706175}
  {\path{arXiv:hep-th/9706175}}, \href
  {https://doi.org/10.1016/S0370-2693(97)00931-3}
  {\path{doi:10.1016/S0370-2693(97)00931-3}}.

\bibitem{Green:1997me}
M.~B. Green, M.~Gutperle, H.-h. Kwon, {Sixteen fermion and related terms in M
  theory on $T^2$}, Phys. Lett. B 421 (1998) 149--161.
\newblock \href {http://arxiv.org/abs/hep-th/9710151}
  {\path{arXiv:hep-th/9710151}}, \href
  {https://doi.org/10.1016/S0370-2693(97)01551-7}
  {\path{doi:10.1016/S0370-2693(97)01551-7}}.

\bibitem{Green:1998by}
M.~B. Green, S.~Sethi, {Supersymmetry constraints on type IIB supergravity},
  Phys. Rev. D 59 (1999) 046006.
\newblock \href {http://arxiv.org/abs/hep-th/9808061}
  {\path{arXiv:hep-th/9808061}}, \href
  {https://doi.org/10.1103/PhysRevD.59.046006}
  {\path{doi:10.1103/PhysRevD.59.046006}}.

\bibitem{Liu:2019ses}
J.~T. Liu, R.~Minasian, {Higher-derivative couplings in string theory:
  five-point contact terms}, Nucl. Phys. B 967 (2021) 115386.
\newblock \href {http://arxiv.org/abs/1912.10974} {\path{arXiv:1912.10974}},
  \href {https://doi.org/10.1016/j.nuclphysb.2021.115386}
  {\path{doi:10.1016/j.nuclphysb.2021.115386}}.

\bibitem{Stieberger:2009rr}
S.~Stieberger, {Constraints on Tree-Level Higher Order Gravitational Couplings
  in Superstring Theory}, Phys. Rev. Lett. 106 (2011) 111601.
\newblock \href {http://arxiv.org/abs/0910.0180} {\path{arXiv:0910.0180}},
  \href {https://doi.org/10.1103/PhysRevLett.106.111601}
  {\path{doi:10.1103/PhysRevLett.106.111601}}.

\bibitem{Policastro:2006vt}
G.~Policastro, D.~Tsimpis, {$R^4$, purified}, Class. Quant. Grav. 23 (2006)
  4753--4780.
\newblock \href {http://arxiv.org/abs/hep-th/0603165}
  {\path{arXiv:hep-th/0603165}}, \href
  {https://doi.org/10.1088/0264-9381/23/14/012}
  {\path{doi:10.1088/0264-9381/23/14/012}}.

\bibitem{Policastro:2008hg}
G.~Policastro, D.~Tsimpis, {A Note on the quartic effective action of type IIB
  superstring}, Class. Quant. Grav. 26 (2009) 125001.
\newblock \href {http://arxiv.org/abs/0812.3138} {\path{arXiv:0812.3138}},
  \href {https://doi.org/10.1088/0264-9381/26/12/125001}
  {\path{doi:10.1088/0264-9381/26/12/125001}}.

\bibitem{Polchinski:1998rr}
J.~Polchinski, {String theory. Vol. 2: Superstring theory and beyond},
  Cambridge Monographs on Mathematical Physics, Cambridge University Press,
  2007.
\newblock \href {https://doi.org/10.1017/CBO9780511618123}
  {\path{doi:10.1017/CBO9780511618123}}.

\bibitem{Green:1981ya}
M.~B. Green, J.~H. Schwarz, {Supersymmetrical Dual String Theory. 3. Loops and
  Renormalization}, Nucl. Phys. B 198 (1982) 441--460.
\newblock \href {https://doi.org/10.1016/0550-3213(82)90334-0}
  {\path{doi:10.1016/0550-3213(82)90334-0}}.

\bibitem{Green:1997tv}
M.~B. Green, M.~Gutperle, {Effects of D instantons}, Nucl. Phys. B 498 (1997)
  195--227.
\newblock \href {http://arxiv.org/abs/hep-th/9701093}
  {\path{arXiv:hep-th/9701093}}, \href
  {https://doi.org/10.1016/S0550-3213(97)00269-1}
  {\path{doi:10.1016/S0550-3213(97)00269-1}}.

\bibitem{Fleig:2015vky}
P.~Fleig, H.~P.~A. Gustafsson, A.~Kleinschmidt, D.~Persson, {Eisenstein series
  and automorphic representations}, Cambridge University Press, 2018.
\newblock \href {http://arxiv.org/abs/1511.04265} {\path{arXiv:1511.04265}}.

\bibitem{Sinha:2002zr}
A.~Sinha, {The G(hat)**4 lambda**16 term in IIB supergravity}, JHEP 08 (2002)
  017.
\newblock \href {http://arxiv.org/abs/hep-th/0207070}
  {\path{arXiv:hep-th/0207070}}, \href
  {https://doi.org/10.1088/1126-6708/2002/08/017}
  {\path{doi:10.1088/1126-6708/2002/08/017}}.

\bibitem{Green:1999pu}
M.~B. Green, H.-h. Kwon, P.~Vanhove, {Two loops in eleven-dimensions}, Phys.
  Rev. D 61 (2000) 104010.
\newblock \href {http://arxiv.org/abs/hep-th/9910055}
  {\path{arXiv:hep-th/9910055}}, \href
  {https://doi.org/10.1103/PhysRevD.61.104010}
  {\path{doi:10.1103/PhysRevD.61.104010}}.

\bibitem{Gaberdiel:1998ui}
M.~R. Gaberdiel, M.~B. Green, {An SL(2, Z) anomaly in IIB supergravity and its
  F theory interpretation}, JHEP 11 (1998) 026.
\newblock \href {http://arxiv.org/abs/hep-th/9810153}
  {\path{arXiv:hep-th/9810153}}, \href
  {https://doi.org/10.1088/1126-6708/1998/11/026}
  {\path{doi:10.1088/1126-6708/1998/11/026}}.

\bibitem{Gross:1986mw}
D.~J. Gross, J.~H. Sloan, {The Quartic Effective Action for the Heterotic
  String}, Nucl. Phys. B 291 (1987) 41--89.
\newblock \href {https://doi.org/10.1016/0550-3213(87)90465-2}
  {\path{doi:10.1016/0550-3213(87)90465-2}}.

\bibitem{Garousi:2020gio}
M.~R. Garousi, {Effective action of type II superstring theories at order
  $\alpha'^{3}$: NS-NS couplings}, JHEP 02 (2021) 157.
\newblock \href {http://arxiv.org/abs/2011.02753} {\path{arXiv:2011.02753}},
  \href {https://doi.org/10.1007/JHEP02(2021)157}
  {\path{doi:10.1007/JHEP02(2021)157}}.

\bibitem{Garousi:2020lof}
M.~R. Garousi, {On NS-NS couplings at order $\alpha'^3$}, Nucl. Phys. B 971
  (2021) 115510.
\newblock \href {http://arxiv.org/abs/2012.15091} {\path{arXiv:2012.15091}},
  \href {https://doi.org/10.1016/j.nuclphysb.2021.115510}
  {\path{doi:10.1016/j.nuclphysb.2021.115510}}.

\bibitem{Moura:2007ks}
F.~Moura, {Type II and heterotic one loop string effective actions in four
  dimensions}, JHEP 06 (2007) 052.
\newblock \href {http://arxiv.org/abs/hep-th/0703026}
  {\path{arXiv:hep-th/0703026}}, \href
  {https://doi.org/10.1088/1126-6708/2007/06/052}
  {\path{doi:10.1088/1126-6708/2007/06/052}}.

\bibitem{Peeters:2003pv}
K.~Peeters, A.~Westerberg, {The Ramond-Ramond sector of string theory beyond
  leading order}, Class. Quant. Grav. 21 (2004) 1643--1666.
\newblock \href {http://arxiv.org/abs/hep-th/0307298}
  {\path{arXiv:hep-th/0307298}}, \href
  {https://doi.org/10.1088/0264-9381/21/6/022}
  {\path{doi:10.1088/0264-9381/21/6/022}}.

\bibitem{Liu:2022bfg}
J.~T. Liu, R.~Minasian, R.~Savelli, A.~Schachner, {Type IIB at eight
  derivatives: insights from Superstrings, Superfields and Superparticles},
  JHEP 08 (2022) 267.
\newblock \href {http://arxiv.org/abs/2205.11530} {\path{arXiv:2205.11530}},
  \href {https://doi.org/10.1007/JHEP08(2022)267}
  {\path{doi:10.1007/JHEP08(2022)267}}.

\bibitem{Richards:2008jg}
D.~M. Richards, {The One-Loop Five-Graviton Amplitude and the Effective
  Action}, JHEP 10 (2008) 042.
\newblock \href {http://arxiv.org/abs/0807.2421} {\path{arXiv:0807.2421}},
  \href {https://doi.org/10.1088/1126-6708/2008/10/042}
  {\path{doi:10.1088/1126-6708/2008/10/042}}.

\bibitem{Richards:2008sa}
D.~M. Richards, {The One-Loop $H^2R^3$ and $H^2$($\Delta H$)$^2R$ Terms in the
  Effective Action}, JHEP 10 (2008) 043.
\newblock \href {http://arxiv.org/abs/0807.3453} {\path{arXiv:0807.3453}},
  \href {https://doi.org/10.1088/1126-6708/2008/10/043}
  {\path{doi:10.1088/1126-6708/2008/10/043}}.

\bibitem{Schwarz:1982jn}
J.~H. Schwarz, {Superstring Theory}, Phys. Rept. 89 (1982) 223--322.
\newblock \href {https://doi.org/10.1016/0370-1573(82)90087-4}
  {\path{doi:10.1016/0370-1573(82)90087-4}}.

\bibitem{Grisaru:1986px}
M.~T. Grisaru, A.~E.~M. van~de Ven, D.~Zanon, {Four Loop beta Function for the
  N=1 and N=2 Supersymmetric Nonlinear Sigma Model in Two-Dimensions}, Phys.
  Lett. B 173 (1986) 423--428.
\newblock \href {https://doi.org/10.1016/0370-2693(86)90408-9}
  {\path{doi:10.1016/0370-2693(86)90408-9}}.

\bibitem{Grisaru:1986dk}
M.~T. Grisaru, A.~E.~M. van~de Ven, D.~Zanon, {Two-Dimensional Supersymmetric
  Sigma Models on Ricci Flat Kahler Manifolds Are Not Finite}, Nucl. Phys. B
  277 (1986) 388--408.
\newblock \href {https://doi.org/10.1016/0550-3213(86)90448-7}
  {\path{doi:10.1016/0550-3213(86)90448-7}}.

\bibitem{Grisaru:1986kw}
M.~T. Grisaru, A.~E.~M. van~de Ven, D.~Zanon, {Four Loop Divergences for the
  N=1 Supersymmetric Nonlinear Sigma Model in Two-Dimensions}, Nucl. Phys. B
  277 (1986) 409--428.
\newblock \href {https://doi.org/10.1016/0550-3213(86)90449-9}
  {\path{doi:10.1016/0550-3213(86)90449-9}}.

\bibitem{Grisaru:1986vi}
M.~T. Grisaru, D.~Zanon, {$\sigma$ Model Superstring Corrections to the
  Einstein-hilbert Action}, Phys. Lett. B 177 (1986) 347--351.
\newblock \href {https://doi.org/10.1016/0370-2693(86)90765-3}
  {\path{doi:10.1016/0370-2693(86)90765-3}}.

\bibitem{Freeman:1986br}
M.~D. Freeman, C.~N. Pope, {Beta Functions and Superstring Compactifications},
  Phys. Lett. B 174 (1986) 48--50.
\newblock \href {https://doi.org/10.1016/0370-2693(86)91127-5}
  {\path{doi:10.1016/0370-2693(86)91127-5}}.

\bibitem{Freeman:1986zh}
M.~D. Freeman, C.~N. Pope, M.~F. Sohnius, K.~S. Stelle, {Higher Order $\sigma$
  Model Counterterms and the Effective Action for Superstrings}, Phys. Lett. B
  178 (1986) 199--204.
\newblock \href {https://doi.org/10.1016/0370-2693(86)91495-4}
  {\path{doi:10.1016/0370-2693(86)91495-4}}.

\bibitem{Sakai:1986bi}
N.~Sakai, Y.~Tanii, {One Loop Amplitudes and Effective Action in Superstring
  Theories}, Nucl. Phys. B 287 (1987) 457.
\newblock \href {https://doi.org/10.1016/0550-3213(87)90114-3}
  {\path{doi:10.1016/0550-3213(87)90114-3}}.

\bibitem{Hull:1994ys}
C.~M. Hull, P.~K. Townsend, {Unity of superstring dualities}, Nucl. Phys. B 438
  (1995) 109--137.
\newblock \href {http://arxiv.org/abs/hep-th/9410167}
  {\path{arXiv:hep-th/9410167}}, \href
  {https://doi.org/10.1016/0550-3213(94)00559-W}
  {\path{doi:10.1016/0550-3213(94)00559-W}}.

\bibitem{Witten:1995ex}
E.~Witten, {String theory dynamics in various dimensions}, Nucl. Phys. B 443
  (1995) 85--126.
\newblock \href {http://arxiv.org/abs/hep-th/9503124}
  {\path{arXiv:hep-th/9503124}}, \href
  {https://doi.org/10.1016/0550-3213(95)00158-O}
  {\path{doi:10.1016/0550-3213(95)00158-O}}.

\bibitem{Sen:1995cj}
A.~Sen, {String string duality conjecture in six-dimensions and charged
  solitonic strings}, Nucl. Phys. B 450 (1995) 103--114.
\newblock \href {http://arxiv.org/abs/hep-th/9504027}
  {\path{arXiv:hep-th/9504027}}, \href
  {https://doi.org/10.1016/0550-3213(95)00320-R}
  {\path{doi:10.1016/0550-3213(95)00320-R}}.

\bibitem{Harvey:1995rn}
J.~A. Harvey, A.~Strominger, {The heterotic string is a soliton}, Nucl. Phys. B
  449 (1995) 535--552, [Erratum: Nucl.Phys.B 458, 456--473 (1996)].
\newblock \href {http://arxiv.org/abs/hep-th/9504047}
  {\path{arXiv:hep-th/9504047}}, \href
  {https://doi.org/10.1016/0550-3213(95)00310-O}
  {\path{doi:10.1016/0550-3213(95)00310-O}}.

\bibitem{Vafa:1995fj}
C.~Vafa, E.~Witten, {A One loop test of string duality}, Nucl. Phys. B 447
  (1995) 261--270.
\newblock \href {http://arxiv.org/abs/hep-th/9505053}
  {\path{arXiv:hep-th/9505053}}, \href
  {https://doi.org/10.1016/0550-3213(95)00280-6}
  {\path{doi:10.1016/0550-3213(95)00280-6}}.

\bibitem{BD1989}
E.~Bergshoeff, M.~de~Roo, {Supersymmetric Chern-simons Terms in
  Ten-dimensions}, Phys. Lett. B 218 (1989) 210--215.
\newblock \href {https://doi.org/10.1016/0370-2693(89)91420-2}
  {\path{doi:10.1016/0370-2693(89)91420-2}}.

\bibitem{Bergshoeff:1989de}
E.~A. Bergshoeff, M.~de~Roo, {The Quartic Effective Action of the Heterotic
  String and Supersymmetry}, Nucl. Phys. B 328 (1989) 439--468.
\newblock \href {https://doi.org/10.1016/0550-3213(89)90336-2}
  {\path{doi:10.1016/0550-3213(89)90336-2}}.

\bibitem{Suelmann:1994vk}
H.~Suelmann, {Effective actions for heterotic string theory}, Int. J. Mod.
  Phys. D 3 (1994) 285--288.
\newblock \href {https://doi.org/10.1142/S0218271894000472}
  {\path{doi:10.1142/S0218271894000472}}.

\bibitem{deRoo:1992zp}
M.~de~Roo, H.~Suelmann, A.~Wiedemann, {The Supersymmetric effective action of
  the heterotic string in ten-dimensions}, Nucl. Phys. B 405 (1993) 326--366.
\newblock \href {http://arxiv.org/abs/hep-th/9210099}
  {\path{arXiv:hep-th/9210099}}, \href
  {https://doi.org/10.1016/0550-3213(93)90550-9}
  {\path{doi:10.1016/0550-3213(93)90550-9}}.

\bibitem{Suelmann:1994qk}
J.~H. Suelmann, {Supersymmetry and string effective actions}, Phd thesis (11
  1994).

\bibitem{Tseytlin:1995fy}
A.~A. Tseytlin, {On SO(32) heterotic type I superstring duality in
  ten-dimensions}, Phys. Lett. B 367 (1996) 84--90.
\newblock \href {http://arxiv.org/abs/hep-th/9510173}
  {\path{arXiv:hep-th/9510173}}, \href
  {https://doi.org/10.1016/0370-2693(95)01452-7}
  {\path{doi:10.1016/0370-2693(95)01452-7}}.

\bibitem{Tseytlin:1995bi}
A.~A. Tseytlin, {Heterotic type I superstring duality and low-energy effective
  actions}, Nucl. Phys. B 467 (1996) 383--398.
\newblock \href {http://arxiv.org/abs/hep-th/9512081}
  {\path{arXiv:hep-th/9512081}}, \href
  {https://doi.org/10.1016/0550-3213(96)00080-6}
  {\path{doi:10.1016/0550-3213(96)00080-6}}.

\bibitem{Bachas:1996bp}
C.~Bachas, E.~Kiritsis, {F(4) terms in N=4 string vacua}, Nucl. Phys. B Proc.
  Suppl. 55 (1997) 194--199.
\newblock \href {http://arxiv.org/abs/hep-th/9611205}
  {\path{arXiv:hep-th/9611205}}, \href
  {https://doi.org/10.1016/S0920-5632(97)00079-0}
  {\path{doi:10.1016/S0920-5632(97)00079-0}}.

\bibitem{Bergshoeff:1988nn}
E.~Bergshoeff, M.~de~Roo, {Supersymmetric Chern-simons Terms in
  Ten-dimensions}, Phys. Lett. B 218 (1989) 210--215.
\newblock \href {https://doi.org/10.1016/0370-2693(89)91420-2}
  {\path{doi:10.1016/0370-2693(89)91420-2}}.

\bibitem{Bergshoeff:1987rb}
E.~Bergshoeff, M.~Rakowski, {An Off-shell Superspace R(2) Action in
  Six-dimensions}, Phys. Lett. B 191 (1987) 399--403.
\newblock \href {https://doi.org/10.1016/0370-2693(87)90629-0}
  {\path{doi:10.1016/0370-2693(87)90629-0}}.

\bibitem{Bento:1986hx}
M.~C. Bento, N.~E. Mavromatos, {Ambiguities in the Low-energy Effective Actions
  of String Theories With the Inclusion of Antisymmetric Tensor and Dilaton
  Fields}, Phys. Lett. B 190 (1987) 105--109.
\newblock \href {https://doi.org/10.1016/0370-2693(87)90847-1}
  {\path{doi:10.1016/0370-2693(87)90847-1}}.

\bibitem{Nunez:1987ig}
C.~A. Nunez, {On the Equivalence Between Beta Functions and Massless Bosonic
  String Fields Equations of Motion}, Phys. Lett. B 193 (1987) 195.
\newblock \href {https://doi.org/10.1016/0370-2693(87)91221-4}
  {\path{doi:10.1016/0370-2693(87)91221-4}}.

\bibitem{Metsaev:1987zx}
R.~R. Metsaev, A.~A. Tseytlin, {Order alpha-prime (Two Loop) Equivalence of the
  String Equations of Motion and the Sigma Model Weyl Invariance Conditions:
  Dependence on the Dilaton and the Antisymmetric Tensor}, Nucl. Phys. B 293
  (1987) 385--419.
\newblock \href {https://doi.org/10.1016/0550-3213(87)90077-0}
  {\path{doi:10.1016/0550-3213(87)90077-0}}.

\bibitem{Ferrara:1996wv}
S.~Ferrara, R.~Minasian, A.~Sagnotti, {Low-energy analysis of M and F theories
  on Calabi-Yau threefolds}, Nucl. Phys. B 474 (1996) 323--342.
\newblock \href {http://arxiv.org/abs/hep-th/9604097}
  {\path{arXiv:hep-th/9604097}}, \href
  {https://doi.org/10.1016/0550-3213(96)00268-4}
  {\path{doi:10.1016/0550-3213(96)00268-4}}.

\bibitem{Riccioni:1998th}
F.~Riccioni, A.~Sagnotti, {Consistent and covariant anomalies in
  six-dimensional supergravity}, Phys. Lett. B 436 (1998) 298--305.
\newblock \href {http://arxiv.org/abs/hep-th/9806129}
  {\path{arXiv:hep-th/9806129}}, \href
  {https://doi.org/10.1016/S0370-2693(98)00846-6}
  {\path{doi:10.1016/S0370-2693(98)00846-6}}.

\bibitem{Wulff:2021fhr}
L.~Wulff, {Completing R$^{4}$ using O(d, d)}, JHEP 08 (2022) 187.
\newblock \href {http://arxiv.org/abs/2111.00018} {\path{arXiv:2111.00018}},
  \href {https://doi.org/10.1007/JHEP08(2022)187}
  {\path{doi:10.1007/JHEP08(2022)187}}.

\bibitem{Cai:1986sa}
Y.~Cai, C.~A. Nunez, {Heterotic String Covariant Amplitudes and Low-energy
  Effective Action}, Nucl. Phys. B 287 (1987) 279.
\newblock \href {https://doi.org/10.1016/0550-3213(87)90106-4}
  {\path{doi:10.1016/0550-3213(87)90106-4}}.

\bibitem{Kikuchi:1986cz}
Y.~Kikuchi, C.~Marzban, {Low-energy Effective Lagrangian of Heterotic String
  Theory}, Phys. Rev. D 35 (1987) 1400.
\newblock \href {https://doi.org/10.1103/PhysRevD.35.1400}
  {\path{doi:10.1103/PhysRevD.35.1400}}.

\bibitem{Kikuchi:1986rk}
Y.~Kikuchi, C.~Marzban, Y.~J. Ng, {Heterotic String Modifications of Einstein's
  and {Yang-Mills}' Actions}, Phys. Lett. B 176 (1986) 57--60.
\newblock \href {https://doi.org/10.1016/0370-2693(86)90924-X}
  {\path{doi:10.1016/0370-2693(86)90924-X}}.

\bibitem{Ellwanger:1988cc}
U.~Ellwanger, J.~Fuchs, M.~G. Schmidt, {The Heterotic $\sigma$ Model With
  Background Gauge Fields}, Nucl. Phys. B 314 (1989) 175.
\newblock \href {https://doi.org/10.1016/0550-3213(89)90117-X}
  {\path{doi:10.1016/0550-3213(89)90117-X}}.

\bibitem{Ellis:1987dc}
J.~R. Ellis, P.~Jetzer, L.~Mizrachi, {One Loop String Corrections to the
  Effective Field Theory}, Nucl. Phys. B 303 (1988) 1--35.
\newblock \href {https://doi.org/10.1016/0550-3213(88)90214-3}
  {\path{doi:10.1016/0550-3213(88)90214-3}}.

\bibitem{Abe:1988cq}
M.~Abe, H.~Kubota, N.~Sakai, {Loop Corrections to the $E(8)$ X $E(8)$ Heterotic
  String Effective Lagrangian}, Nucl. Phys. B 306 (1988) 405--424.
\newblock \href {https://doi.org/10.1016/0550-3213(88)90699-2}
  {\path{doi:10.1016/0550-3213(88)90699-2}}.

\bibitem{Lerche:1988zy}
W.~Lerche, {Elliptic Index and Superstring Effective Actions}, Nucl. Phys. B
  308 (1988) 102--126.
\newblock \href {https://doi.org/10.1016/0550-3213(88)90044-2}
  {\path{doi:10.1016/0550-3213(88)90044-2}}.

\bibitem{deRoo:1992sm}
M.~de~Roo, H.~Suelmann, A.~Wiedemann, {Supersymmetric R**4 actions in
  ten-dimensions}, Phys. Lett. B 280 (1992) 39--46.
\newblock \href {https://doi.org/10.1016/0370-2693(92)90769-Z}
  {\path{doi:10.1016/0370-2693(92)90769-Z}}.

\bibitem{Becker:2006dvp}
K.~Becker, M.~Becker, J.~H. Schwarz, {String theory and M-theory: A modern
  introduction}, Cambridge University Press, 2006.
\newblock \href {https://doi.org/10.1017/CBO9780511816086}
  {\path{doi:10.1017/CBO9780511816086}}.

\bibitem{Candelas:1985en}
P.~Candelas, G.~T. Horowitz, A.~Strominger, E.~Witten, {Vacuum configurations
  for superstrings}, Nucl. Phys. B 258 (1985) 46--74.
\newblock \href {https://doi.org/10.1016/0550-3213(85)90602-9}
  {\path{doi:10.1016/0550-3213(85)90602-9}}.

\bibitem{Candelas:1986tz}
P.~Candelas, M.~D. Freeman, C.~N. Pope, M.~F. Sohnius, K.~S. Stelle, {Higher
  Order Corrections to Supersymmetry and Compactifications of the Heterotic
  String}, Phys. Lett. B 177 (1986) 341--346.
\newblock \href {https://doi.org/10.1016/0370-2693(86)90764-1}
  {\path{doi:10.1016/0370-2693(86)90764-1}}.

\bibitem{BdR90}
E.~A. Bergshoeff, M.~de~Roo, {Duality transformations of string effective
  actions}, Phys. Lett. B 249 (1990) 27--34.
\newblock \href {https://doi.org/10.1016/0370-2693(90)90522-8}
  {\path{doi:10.1016/0370-2693(90)90522-8}}.

\bibitem{Chang:2022urm}
H.-Y. Chang, E.~Sezgin, Y.~Tanii, {Dualization of higher derivative heterotic
  supergravities in 6D and 10D}, JHEP 10 (2022) 062.
\newblock \href {http://arxiv.org/abs/2209.03981} {\path{arXiv:2209.03981}},
  \href {https://doi.org/10.1007/JHEP10(2022)062}
  {\path{doi:10.1007/JHEP10(2022)062}}.

\bibitem{Saulina:1996vn}
N.~A. Saulina, M.~V. Terentev, K.~N. Zyablyuk, {Five-brane Lagrangian with loop
  corrections in field theory limit}, Int. J. Mod. Phys. A 12 (1997)
  4559--4580.
\newblock \href {http://arxiv.org/abs/hep-th/9607079}
  {\path{arXiv:hep-th/9607079}}, \href
  {https://doi.org/10.1142/S0217751X97002462}
  {\path{doi:10.1142/S0217751X97002462}}.

\bibitem{Duff:1990wv}
M.~J. Duff, J.~X. Lu, {Elementary five-brane solutions of D = 10 supergravity},
  Nucl. Phys. B 354 (1991) 141--153.
\newblock \href {https://doi.org/10.1016/0550-3213(91)90180-6}
  {\path{doi:10.1016/0550-3213(91)90180-6}}.

\bibitem{Bonora:1986ix}
L.~Bonora, P.~Pasti, M.~Tonin, {Superspace Formulation of 10-$D$ {SUGRA}+{SYM}
  Theory a La Green-schwarz}, Phys. Lett. B 188 (1987) 335.
\newblock \href {https://doi.org/10.1016/0370-2693(87)91392-X}
  {\path{doi:10.1016/0370-2693(87)91392-X}}.

\bibitem{Bonora:1987xn}
L.~Bonora, M.~Bregola, K.~Lechner, P.~Pasti, M.~Tonin, {Anomaly Free
  Supergravity and Superyang-mills Theories in Ten-dimensions}, Nucl. Phys. B
  296 (1988) 877--901.
\newblock \href {https://doi.org/10.1016/0550-3213(88)90402-6}
  {\path{doi:10.1016/0550-3213(88)90402-6}}.

\bibitem{DAuria:1987tdr}
R.~D'Auria, P.~Fre, M.~Raciti, F.~Riva, {Anomaly Free Supergravity in $D=10$.
  1. The Bianchi Identities and the Bosonic Lagrangian}, Int. J. Mod. Phys. A 3
  (1988) 953.
\newblock \href {https://doi.org/10.1142/S0217751X88000436}
  {\path{doi:10.1142/S0217751X88000436}}.

\bibitem{Raciti:1989je}
M.~Raciti, F.~Riva, D.~Zanon, {Perturbative Approach to $D=10$ Superspace
  Supergravity With a Lorentz {Chern-Simons} Form}, Phys. Lett. B 227 (1989)
  118--123.
\newblock \href {https://doi.org/10.1016/0370-2693(89)91292-6}
  {\path{doi:10.1016/0370-2693(89)91292-6}}.

\bibitem{Bonora:1990mt}
L.~Bonora, K.~Lechner, M.~Bregola, P.~Pasti, M.~Tonin, {A Discussion of the
  constraints in N=1 SUGRA-SYM in 10-D}, Int. J. Mod. Phys. A 5 (1990)
  461--477.
\newblock \href {https://doi.org/10.1142/S0217751X90000222}
  {\path{doi:10.1142/S0217751X90000222}}.

\bibitem{Bonora:1992tx}
L.~Bonora, et~al., {Some remarks on the supersymmetrization of the Lorentz
  Chern-Simons form in D = 10 N=1 supergravity theories}, Phys. Lett. B 277
  (1992) 306--312.
\newblock \href {https://doi.org/10.1016/0370-2693(92)90751-O}
  {\path{doi:10.1016/0370-2693(92)90751-O}}.

\bibitem{Fre:1991ef}
P.~Fre, I.~Pesando, {Supersymmetrization of the Lorentz Chern-Simons term in D
  = 10}, in: {Strings and Symmetries 1991}, 1991.

\bibitem{Pesando:1992pa}
I.~Pesando, {Completion of the ten-dimensional anomaly free supergravity
  program: The Field equations}, Class. Quant. Grav. 9 (1992) 823--866.
\newblock \href {https://doi.org/10.1088/0264-9381/9/4/004}
  {\path{doi:10.1088/0264-9381/9/4/004}}.

\bibitem{Lechner:2008uz}
K.~Lechner, M.~Tonin, {Superspace formulations of ten-dimensional
  supergravity}, JHEP 06 (2008) 021.
\newblock \href {http://arxiv.org/abs/0802.3869} {\path{arXiv:0802.3869}},
  \href {https://doi.org/10.1088/1126-6708/2008/06/021}
  {\path{doi:10.1088/1126-6708/2008/06/021}}.

\bibitem{Bellucci:1988ff}
S.~Bellucci, S.~J. Gates, Jr., {$D=10$, $N=1$ Superspace Supergravity and the
  Lorentz Chern-simons Form}, Phys. Lett. B 208 (1988) 456--462.
\newblock \href {https://doi.org/10.1016/0370-2693(88)90647-8}
  {\path{doi:10.1016/0370-2693(88)90647-8}}.

\bibitem{Bellucci:1990fa}
S.~Bellucci, D.~A. Depireux, S.~J. Gates, Jr., {Consistent and Universal
  Inclusion of the Lorentz {Chern-Simons} Form in $D=10$, $N=1$ Supergravity
  Theories}, Phys. Lett. B 238 (1990) 315--322.
\newblock \href {https://doi.org/10.1016/0370-2693(90)91741-S}
  {\path{doi:10.1016/0370-2693(90)91741-S}}.

\bibitem{Bellucci:2006cx}
S.~Bellucci, D.~O'Reilly, {Non-minimal string corrections and supergravity},
  Phys. Rev. D 73 (2006) 065009.
\newblock \href {http://arxiv.org/abs/hep-th/0603033}
  {\path{arXiv:hep-th/0603033}}, \href
  {https://doi.org/10.1103/PhysRevD.73.065009}
  {\path{doi:10.1103/PhysRevD.73.065009}}.

\bibitem{OReilly:2006eeg}
D.~O'Reilly, {String corrected supergravity: A Complete and consistent
  non-minimal solution} (11 2006).
\newblock \href {http://arxiv.org/abs/hep-th/0611068}
  {\path{arXiv:hep-th/0611068}}.

\bibitem{Nilsson:1986cz}
B.~E.~W. Nilsson, {Off-shell $d=10$, $N=1$ Poincare Supergravity and the
  Embeddibility of Higher Derivative Field Theories in Superspace}, Phys. Lett.
  B 175 (1986) 319--324.
\newblock \href {https://doi.org/10.1016/0370-2693(86)90863-4}
  {\path{doi:10.1016/0370-2693(86)90863-4}}.

\bibitem{Nilsson:1986rh}
B.~E.~W. Nilsson, A.~K. Tollsten, {Supersymmetrization of Zeta (3) (R $\mu \nu
  \rho \sigma$)**4 in Superstring Theories}, Phys. Lett. B 181 (1986) 63--66.
\newblock \href {https://doi.org/10.1016/0370-2693(86)91255-4}
  {\path{doi:10.1016/0370-2693(86)91255-4}}.

\bibitem{Candiello:1994ew}
A.~Candiello, K.~Lechner, {The Supersymmetric version of the Green-Schwarz
  anomaly cancellation mechanism}, Phys. Lett. B 332 (1994) 71--76.
\newblock \href {http://arxiv.org/abs/hep-th/9404095}
  {\path{arXiv:hep-th/9404095}}, \href
  {https://doi.org/10.1016/0370-2693(94)90860-5}
  {\path{doi:10.1016/0370-2693(94)90860-5}}.

\bibitem{Howe:2008vb}
P.~S. Howe, {Heterotic supergeometry revisited}, in: {Gravity, Supersymmetry
  and Branes: A Meeting in Celebration of Kellogg Stelle's 60th Birthday},
  2008.
\newblock \href {http://arxiv.org/abs/0805.2893} {\path{arXiv:0805.2893}}.

\bibitem{Lechner:2010ti}
K.~Lechner, {Quantum properties of the heterotic five-brane}, Phys. Lett. B 693
  (2010) 323--329.
\newblock \href {http://arxiv.org/abs/1005.5719} {\path{arXiv:1005.5719}},
  \href {https://doi.org/10.1016/j.physletb.2010.08.041}
  {\path{doi:10.1016/j.physletb.2010.08.041}}.

\bibitem{Lechner:1987ip}
K.~Lechner, P.~Pasti, M.~Tonin, {Anomaly Free {SUGRA} and the R**4 Superstring
  Term}, Mod. Phys. Lett. A 2 (1987) 929.
\newblock \href {https://doi.org/10.1142/S021773238700118X}
  {\path{doi:10.1142/S021773238700118X}}.

\bibitem{Gates:1985wh}
S.~J. Gates, Jr., H.~Nishino, {Manifestly Supersymmetric O ($\alpha^\prime$)
  Superstring Corrections in New $D=10$, $N=1$ Supergravity {Yang-Mills}
  Theory}, Phys. Lett. B 173 (1986) 52--58.
\newblock \href {https://doi.org/10.1016/0370-2693(86)91229-3}
  {\path{doi:10.1016/0370-2693(86)91229-3}}.

\bibitem{Gates:1986is}
S.~J. Gates, Jr., S.~Vashakidze, {On $D=10$, $N=1$ Supersymmetry, Superspace
  Geometry and Superstring Effects}, Nucl. Phys. B 291 (1987) 172.
\newblock \href {https://doi.org/10.1016/0550-3213(87)90470-6}
  {\path{doi:10.1016/0550-3213(87)90470-6}}.

\bibitem{Gates:1986tj}
S.~J. Gates, Jr., H.~Nishino, {On D = 10, N=1 Supersymmetry, Superspace
  Geometry and Superstring Effects. 2.}, Nucl. Phys. B 291 (1987) 205.
\newblock \href {https://doi.org/10.1016/0550-3213(87)90471-8}
  {\path{doi:10.1016/0550-3213(87)90471-8}}.

\bibitem{Nishino:1986mj}
H.~Nishino, {General form of string corrections to supersymmetry transformation
  in D = 10, N=1 supergravity}, Phys. Lett. B 188 (1987) 437--441.
\newblock \href {https://doi.org/10.1016/0370-2693(87)91644-3}
  {\path{doi:10.1016/0370-2693(87)91644-3}}.

\bibitem{Nishino:1990ky}
H.~Nishino, {Superstring corrections in simplest constraint set for D = 10, N=1
  superspace}, Phys. Lett. B 258 (1991) 104--110.
\newblock \href {https://doi.org/10.1016/0370-2693(91)91216-I}
  {\path{doi:10.1016/0370-2693(91)91216-I}}.

\bibitem{Terentev:1993wm}
M.~V. Terentev, {Dynamical equations for superstring corrections in D = 10, N=1
  superspace}, Phys. Lett. B 313 (1993) 351--356.
\newblock \href {https://doi.org/10.1016/0370-2693(93)90003-Z}
  {\path{doi:10.1016/0370-2693(93)90003-Z}}.

\bibitem{Terentev:1994br}
M.~V. Terentev, {Dual supergravity in D = 10, N=1 superspace with tree level
  superstring corrections}, Phys. Lett. B 325 (1994) 96--102.
\newblock \href {http://arxiv.org/abs/hep-th/9406152}
  {\path{arXiv:hep-th/9406152}}, \href
  {https://doi.org/10.1016/0370-2693(94)90077-9}
  {\path{doi:10.1016/0370-2693(94)90077-9}}.

\bibitem{Zyablyuk:1994xk}
K.~N. Zyablyuk, {On the Lagrangian of N=1, D = 10 dual supergravity}, Phys.
  Atom. Nucl. 58 (1995) 1425--1429.
\newblock \href {http://arxiv.org/abs/hep-ph/9411240}
  {\path{arXiv:hep-ph/9411240}}.

\bibitem{Saulina:1995eq}
N.~A. Saulina, M.~V. Terentev, K.~N. Zyablyuk, {Duality in N=1, D = 10
  superspace and supergravity with tree level superstring corrections}, Phys.
  Lett. B 366 (1996) 134--140.
\newblock \href {http://arxiv.org/abs/hep-th/9507033}
  {\path{arXiv:hep-th/9507033}}, \href
  {https://doi.org/10.1016/0370-2693(95)01338-5}
  {\path{doi:10.1016/0370-2693(95)01338-5}}.

\bibitem{DAuria:1987qjh}
R.~D'Auria, P.~Fre, {Duality in Superspace and Anomaly Free Supergravity: Some
  Remarks}, Mod. Phys. Lett. A 3 (1988) 673.
\newblock \href {https://doi.org/10.1142/S0217732388000817}
  {\path{doi:10.1142/S0217732388000817}}.

\bibitem{Chang:2023pss}
H.-Y. Chang, E.~Sezgin, Y.~Tanii, {Higher derivative couplings of
  hypermultiplets}, JHEP 06 (2023) 172.
\newblock \href {http://arxiv.org/abs/2304.06073} {\path{arXiv:2304.06073}},
  \href {https://doi.org/10.1007/JHEP06(2023)172}
  {\path{doi:10.1007/JHEP06(2023)172}}.

\bibitem{Fontanella:2019avn}
A.~Fontanella, T.~Ort\'\i{}n, {On the supersymmetric solutions of the Heterotic
  Superstring effective action}, JHEP 06 (2020) 106, [Erratum: JHEP 10, 130
  (2021)].
\newblock \href {http://arxiv.org/abs/1910.08496} {\path{arXiv:1910.08496}},
  \href {https://doi.org/10.1007/JHEP10(2021)130}
  {\path{doi:10.1007/JHEP10(2021)130}}.

\bibitem{Kallosh:1993wx}
R.~Kallosh, T.~Ortin, {Killing spinor identities} (6 1993).
\newblock \href {http://arxiv.org/abs/hep-th/9306085}
  {\path{arXiv:hep-th/9306085}}.

\bibitem{Cano:2018qev}
P.~A. Cano, P.~Meessen, T.~Ort\'\i{}n, P.~F. Ram\'\i{}rez, {$\alpha'$-corrected
  black holes in String Theory}, JHEP 05 (2018) 110.
\newblock \href {http://arxiv.org/abs/1803.01919} {\path{arXiv:1803.01919}},
  \href {https://doi.org/10.1007/JHEP05(2018)110}
  {\path{doi:10.1007/JHEP05(2018)110}}.

\bibitem{Cano:2019oma}
P.~A. Cano, T.~Ort\'\i{}n, P.~F. Ramirez, {On the extremality bound of stringy
  black holes}, JHEP 02 (2020) 175.
\newblock \href {http://arxiv.org/abs/1909.08530} {\path{arXiv:1909.08530}},
  \href {https://doi.org/10.1007/JHEP02(2020)175}
  {\path{doi:10.1007/JHEP02(2020)175}}.

\bibitem{Cano:2019ycn}
P.~A. Cano, S.~Chimento, R.~Linares, T.~Ort\'\i{}n, P.~F. Ram\'\i{}rez,
  {$\alpha'$ corrections of Reissner-Nordstr\"om black holes}, JHEP 02 (2020)
  031.
\newblock \href {http://arxiv.org/abs/1910.14324} {\path{arXiv:1910.14324}},
  \href {https://doi.org/10.1007/JHEP02(2020)031}
  {\path{doi:10.1007/JHEP02(2020)031}}.

\bibitem{Cheung:2018cwt}
C.~Cheung, J.~Liu, G.~N. Remmen, {Proof of the Weak Gravity Conjecture from
  Black Hole Entropy}, JHEP 10 (2018) 004.
\newblock \href {http://arxiv.org/abs/1801.08546} {\path{arXiv:1801.08546}},
  \href {https://doi.org/10.1007/JHEP10(2018)004}
  {\path{doi:10.1007/JHEP10(2018)004}}.

\bibitem{Eloy:2020dko}
C.~Eloy, O.~Hohm, H.~Samtleben, {Duality Invariance and Higher Derivatives},
  Phys. Rev. D 101~(12) (2020) 126018.
\newblock \href {http://arxiv.org/abs/2004.13140} {\path{arXiv:2004.13140}},
  \href {https://doi.org/10.1103/PhysRevD.101.126018}
  {\path{doi:10.1103/PhysRevD.101.126018}}.

\bibitem{Chang:2021tsj}
H.-Y. Chang, E.~Sezgin, Y.~Tanii, {Dimensional reduction of higher derivative
  heterotic supergravity}, JHEP 03 (2022) 081.
\newblock \href {http://arxiv.org/abs/2110.13163} {\path{arXiv:2110.13163}},
  \href {https://doi.org/10.1007/JHEP03(2022)081}
  {\path{doi:10.1007/JHEP03(2022)081}}.

\bibitem{Bergshoeff:1985mr}
E.~Bergshoeff, I.~G. Koh, E.~Sezgin, {{Yang-Mills} / Einstein Supergravity in
  Seven-dimensions}, Phys. Rev. D 32 (1985) 1353--1357.
\newblock \href {https://doi.org/10.1103/PhysRevD.32.1353}
  {\path{doi:10.1103/PhysRevD.32.1353}}.

\bibitem{Romans:1985tw}
L.~J. Romans, {The F(4) Gauged Supergravity in Six-dimensions}, Nucl. Phys. B
  269 (1986) 691.
\newblock \href {https://doi.org/10.1016/0550-3213(86)90517-1}
  {\path{doi:10.1016/0550-3213(86)90517-1}}.

\bibitem{Giani:1984dw}
F.~Giani, M.~Pernici, P.~van Nieuwenhuizen, {Gauged N=4 d = 6 Supergravity},
  Phys. Rev. D 30 (1984) 1680.
\newblock \href {https://doi.org/10.1103/PhysRevD.30.1680}
  {\path{doi:10.1103/PhysRevD.30.1680}}.

\bibitem{Romans:1986er}
L.~J. Romans, {Selfduality for Interacting Fields: Covariant Field Equations
  for Six-dimensional Chiral Supergravities}, Nucl. Phys. B 276 (1986) 71.
\newblock \href {https://doi.org/10.1016/0550-3213(86)90016-7}
  {\path{doi:10.1016/0550-3213(86)90016-7}}.

\bibitem{Bergshoeff:1999db}
E.~Bergshoeff, E.~Sezgin, A.~Van~Proeyen, {(2,0) tensor multiplets and
  conformal supergravity in D = 6}, Class. Quant. Grav. 16 (1999) 3193--3206.
\newblock \href {http://arxiv.org/abs/hep-th/9904085}
  {\path{arXiv:hep-th/9904085}}, \href
  {https://doi.org/10.1088/0264-9381/16/10/311}
  {\path{doi:10.1088/0264-9381/16/10/311}}.

\bibitem{Riccioni:1997np}
F.~Riccioni, {Tensor multiplets in six-dimensional (2,0) supergravity}, Phys.
  Lett. B 422 (1998) 126--134.
\newblock \href {http://arxiv.org/abs/hep-th/9712176}
  {\path{arXiv:hep-th/9712176}}, \href
  {https://doi.org/10.1016/S0370-2693(98)00070-7}
  {\path{doi:10.1016/S0370-2693(98)00070-7}}.

\bibitem{Bergshoeff:2012ax}
E.~Bergshoeff, F.~Coomans, E.~Sezgin, A.~Van~Proeyen, {Higher Derivative
  Extension of 6D Chiral Gauged Supergravity}, JHEP 07 (2012) 011.
\newblock \href {http://arxiv.org/abs/1203.2975} {\path{arXiv:1203.2975}},
  \href {https://doi.org/10.1007/JHEP07(2012)011}
  {\path{doi:10.1007/JHEP07(2012)011}}.

\bibitem{Bergshoeff:1986wc}
E.~Bergshoeff, A.~Salam, E.~Sezgin, {Supersymmetric R**2 Actions, Conformal
  Invariance and Lorentz Chern-simons Term in Six-dimensions and
  Ten-dimensions}, Nucl. Phys. B 279 (1987) 659--683.
\newblock \href {https://doi.org/10.1016/0550-3213(87)90015-0}
  {\path{doi:10.1016/0550-3213(87)90015-0}}.

\bibitem{Butter:2016qkx}
D.~Butter, S.~M. Kuzenko, J.~Novak, S.~Theisen, {Invariants for minimal
  conformal supergravity in six dimensions}, JHEP 12 (2016) 072.
\newblock \href {http://arxiv.org/abs/1606.02921} {\path{arXiv:1606.02921}},
  \href {https://doi.org/10.1007/JHEP12(2016)072}
  {\path{doi:10.1007/JHEP12(2016)072}}.

\bibitem{Butter:2017jqu}
D.~Butter, J.~Novak, G.~Tartaglino-Mazzucchelli, {The component structure of
  conformal supergravity invariants in six dimensions}, JHEP 05 (2017) 133.
\newblock \href {http://arxiv.org/abs/1701.08163} {\path{arXiv:1701.08163}},
  \href {https://doi.org/10.1007/JHEP05(2017)133}
  {\path{doi:10.1007/JHEP05(2017)133}}.

\bibitem{Novak:2017wqc}
J.~Novak, M.~Ozkan, Y.~Pang, G.~Tartaglino-Mazzucchelli, {Gauss-Bonnet
  supergravity in six dimensions}, Phys. Rev. Lett. 119~(11) (2017) 111602.
\newblock \href {http://arxiv.org/abs/1706.09330} {\path{arXiv:1706.09330}},
  \href {https://doi.org/10.1103/PhysRevLett.119.111602}
  {\path{doi:10.1103/PhysRevLett.119.111602}}.

\bibitem{Butter:2018wss}
D.~Butter, J.~Novak, M.~Ozkan, Y.~Pang, G.~Tartaglino-Mazzucchelli, {Curvature
  squared invariants in six-dimensional ${\cal N} = (1,0)$ supergravity}, JHEP
  04 (2019) 013.
\newblock \href {http://arxiv.org/abs/1808.00459} {\path{arXiv:1808.00459}},
  \href {https://doi.org/10.1007/JHEP04(2019)013}
  {\path{doi:10.1007/JHEP04(2019)013}}.

\bibitem{DallAgata:1997yqq}
G.~Dall'Agata, K.~Lechner, {N=1, D = 6 supergravity: Duality and nonminimal
  couplings}, Nucl. Phys. B 511 (1998) 326--352.
\newblock \href {http://arxiv.org/abs/hep-th/9707236}
  {\path{arXiv:hep-th/9707236}}, \href
  {https://doi.org/10.1016/S0550-3213(97)00719-0}
  {\path{doi:10.1016/S0550-3213(97)00719-0}}.

\bibitem{Sen:1991zi}
A.~Sen, {O(d) x O(d) symmetry of the space of cosmological solutions in string
  theory, scale factor duality and two-dimensional black holes}, Phys. Lett. B
  271 (1991) 295--300.
\newblock \href {https://doi.org/10.1016/0370-2693(91)90090-D}
  {\path{doi:10.1016/0370-2693(91)90090-D}}.

\bibitem{Maharana:1992my}
J.~Maharana, J.~H. Schwarz, {Noncompact symmetries in string theory}, Nucl.
  Phys. B 390 (1993) 3--32.
\newblock \href {http://arxiv.org/abs/hep-th/9207016}
  {\path{arXiv:hep-th/9207016}}, \href
  {https://doi.org/10.1016/0550-3213(93)90387-5}
  {\path{doi:10.1016/0550-3213(93)90387-5}}.

\bibitem{Ortin:2020xdm}
T.~Ortin, {O(n, n) invariance and Wald entropy formula in the Heterotic
  Superstring effective action at first order in $\alpha'$}, JHEP 01 (2021)
  187.
\newblock \href {http://arxiv.org/abs/2005.14618} {\path{arXiv:2005.14618}},
  \href {https://doi.org/10.1007/JHEP01(2021)187}
  {\path{doi:10.1007/JHEP01(2021)187}}.

\bibitem{Hohm:2013jaa}
O.~Hohm, W.~Siegel, B.~Zwiebach, {Doubled $\alpha'$-geometry}, JHEP 02 (2014)
  065.
\newblock \href {http://arxiv.org/abs/1306.2970} {\path{arXiv:1306.2970}},
  \href {https://doi.org/10.1007/JHEP02(2014)065}
  {\path{doi:10.1007/JHEP02(2014)065}}.

\bibitem{Hohm:2014xsa}
O.~Hohm, B.~Zwiebach, {Double field theory at order $\alpha'$}, JHEP 11 (2014)
  075.
\newblock \href {http://arxiv.org/abs/1407.3803} {\path{arXiv:1407.3803}},
  \href {https://doi.org/10.1007/JHEP11(2014)075}
  {\path{doi:10.1007/JHEP11(2014)075}}.

\bibitem{Marques:2015vua}
D.~Marques, C.~A. Nunez, {T-duality and \ensuremath{\alpha}'-corrections}, JHEP
  10 (2015) 084.
\newblock \href {http://arxiv.org/abs/1507.00652} {\path{arXiv:1507.00652}},
  \href {https://doi.org/10.1007/JHEP10(2015)084}
  {\path{doi:10.1007/JHEP10(2015)084}}.

\bibitem{Seiberg:1988pf}
N.~Seiberg, {Observations on the Moduli Space of Superconformal Field
  Theories}, Nucl. Phys. B 303 (1988) 286--304.
\newblock \href {https://doi.org/10.1016/0550-3213(88)90183-6}
  {\path{doi:10.1016/0550-3213(88)90183-6}}.

\bibitem{Aspinwall:1994rg}
P.~S. Aspinwall, D.~R. Morrison, {String theory on K3 surfaces}, AMS/IP Stud.
  Adv. Math. 1 (1996) 703--716.
\newblock \href {http://arxiv.org/abs/hep-th/9404151}
  {\path{arXiv:hep-th/9404151}}.

\bibitem{Forste:1996yd}
S.~Forste, J.~Louis, {Duality in string theory}, Nucl. Phys. B Proc. Suppl. 61
  (1998) 3--22.
\newblock \href {http://arxiv.org/abs/hep-th/9612192}
  {\path{arXiv:hep-th/9612192}}, \href
  {https://doi.org/10.1016/S0920-5632(97)00516-1}
  {\path{doi:10.1016/S0920-5632(97)00516-1}}.

\bibitem{Chow:2019win}
D.~D.~K. Chow, Y.~Pang, {Rotating Strings in Six-Dimensional Higher-Derivative
  Supergravity}, Phys. Rev. D 100~(10) (2019) 106004.
\newblock \href {http://arxiv.org/abs/1906.07426} {\path{arXiv:1906.07426}},
  \href {https://doi.org/10.1103/PhysRevD.100.106004}
  {\path{doi:10.1103/PhysRevD.100.106004}}.

\bibitem{Pang:2019qwq}
Y.~Pang, {Attractor mechanism and nonrenormalization theorem in 6D (1, 0)
  supergravity}, Phys. Rev. D 103~(2) (2021) 026018.
\newblock \href {http://arxiv.org/abs/1910.10192} {\path{arXiv:1910.10192}},
  \href {https://doi.org/10.1103/PhysRevD.103.026018}
  {\path{doi:10.1103/PhysRevD.103.026018}}.

\bibitem{Gutowski:2003rg}
J.~B. Gutowski, D.~Martelli, H.~S. Reall, {All Supersymmetric solutions of
  minimal supergravity in six- dimensions}, Class. Quant. Grav. 20 (2003)
  5049--5078.
\newblock \href {http://arxiv.org/abs/hep-th/0306235}
  {\path{arXiv:hep-th/0306235}}, \href
  {https://doi.org/10.1088/0264-9381/20/23/008}
  {\path{doi:10.1088/0264-9381/20/23/008}}.

\bibitem{Ma:2021opb}
L.~Ma, Y.~Pang, H.~L\"u, {\ensuremath{\alpha}'-corrections to near extremal
  dyonic strings and weak gravity conjecture}, JHEP 01 (2022) 157.
\newblock \href {http://arxiv.org/abs/2110.03129} {\path{arXiv:2110.03129}},
  \href {https://doi.org/10.1007/JHEP01(2022)157}
  {\path{doi:10.1007/JHEP01(2022)157}}.

\bibitem{Ma:2022nwq}
L.~Ma, Y.~Pang, H.~Lu, {Improved Wald formalism and first law of dyonic black
  strings with mixed Chern-Simons terms}, JHEP 10 (2022) 142.
\newblock \href {http://arxiv.org/abs/2202.08290} {\path{arXiv:2202.08290}},
  \href {https://doi.org/10.1007/JHEP10(2022)142}
  {\path{doi:10.1007/JHEP10(2022)142}}.

\bibitem{Ma:2022gtm}
L.~Ma, Y.~Pang, H.~L\"u, {Negative corrections to black hole entropy from
  string theory}, Sci. China Phys. Mech. Astron. 66~(12) (2023) 121011.
\newblock \href {http://arxiv.org/abs/2212.03262} {\path{arXiv:2212.03262}},
  \href {https://doi.org/10.1007/s11433-023-2257-6}
  {\path{doi:10.1007/s11433-023-2257-6}}.

\bibitem{Pang:2012xs}
Y.~Pang, C.~N. Pope, E.~Sezgin, {Spectrum of Higher Derivative 6D Chiral
  Supergravity on Minkowski $ x S^2$}, JHEP 10 (2012) 154.
\newblock \href {http://arxiv.org/abs/1204.1060} {\path{arXiv:1204.1060}},
  \href {https://doi.org/10.1007/JHEP10(2012)154}
  {\path{doi:10.1007/JHEP10(2012)154}}.

\bibitem{Butter:2014xxa}
D.~Butter, S.~M. Kuzenko, J.~Novak, G.~Tartaglino-Mazzucchelli, {Conformal
  supergravity in five dimensions: New approach and applications}, JHEP 02
  (2015) 111.
\newblock \href {http://arxiv.org/abs/1410.8682} {\path{arXiv:1410.8682}},
  \href {https://doi.org/10.1007/JHEP02(2015)111}
  {\path{doi:10.1007/JHEP02(2015)111}}.

\bibitem{Gold:2023dfe}
G.~Gold, J.~Hutomo, S.~Khandelwal, G.~Tartaglino-Mazzucchelli,
  {Curvature-squared invariants of minimal five-dimensional supergravity from
  superspace}, Phys. Rev. D 107~(10) (2023) 106013.
\newblock \href {http://arxiv.org/abs/2302.14295} {\path{arXiv:2302.14295}},
  \href {https://doi.org/10.1103/PhysRevD.107.106013}
  {\path{doi:10.1103/PhysRevD.107.106013}}.

\bibitem{Gold:2023ykx}
G.~Gold, J.~Hutomo, S.~Khandelwal, G.~Tartaglino-Mazzucchelli, {Components of
  curvature-squared invariants of minimal supergravity in five dimensions} (11
  2023).
\newblock \href {http://arxiv.org/abs/2311.00679} {\path{arXiv:2311.00679}}.

\bibitem{Bergshoeff:2011xn}
E.~A. Bergshoeff, J.~Rosseel, E.~Sezgin, {Off-shell D=5, N=2 Riemann Squared
  Supergravity}, Class. Quant. Grav. 28 (2011) 225016.
\newblock \href {http://arxiv.org/abs/1107.2825} {\path{arXiv:1107.2825}},
  \href {https://doi.org/10.1088/0264-9381/28/22/225016}
  {\path{doi:10.1088/0264-9381/28/22/225016}}.

\bibitem{VanProeyen:1999ni}
A.~Van~Proeyen, {Tools for supersymmetry}, Ann. U. Craiova Phys. 9~(I) (1999)
  1--48.
\newblock \href {http://arxiv.org/abs/hep-th/9910030}
  {\path{arXiv:hep-th/9910030}}.

\bibitem{Kugo:2000hn}
T.~Kugo, K.~Ohashi, {Supergravity tensor calculus in 5-D from 6-D}, Prog.
  Theor. Phys. 104 (2000) 835--865.
\newblock \href {http://arxiv.org/abs/hep-ph/0006231}
  {\path{arXiv:hep-ph/0006231}}, \href {https://doi.org/10.1143/PTP.104.835}
  {\path{doi:10.1143/PTP.104.835}}.

\bibitem{Bergshoeff:2002qk}
E.~Bergshoeff, S.~Cucu, T.~De~Wit, J.~Gheerardyn, R.~Halbersma, S.~Vandoren,
  A.~Van~Proeyen, {Superconformal N=2, D = 5 matter with and without actions},
  JHEP 10 (2002) 045.
\newblock \href {http://arxiv.org/abs/hep-th/0205230}
  {\path{arXiv:hep-th/0205230}}, \href
  {https://doi.org/10.1088/1126-6708/2002/10/045}
  {\path{doi:10.1088/1126-6708/2002/10/045}}.

\bibitem{Gunaydin:1983bi}
M.~Gunaydin, G.~Sierra, P.~K. Townsend, {The Geometry of N=2 Maxwell-Einstein
  Supergravity and Jordan Algebras}, Nucl. Phys. B 242 (1984) 244--268.
\newblock \href {https://doi.org/10.1016/0550-3213(84)90142-1}
  {\path{doi:10.1016/0550-3213(84)90142-1}}.

\bibitem{Gunaydin:1984ak}
M.~Gunaydin, G.~Sierra, P.~K. Townsend, {Gauging the d = 5 Maxwell-Einstein
  Supergravity Theories: More on Jordan Algebras}, Nucl. Phys. B 253 (1985)
  573.
\newblock \href {https://doi.org/10.1016/0550-3213(85)90547-4}
  {\path{doi:10.1016/0550-3213(85)90547-4}}.

\bibitem{Hanaki:2006pj}
K.~Hanaki, K.~Ohashi, Y.~Tachikawa, {Supersymmetric Completion of an R**2 term
  in Five-dimensional Supergravity}, Prog. Theor. Phys. 117 (2007) 533.
\newblock \href {http://arxiv.org/abs/hep-th/0611329}
  {\path{arXiv:hep-th/0611329}}, \href {https://doi.org/10.1143/PTP.117.533}
  {\path{doi:10.1143/PTP.117.533}}.

\bibitem{Cremonini:2009sy}
S.~Cremonini, K.~Hanaki, J.~T. Liu, P.~Szepietowski, {Higher derivative effects
  on eta/s at finite chemical potential}, Phys. Rev. D 80 (2009) 025002.
\newblock \href {http://arxiv.org/abs/0903.3244} {\path{arXiv:0903.3244}},
  \href {https://doi.org/10.1103/PhysRevD.80.025002}
  {\path{doi:10.1103/PhysRevD.80.025002}}.

\bibitem{Gold:2023ymc}
G.~Gold, J.~Hutomo, S.~Khandelwal, M.~Ozkan, Y.~Pang,
  G.~Tartaglino-Mazzucchelli, {All Gauged Curvature-Squared Supergravities in
  Five Dimensions}, Phys. Rev. Lett. 131~(25) (2023) 251603.
\newblock \href {http://arxiv.org/abs/2309.07637} {\path{arXiv:2309.07637}},
  \href {https://doi.org/10.1103/PhysRevLett.131.251603}
  {\path{doi:10.1103/PhysRevLett.131.251603}}.

\bibitem{Ozkan:2013uk}
M.~Ozkan, Y.~Pang, {Supersymmetric Completion of Gauss-Bonnet Combination in
  Five Dimensions}, JHEP 03 (2013) 158, [Erratum: JHEP 07, 152 (2013)].
\newblock \href {http://arxiv.org/abs/1301.6622} {\path{arXiv:1301.6622}},
  \href {https://doi.org/10.1007/JHEP07(2013)152}
  {\path{doi:10.1007/JHEP07(2013)152}}.

\bibitem{Liu:2022sew}
J.~T. Liu, R.~J. Saskowski, {Four-derivative corrections to minimal gauged
  supergravity in five dimensions}, JHEP 05 (2022) 171.
\newblock \href {http://arxiv.org/abs/2201.04690} {\path{arXiv:2201.04690}},
  \href {https://doi.org/10.1007/JHEP05(2022)171}
  {\path{doi:10.1007/JHEP05(2022)171}}.

\bibitem{Cassani:2022lrk}
D.~Cassani, A.~Ruip\'erez, E.~Turetta, {Corrections to AdS$_{5}$ black hole
  thermodynamics from higher-derivative supergravity}, JHEP 11 (2022) 059.
\newblock \href {http://arxiv.org/abs/2208.01007} {\path{arXiv:2208.01007}},
  \href {https://doi.org/10.1007/JHEP11(2022)059}
  {\path{doi:10.1007/JHEP11(2022)059}}.

\bibitem{Bonetti:2018lfb}
F.~Bonetti, D.~Klemm, W.~A. Sabra, P.~Sloane, {Spinorial geometry, off-shell
  Killing spinor identities and higher derivative 5D supergravities}, JHEP 08
  (2018) 121.
\newblock \href {http://arxiv.org/abs/1806.04108} {\path{arXiv:1806.04108}},
  \href {https://doi.org/10.1007/JHEP08(2018)121}
  {\path{doi:10.1007/JHEP08(2018)121}}.

\bibitem{Gauntlett:2002nw}
J.~P. Gauntlett, J.~B. Gutowski, C.~M. Hull, S.~Pakis, H.~S. Reall, {All
  supersymmetric solutions of minimal supergravity in five- dimensions}, Class.
  Quant. Grav. 20 (2003) 4587--4634.
\newblock \href {http://arxiv.org/abs/hep-th/0209114}
  {\path{arXiv:hep-th/0209114}}, \href
  {https://doi.org/10.1088/0264-9381/20/21/005}
  {\path{doi:10.1088/0264-9381/20/21/005}}.

\bibitem{Baggio:2014hua}
M.~Baggio, N.~Halmagyi, D.~R. Mayerson, D.~Robbins, B.~Wecht, {Higher
  Derivative Corrections and Central Charges from Wrapped M5-branes}, JHEP 12
  (2014) 042.
\newblock \href {http://arxiv.org/abs/1408.2538} {\path{arXiv:1408.2538}},
  \href {https://doi.org/10.1007/JHEP12(2014)042}
  {\path{doi:10.1007/JHEP12(2014)042}}.

\bibitem{Castro:2007sd}
A.~Castro, J.~L. Davis, P.~Kraus, F.~Larsen, {5D attractors with higher
  derivatives}, JHEP 04 (2007) 091.
\newblock \href {http://arxiv.org/abs/hep-th/0702072}
  {\path{arXiv:hep-th/0702072}}, \href
  {https://doi.org/10.1088/1126-6708/2007/04/091}
  {\path{doi:10.1088/1126-6708/2007/04/091}}.

\bibitem{Kraus:2005zm}
P.~Kraus, F.~Larsen, {Holographic gravitational anomalies}, JHEP 01 (2006) 022.
\newblock \href {http://arxiv.org/abs/hep-th/0508218}
  {\path{arXiv:hep-th/0508218}}, \href
  {https://doi.org/10.1088/1126-6708/2006/01/022}
  {\path{doi:10.1088/1126-6708/2006/01/022}}.

\bibitem{Maldacena:1997de}
J.~M. Maldacena, A.~Strominger, E.~Witten, {Black hole entropy in M theory},
  JHEP 12 (1997) 002.
\newblock \href {http://arxiv.org/abs/hep-th/9711053}
  {\path{arXiv:hep-th/9711053}}, \href
  {https://doi.org/10.1088/1126-6708/1997/12/002}
  {\path{doi:10.1088/1126-6708/1997/12/002}}.

\bibitem{Breckenridge:1996is}
J.~C. Breckenridge, R.~C. Myers, A.~W. Peet, C.~Vafa, {D-branes and spinning
  black holes}, Phys. Lett. B 391 (1997) 93--98.
\newblock \href {http://arxiv.org/abs/hep-th/9602065}
  {\path{arXiv:hep-th/9602065}}, \href
  {https://doi.org/10.1016/S0370-2693(96)01460-8}
  {\path{doi:10.1016/S0370-2693(96)01460-8}}.

\bibitem{deWit:2009de}
B.~de~Wit, S.~Katmadas, {Near-Horizon Analysis of D=5 BPS Black Holes and
  Rings}, JHEP 02 (2010) 056.
\newblock \href {http://arxiv.org/abs/0910.4907} {\path{arXiv:0910.4907}},
  \href {https://doi.org/10.1007/JHEP02(2010)056}
  {\path{doi:10.1007/JHEP02(2010)056}}.

\bibitem{Castro:2007hc}
A.~Castro, J.~L. Davis, P.~Kraus, F.~Larsen, {5D Black Holes and Strings with
  Higher Derivatives}, JHEP 06 (2007) 007.
\newblock \href {http://arxiv.org/abs/hep-th/0703087}
  {\path{arXiv:hep-th/0703087}}, \href
  {https://doi.org/10.1088/1126-6708/2007/06/007}
  {\path{doi:10.1088/1126-6708/2007/06/007}}.

\bibitem{Castro:2007ci}
A.~Castro, J.~L. Davis, P.~Kraus, F.~Larsen, {Precision Entropy of Spinning
  Black Holes}, JHEP 09 (2007) 003.
\newblock \href {http://arxiv.org/abs/0705.1847} {\path{arXiv:0705.1847}},
  \href {https://doi.org/10.1088/1126-6708/2007/09/003}
  {\path{doi:10.1088/1126-6708/2007/09/003}}.

\bibitem{Kraus:2005vz}
P.~Kraus, F.~Larsen, {Microscopic black hole entropy in theories with higher
  derivatives}, JHEP 09 (2005) 034.
\newblock \href {http://arxiv.org/abs/hep-th/0506176}
  {\path{arXiv:hep-th/0506176}}, \href
  {https://doi.org/10.1088/1126-6708/2005/09/034}
  {\path{doi:10.1088/1126-6708/2005/09/034}}.

\bibitem{Cecotti:1985nf}
S.~Cecotti, S.~Ferrara, L.~Girardello, M.~Porrati, {Lorentz Chern-simons Terms
  in $N=1$ Four-dimensional Supergravity Consistent With Supersymmetry and
  String Compactification}, Phys. Lett. B 164 (1985) 46--50.
\newblock \href {https://doi.org/10.1016/0370-2693(85)90028-0}
  {\path{doi:10.1016/0370-2693(85)90028-0}}.

\bibitem{Cecotti:1985mf}
S.~Cecotti, S.~Ferrara, L.~Girardello, M.~Porrati, A.~Pasquinucci, {Matter
  Coupling in Higher Derivative Supergravity}, Phys. Rev. D 33 (1986) 2504.
\newblock \href {https://doi.org/10.1103/PhysRevD.33.2504}
  {\path{doi:10.1103/PhysRevD.33.2504}}.

\bibitem{Cecotti:1987mr}
S.~Cecotti, S.~Ferrara, L.~Girardello, A.~Pasquinucci, M.~Porrati, {Matter
  Coupled Supergravity With {Gauss-Bonnet} Invariants: Component Lagrangian and
  Supersymmetry Breaking}, Int. J. Mod. Phys. A 3 (1988) 1675--1733.
\newblock \href {https://doi.org/10.1142/S0217751X88000734}
  {\path{doi:10.1142/S0217751X88000734}}.

\bibitem{Freedman:2011uc}
D.~Z. Freedman, E.~Tonni, {The $D^{2k}R^{4}$ Invariants of $\mathcal{N}$=8
  Supergravity}, JHEP 04 (2011) 006.
\newblock \href {http://arxiv.org/abs/1101.1672} {\path{arXiv:1101.1672}},
  \href {https://doi.org/10.1007/JHEP04(2011)006}
  {\path{doi:10.1007/JHEP04(2011)006}}.

\bibitem{Bianchi:2008pu}
M.~Bianchi, H.~Elvang, D.~Z. Freedman, {Generating Tree Amplitudes in N=4 SYM
  and N = 8 SG}, JHEP 09 (2008) 063.
\newblock \href {http://arxiv.org/abs/0805.0757} {\path{arXiv:0805.0757}},
  \href {https://doi.org/10.1088/1126-6708/2008/09/063}
  {\path{doi:10.1088/1126-6708/2008/09/063}}.

\bibitem{Kallosh:1980fi}
R.~E. Kallosh, {Counterterms in extended supergravities}, Phys. Lett. B 99
  (1981) 122--127.
\newblock \href {https://doi.org/10.1016/0370-2693(81)90964-3}
  {\path{doi:10.1016/0370-2693(81)90964-3}}.

\bibitem{Elvang:2010jv}
H.~Elvang, D.~Z. Freedman, M.~Kiermaier, {A simple approach to counterterms in
  N=8 supergravity}, JHEP 11 (2010) 016.
\newblock \href {http://arxiv.org/abs/1003.5018} {\path{arXiv:1003.5018}},
  \href {https://doi.org/10.1007/JHEP11(2010)016}
  {\path{doi:10.1007/JHEP11(2010)016}}.

\bibitem{Beisert:2010jx}
N.~Beisert, H.~Elvang, D.~Z. Freedman, M.~Kiermaier, A.~Morales, S.~Stieberger,
  {E7(7) constraints on counterterms in N=8 supergravity}, Phys. Lett. B 694
  (2011) 265--271.
\newblock \href {http://arxiv.org/abs/1009.1643} {\path{arXiv:1009.1643}},
  \href {https://doi.org/10.1016/j.physletb.2010.09.069}
  {\path{doi:10.1016/j.physletb.2010.09.069}}.

\bibitem{Bossard:2010dq}
G.~Bossard, C.~Hillmann, H.~Nicolai, {E7(7) symmetry in perturbatively
  quantised N=8 supergravity}, JHEP 12 (2010) 052.
\newblock \href {http://arxiv.org/abs/1007.5472} {\path{arXiv:1007.5472}},
  \href {https://doi.org/10.1007/JHEP12(2010)052}
  {\path{doi:10.1007/JHEP12(2010)052}}.

\bibitem{Arkani-Hamed:2008owk}
N.~Arkani-Hamed, F.~Cachazo, J.~Kaplan, {What is the Simplest Quantum Field
  Theory?}, JHEP 09 (2010) 016.
\newblock \href {http://arxiv.org/abs/0808.1446} {\path{arXiv:0808.1446}},
  \href {https://doi.org/10.1007/JHEP09(2010)016}
  {\path{doi:10.1007/JHEP09(2010)016}}.

\bibitem{Kallosh:2008rr}
R.~Kallosh, T.~Kugo, {The Footprint of E(7(7)) amplitudes of N=8 supergravity},
  JHEP 01 (2009) 072.
\newblock \href {http://arxiv.org/abs/0811.3414} {\path{arXiv:0811.3414}},
  \href {https://doi.org/10.1088/1126-6708/2009/01/072}
  {\path{doi:10.1088/1126-6708/2009/01/072}}.

\bibitem{Green:2010wi}
M.~B. Green, J.~G. Russo, P.~Vanhove, {Automorphic properties of low energy
  string amplitudes in various dimensions}, Phys. Rev. D 81 (2010) 086008.
\newblock \href {http://arxiv.org/abs/1001.2535} {\path{arXiv:1001.2535}},
  \href {https://doi.org/10.1103/PhysRevD.81.086008}
  {\path{doi:10.1103/PhysRevD.81.086008}}.

\bibitem{Pioline:2015yea}
B.~Pioline, {D$^{6}$\ensuremath{\mathscr{R}}$^{4}$ amplitudes in various
  dimensions}, JHEP 04 (2015) 057.
\newblock \href {http://arxiv.org/abs/1502.03377} {\path{arXiv:1502.03377}},
  \href {https://doi.org/10.1007/JHEP04(2015)057}
  {\path{doi:10.1007/JHEP04(2015)057}}.

\bibitem{Bern:2009kd}
Z.~Bern, J.~J. Carrasco, L.~J. Dixon, H.~Johansson, R.~Roiban, {The Ultraviolet
  Behavior of N=8 Supergravity at Four Loops}, Phys. Rev. Lett. 103 (2009)
  081301.
\newblock \href {http://arxiv.org/abs/0905.2326} {\path{arXiv:0905.2326}},
  \href {https://doi.org/10.1103/PhysRevLett.103.081301}
  {\path{doi:10.1103/PhysRevLett.103.081301}}.

\bibitem{Cremmer:1977tc}
E.~Cremmer, J.~Scherk, {Algebraic Simplifications in Supergravity Theories},
  Nucl. Phys. B 127 (1977) 259--268.
\newblock \href {https://doi.org/10.1016/0550-3213(77)90214-0}
  {\path{doi:10.1016/0550-3213(77)90214-0}}.

\bibitem{Das:1977uy}
A.~K. Das, {SO(4) Invariant Extended Supergravity}, Phys. Rev. D 15 (1977)
  2805.
\newblock \href {https://doi.org/10.1103/PhysRevD.15.2805}
  {\path{doi:10.1103/PhysRevD.15.2805}}.

\bibitem{Cremmer:1977tt}
E.~Cremmer, J.~Scherk, S.~Ferrara, {SU(4) Invariant Supergravity Theory}, Phys.
  Lett. B 74 (1978) 61--64.
\newblock \href {https://doi.org/10.1016/0370-2693(78)90060-6}
  {\path{doi:10.1016/0370-2693(78)90060-6}}.

\bibitem{Fradkin:1985am}
E.~S. Fradkin, A.~A. Tseytlin, {Conformal supergravity}, Phys. Rept. 119 (1985)
  233--362.
\newblock \href {https://doi.org/10.1016/0370-1573(85)90138-3}
  {\path{doi:10.1016/0370-1573(85)90138-3}}.

\bibitem{Berkovits:2004jj}
N.~Berkovits, E.~Witten, {Conformal supergravity in twistor-string theory},
  JHEP 08 (2004) 009.
\newblock \href {http://arxiv.org/abs/hep-th/0406051}
  {\path{arXiv:hep-th/0406051}}, \href
  {https://doi.org/10.1088/1126-6708/2004/08/009}
  {\path{doi:10.1088/1126-6708/2004/08/009}}.

\bibitem{Butter:2016mtk}
D.~Butter, F.~Ciceri, B.~de~Wit, B.~Sahoo, {Construction of all N=4 conformal
  supergravities}, Phys. Rev. Lett. 118~(8) (2017) 081602.
\newblock \href {http://arxiv.org/abs/1609.09083} {\path{arXiv:1609.09083}},
  \href {https://doi.org/10.1103/PhysRevLett.118.081602}
  {\path{doi:10.1103/PhysRevLett.118.081602}}.

\bibitem{Butter:2019edc}
D.~Butter, F.~Ciceri, B.~Sahoo, {$N=4$ conformal supergravity: the complete
  actions}, JHEP 01 (2020) 029.
\newblock \href {http://arxiv.org/abs/1910.11874} {\path{arXiv:1910.11874}},
  \href {https://doi.org/10.1007/JHEP01(2020)029}
  {\path{doi:10.1007/JHEP01(2020)029}}.

\bibitem{Howe:1981qj}
P.~S. Howe, K.~S. Stelle, P.~K. Townsend, {Supercurrents}, Nucl. Phys. B 192
  (1981) 332--352.
\newblock \href {https://doi.org/10.1016/0550-3213(81)90429-6}
  {\path{doi:10.1016/0550-3213(81)90429-6}}.

\bibitem{Hegde:2021rte}
S.~Hegde, M.~Mishra, B.~Sahoo, {N = 3 conformal supergravity in four
  dimensions}, JHEP 04 (2022) 001.
\newblock \href {http://arxiv.org/abs/2104.07453} {\path{arXiv:2104.07453}},
  \href {https://doi.org/10.1007/JHEP04(2022)001}
  {\path{doi:10.1007/JHEP04(2022)001}}.

\bibitem{Hegde:2022wnb}
S.~Hegde, M.~Mishra, D.~Mukherjee, B.~Sahoo, {Higher derivative invariants in
  four dimensional \ensuremath{\mathscr{N}} = 3 Poincar\'e supergravity}, JHEP
  02 (2023) 145.
\newblock \href {http://arxiv.org/abs/2211.06628} {\path{arXiv:2211.06628}},
  \href {https://doi.org/10.1007/JHEP02(2023)145}
  {\path{doi:10.1007/JHEP02(2023)145}}.

\bibitem{deWit:1980lyi}
B.~de~Wit, J.~W. van Holten, A.~Van~Proeyen, {Structure of N=2 Supergravity},
  Nucl. Phys. B 184 (1981) 77, [Erratum: Nucl.Phys.B 222, 516 (1983)].
\newblock \href {https://doi.org/10.1016/0550-3213(83)90548-5}
  {\path{doi:10.1016/0550-3213(83)90548-5}}.

\bibitem{deWit:2006gn}
B.~de~Wit, F.~Saueressig, {Off-shell N=2 tensor supermultiplets}, JHEP 09
  (2006) 062.
\newblock \href {http://arxiv.org/abs/hep-th/0606148}
  {\path{arXiv:hep-th/0606148}}, \href
  {https://doi.org/10.1088/1126-6708/2006/09/062}
  {\path{doi:10.1088/1126-6708/2006/09/062}}.

\bibitem{Bobev:2021oku}
N.~Bobev, A.~M. Charles, K.~Hristov, V.~Reys, {Higher-derivative supergravity,
  AdS$_{4}$ holography, and black holes}, JHEP 08 (2021) 173.
\newblock \href {http://arxiv.org/abs/2106.04581} {\path{arXiv:2106.04581}},
  \href {https://doi.org/10.1007/JHEP08(2021)173}
  {\path{doi:10.1007/JHEP08(2021)173}}.

\bibitem{Butter:2013lta}
D.~Butter, B.~de~Wit, S.~M. Kuzenko, I.~Lodato, {New higher-derivative
  invariants in N=2 supergravity and the Gauss-Bonnet term}, JHEP 12 (2013)
  062.
\newblock \href {http://arxiv.org/abs/1307.6546} {\path{arXiv:1307.6546}},
  \href {https://doi.org/10.1007/JHEP12(2013)062}
  {\path{doi:10.1007/JHEP12(2013)062}}.

\bibitem{Charles:2016wjs}
A.~M. Charles, F.~Larsen, {Kerr-Newman Black Holes with String Corrections},
  JHEP 10 (2016) 142.
\newblock \href {http://arxiv.org/abs/1605.07622} {\path{arXiv:1605.07622}},
  \href {https://doi.org/10.1007/JHEP10(2016)142}
  {\path{doi:10.1007/JHEP10(2016)142}}.

\bibitem{Charles:2017dbr}
A.~M. Charles, F.~Larsen, D.~R. Mayerson, {Non-Renormalization For
  Non-Supersymmetric Black Holes}, JHEP 08 (2017) 048.
\newblock \href {http://arxiv.org/abs/1702.08458} {\path{arXiv:1702.08458}},
  \href {https://doi.org/10.1007/JHEP08(2017)048}
  {\path{doi:10.1007/JHEP08(2017)048}}.

\bibitem{Bobev:2020egg}
N.~Bobev, A.~M. Charles, K.~Hristov, V.~Reys, {The Unreasonable Effectiveness
  of Higher-Derivative Supergravity in AdS$_4$ Holography}, Phys. Rev. Lett.
  125~(13) (2020) 131601.
\newblock \href {http://arxiv.org/abs/2006.09390} {\path{arXiv:2006.09390}},
  \href {https://doi.org/10.1103/PhysRevLett.125.131601}
  {\path{doi:10.1103/PhysRevLett.125.131601}}.

\bibitem{Alday:2021ymb}
L.~F. Alday, S.~M. Chester, H.~Raj, {ABJM at strong coupling from M-theory,
  localization, and Lorentzian inversion}, JHEP 02 (2022) 005.
\newblock \href {http://arxiv.org/abs/2107.10274} {\path{arXiv:2107.10274}},
  \href {https://doi.org/10.1007/JHEP02(2022)005}
  {\path{doi:10.1007/JHEP02(2022)005}}.

\bibitem{Alday:2022rly}
L.~F. Alday, S.~M. Chester, H.~Raj, {M-theory on AdS$_{4}$ $\times$ S$^{7}$ at
  1-loop and beyond}, JHEP 11 (2022) 091.
\newblock \href {http://arxiv.org/abs/2207.11138} {\path{arXiv:2207.11138}},
  \href {https://doi.org/10.1007/JHEP11(2022)091}
  {\path{doi:10.1007/JHEP11(2022)091}}.

\bibitem{Moura:2002ip}
F.~Moura, {Four-dimensional 'old minimal' N=2 supersymmetrization of R**4},
  JHEP 07 (2003) 057.
\newblock \href {http://arxiv.org/abs/hep-th/0212271}
  {\path{arXiv:hep-th/0212271}}, \href
  {https://doi.org/10.1088/1126-6708/2003/07/057}
  {\path{doi:10.1088/1126-6708/2003/07/057}}.

\bibitem{Mishra:2020jlc}
M.~Mishra, B.~Sahoo, {Curvature squared action in four dimensional $N = 2$
  supergravity using the dilaton Weyl multiplet}, JHEP 04 (2021) 027.
\newblock \href {http://arxiv.org/abs/2012.03760} {\path{arXiv:2012.03760}},
  \href {https://doi.org/10.1007/JHEP04(2021)027}
  {\path{doi:10.1007/JHEP04(2021)027}}.

\bibitem{Wald:1993nt}
R.~M. Wald, {Black hole entropy is the Noether charge}, Phys. Rev. D 48~(8)
  (1993) R3427--R3431.
\newblock \href {http://arxiv.org/abs/gr-qc/9307038}
  {\path{arXiv:gr-qc/9307038}}, \href
  {https://doi.org/10.1103/PhysRevD.48.R3427}
  {\path{doi:10.1103/PhysRevD.48.R3427}}.

\bibitem{LopesCardoso:1999fsj}
G.~Lopes~Cardoso, B.~de~Wit, T.~Mohaupt, {Macroscopic entropy formulae and
  nonholomorphic corrections for supersymmetric black holes}, Nucl. Phys. B 567
  (2000) 87--110.
\newblock \href {http://arxiv.org/abs/hep-th/9906094}
  {\path{arXiv:hep-th/9906094}}, \href
  {https://doi.org/10.1016/S0550-3213(99)00560-X}
  {\path{doi:10.1016/S0550-3213(99)00560-X}}.

\bibitem{Sahoo:2006rp}
B.~Sahoo, A.~Sen, {Higher derivative corrections to non-supersymmetric extremal
  black holes in N=2 supergravity}, JHEP 09 (2006) 029.
\newblock \href {http://arxiv.org/abs/hep-th/0603149}
  {\path{arXiv:hep-th/0603149}}, \href
  {https://doi.org/10.1088/1126-6708/2006/09/029}
  {\path{doi:10.1088/1126-6708/2006/09/029}}.

\bibitem{Behrndt:1998eq}
K.~Behrndt, G.~Lopes~Cardoso, B.~de~Wit, D.~Lust, T.~Mohaupt, W.~A. Sabra,
  {Higher order black hole solutions in N=2 supergravity and Calabi-Yau string
  backgrounds}, Phys. Lett. B 429 (1998) 289--296.
\newblock \href {http://arxiv.org/abs/hep-th/9801081}
  {\path{arXiv:hep-th/9801081}}, \href
  {https://doi.org/10.1016/S0370-2693(98)00413-4}
  {\path{doi:10.1016/S0370-2693(98)00413-4}}.

\bibitem{LopesCardoso:1998tkj}
G.~Lopes~Cardoso, B.~de~Wit, T.~Mohaupt, {Corrections to macroscopic
  supersymmetric black hole entropy}, Phys. Lett. B 451 (1999) 309--316.
\newblock \href {http://arxiv.org/abs/hep-th/9812082}
  {\path{arXiv:hep-th/9812082}}, \href
  {https://doi.org/10.1016/S0370-2693(99)00227-0}
  {\path{doi:10.1016/S0370-2693(99)00227-0}}.

\bibitem{LopesCardoso:1999cv}
G.~Lopes~Cardoso, B.~de~Wit, T.~Mohaupt, {Deviations from the area law for
  supersymmetric black holes}, Fortsch. Phys. 48 (2000) 49--64.
\newblock \href {http://arxiv.org/abs/hep-th/9904005}
  {\path{arXiv:hep-th/9904005}}.

\bibitem{LopesCardoso:1999xn}
G.~Lopes~Cardoso, B.~de~Wit, T.~Mohaupt, {Area law corrections from state
  counting and supergravity}, Class. Quant. Grav. 17 (2000) 1007--1015.
\newblock \href {http://arxiv.org/abs/hep-th/9910179}
  {\path{arXiv:hep-th/9910179}}, \href
  {https://doi.org/10.1088/0264-9381/17/5/310}
  {\path{doi:10.1088/0264-9381/17/5/310}}.

\bibitem{LopesCardoso:2000qm}
G.~Lopes~Cardoso, B.~de~Wit, J.~Kappeli, T.~Mohaupt, {Stationary BPS solutions
  in N=2 supergravity with R**2 interactions}, JHEP 12 (2000) 019.
\newblock \href {http://arxiv.org/abs/hep-th/0009234}
  {\path{arXiv:hep-th/0009234}}, \href
  {https://doi.org/10.1088/1126-6708/2000/12/019}
  {\path{doi:10.1088/1126-6708/2000/12/019}}.

\bibitem{LopesCardoso:2000fp}
G.~Lopes~Cardoso, B.~de~Wit, J.~Kappeli, T.~Mohaupt, {Examples of stationary
  BPS solutions in N=2 supergravity theories with R**2 interactions}, Fortsch.
  Phys. 49 (2001) 557--563.
\newblock \href {http://arxiv.org/abs/hep-th/0012232}
  {\path{arXiv:hep-th/0012232}}, \href
  {https://doi.org/10.1002/1521-3978(200105)49:4/6<557::AID-PROP557>3.0.CO;2-2}
  {\path{doi:10.1002/1521-3978(200105)49:4/6<557::AID-PROP557>3.0.CO;2-2}}.

\bibitem{Vafa:1997gr}
C.~Vafa, {Black holes and Calabi-Yau threefolds}, Adv. Theor. Math. Phys. 2
  (1998) 207--218.
\newblock \href {http://arxiv.org/abs/hep-th/9711067}
  {\path{arXiv:hep-th/9711067}}, \href
  {https://doi.org/10.4310/ATMP.1998.v2.n1.a8}
  {\path{doi:10.4310/ATMP.1998.v2.n1.a8}}.

\bibitem{Hubeny:2004ji}
V.~Hubeny, A.~Maloney, M.~Rangamani, {String-corrected black holes}, JHEP 05
  (2005) 035.
\newblock \href {http://arxiv.org/abs/hep-th/0411272}
  {\path{arXiv:hep-th/0411272}}, \href
  {https://doi.org/10.1088/1126-6708/2005/05/035}
  {\path{doi:10.1088/1126-6708/2005/05/035}}.

\bibitem{Dabholkar:2004dq}
A.~Dabholkar, R.~Kallosh, A.~Maloney, {A Stringy cloak for a classical
  singularity}, JHEP 12 (2004) 059.
\newblock \href {http://arxiv.org/abs/hep-th/0410076}
  {\path{arXiv:hep-th/0410076}}, \href
  {https://doi.org/10.1088/1126-6708/2004/12/059}
  {\path{doi:10.1088/1126-6708/2004/12/059}}.

\bibitem{Chimento:2018kop}
S.~Chimento, P.~Meessen, T.~Ortin, P.~F. Ramirez, A.~Ruiperez, {On a family of
  $\alpha'$-corrected solutions of the Heterotic Superstring effective action},
  JHEP 07 (2018) 080.
\newblock \href {http://arxiv.org/abs/1803.04463} {\path{arXiv:1803.04463}},
  \href {https://doi.org/10.1007/JHEP07(2018)080}
  {\path{doi:10.1007/JHEP07(2018)080}}.

\bibitem{Cano:2018brq}
P.~A. Cano, S.~Chimento, P.~Meessen, T.~Ort\'\i{}n, P.~F. Ram\'\i{}rez,
  A.~Ruip\'erez, {Beyond the near-horizon limit: Stringy corrections to
  Heterotic Black Holes}, JHEP 02 (2019) 192.
\newblock \href {http://arxiv.org/abs/1808.03651} {\path{arXiv:1808.03651}},
  \href {https://doi.org/10.1007/JHEP02(2019)192}
  {\path{doi:10.1007/JHEP02(2019)192}}.

\bibitem{Cano:2018hut}
P.~A. Cano, P.~F. Ram\'\i{}rez, A.~Ruip\'erez, {The small black hole illusion},
  JHEP 03 (2020) 115.
\newblock \href {http://arxiv.org/abs/1808.10449} {\path{arXiv:1808.10449}},
  \href {https://doi.org/10.1007/JHEP03(2020)115}
  {\path{doi:10.1007/JHEP03(2020)115}}.

\bibitem{Cano:2021dyy}
P.~A. Cano, A.~Murcia, P.~F. Ram\'\i{}rez, A.~Ruip\'erez, {On small black
  holes, KK monopoles and solitonic 5-branes}, JHEP 05 (2021) 272.
\newblock \href {http://arxiv.org/abs/2102.04476} {\path{arXiv:2102.04476}},
  \href {https://doi.org/10.1007/JHEP05(2021)272}
  {\path{doi:10.1007/JHEP05(2021)272}}.

\bibitem{Massai:2023cis}
S.~Massai, A.~Ruip\'erez, M.~Zatti, {Revisiting $\alpha'$ corrections to
  heterotic two-charge black holes} (11 2023).
\newblock \href {http://arxiv.org/abs/2311.03308} {\path{arXiv:2311.03308}}.

\bibitem{Muller:1985vga}
M.~Muller, {Supergravity in U(1) Superspace With a Two Form Gauge Potential},
  Nucl. Phys. B 264 (1986) 292--316.
\newblock \href {https://doi.org/10.1016/0550-3213(86)90484-0}
  {\path{doi:10.1016/0550-3213(86)90484-0}}.

\bibitem{deWit:1978ww}
B.~de~Wit, P.~van Nieuwenhuizen, {The Auxiliary Field Structure in Chirally
  Extended Supergravity}, Nucl. Phys. B 139 (1978) 216--220.
\newblock \href {https://doi.org/10.1016/0550-3213(78)90188-8}
  {\path{doi:10.1016/0550-3213(78)90188-8}}.

\bibitem{Ferrara:1978em}
S.~Ferrara, P.~van Nieuwenhuizen, {The Auxiliary Fields of Supergravity}, Phys.
  Lett. B 74 (1978) 333.
\newblock \href {https://doi.org/10.1016/0370-2693(78)90670-6}
  {\path{doi:10.1016/0370-2693(78)90670-6}}.

\bibitem{Stelle:1978ye}
K.~S. Stelle, P.~C. West, {Minimal Auxiliary Fields for Supergravity}, Phys.
  Lett. B 74 (1978) 330--332.
\newblock \href {https://doi.org/10.1016/0370-2693(78)90669-X}
  {\path{doi:10.1016/0370-2693(78)90669-X}}.

\bibitem{Sohnius:1981tp}
M.~F. Sohnius, P.~C. West, {An Alternative Minimal Off-Shell Version of N=1
  Supergravity}, Phys. Lett. B 105 (1981) 353--357.
\newblock \href {https://doi.org/10.1016/0370-2693(81)90778-4}
  {\path{doi:10.1016/0370-2693(81)90778-4}}.

\bibitem{LeDu:1997us}
R.~Le~Du, {Higher derivative supergravity in U(1) superspace}, Eur. Phys. J. C
  5 (1998) 181--187.
\newblock \href {http://arxiv.org/abs/hep-th/9706058}
  {\path{arXiv:hep-th/9706058}}, \href {https://doi.org/10.1007/s100520050260}
  {\path{doi:10.1007/s100520050260}}.

\bibitem{deRoo:1990zm}
M.~de~Roo, A.~Wiedemann, E.~Zijlstra, {The Construction of $R^{2}$ Actions in
  $D=4$, $N=1$ Supergravity}, Class. Quant. Grav. 7 (1990) 1181--1196.
\newblock \href {https://doi.org/10.1088/0264-9381/7/7/014}
  {\path{doi:10.1088/0264-9381/7/7/014}}.

\bibitem{Freedman:1976uk}
D.~Z. Freedman, {Supergravity with Axial Gauge Invariance}, Phys. Rev. D 15
  (1977) 1173.
\newblock \href {https://doi.org/10.1103/PhysRevD.15.1173}
  {\path{doi:10.1103/PhysRevD.15.1173}}.

\bibitem{Lu:2011mw}
H.~Lu, C.~N. Pope, E.~Sezgin, L.~Wulff, {Critical and Non-Critical
  Einstein-Weyl Supergravity}, JHEP 10 (2011) 131.
\newblock \href {http://arxiv.org/abs/1107.2480} {\path{arXiv:1107.2480}},
  \href {https://doi.org/10.1007/JHEP10(2011)131}
  {\path{doi:10.1007/JHEP10(2011)131}}.

\bibitem{Lu:2011ks}
H.~Lu, Y.~Pang, C.~N. Pope, {Conformal Gravity and Extensions of Critical
  Gravity}, Phys. Rev. D 84 (2011) 064001.
\newblock \href {http://arxiv.org/abs/1106.4657} {\path{arXiv:1106.4657}},
  \href {https://doi.org/10.1103/PhysRevD.84.064001}
  {\path{doi:10.1103/PhysRevD.84.064001}}.

\bibitem{Cremmer:1982en}
E.~Cremmer, S.~Ferrara, L.~Girardello, A.~Van~Proeyen, {Yang-Mills Theories
  with Local Supersymmetry: Lagrangian, Transformation Laws and SuperHiggs
  Effect}, Nucl. Phys. B 212 (1983) 413.
\newblock \href {https://doi.org/10.1016/0550-3213(83)90679-X}
  {\path{doi:10.1016/0550-3213(83)90679-X}}.

\bibitem{Koehn:2012ar}
M.~Koehn, J.-L. Lehners, B.~A. Ovrut, {Higher-Derivative Chiral Superfield
  Actions Coupled to N=1 Supergravity}, Phys. Rev. D 86 (2012) 085019.
\newblock \href {http://arxiv.org/abs/1207.3798} {\path{arXiv:1207.3798}},
  \href {https://doi.org/10.1103/PhysRevD.86.085019}
  {\path{doi:10.1103/PhysRevD.86.085019}}.

\bibitem{Farakos:2012qu}
F.~Farakos, A.~Kehagias, {Emerging Potentials in Higher-Derivative Gauged
  Chiral Models Coupled to N=1 Supergravity}, JHEP 11 (2012) 077.
\newblock \href {http://arxiv.org/abs/1207.4767} {\path{arXiv:1207.4767}},
  \href {https://doi.org/10.1007/JHEP11(2012)077}
  {\path{doi:10.1007/JHEP11(2012)077}}.

\bibitem{Moura:2002ft}
F.~Moura, {Four-dimensional R**4 superinvariants through gauge completion},
  JHEP 08 (2002) 038.
\newblock \href {http://arxiv.org/abs/hep-th/0206119}
  {\path{arXiv:hep-th/0206119}}, \href
  {https://doi.org/10.1088/1126-6708/2002/08/038}
  {\path{doi:10.1088/1126-6708/2002/08/038}}.

\bibitem{Starobinsky:1980te}
A.~A. Starobinsky, {A New Type of Isotropic Cosmological Models Without
  Singularity}, Phys. Lett. B 91 (1980) 99--102.
\newblock \href {https://doi.org/10.1016/0370-2693(80)90670-X}
  {\path{doi:10.1016/0370-2693(80)90670-X}}.

\bibitem{DeFelice:2010aj}
A.~De~Felice, S.~Tsujikawa, {f(R) theories}, Living Rev. Rel. 13 (2010) 3.
\newblock \href {http://arxiv.org/abs/1002.4928} {\path{arXiv:1002.4928}},
  \href {https://doi.org/10.12942/lrr-2010-3} {\path{doi:10.12942/lrr-2010-3}}.

\bibitem{Cecotti:1987sa}
S.~Cecotti, {Higher derivative supergravity is equivalent to standard
  supergravity coupled to matter. 1.}, Phys. Lett. B 190 (1987) 86--92.
\newblock \href {https://doi.org/10.1016/0370-2693(87)90844-6}
  {\path{doi:10.1016/0370-2693(87)90844-6}}.

\bibitem{Farakos:2013cqa}
F.~Farakos, A.~Kehagias, A.~Riotto, {On the Starobinsky Model of Inflation from
  Supergravity}, Nucl. Phys. B 876 (2013) 187--200.
\newblock \href {http://arxiv.org/abs/1307.1137} {\path{arXiv:1307.1137}},
  \href {https://doi.org/10.1016/j.nuclphysb.2013.08.005}
  {\path{doi:10.1016/j.nuclphysb.2013.08.005}}.

\bibitem{Ozkan:2014cua}
M.~Ozkan, Y.~Pang, {$R^n$ Extension of Starobinsky Model in Old Minimal
  Supergravity}, Class. Quant. Grav. 31 (2014) 205004.
\newblock \href {http://arxiv.org/abs/1402.5427} {\path{arXiv:1402.5427}},
  \href {https://doi.org/10.1088/0264-9381/31/20/205004}
  {\path{doi:10.1088/0264-9381/31/20/205004}}.

\bibitem{Kallosh:2013lkr}
R.~Kallosh, A.~Linde, {Superconformal generalizations of the Starobinsky
  model}, JCAP 06 (2013) 028.
\newblock \href {http://arxiv.org/abs/1306.3214} {\path{arXiv:1306.3214}},
  \href {https://doi.org/10.1088/1475-7516/2013/06/028}
  {\path{doi:10.1088/1475-7516/2013/06/028}}.

\bibitem{Antoniadis:2014oya}
I.~Antoniadis, E.~Dudas, S.~Ferrara, A.~Sagnotti, {The
  Volkov\textendash{}Akulov\textendash{}Starobinsky supergravity}, Phys. Lett.
  B 733 (2014) 32--35.
\newblock \href {http://arxiv.org/abs/1403.3269} {\path{arXiv:1403.3269}},
  \href {https://doi.org/10.1016/j.physletb.2014.04.015}
  {\path{doi:10.1016/j.physletb.2014.04.015}}.


\bibitem{Rocek:1978nb}
M.~Rocek, {Linearizing the Volkov-Akulov Model}, Phys. Rev. Lett. 41 (1978)
  451--453.
\newblock \href {https://doi.org/10.1103/PhysRevLett.41.451}
  {\path{doi:10.1103/PhysRevLett.41.451}}.

\bibitem{Lindstrom:1979kq}
U.~Lindstrom, M.~Rocek, {CONSTRAINED LOCAL SUPERFIELDS}, Phys. Rev. D 19 (1979)
  2300--2303.
\newblock \href {https://doi.org/10.1103/PhysRevD.19.2300}
  {\path{doi:10.1103/PhysRevD.19.2300}}.

\bibitem{Casalbuoni:1988xh}
R.~Casalbuoni, S.~De~Curtis, D.~Dominici, F.~Feruglio, R.~Gatto, {Nonlinear
  Realization of Supersymmetry Algebra From Supersymmetric Constraint}, Phys.
  Lett. B 220 (1989) 569--575.
\newblock \href {https://doi.org/10.1016/0370-2693(89)90788-0}
  {\path{doi:10.1016/0370-2693(89)90788-0}}.

\bibitem{Ferrara:2013kca}
S.~Ferrara, R.~Kallosh, A.~Linde, M.~Porrati, {Higher Order Corrections in
  Minimal Supergravity Models of Inflation}, JCAP 11 (2013) 046.
\newblock \href {http://arxiv.org/abs/1309.1085} {\path{arXiv:1309.1085}},
  \href {https://doi.org/10.1088/1475-7516/2013/11/046}
  {\path{doi:10.1088/1475-7516/2013/11/046}}.

\bibitem{Samtleben:2012gy}
H.~Samtleben, D.~Tsimpis, {Rigid supersymmetric theories in 4d Riemannian
  space}, JHEP 05 (2012) 132.
\newblock \href {http://arxiv.org/abs/1203.3420} {\path{arXiv:1203.3420}},
  \href {https://doi.org/10.1007/JHEP05(2012)132}
  {\path{doi:10.1007/JHEP05(2012)132}}.

\bibitem{Dumitrescu:2012ha}
T.~T. Dumitrescu, G.~Festuccia, N.~Seiberg, {Exploring Curved Superspace}, JHEP
  08 (2012) 141.
\newblock \href {http://arxiv.org/abs/1205.1115} {\path{arXiv:1205.1115}},
  \href {https://doi.org/10.1007/JHEP08(2012)141}
  {\path{doi:10.1007/JHEP08(2012)141}}.

\bibitem{Kehagias:1997cq}
A.~Kehagias, H.~Partouche, {On the exact quartic effective action for the type
  IIB superstring}, Phys. Lett. B 422 (1998) 109--116.
\newblock \href {http://arxiv.org/abs/hep-th/9710023}
  {\path{arXiv:hep-th/9710023}}, \href
  {https://doi.org/10.1016/S0370-2693(97)01430-5}
  {\path{doi:10.1016/S0370-2693(97)01430-5}}.

\bibitem{Klare:2012gn}
C.~Klare, A.~Tomasiello, A.~Zaffaroni, {Supersymmetry on Curved Spaces and
  Holography}, JHEP 08 (2012) 061.
\newblock \href {http://arxiv.org/abs/1205.1062} {\path{arXiv:1205.1062}},
  \href {https://doi.org/10.1007/JHEP08(2012)061}
  {\path{doi:10.1007/JHEP08(2012)061}}.

\bibitem{Klare:2013dka}
C.~Klare, A.~Zaffaroni, {Extended Supersymmetry on Curved Spaces}, JHEP 10
  (2013) 218.
\newblock \href {http://arxiv.org/abs/1308.1102} {\path{arXiv:1308.1102}},
  \href {https://doi.org/10.1007/JHEP10(2013)218}
  {\path{doi:10.1007/JHEP10(2013)218}}.

\bibitem{Butter:2015tra}
D.~Butter, G.~Inverso, I.~Lodato, {Rigid 4D $ \mathcal{N}=2 $ supersymmetric
  backgrounds and actions}, JHEP 09 (2015) 088.
\newblock \href {http://arxiv.org/abs/1505.03500} {\path{arXiv:1505.03500}},
  \href {https://doi.org/10.1007/JHEP09(2015)088}
  {\path{doi:10.1007/JHEP09(2015)088}}.

\bibitem{Gupta:2012cy}
R.~K. Gupta, S.~Murthy, {All solutions of the localization equations for N=2
  quantum black hole entropy}, JHEP 02 (2013) 141.
\newblock \href {http://arxiv.org/abs/1208.6221} {\path{arXiv:1208.6221}},
  \href {https://doi.org/10.1007/JHEP02(2013)141}
  {\path{doi:10.1007/JHEP02(2013)141}}.

\bibitem{Festuccia:2011ws}
G.~Festuccia, N.~Seiberg, {Rigid Supersymmetric Theories in Curved Superspace},
  JHEP 06 (2011) 114.
\newblock \href {http://arxiv.org/abs/1105.0689} {\path{arXiv:1105.0689}},
  \href {https://doi.org/10.1007/JHEP06(2011)114}
  {\path{doi:10.1007/JHEP06(2011)114}}.

\bibitem{Liu:2012bi}
J.~T. Liu, L.~A. Pando~Zayas, D.~Reichmann, {Rigid Supersymmetric Backgrounds
  of Minimal Off-Shell Supergravity}, JHEP 10 (2012) 034.
\newblock \href {http://arxiv.org/abs/1207.2785} {\path{arXiv:1207.2785}},
  \href {https://doi.org/10.1007/JHEP10(2012)034}
  {\path{doi:10.1007/JHEP10(2012)034}}.

\bibitem{Lu:2012am}
H.~Lu, Z.-L. Wang, {Supersymmetric Asymptotic AdS and Lifshitz Solutions in
  Einstein-Weyl and Conformal Supergravities}, JHEP 08 (2012) 012.
\newblock \href {http://arxiv.org/abs/1205.2092} {\path{arXiv:1205.2092}},
  \href {https://doi.org/10.1007/JHEP08(2012)012}
  {\path{doi:10.1007/JHEP08(2012)012}}.

\bibitem{Lu:2012cz}
H.~Lu, C.~N. Pope, {Gyrating Schrodinger Geometries and Non-Relativistic Field
  Theories}, Phys. Rev. D 86 (2012) 061501.
\newblock \href {http://arxiv.org/abs/1206.6510} {\path{arXiv:1206.6510}},
  \href {https://doi.org/10.1103/PhysRevD.86.061501}
  {\path{doi:10.1103/PhysRevD.86.061501}}.

\bibitem{Liu:2012mh}
H.-S. Liu, H.~L\"u, Y.~Pang, C.~N. Pope, {Supersymmetric Solutions in
  Four-Dimensional Off-Shell Curvature-Squared Supergravity}, Phys. Rev. D
  87~(6) (2013) 065014.
\newblock \href {http://arxiv.org/abs/1209.6065} {\path{arXiv:1209.6065}},
  \href {https://doi.org/10.1103/PhysRevD.87.065014}
  {\path{doi:10.1103/PhysRevD.87.065014}}.

\bibitem{Cassani:2012ri}
D.~Cassani, C.~Klare, D.~Martelli, A.~Tomasiello, A.~Zaffaroni, {Supersymmetry
  in Lorentzian Curved Spaces and Holography}, Commun. Math. Phys. 327 (2014)
  577--602.
\newblock \href {http://arxiv.org/abs/1207.2181} {\path{arXiv:1207.2181}},
  \href {https://doi.org/10.1007/s00220-014-1983-3}
  {\path{doi:10.1007/s00220-014-1983-3}}.

\bibitem{Marcus:1983hb}
N.~Marcus, J.~H. Schwarz, {Three-Dimensional Supergravity Theories}, Nucl.
  Phys. B 228 (1983) 145.
\newblock \href {https://doi.org/10.1016/0550-3213(83)90402-9}
  {\path{doi:10.1016/0550-3213(83)90402-9}}.

\bibitem{deWit:1992psp}
B.~de~Wit, A.~K. Tollsten, H.~Nicolai, {Locally supersymmetric D = 3 nonlinear
  sigma models}, Nucl. Phys. B 392 (1993) 3--38.
\newblock \href {http://arxiv.org/abs/hep-th/9208074}
  {\path{arXiv:hep-th/9208074}}, \href
  {https://doi.org/10.1016/0550-3213(93)90195-U}
  {\path{doi:10.1016/0550-3213(93)90195-U}}.

\bibitem{deWit:2003ja}
B.~de~Wit, I.~Herger, H.~Samtleben, {Gauged locally supersymmetric D = 3
  nonlinear sigma models}, Nucl. Phys. B 671 (2003) 175--216.
\newblock \href {http://arxiv.org/abs/hep-th/0307006}
  {\path{arXiv:hep-th/0307006}}, \href
  {https://doi.org/10.1016/j.nuclphysb.2003.08.022}
  {\path{doi:10.1016/j.nuclphysb.2003.08.022}}.

\bibitem{Bergshoeff:2010ui}
E.~A. Bergshoeff, O.~Hohm, J.~Rosseel, P.~K. Townsend, {On Maximal Massive 3D
  Supergravity}, Class. Quant. Grav. 27 (2010) 235012.
\newblock \href {http://arxiv.org/abs/1007.4075} {\path{arXiv:1007.4075}},
  \href {https://doi.org/10.1088/0264-9381/27/23/235012}
  {\path{doi:10.1088/0264-9381/27/23/235012}}.

\bibitem{Gupta:2007th}
R.~K. Gupta, A.~Sen, {Consistent Truncation to Three Dimensional
  (Super-)gravity}, JHEP 03 (2008) 015.
\newblock \href {http://arxiv.org/abs/0710.4177} {\path{arXiv:0710.4177}},
  \href {https://doi.org/10.1088/1126-6708/2008/03/015}
  {\path{doi:10.1088/1126-6708/2008/03/015}}.

\bibitem{Bergshoeff:2010mf}
E.~A. Bergshoeff, O.~Hohm, J.~Rosseel, E.~Sezgin, P.~K. Townsend, {More on
  Massive 3D Supergravity}, Class. Quant. Grav. 28 (2011) 015002.
\newblock \href {http://arxiv.org/abs/1005.3952} {\path{arXiv:1005.3952}},
  \href {https://doi.org/10.1088/0264-9381/28/1/015002}
  {\path{doi:10.1088/0264-9381/28/1/015002}}.

\bibitem{Bergshoeff:2014ida}
E.~Bergshoeff, M.~Ozkan, {3D Born-Infeld Gravity and Supersymmetry}, JHEP 08
  (2014) 149.
\newblock \href {http://arxiv.org/abs/1405.6212} {\path{arXiv:1405.6212}},
  \href {https://doi.org/10.1007/JHEP08(2014)149}
  {\path{doi:10.1007/JHEP08(2014)149}}.

\bibitem{Andringa:2009yc}
R.~Andringa, E.~A. Bergshoeff, M.~de~Roo, O.~Hohm, E.~Sezgin, P.~K. Townsend,
  {Massive 3D Supergravity}, Class. Quant. Grav. 27 (2010) 025010.
\newblock \href {http://arxiv.org/abs/0907.4658} {\path{arXiv:0907.4658}},
  \href {https://doi.org/10.1088/0264-9381/27/2/025010}
  {\path{doi:10.1088/0264-9381/27/2/025010}}.

\bibitem{Bergshoeff:2015zga}
E.~Bergshoeff, W.~Merbis, A.~J. Routh, P.~K. Townsend, {The Third Way to 3D
  Gravity}, Int. J. Mod. Phys. D 24~(12) (2015) 1544015.
\newblock \href {http://arxiv.org/abs/1506.05949} {\path{arXiv:1506.05949}},
  \href {https://doi.org/10.1142/S0218271815440150}
  {\path{doi:10.1142/S0218271815440150}}.

\bibitem{Ozkan:2018cxj}
M.~\"Ozkan, Y.~Pang, P.~K. Townsend, {Exotic Massive 3D Gravity}, JHEP 08
  (2018) 035.
\newblock \href {http://arxiv.org/abs/1806.04179} {\path{arXiv:1806.04179}},
  \href {https://doi.org/10.1007/JHEP08(2018)035}
  {\path{doi:10.1007/JHEP08(2018)035}}.

\bibitem{Deger:2022gim}
N.~S. Deger, M.~Geiller, J.~Rosseel, H.~Samtleben, {Minimal Massive
  Supergravity}, Phys. Rev. Lett. 129~(17) (2022) 171601.
\newblock \href {http://arxiv.org/abs/2206.00675} {\path{arXiv:2206.00675}},
  \href {https://doi.org/10.1103/PhysRevLett.129.171601}
  {\path{doi:10.1103/PhysRevLett.129.171601}}.

\bibitem{Deger:2023eah}
N.~S. Deger, M.~Geiller, J.~Rosseel, H.~Samtleben, {Minimal massive
  supergravity and new theories of massive gravity}, Phys. Rev. D 109~(8)
  (2024) 086014.
\newblock \href {http://arxiv.org/abs/2312.12387} {\path{arXiv:2312.12387}},
  \href {https://doi.org/10.1103/PhysRevD.109.086014}
  {\path{doi:10.1103/PhysRevD.109.086014}}.

\bibitem{Bergshoeff:2014pca}
E.~Bergshoeff, O.~Hohm, W.~Merbis, A.~J. Routh, P.~K. Townsend, {Minimal
  Massive 3D Gravity}, Class. Quant. Grav. 31 (2014) 145008.
\newblock \href {http://arxiv.org/abs/1404.2867} {\path{arXiv:1404.2867}},
  \href {https://doi.org/10.1088/0264-9381/31/14/145008}
  {\path{doi:10.1088/0264-9381/31/14/145008}}.

\bibitem{Alkac:2014hwa}
G.~Alka\c{c}, L.~Basanisi, E.~A. Bergshoeff, M.~Ozkan, E.~Sezgin, {Massive $
  \mathcal{N} $ = 2 supergravity in three dimensions}, JHEP 02 (2015) 125.
\newblock \href {http://arxiv.org/abs/1412.3118} {\path{arXiv:1412.3118}},
  \href {https://doi.org/10.1007/JHEP02(2015)125}
  {\path{doi:10.1007/JHEP02(2015)125}}.

\bibitem{Lauf:2016sac}
F.~Lauf, I.~Sachs, {Topologically massive gravity with extended supersymmetry},
  Phys. Rev. D 94~(6) (2016) 065028.
\newblock \href {http://arxiv.org/abs/1605.00103} {\path{arXiv:1605.00103}},
  \href {https://doi.org/10.1103/PhysRevD.94.065028}
  {\path{doi:10.1103/PhysRevD.94.065028}}.

\bibitem{Butter:2013rba}
D.~Butter, S.~M. Kuzenko, J.~Novak, G.~Tartaglino-Mazzucchelli, {Conformal
  supergravity in three dimensions: Off-shell actions}, JHEP 10 (2013) 073.
\newblock \href {http://arxiv.org/abs/1306.1205} {\path{arXiv:1306.1205}},
  \href {https://doi.org/10.1007/JHEP10(2013)073}
  {\path{doi:10.1007/JHEP10(2013)073}}.

\bibitem{Li:2008dq}
W.~Li, W.~Song, A.~Strominger, {Chiral Gravity in Three Dimensions}, JHEP 04
  (2008) 082.
\newblock \href {http://arxiv.org/abs/0801.4566} {\path{arXiv:0801.4566}},
  \href {https://doi.org/10.1088/1126-6708/2008/04/082}
  {\path{doi:10.1088/1126-6708/2008/04/082}}.

\bibitem{Lauf:2017nfa}
F.~Lauf, I.~Sachs, {Complete superspace classification of three-dimensional
  Chern-Simons-matter theories coupled to supergravity}, JHEP 02 (2018) 154.
\newblock \href {http://arxiv.org/abs/1709.01461} {\path{arXiv:1709.01461}},
  \href {https://doi.org/10.1007/JHEP02(2018)154}
  {\path{doi:10.1007/JHEP02(2018)154}}.

\bibitem{Chu:2009gi}
X.~Chu, B.~E.~W. Nilsson, {Three-dimensional topologically gauged N=6 ABJM type
  theories}, JHEP 06 (2010) 057.
\newblock \href {http://arxiv.org/abs/0906.1655} {\path{arXiv:0906.1655}},
  \href {https://doi.org/10.1007/JHEP06(2010)057}
  {\path{doi:10.1007/JHEP06(2010)057}}.

\bibitem{Nishimura:2013poa}
M.~Nishimura, Y.~Tanii, {N=6 conformal supergravity in three dimensions}, JHEP
  10 (2013) 123.
\newblock \href {http://arxiv.org/abs/1308.3960} {\path{arXiv:1308.3960}},
  \href {https://doi.org/10.1007/JHEP10(2013)123}
  {\path{doi:10.1007/JHEP10(2013)123}}.

\bibitem{Kuzenko:2013vha}
S.~M. Kuzenko, J.~Novak, G.~Tartaglino-Mazzucchelli, {N=6 superconformal
  gravity in three dimensions from superspace}, JHEP 01 (2014) 121.
\newblock \href {http://arxiv.org/abs/1308.5552} {\path{arXiv:1308.5552}},
  \href {https://doi.org/10.1007/JHEP01(2014)121}
  {\path{doi:10.1007/JHEP01(2014)121}}.

\bibitem{Schnabl:2008wj}
M.~Schnabl, Y.~Tachikawa, {Classification of N=6 superconformal theories of
  ABJM type}, JHEP 09 (2010) 103.
\newblock \href {http://arxiv.org/abs/0807.1102} {\path{arXiv:0807.1102}},
  \href {https://doi.org/10.1007/JHEP09(2010)103}
  {\path{doi:10.1007/JHEP09(2010)103}}.

\bibitem{Bergshoeff:2008bh}
E.~A. Bergshoeff, O.~Hohm, D.~Roest, H.~Samtleben, E.~Sezgin, {The
  Superconformal Gaugings in Three Dimensions}, JHEP 09 (2008) 101.
\newblock \href {http://arxiv.org/abs/0807.2841} {\path{arXiv:0807.2841}},
  \href {https://doi.org/10.1088/1126-6708/2008/09/101}
  {\path{doi:10.1088/1126-6708/2008/09/101}}.

\bibitem{Bagger:2008se}
J.~Bagger, N.~Lambert, {Three-Algebras and N=6 Chern-Simons Gauge Theories},
  Phys. Rev. D 79 (2009) 025002.
\newblock \href {http://arxiv.org/abs/0807.0163} {\path{arXiv:0807.0163}},
  \href {https://doi.org/10.1103/PhysRevD.79.025002}
  {\path{doi:10.1103/PhysRevD.79.025002}}.

\bibitem{Nishimura:2012jh}
M.~Nishimura, Y.~Tanii, {Coupling of the BLG theory to a conformal supergravity
  background}, JHEP 01 (2013) 120.
\newblock \href {http://arxiv.org/abs/1206.5388} {\path{arXiv:1206.5388}},
  \href {https://doi.org/10.1007/JHEP01(2013)120}
  {\path{doi:10.1007/JHEP01(2013)120}}.

\bibitem{Bagger:2007jr}
J.~Bagger, N.~Lambert, {Gauge symmetry and supersymmetry of multiple
  M2-branes}, Phys. Rev. D 77 (2008) 065008.
\newblock \href {http://arxiv.org/abs/0711.0955} {\path{arXiv:0711.0955}},
  \href {https://doi.org/10.1103/PhysRevD.77.065008}
  {\path{doi:10.1103/PhysRevD.77.065008}}.

\bibitem{Gustavsson:2007vu}
A.~Gustavsson, {Algebraic structures on parallel M2-branes}, Nucl. Phys. B 811
  (2009) 66--76.
\newblock \href {http://arxiv.org/abs/0709.1260} {\path{arXiv:0709.1260}},
  \href {https://doi.org/10.1016/j.nuclphysb.2008.11.014}
  {\path{doi:10.1016/j.nuclphysb.2008.11.014}}.

\bibitem{Gran:2008qx}
U.~Gran, B.~E.~W. Nilsson, {Three-dimensional $N=8$ superconformal gravity and
  its coupling to BLG M2-branes}, JHEP 03 (2009) 074.
\newblock \href {http://arxiv.org/abs/0809.4478} {\path{arXiv:0809.4478}},
  \href {https://doi.org/10.1088/1126-6708/2009/03/074}
  {\path{doi:10.1088/1126-6708/2009/03/074}}.

\bibitem{Gran:2012mg}
U.~Gran, J.~Greitz, P.~S. Howe, B.~E.~W. Nilsson, {Topologically gauged
  superconformal Chern-Simons matter theories}, JHEP 12 (2012) 046.
\newblock \href {http://arxiv.org/abs/1204.2521} {\path{arXiv:1204.2521}},
  \href {https://doi.org/10.1007/JHEP12(2012)046}
  {\path{doi:10.1007/JHEP12(2012)046}}.

\bibitem{Eloy:2022vsq}
C.~Eloy, O.~Hohm, H.~Samtleben, {U duality and \ensuremath{\alpha}' corrections
  in three dimensions}, Phys. Rev. D 108~(2) (2023) 026015.
\newblock \href {http://arxiv.org/abs/2211.16358} {\path{arXiv:2211.16358}},
  \href {https://doi.org/10.1103/PhysRevD.108.026015}
  {\path{doi:10.1103/PhysRevD.108.026015}}.

\bibitem{Kaloper:1993fg}
N.~Kaloper, {Topological mass generation in three-dimensional string theory},
  Phys. Lett. B 320 (1994) 16--20.
\newblock \href {http://arxiv.org/abs/hep-th/9310011}
  {\path{arXiv:hep-th/9310011}}, \href
  {https://doi.org/10.1016/0370-2693(94)90817-6}
  {\path{doi:10.1016/0370-2693(94)90817-6}}.

\bibitem{Alkac:2015lma}
G.~Alkac, L.~Basanisi, E.~A. Bergshoeff, D.~O. Devecio\u{g}lu, M.~Ozkan,
  {Supersymmetric backgrounds and black holes in $
  \mathcal{N}=\left(1,\;1\right) $ cosmological new massive supergravity}, JHEP
  10 (2015) 141.
\newblock \href {http://arxiv.org/abs/1507.06928} {\path{arXiv:1507.06928}},
  \href {https://doi.org/10.1007/JHEP10(2015)141}
  {\path{doi:10.1007/JHEP10(2015)141}}.

\bibitem{Deger:2018kur}
N.~S. Deger, Z.~Nazari, O.~Sarioglu, {Supersymmetric solutions of N=(1,1)
  general massive supergravity}, Phys. Rev. D 97~(10) (2018) 106022.
\newblock \href {http://arxiv.org/abs/1803.06926} {\path{arXiv:1803.06926}},
  \href {https://doi.org/10.1103/PhysRevD.97.106022}
  {\path{doi:10.1103/PhysRevD.97.106022}}.

\bibitem{Sarioglu:2011vz}
O.~Sarioglu, {Stationary Lifshitz black holes of $R^2$-corrected gravity
  theory}, Phys. Rev. D 84 (2011) 127501.
\newblock \href {http://arxiv.org/abs/1109.4721} {\path{arXiv:1109.4721}},
  \href {https://doi.org/10.1103/PhysRevD.84.127501}
  {\path{doi:10.1103/PhysRevD.84.127501}}.

\bibitem{Gibbons:2008vi}
G.~W. Gibbons, C.~N. Pope, E.~Sezgin, {The General Supersymmetric Solution of
  Topologically Massive Supergravity}, Class. Quant. Grav. 25 (2008) 205005.
\newblock \href {http://arxiv.org/abs/0807.2613} {\path{arXiv:0807.2613}},
  \href {https://doi.org/10.1088/0264-9381/25/20/205005}
  {\path{doi:10.1088/0264-9381/25/20/205005}}.

\bibitem{Deger:2013yla}
N.~S. Deger, A.~Kaya, H.~Samtleben, E.~Sezgin, {Supersymmetric Warped AdS in
  Extended Topologically Massive Supergravity}, Nucl. Phys. B 884 (2014)
  106--124.
\newblock \href {http://arxiv.org/abs/1311.4583} {\path{arXiv:1311.4583}},
  \href {https://doi.org/10.1016/j.nuclphysb.2014.04.011}
  {\path{doi:10.1016/j.nuclphysb.2014.04.011}}.

\bibitem{Deger:2016vrn}
N.~S. Deger, G.~Moutsopoulos, {Supersymmetric solutions of $N=(2,0)$
  Topologically Massive Supergravity}, Class. Quant. Grav. 33~(15) (2016)
  155006.
\newblock \href {http://arxiv.org/abs/1602.07263} {\path{arXiv:1602.07263}},
  \href {https://doi.org/10.1088/0264-9381/33/15/155006}
  {\path{doi:10.1088/0264-9381/33/15/155006}}.

\bibitem{Nilsson:2013fya}
B.~E.~W. Nilsson, {Critical solutions of topologically gauged N = 8 CFTs in
  three dimensions}, JHEP 04 (2014) 107.
\newblock \href {http://arxiv.org/abs/1304.2270} {\path{arXiv:1304.2270}},
  \href {https://doi.org/10.1007/JHEP04(2014)107}
  {\path{doi:10.1007/JHEP04(2014)107}}.

\bibitem{Strathdee:1986jr}
J.~A. Strathdee, {Extended Poincar\'e supersymmetry}, Int. J. Mod. Phys. A 2
  (1987) 273.
\newblock \href {https://doi.org/10.1142/S0217751X87000120}
  {\path{doi:10.1142/S0217751X87000120}}.

\bibitem{Mertens:2022irh}
T.~G. Mertens, G.~J. Turiaci, {Solvable models of quantum black holes: a review
  on Jackiw\textendash{}Teitelboim gravity}, Living Rev. Rel. 26~(1) (2023) 4.
\newblock \href {http://arxiv.org/abs/2210.10846} {\path{arXiv:2210.10846}},
  \href {https://doi.org/10.1007/s41114-023-00046-1}
  {\path{doi:10.1007/s41114-023-00046-1}}.

\bibitem{Chamseddine:1991fg}
A.~H. Chamseddine, {Superstrings in arbitrary dimensions}, Phys. Lett. B 258
  (1991) 97--103.
\newblock \href {https://doi.org/10.1016/0370-2693(91)91215-H}
  {\path{doi:10.1016/0370-2693(91)91215-H}}.

\bibitem{Hindawi:1995fy}
A.~Hindawi, B.~A. Ovrut, D.~Waldram, {Two-dimensional higher derivative
  supergravity and a new mechanism for supersymmetry breaking}, Nucl. Phys. B
  471 (1996) 409--429.
\newblock \href {http://arxiv.org/abs/hep-th/9509174}
  {\path{arXiv:hep-th/9509174}}, \href
  {https://doi.org/10.1016/0550-3213(96)00169-1}
  {\path{doi:10.1016/0550-3213(96)00169-1}}.

\bibitem{Hindawi:1995qa}
A.~Hindawi, B.~A. Ovrut, D.~Waldram, {Four-dimensional higher derivative
  supergravity and spontaneous supersymmetry breaking}, Nucl. Phys. B 476
  (1996) 175--199.
\newblock \href {http://arxiv.org/abs/hep-th/9511223}
  {\path{arXiv:hep-th/9511223}}, \href
  {https://doi.org/10.1016/0550-3213(96)00281-7}
  {\path{doi:10.1016/0550-3213(96)00281-7}}.

\bibitem{Gates:2000fj}
S.~J. Gates, Jr., S.~Gukov, E.~Witten, {Two two-dimensional supergravity
  theories from Calabi-Yau four folds}, Nucl. Phys. B 584 (2000) 109--148.
\newblock \href {http://arxiv.org/abs/hep-th/0005120}
  {\path{arXiv:hep-th/0005120}}, \href
  {https://doi.org/10.1016/S0550-3213(00)00374-6}
  {\path{doi:10.1016/S0550-3213(00)00374-6}}.

\bibitem{Haack:2000di}
M.~Haack, J.~Louis, M.~Marquart, {Type IIA and heterotic string vacua in D =
  2}, Nucl. Phys. B 598 (2001) 30--56.
\newblock \href {http://arxiv.org/abs/hep-th/0011075}
  {\path{arXiv:hep-th/0011075}}, \href
  {https://doi.org/10.1016/S0550-3213(00)00786-0}
  {\path{doi:10.1016/S0550-3213(00)00786-0}}.

\bibitem{Julia:1980gr}
B.~Julia, {Group Disintegrations}, Conf. Proc. C 8006162 (1980) 331--350.

\bibitem{Nicolai:1998gi}
H.~Nicolai, H.~Samtleben, {Integrability and canonical structure of d = 2, N=16
  supergravity}, Nucl. Phys. B 533 (1998) 210--242.
\newblock \href {http://arxiv.org/abs/hep-th/9804152}
  {\path{arXiv:hep-th/9804152}}, \href
  {https://doi.org/10.1016/S0550-3213(98)00496-9}
  {\path{doi:10.1016/S0550-3213(98)00496-9}}.

\bibitem{Nicolai:1987kz}
H.~Nicolai, {The Integrability of $N=16$ Supergravity}, Phys. Lett. B 194
  (1987) 402.
\newblock \href {https://doi.org/10.1016/0370-2693(87)91072-0}
  {\path{doi:10.1016/0370-2693(87)91072-0}}.

\bibitem{Nicolai:1988jb}
H.~Nicolai, N.~P. Warner, {The Structure of $N=16$ Supergravity in
  Two-dimensions}, Commun. Math. Phys. 125 (1989) 369.
\newblock \href {https://doi.org/10.1007/BF01218408}
  {\path{doi:10.1007/BF01218408}}.

\bibitem{Codina:2021cxh}
T.~Codina, O.~Hohm, D.~Marques, {General string cosmologies at order
  $\alpha'^3$}, Phys. Rev. D 104~(10) (2021) 106007.
\newblock \href {http://arxiv.org/abs/2107.00053} {\path{arXiv:2107.00053}},
  \href {https://doi.org/10.1103/PhysRevD.104.106007}
  {\path{doi:10.1103/PhysRevD.104.106007}}.

\bibitem{Garousi:2021ikb}
M.~R. Garousi, {O(9,9) symmetry of NS-NS couplings at order
  \ensuremath{\alpha}'3}, Phys. Rev. D 104~(6) (2021) 066013.
\newblock \href {http://arxiv.org/abs/2105.07598} {\path{arXiv:2105.07598}},
  \href {https://doi.org/10.1103/PhysRevD.104.066013}
  {\path{doi:10.1103/PhysRevD.104.066013}}.

\bibitem{David:2021jqn}
M.~David, J.~T. Liu, {T duality and hints of generalized geometry in string
  \ensuremath{\alpha}' corrections}, Phys. Rev. D 106~(10) (2022) 106008.
\newblock \href {http://arxiv.org/abs/2108.04370} {\path{arXiv:2108.04370}},
  \href {https://doi.org/10.1103/PhysRevD.106.106008}
  {\path{doi:10.1103/PhysRevD.106.106008}}.

\bibitem{Damour:2005zb}
T.~Damour, H.~Nicolai, {Higher order M theory corrections and the Kac-Moody
  algebra E(10)}, Class. Quant. Grav. 22 (2005) 2849--2880.
\newblock \href {http://arxiv.org/abs/hep-th/0504153}
  {\path{arXiv:hep-th/0504153}}, \href
  {https://doi.org/10.1088/0264-9381/22/14/003}
  {\path{doi:10.1088/0264-9381/22/14/003}}.

\bibitem{Damour:2005zs}
T.~Damour, A.~Kleinschmidt, H.~Nicolai, {Hidden symmetries and the fermionic
  sector of eleven-dimensional supergravity}, Phys. Lett. B 634 (2006)
  319--324.
\newblock \href {http://arxiv.org/abs/hep-th/0512163}
  {\path{arXiv:hep-th/0512163}}, \href
  {https://doi.org/10.1016/j.physletb.2006.01.015}
  {\path{doi:10.1016/j.physletb.2006.01.015}}.

\bibitem{Damour:2006ez}
T.~Damour, A.~Hanany, M.~Henneaux, A.~Kleinschmidt, H.~Nicolai, {Curvature
  corrections and Kac-Moody compatibility conditions}, Gen. Rel. Grav. 38
  (2006) 1507--1528.
\newblock \href {http://arxiv.org/abs/hep-th/0604143}
  {\path{arXiv:hep-th/0604143}}, \href
  {https://doi.org/10.1007/s10714-006-0317-y}
  {\path{doi:10.1007/s10714-006-0317-y}}.

\bibitem{Fleig:2012xa}
P.~Fleig, A.~Kleinschmidt, {Eisenstein series for infinite-dimensional
  U-duality groups}, JHEP 06 (2012) 054.
\newblock \href {http://arxiv.org/abs/1204.3043} {\path{arXiv:1204.3043}},
  \href {https://doi.org/10.1007/JHEP06(2012)054}
  {\path{doi:10.1007/JHEP06(2012)054}}.

\bibitem{Fleig:2013psa}
P.~Fleig, A.~Kleinschmidt, D.~Persson, {Fourier expansions of Kac-Moody
  Eisenstein series and degenerate Whittaker vectors}, Commun. Num. Theor.
  Phys. 08 (2014) 41--100.
\newblock \href {http://arxiv.org/abs/1312.3643} {\path{arXiv:1312.3643}},
  \href {https://doi.org/10.4310/CNTP.2014.v8.n1.a2}
  {\path{doi:10.4310/CNTP.2014.v8.n1.a2}}.

\bibitem{Pioline:1998mn}
B.~Pioline, {A Note on nonperturbative $R^4$ couplings}, Phys. Lett. B 431
  (1998) 73--76.
\newblock \href {http://arxiv.org/abs/hep-th/9804023}
  {\path{arXiv:hep-th/9804023}}, \href
  {https://doi.org/10.1016/S0370-2693(98)00554-1}
  {\path{doi:10.1016/S0370-2693(98)00554-1}}.

\bibitem{Basu:2006cs}
A.~Basu, {The $D^{10} R^4$ term in type IIB string theory}, Phys. Lett. B 648
  (2007) 378--382.
\newblock \href {http://arxiv.org/abs/hep-th/0610335}
  {\path{arXiv:hep-th/0610335}}, \href
  {https://doi.org/10.1016/j.physletb.2007.03.024}
  {\path{doi:10.1016/j.physletb.2007.03.024}}.

\bibitem{Bossard:2014lra}
G.~Bossard, V.~Verschinin, {Minimal unitary representations from
  supersymmetry}, JHEP 10 (2014) 008.
\newblock \href {http://arxiv.org/abs/1406.5527} {\path{arXiv:1406.5527}},
  \href {https://doi.org/10.1007/JHEP10(2014)008}
  {\path{doi:10.1007/JHEP10(2014)008}}.

\bibitem{Basu:2014hsa}
A.~Basu, {The $D^{6}R^{4}$ term from three loop maximal supergravity}, Class.
  Quant. Grav. 31~(24) (2014) 245002.
\newblock \href {http://arxiv.org/abs/1407.0535} {\path{arXiv:1407.0535}},
  \href {https://doi.org/10.1088/0264-9381/31/24/245002}
  {\path{doi:10.1088/0264-9381/31/24/245002}}.

\bibitem{Bossard:2014aea}
G.~Bossard, V.~Verschinin, {$\mathcal{E} \nabla^4 R^4$ type invariants and
  their gradient expansion}, JHEP 03 (2015) 089.
\newblock \href {http://arxiv.org/abs/1411.3373} {\path{arXiv:1411.3373}},
  \href {https://doi.org/10.1007/JHEP03(2015)089}
  {\path{doi:10.1007/JHEP03(2015)089}}.

\bibitem{Wang:2015jna}
Y.~Wang, X.~Yin, {Constraining Higher Derivative Supergravity with Scattering
  Amplitudes}, Phys. Rev. D 92~(4) (2015) 041701.
\newblock \href {http://arxiv.org/abs/1502.03810} {\path{arXiv:1502.03810}},
  \href {https://doi.org/10.1103/PhysRevD.92.041701}
  {\path{doi:10.1103/PhysRevD.92.041701}}.

\bibitem{Wang:2015aua}
Y.~Wang, X.~Yin, {Supervertices and Non-renormalization Conditions in Maximal
  Supergravity Theories} (5 2015).
\newblock \href {http://arxiv.org/abs/1505.05861} {\path{arXiv:1505.05861}}.

\bibitem{Bossard:2015uga}
G.~Bossard, V.~Verschinin, {The two \ensuremath{\nabla}$^{6}$ R$^{4}$ type
  invariants and their higher order generalisation}, JHEP 07 (2015) 154.
\newblock \href {http://arxiv.org/abs/1503.04230} {\path{arXiv:1503.04230}},
  \href {https://doi.org/10.1007/JHEP07(2015)154}
  {\path{doi:10.1007/JHEP07(2015)154}}.

\bibitem{Green:2010kv}
M.~B. Green, S.~D. Miller, J.~G. Russo, P.~Vanhove, {Eisenstein series for
  higher-rank groups and string theory amplitudes}, Commun. Num. Theor. Phys. 4
  (2010) 551--596.
\newblock \href {http://arxiv.org/abs/1004.0163} {\path{arXiv:1004.0163}},
  \href {https://doi.org/10.4310/CNTP.2010.v4.n3.a2}
  {\path{doi:10.4310/CNTP.2010.v4.n3.a2}}.

\bibitem{Pioline:2010kb}
B.~Pioline, {$R^4$ couplings and automorphic unipotent representations}, JHEP
  03 (2010) 116.
\newblock \href {http://arxiv.org/abs/1001.3647} {\path{arXiv:1001.3647}},
  \href {https://doi.org/10.1007/JHEP03(2010)116}
  {\path{doi:10.1007/JHEP03(2010)116}}.

\bibitem{Green:2011vz}
M.~B. Green, S.~D. Miller, P.~Vanhove, {Small representations, string
  instantons, and Fourier modes of Eisenstein series}, J. Number Theor. 146
  (2015) 187--309.
\newblock \href {http://arxiv.org/abs/1111.2983} {\path{arXiv:1111.2983}},
  \href {https://doi.org/10.1016/j.jnt.2013.05.018}
  {\path{doi:10.1016/j.jnt.2013.05.018}}.

\bibitem{Green:2010sp}
M.~B. Green, J.~G. Russo, P.~Vanhove, {String theory dualities and supergravity
  divergences}, JHEP 06 (2010) 075.
\newblock \href {http://arxiv.org/abs/1002.3805} {\path{arXiv:1002.3805}},
  \href {https://doi.org/10.1007/JHEP06(2010)075}
  {\path{doi:10.1007/JHEP06(2010)075}}.

\bibitem{Bossard:2015oxa}
G.~Bossard, A.~Kleinschmidt, {Supergravity divergences, supersymmetry and
  automorphic forms}, JHEP 08 (2015) 102.
\newblock \href {http://arxiv.org/abs/1506.00657} {\path{arXiv:1506.00657}},
  \href {https://doi.org/10.1007/JHEP08(2015)102}
  {\path{doi:10.1007/JHEP08(2015)102}}.

\bibitem{Obers:1999um}
N.~A. Obers, B.~Pioline, {Eisenstein series and string thresholds}, Commun.
  Math. Phys. 209 (2000) 275--324.
\newblock \href {http://arxiv.org/abs/hep-th/9903113}
  {\path{arXiv:hep-th/9903113}}, \href {https://doi.org/10.1007/s002200050022}
  {\path{doi:10.1007/s002200050022}}.

\bibitem{Kiritsis:1997em}
E.~Kiritsis, B.~Pioline, {On R**4 threshold corrections in IIb string theory
  and (p, q) string instantons}, Nucl. Phys. B 508 (1997) 509--534.
\newblock \href {http://arxiv.org/abs/hep-th/9707018}
  {\path{arXiv:hep-th/9707018}}, \href
  {https://doi.org/10.1016/S0550-3213(97)00645-7}
  {\path{doi:10.1016/S0550-3213(97)00645-7}}.

\bibitem{Green:1997di}
M.~B. Green, P.~Vanhove, {D instantons, strings and M theory}, Phys. Lett. B
  408 (1997) 122--134.
\newblock \href {http://arxiv.org/abs/hep-th/9704145}
  {\path{arXiv:hep-th/9704145}}, \href
  {https://doi.org/10.1016/S0370-2693(97)00785-5}
  {\path{doi:10.1016/S0370-2693(97)00785-5}}.

\bibitem{Basu:2007ru}
A.~Basu, {The $D^4 R^4$ term in type IIB string theory on $T^2$ and U-duality},
  Phys. Rev. D 77 (2008) 106003.
\newblock \href {http://arxiv.org/abs/0708.2950} {\path{arXiv:0708.2950}},
  \href {https://doi.org/10.1103/PhysRevD.77.106003}
  {\path{doi:10.1103/PhysRevD.77.106003}}.

\bibitem{Bossard:2015foa}
G.~Bossard, A.~Kleinschmidt, {Loops in exceptional field theory}, JHEP 01
  (2016) 164.
\newblock \href {http://arxiv.org/abs/1510.07859} {\path{arXiv:1510.07859}},
  \href {https://doi.org/10.1007/JHEP01(2016)164}
  {\path{doi:10.1007/JHEP01(2016)164}}.

\bibitem{Green:2008uj}
M.~B. Green, J.~G. Russo, P.~Vanhove, {Low energy expansion of the
  four-particle genus-one amplitude in type II superstring theory}, JHEP 02
  (2008) 020.
\newblock \href {http://arxiv.org/abs/0801.0322} {\path{arXiv:0801.0322}},
  \href {https://doi.org/10.1088/1126-6708/2008/02/020}
  {\path{doi:10.1088/1126-6708/2008/02/020}}.

\bibitem{DHoker:2015gmr}
E.~D'Hoker, M.~B. Green, P.~Vanhove, {On the modular structure of the genus-one
  Type II superstring low energy expansion}, JHEP 08 (2015) 041.
\newblock \href {http://arxiv.org/abs/1502.06698} {\path{arXiv:1502.06698}},
  \href {https://doi.org/10.1007/JHEP08(2015)041}
  {\path{doi:10.1007/JHEP08(2015)041}}.

\bibitem{Bossard:2020xod}
G.~Bossard, A.~Kleinschmidt, B.~Pioline, {1/8-BPS Couplings and Exceptional
  Automorphic Functions}, SciPost Phys. 8~(4) (2020) 054.
\newblock \href {http://arxiv.org/abs/2001.05562} {\path{arXiv:2001.05562}},
  \href {https://doi.org/10.21468/SciPostPhys.8.4.054}
  {\path{doi:10.21468/SciPostPhys.8.4.054}}.

\bibitem{Bern:2000fm}
Z.~Bern, L.~J. Dixon, D.~Dunbar, B.~Julia, M.~Perelstein, J.~Rozowsky,
  D.~Seminara, M.~Trigiante, {Counterterms in supergravity}, PoS tmr2000 (2000)
  017.
\newblock \href {http://arxiv.org/abs/hep-th/0012230}
  {\path{arXiv:hep-th/0012230}}, \href {https://doi.org/10.22323/1.006.0017}
  {\path{doi:10.22323/1.006.0017}}.

\bibitem{Bern:2014sna}
Z.~Bern, S.~Davies, T.~Dennen, {Enhanced ultraviolet cancellations in $\mathcal
  N=5$ supergravity at four loops}, Phys. Rev. D 90~(10) (2014) 105011.
\newblock \href {http://arxiv.org/abs/1409.3089} {\path{arXiv:1409.3089}},
  \href {https://doi.org/10.1103/PhysRevD.90.105011}
  {\path{doi:10.1103/PhysRevD.90.105011}}.

\bibitem{Bern:2013uka}
Z.~Bern, S.~Davies, T.~Dennen, A.~V. Smirnov, V.~A. Smirnov, {Ultraviolet
  Properties of N=4 Supergravity at Four Loops}, Phys. Rev. Lett. 111~(23)
  (2013) 231302.
\newblock \href {http://arxiv.org/abs/1309.2498} {\path{arXiv:1309.2498}},
  \href {https://doi.org/10.1103/PhysRevLett.111.231302}
  {\path{doi:10.1103/PhysRevLett.111.231302}}.

\bibitem{Bern:2018jmv}
Z.~Bern, J.~J. Carrasco, W.-M. Chen, A.~Edison, H.~Johansson,
  J.~Parra-Martinez, R.~Roiban, M.~Zeng, {Ultraviolet Properties of $\mathcal N
  = 8$ Supergravity at Five Loops}, Phys. Rev. D 98~(8) (2018) 086021.
\newblock \href {http://arxiv.org/abs/1804.09311} {\path{arXiv:1804.09311}},
  \href {https://doi.org/10.1103/PhysRevD.98.086021}
  {\path{doi:10.1103/PhysRevD.98.086021}}.

\bibitem{Dunbar:1999nj}
D.~C. Dunbar, B.~Julia, D.~Seminara, M.~Trigiante, {Counterterms in type I
  supergravities}, JHEP 01 (2000) 046.
\newblock \href {http://arxiv.org/abs/hep-th/9911158}
  {\path{arXiv:hep-th/9911158}}, \href
  {https://doi.org/10.1088/1126-6708/2000/01/046}
  {\path{doi:10.1088/1126-6708/2000/01/046}}.

\bibitem{Bern:2012gh}
Z.~Bern, S.~Davies, T.~Dennen, Y.-t. Huang, {Ultraviolet Cancellations in
  Half-Maximal Supergravity as a Consequence of the Double-Copy Structure},
  Phys. Rev. D 86 (2012) 105014.
\newblock \href {http://arxiv.org/abs/1209.2472} {\path{arXiv:1209.2472}},
  \href {https://doi.org/10.1103/PhysRevD.86.105014}
  {\path{doi:10.1103/PhysRevD.86.105014}}.

\bibitem{Bern:2008pv}
Z.~Bern, J.~J.~M. Carrasco, L.~J. Dixon, H.~Johansson, R.~Roiban, {Manifest
  Ultraviolet Behavior for the Three-Loop Four-Point Amplitude of N=8
  Supergravity}, Phys. Rev. D 78 (2008) 105019.
\newblock \href {http://arxiv.org/abs/0808.4112} {\path{arXiv:0808.4112}},
  \href {https://doi.org/10.1103/PhysRevD.78.105019}
  {\path{doi:10.1103/PhysRevD.78.105019}}.

\bibitem{Bossard:2010bd}
G.~Bossard, P.~S. Howe, K.~S. Stelle, {On duality symmetries of supergravity
  invariants}, JHEP 01 (2011) 020.
\newblock \href {http://arxiv.org/abs/1009.0743} {\path{arXiv:1009.0743}},
  \href {https://doi.org/10.1007/JHEP01(2011)020}
  {\path{doi:10.1007/JHEP01(2011)020}}.

\bibitem{Bossard:2011tq}
G.~Bossard, P.~S. Howe, K.~S. Stelle, P.~Vanhove, {The vanishing volume of D=4
  superspace}, Class. Quant. Grav. 28 (2011) 215005.
\newblock \href {http://arxiv.org/abs/1105.6087} {\path{arXiv:1105.6087}},
  \href {https://doi.org/10.1088/0264-9381/28/21/215005}
  {\path{doi:10.1088/0264-9381/28/21/215005}}.

\bibitem{Hartwell:1994rp}
G.~G. Hartwell, P.~S. Howe, {(N, p, q) harmonic superspace}, Int. J. Mod. Phys.
  A 10 (1995) 3901--3920.
\newblock \href {http://arxiv.org/abs/hep-th/9412147}
  {\path{arXiv:hep-th/9412147}}, \href
  {https://doi.org/10.1142/S0217751X95001820}
  {\path{doi:10.1142/S0217751X95001820}}.

\bibitem{Drummond:2003ex}
J.~M. Drummond, P.~J. Heslop, P.~S. Howe, S.~F. Kerstan, {Integral invariants
  in N=4 SYM and the effective action for coincident D-branes}, JHEP 08 (2003)
  016.
\newblock \href {http://arxiv.org/abs/hep-th/0305202}
  {\path{arXiv:hep-th/0305202}}, \href
  {https://doi.org/10.1088/1126-6708/2003/08/016}
  {\path{doi:10.1088/1126-6708/2003/08/016}}.

\bibitem{Kallosh:2023pkr}
R.~Kallosh, {Superinvariants Below Critical Loop Order} (12 2023).
\newblock \href {http://arxiv.org/abs/2312.06794} {\path{arXiv:2312.06794}}.

\bibitem{Hofman:2008ar}
D.~M. Hofman, J.~Maldacena, {Conformal collider physics: Energy and charge
  correlations}, JHEP 05 (2008) 012.
\newblock \href {http://arxiv.org/abs/0803.1467} {\path{arXiv:0803.1467}},
  \href {https://doi.org/10.1088/1126-6708/2008/05/012}
  {\path{doi:10.1088/1126-6708/2008/05/012}}.

\bibitem{Buchel:2009tt}
A.~Buchel, R.~C. Myers, {Causality of Holographic Hydrodynamics}, JHEP 08
  (2009) 016.
\newblock \href {http://arxiv.org/abs/0906.2922} {\path{arXiv:0906.2922}},
  \href {https://doi.org/10.1088/1126-6708/2009/08/016}
  {\path{doi:10.1088/1126-6708/2009/08/016}}.

\bibitem{deBoer:2009pn}
J.~de~Boer, M.~Kulaxizi, A.~Parnachev, {AdS(7)/CFT(6), Gauss-Bonnet Gravity,
  and Viscosity Bound}, JHEP 03 (2010) 087.
\newblock \href {http://arxiv.org/abs/0910.5347} {\path{arXiv:0910.5347}},
  \href {https://doi.org/10.1007/JHEP03(2010)087}
  {\path{doi:10.1007/JHEP03(2010)087}}.

\bibitem{Camanho:2009vw}
X.~O. Camanho, J.~D. Edelstein, {Causality constraints in AdS/CFT from
  conformal collider physics and Gauss-Bonnet gravity}, JHEP 04 (2010) 007.
\newblock \href {http://arxiv.org/abs/0911.3160} {\path{arXiv:0911.3160}},
  \href {https://doi.org/10.1007/JHEP04(2010)007}
  {\path{doi:10.1007/JHEP04(2010)007}}.

\bibitem{Camanho:2014apa}
X.~O. Camanho, J.~D. Edelstein, J.~Maldacena, A.~Zhiboedov, {Causality
  Constraints on Corrections to the Graviton Three-Point Coupling}, JHEP 02
  (2016) 020.
\newblock \href {http://arxiv.org/abs/1407.5597} {\path{arXiv:1407.5597}},
  \href {https://doi.org/10.1007/JHEP02(2016)020}
  {\path{doi:10.1007/JHEP02(2016)020}}.

\bibitem{Bellazzini:2015cra}
B.~Bellazzini, C.~Cheung, G.~N. Remmen, {Quantum Gravity Constraints from
  Unitarity and Analyticity}, Phys. Rev. D 93~(6) (2016) 064076.
\newblock \href {http://arxiv.org/abs/1509.00851} {\path{arXiv:1509.00851}},
  \href {https://doi.org/10.1103/PhysRevD.93.064076}
  {\path{doi:10.1103/PhysRevD.93.064076}}.

\bibitem{Cheung:2016wjt}
C.~Cheung, G.~N. Remmen, {Positivity of Curvature-Squared Corrections in
  Gravity}, Phys. Rev. Lett. 118~(5) (2017) 051601.
\newblock \href {http://arxiv.org/abs/1608.02942} {\path{arXiv:1608.02942}},
  \href {https://doi.org/10.1103/PhysRevLett.118.051601}
  {\path{doi:10.1103/PhysRevLett.118.051601}}.

\bibitem{Arkani-Hamed:2020blm}
N.~Arkani-Hamed, T.-C. Huang, Y.-t. Huang, {The EFT-Hedron}, JHEP 05 (2021)
  259.
\newblock \href {http://arxiv.org/abs/2012.15849} {\path{arXiv:2012.15849}},
  \href {https://doi.org/10.1007/JHEP05(2021)259}
  {\path{doi:10.1007/JHEP05(2021)259}}.

\bibitem{Caron-Huot:2021rmr}
S.~Caron-Huot, D.~Mazac, L.~Rastelli, D.~Simmons-Duffin, {Sharp boundaries for
  the swampland}, JHEP 07 (2021) 110.
\newblock \href {http://arxiv.org/abs/2102.08951} {\path{arXiv:2102.08951}},
  \href {https://doi.org/10.1007/JHEP07(2021)110}
  {\path{doi:10.1007/JHEP07(2021)110}}.

\bibitem{Caron-Huot:2021enk}
S.~Caron-Huot, D.~Mazac, L.~Rastelli, D.~Simmons-Duffin, {AdS bulk locality
  from sharp CFT bounds}, JHEP 11 (2021) 164.
\newblock \href {http://arxiv.org/abs/2106.10274} {\path{arXiv:2106.10274}},
  \href {https://doi.org/10.1007/JHEP11(2021)164}
  {\path{doi:10.1007/JHEP11(2021)164}}.

\bibitem{Caron-Huot:2022jli}
S.~Caron-Huot, Y.-Z. Li, J.~Parra-Martinez, D.~Simmons-Duffin, {Graviton
  partial waves and causality in higher dimensions}, Phys. Rev. D 108~(2)
  (2023) 026007.
\newblock \href {http://arxiv.org/abs/2205.01495} {\path{arXiv:2205.01495}},
  \href {https://doi.org/10.1103/PhysRevD.108.026007}
  {\path{doi:10.1103/PhysRevD.108.026007}}.

\bibitem{Arkani-Hamed:2023jwn}
N.~Arkani-Hamed, C.~Cheung, C.~Figueiredo, G.~N. Remmen, {Multiparticle
  Factorization and the Rigidity of String Theory}, Phys. Rev. Lett. 132~(9)
  (2024) 091601.
\newblock \href {http://arxiv.org/abs/2312.07652} {\path{arXiv:2312.07652}},
  \href {https://doi.org/10.1103/PhysRevLett.132.091601}
  {\path{doi:10.1103/PhysRevLett.132.091601}}.

\end{thebibliography}
\end{document}